\documentclass[letterpaper,american,english,prl,simglecolumn,amsmath,amssymb,showpacs,superscriptaddress]{revtex4-1}
\usepackage[T1]{fontenc}
\usepackage{babel}
\usepackage{prettyref}
\usepackage{refstyle}
\usepackage{amsbsy}
\usepackage{amstext}
\usepackage{graphicx}
\usepackage{feyn}

\makeatletter

\AtBeginDocument{\providecommand\secref[1]{\ref{sec:#1}}}
\AtBeginDocument{\providecommand\subref[1]{\ref{sub:#1}}}
\AtBeginDocument{\providecommand\eqref[1]{\ref{eq:#1}}}
\AtBeginDocument{\providecommand\tabref[1]{\ref{tab:#1}}}
\AtBeginDocument{\providecommand\figref[1]{\ref{fig:#1}}}
\AtBeginDocument{\providecommand\enuref[1]{\ref{enu:#1}}}
\pdfpageheight\paperheight
\pdfpagewidth\paperwidth

\providecommand{\tabularnewline}{\\}
\newcommand{\lyxdot}{.}

\RS@ifundefined{subsecref}
  {\newref{subsec}{name = \RSsectxt}}
  {}
\RS@ifundefined{thmref}
  {\def\RSthmtxt{theorem~}\newref{thm}{name = \RSthmtxt}}
  {}
\RS@ifundefined{lemref}
  {\def\RSlemtxt{lemma~}\newref{lem}{name = \RSlemtxt}}
  {}

\usepackage{MnSymbol}  
\usepackage{simplewick}  
\usepackage{feynmp-auto}

\usepackage{ifthen}
\newboolean{includesolutions}
\newboolean{lecture}

\newrefformat{cap}{\hyperref[#1]{Figure~\ref{#1}}}
\newrefformat{fig}{\hyperref[#1]{Figure~\ref{#1}}}
\newrefformat{tab}{\hyperref[#1]{Table ~\ref{#1}}}
\newrefformat{sec}{\hyperref[#1]{Section~\ref{#1}}}
\newrefformat{sub}{\hyperref[#1]{Section~\ref{#1}}}
\newrefformat{cha}{\hyperref[#1]{Chapter~\ref{#1}}}

\newif\ifContLineOne
\newif\ifContLineTwo
\newif\ifContLineThree

\def\conC#1{\vbox{\ialign{##\crcr
  \ifContLineThree\hrulefill\else\vphantom{\hrulefill}\fi\crcr
  \noalign{\kern3.2pt\nointerlineskip}
  \ifContLineTwo\hrulefill\else\vphantom{\hrulefill}\fi\crcr
  \noalign{\kern3.2pt\nointerlineskip}
  \ifContLineOne\hrulefill\else\vphantom{\hrulefill}\fi\crcr
  \noalign{\nointerlineskip}
  $\hfil\textstyle{\vbox to 14pt{}#1}\hfil$\crcr}}}

\def\DrawLeg#1#2{
  \kern-.2pt              
  \dimen2 =#1             
  \advance\dimen2 by 2pt  
  \dimen3 = 10.6pt        
  \dimen4 =3.6pt          
  \advance\dimen3 by -\dimen2 
  \multiply\dimen4 by #2
  \advance\dimen3 by \dimen4
  \raise\dimen2 \hbox{\vrule height\dimen3 width .4pt} 
  \kern-.2pt}             

\def\begC#1#2{\setbox0 =\hbox{$\textstyle{#2}$}
  \dimen0=.5\wd0 \dimen1=\ht0
  \conC{\hskip\dimen0}
  \count255=#1
  \ifnum\count255 =1 \ContLineOnetrue\else
  \ifnum\count255 =2 \ContLineTwotrue\else
  \ifnum\count255 =3 \ContLineThreetrue\fi\fi\fi
  \DrawLeg{\dimen1}{\count255}
  \conC{\hskip\dimen0}
  \kern-\dimen0\kern-\dimen0 \box0}

\def\endC#1#2{\setbox0 =\hbox{$\textstyle{#2}$}
  \dimen0=.5\wd0 \dimen1=\ht0
  \conC{\hskip\dimen0}
  \count255=#1
  \ifnum\count255 =1 \ContLineOnefalse\else
  \ifnum\count255 =2 \ContLineTwofalse\else
  \ifnum\count255 =3 \ContLineThreefalse\fi\fi\fi
  \DrawLeg{\dimen1}{\count255}
  \conC{\hskip\dimen0}
  \kern-\dimen0\kern-\dimen0 \box0}

\makeatother

\begin{document}
\global\long\def\D{\mathcal{D}}%
\global\long\def\bx{\mathbf{x}}%
\global\long\def\bl{\mathbf{l}}%
\global\long\def\bh{\mathbf{h}}%
\global\long\def\bJ{\mathbf{J}}%
\global\long\def\N{\mathcal{N}}%
\global\long\def\hh{\hat{h}}%
\global\long\def\bhh{\mathbf{\hh}}%
\global\long\def\T{\mathrm{T}}%
\global\long\def\by{\mathrm{\mathbf{y}}}%
\global\long\def\diag{\mathrm{diag}}%
\global\long\def\Ftr#1#2{\mathcal{F}\left[#1\right]\left(#2\right)}%
\global\long\def\iFtr#1#2{\mathfrak{\mathcal{F}^{-1}}\left[#1\right]\left(#2\right)}%
\global\long\def\D{\mathcal{D}}%
\global\long\def\T{\mathrm{T}}%
\global\long\def\Gammafl{\Gamma_{\mathrm{fl}}}%
\global\long\def\gammafl{\gamma_{\mathrm{fl}}}%
\global\long\def\E#1{\left\langle #1\right\rangle }%

\global\long\def\D{\mathcal{D}}%
\global\long\def\J{\mathbf{J}}%
\global\long\def\one{\mathbf{1}}%
\global\long\def\e{\mathbf{e}}%
\global\long\def\Cpp{\mathcal{K}_{\phi^{\prime}\phi^{\prime}}^{(0)}}%
\global\long\def\CCpp{C_{\phi^{\prime}\phi^{\prime}}^{(0)}}%
\global\long\def\Cppj{\mathcal{K}_{\phi_{j}^{\prime}\phi_{j}^{\prime}}^{(0)}}%
\global\long\def\tx{\tilde{x}}%
\global\long\def\xo{x^{(0)}}%
\global\long\def\xii{x^{(1)}}%
\global\long\def\txi{\tilde{x}^{(1)}}%
\global\long\def\bx{\mathbf{x}}%
\global\long\def\tbx{\tilde{\mathbf{x}}}%
\global\long\def\bl{\mathbf{j}}%
\global\long\def\tbj{\tilde{\mathbf{j}}}%
\global\long\def\bk{\mathbf{k}}%
\global\long\def\tbk{\tilde{\mathbf{k}}}%
\global\long\def\bh{\mathbf{h}}%
\global\long\def\bJ{\mathbf{J}}%
\global\long\def\bN{\mathcal{N}}%
\global\long\def\bH{\mathbf{H}}%
\global\long\def\bK{\mathbf{K}}%
\global\long\def\bxo{\bx^{(0)}}%
\global\long\def\tbxo{\tilde{\bx}^{(0)}}%
\global\long\def\bxi{\bx^{(1)}}%
\global\long\def\tbxi{\tilde{\bx}^{(1)}}%
\global\long\def\tbxi{\tbx^{(1)}}%
\global\long\def\tpsi{\tilde{\psi}}%
\global\long\def\Cxi{C_{x^{(1)}x^{(1)}}}%
\global\long\def\bxxi{\mathbf{\xi}}%
\global\long\def\N{\mathcal{N}}%
\global\long\def\bW{\mathbf{W}}%
\global\long\def\bon{\mathbf{1}}%
\global\long\def\tj{\tilde{j}}%
\global\long\def\tJ{\tilde{J}}%
\global\long\def\Z{\mathcal{Z}}%
\global\long\def\SOM{S_{\mathrm{OM}}}%
\global\long\def\SMSR{S_{\mathrm{MSR}}}%

\global\long\def\cX{\mathcal{X}}%
\global\long\def\cK{\mathcal{K}}%
\global\long\def\cF{\mathcal{F}}%
\global\long\def\tr{\mathrm{tr}}%

\global\long\def\mcC{\mathcal{C}}%
\global\long\def\tX{\tilde{X}}%
\global\long\def\ty{\tilde{y}}%
\global\long\def\tQ{\tilde{Q}}%
\global\long\def\tk{\tilde{k}}%
\global\long\def\bR{\mathbf{R}}%
\global\long\def\bQ{\mathbf{Q}}%
\global\long\def\tbQ{\mathbf{\tilde{Q}}}%

\global\long\def\bX{\mathbf{\mathbf{X}}}%
\global\long\def\tbX{\tilde{\mathbf{\mathbf{X}}}}%
\global\long\def\bW{\mathbf{\mathbf{W}}}%
\global\long\def\bXi{\boldsymbol{\Xi}}%
\global\long\def\bC{\boldsymbol{C}}%

\title{Statistical field theory for neural networks}
\author{Moritz Helias}
\affiliation{Institute of Neuroscience and Medicine (INM-6) and Institute for Advanced
Simulation (IAS-6) and JARA BRAIN Institute I, Jülich Research Centre,
Jülich, Germany}
\affiliation{Department of Physics, Faculty 1, RWTH Aachen University, Aachen,
Germany}
\author{David Dahmen}
\affiliation{Institute of Neuroscience and Medicine (INM-6) and Institute for Advanced
Simulation (IAS-6) and JARA BRAIN Institute I, Jülich Research Centre,
Jülich, Germany}
\date{\today}
\begin{abstract}
Many qualitative features of the emerging collective dynamics in neuronal
networks, such as correlated activity, stability, response to inputs,
chaotic and regular behavior, can be understood in models that are
accessible to a treatment in statistical mechanics, or, more precisely,
statistical field theory. These notes attempt at a self-contained
introduction into these methods, explained on the example of neural
networks of rate units or binary spins. In particular we will focus
on a relevant class of systems that have quenched (time independent)
disorder, mostly arising from random synaptic couplings between neurons.

The presentation consists of three parts. First we introduce fundamental
notions of probabilities, moments, cumulants, and their relation by
the linked cluster theorem, of which Wick's theorem is the most important
special case. The graphical formulation of perturbation theory with
the help of Feynman diagrams will be reviewed in the statistical setting.
The second part extends these concepts to dynamics, in particular
stochastic differential equations in the Ito-formulation, treated
in the Martin-Siggia-Rose-De Dominicis-Janssen path integral formalism.
Employing concepts from disordered systems, we will study networks
with random connectivity and derive their self-consistent dynamic
mean-field theory. We will employ this formalism to explain the statistics
of the fluctuations in these networks and the emergence of different
phases with regular and chaotic dynamics and also cover a recent extension
of the model to stochastic units. The last part introduces as more
advanced concepts the effective action, vertex functions, and the
loopwise expansion. The use of these tools is illustrated in systematic
derivations of self-consistency equations that are grounded on and
going beyond the mean-field approximation. We will illustrate these
methods on the example of the pairwise maximum entropy (Ising spin)
model, including the recently-found diagrammatic derivation of the
Thouless-Anderson-Palmer mean field theory.

\vfill{}
\end{abstract}
\maketitle
\tableofcontents{}

\vfill{}

\pagebreak{}

\setboolean{includesolutions}{false}\setboolean{lecture}{false}

\section{Introduction\label{sec:Introduction}}

The organization of the outer shell of the mammalian brain, the cerebral
cortex, extends over a wide range of spatial scales, from fine-scale
specificity of the connectivity between small assemblies of neurons
\citep{Yoshimura05_1552} to hierarchically organized networks of
entire cortical areas \citep{Schmidt16_arxiv_v3}. These neuronal
networks share many features with interacting many particle systems
in physics. While the single neuron dynamics is rather simple, interesting
behavior of networks arises from the interaction of these many components.
As a result, the activity in the electrically active tissue of the
brain exhibits is correlated on a multitude of spatial and temporal
scales.

Understanding the processes that take place in brain, we face a fundamental
problem: We want to infer the behavior of these networks and identify
the mechanisms that process information from the observation of a
very limited number of measurements. In addition, each available measurement
comes with its characteristic constraints. Recordings from single
neurons have a high temporal resolution, but obviously enforce a serious
sub-sampling. Today, it is possible to record from hundreds of neurons
in parallel. Still this is only a tiny fraction of the number of cells
believed to form the fundamental building blocks of the brain \citep{Mountcastle97_701}.
Alternatively, recordings of the local field potential measure a mesoscopic
collective signal, the superposition of hundreds of thousands to millions
of neurons \citep{Nunez06}. But this signal has a moderate temporal
resolution and it does not allow us to reconstruct the activities
of individual neurons from which it is composed.

A way around this dilemma is to build models, constituted of the elementary
building blocks, neurons connected and interacting by synapses. These
models then enable us to bridge from the microscopic neuronal dynamics
to the mesoscopic or macroscopic measurements and, in the optimal
case, allow us to constrain the regimes of operation of these networks
on the microscopic scale. It is a basic biophysical property that
single cells receive on the order of thousands of synaptic inputs.
This property may on the one hand seem daunting. On the other hand
this superposition of many small input signals typically allows the
application of the law of large numbers. If the connectivity in such
networks is moreover homogeneous on a statistical level, a successful
route to understanding the collective dynamics is by means of population
equations \citep{Bressloff12}.

Such descriptions are, however, only rarely formally justified from
the underlying microscopic behavior. These phenomenological models
present effective equations of motion for the dynamics on the macroscopic
level of neuronal networks. Typically, intuitive ``mean-field''
considerations are employed, performing a coarse graining in space
by collecting a set of neurons in a close vicinity into larger groups
described in terms of their average activity. Often this spatial coarse
graining is accompanied by a temporal coarse graining, replacing the
pulsed coupling among neurons by a temporally smooth interaction (see
e.g. \citet{Bressloff12} for a recent review, esp. section 2 and
\citet{Ermentrout10}). The resulting descriptions are often referred
to as \textquotedblleft rate models\textquotedblright , sometimes
also as \textquotedblleft mean-field models\textquotedblright . The
conceptual step from the microscopic dynamics to the effective macroscopic
description is conceptually difficult. This step therefore often
requires considerable intuition to include the important parts and
there is little control as to which effects are captured and which
are not. One might say this approach so far lacks systematics: It
is not based on a classification scheme that allows us to identify
which constituents of the original microscopic dynamics enter the
approximate expressions and which have to be left out. The lack of
systematics prohibits the assessment of their consistency: It is unclear
if all terms of a certain order of approximation are contained in
the coarse-grained description. While mean-field approaches in their
simplest form neglect fluctuations, the latter are important to explain
the in-vivo like irregular \citep{Softky93,Vreeswijk96,Vreeswijk98,Amit-1997_373}
and oscillating activity in cortex \citep{Brunel99,Brunel00_183,Brunel03a}.
The attempt to include fluctuations into mean-field approaches has
so far been performed in a semi-systematic manner, based on linear
response theory around a mean-field solution \citep{Ginzburg94,Renart10_587,Pernice11_e1002059,Pernice12_031916,Trousdale12_e1002408,Tetzlaff12_e1002596,Helias13_023002}.

To overcome the problems of current approaches based on mean-field
theory or ad-hoc approximations, a natural choice for the formulation
of a theory of fluctuating activity of cortical networks is in the
language of classical stochastic fields, as pioneered by \citet{Buice07_051919,Buice09_377}.
Functional or path integral formulations are ubiquitously employed
throughout many fields of physics, from particle physics to condensed
matter \citep[see e.g. ][]{ZinnJustin96}, but are still rare in theoretical
neuroscience \citep[see][for recent reviews]{Chow15,Hertz16_033001,Schuecker16b_arxiv}.
Such formulations not only provide compact representations of the
physical content of a theory, for example in terms of Feynman diagrams
or vertex functions, but also come with a rich set of systematic approximation
schemes, such as perturbation theory and loop-wise expansion \citep{NegeleOrland98,ZinnJustin96}.
In combination with renormalization methods \citep{Wilson74_75,Wilson75_773}
and, more recently, the functional renormalization group (\citet{WETTERICH93_90},
reviewed in \citet{Berges02_223,Gies06,Metzner12_299}), the framework
can tackle one of the hardest problems in physics, collective behavior
that emerges from the interaction between phenomena on a multitude
of scales spanning several orders of magnitude. It is likely that
in an analogous way that multi-scale dynamics of neuronal networks
can be treated, but corresponding developments are just about to start
\citep{Buice10_377,Steyn-Ross16_022402}.

From the physics perspective, the research in theoretical solid state
physics is often motivated by the development and characterization
of new materials, quantum states, and quantum devices. In these systems
an important microscopic interaction that gives rise to a wealth of
phenomena is the Coulomb interaction: It is reciprocal or symmetric,
instantaneous, and continuously present over time. The interaction
in neuronal systems, in contrast, is directed or asymmetric, delayed,
and is mediated by temporally short pulses. In this view, a neuronal
network can be considered as an exotic physical system, that promises
phenomena hitherto unknown from solid state systems with Coulomb interaction.
Formulating neuronal networks in the language of field theory, which
has brought many insights into collective phenomena in solid state
physics, therefore promises to open the exotic physical system of
the brain to investigations on a similarly informative level.

Historically, the idea of a mean-field theory for neuronal networks
was brought into the field by experts who had a background in disordered
systems, such as spin glasses. In the physics literature the term
mean-field approximation indeed refers to at least two slightly different
approximations. Often it is understood in the sense of Curie-Weiss
mean-field theory of ferromagnetism. Here it is a saddle point approximation
in the local order parameters, each corresponding to one of the spins
in the original system \citep[i.p. section 4.3]{NegeleOrland98}.
To lowest order, the so called tree level or mean-field approximation,
fluctuations of the order parameter are neglected altogether. Corrections
within this so-called loopwise expansion contain fluctuations of the
order parameter around the mean. The other use of the term mean-field
theory, to our knowledge, originates in the spin glass literature
\citep{Kirkpatrick1978}: Their equation 2.17 for the magnetization
$m$ resembles the Curie-Weiss mean-field equation as described before.
A crucial difference is, though, the presence of the Gaussian variable
$z$, which contains fluctuations. Hence the theory, termed \textquotedbl a
novel kind of mean-field theory\textquotedbl{} by the authors, contains
fluctuations. The reason for the difference formally results from
a saddle point approximation performed on the auxiliary field $q$
instead of the local spin-specific order parameter for each spin as
in the Curie-Weiss mean field theory. The auxiliary field only appears
in the partition function of the system after averaging over the quenched
disorder, the frozen and disordered couplings $J$ between spins.

In the same spirit, the work by \citet{Sompolinsky82_6860} obtained
a mean-field equation that reduces the dynamics in a spin glass to
the equation of a single spin embedded into a fluctuating field, whose
statistics is determined self-consistently (see their equation 3.5).
This saddle point approximation of the auxiliary field is sometimes
also called \textquotedbl dynamic mean-field theory\textquotedbl ,
because the statistics of the field is described by a time-lag-dependent
autocorrelation function. By the seminal work of \citet{Sompolinsky88_259}
on a deterministic network of nonlinear rate units (see their eqs.
(3) and (4)), this technique entered neuroscience. The presentation
of the latter work, however, spared many details of the actual derivation,
so that the logical origin of this mean field theory is hard to see
from the published work. The result, the reduction of the disordered
network to an equation of motion of a single unit in the background
of a Gaussian fluctuating field with self-consistently determined
statistics, has since found entry into many subsequent studies. The
seminal work by \citet{Amit-1997_373} presents the analogue approach
for spiking neuron models, for which to date a more formal derivation
as in the case of rate models is lacking. The counterpart for binary
model neurons \citep{Vreeswijk96,Vreeswijk98} follows conceptually
the same view.

Unfortunately, the formal origins of these very successful and influential
approaches have only sparsely found entry into neuroscience until
today. It is likely that part of the reason is the number of formal
concepts that need to be introduced prior to making this approach
comprehensible. Another reason is the lack of introductory texts into
the topic and and the unfortunate fact that seminal papers, such as
\citep{Sompolinsky88_259}, have appeared in journals with tight page
constraints, so that the functional methods, by which the results
were obtained, were necessarily skipped to cater for a broad audience.
As a consequence, a whole stream of literature has used the outcome
of the mean-field reduction as the very starting point without going
back to the roots of the original work. Recently, an attempt has been
made to re-derive the old results using the original functional methods
\citep{Schuecker16b_arxiv}. A detailed version by the original authors
of \citep{Sompolinsky88_259} just became available \citep{Crisanti18_062120}.

These notes present the formal developments of statistical field theory
to an extent that puts the reader in the position to understand the
aforementioned works and to extend them towards novel questions arising
in theoretical neuroscience. Compared to most text books on statistical
field theory, we here chose a different approach: We aim at separating
the conceptual difficulties from the mathematical complications of
infinite dimensions. For this reason the material is separated into
two parts. The first part focuses on the introduction of the conceptually
challenging topics of field theory. We start with a purely stochastic
view point, introducing probability distributions and their respective
descriptions in terms of moments, cumulants, and generating functions.
We exemplify all methods on joint distributions of $N$ scalar real
valued random variables, instead of treating infinite-dimensional
problems right from the beginning. One could call this the field theory
of $N$ numbers, or zero-dimensional fields. This step is, however,
not only done for didactic purposes. Indeed, the pairwise maximum
entropy model, or Ising spin system, can be treated within this framework.
Didactically, this approach allows us to focus on the concepts, which
are challenging enough, without the need of advanced mathematical
tools: We only require elementary tools from analysis and algebra.
Also standard methods that are the foundation of contemporary theoretical
physics, such as diagrammatic perturbation theory, loopwise or background-field
expansions, and the effective action will be introduced within this
framework. Within the first part we will throughout highlight the
connection between the concepts of statistics, such as probabilities,
moments, cumulants and the corresponding counterparts appearing in
the literature of field theory, such as the action, green's functions,
connected Green's functions.

After these conceptual steps, the second part focuses on time-dependent
systems. We will here introduce the functional formalism of classical
systems pioneered by Martin, Siggia and Rose \citep{Martin73} and
further developed by De Dominicis \citep{dedominicis1976_247,DeDomincis78_353}
and Janssen \citep{janssen1976_377}. This development in the mid
seventies arose from the observation that elaborated methods existed
for quantum systems, which were unavailable to stochastic classical
systems. Based on the ideas by \citet{DeDomincis78_353}, we will
then apply these methods to networks with random connectivity, making
use of the randomness of their connectivity to introduce quenched
averages of the moment generating functional and its treatment in
the large $N$ limit by auxiliary fields \citep{Moshe03} to derive
the seminal theory by \citep{Sompolinsky88_259}, which provides the
starting point for many current works. We will then present some examples
of extensions of their work to current questions in theoretical neuroscience
\citep{Dahmen15_1605,Schuecker17_arxiv_v3,Mastrogiuseppe17_e1005498,Marti18_062314}.

The material collected here arose from a lecture held at the RWTH
university in Aachen in the winter terms 2016-2018. Parts of the material
have been presented in different form at the aCNS summer school in
G\"ottigen 2016 and the latter part, namely sections \secref{Martin-Siggia-Rose-De-Dominicis}
and \secref{Sompolinsky-Crisanti-Sommers-theory} on the sparks workshop
2016 in G\"ottingen. In parts the material has been developed within
the PhD theses of David Dahmen, Jannis Sch\"ucker, Sven Goedeke,
and Tobias K\"uhn, to whom we are very grateful.

\section{Probabilities, moments, cumulants\label{sec:Probabilities-moments-cumulant}}

\subsection{Probabilities, observables, and moments\label{sec:Probabilities}}

Assume we want to describe some physical system. Let us further assume
the state of the system is denoted as $x\in\mathbb{R}^{N}$. Imagine,
for example, the activity of $N$ neurons at a given time point. Or
the activity of a single neuron at $N$ different time points. We
can make observations of the system that are functions $f(x)\in\mathbb{R}$
of the state of the system. Often we are repeating our measurements,
either over different trials or we average the observable in a stationary
system over time. It is therefore useful to describe the system in
terms of the density
\begin{align*}
p(y) & =\lim_{\epsilon\to0}\,\frac{1}{\Pi_{i}\epsilon_{i}}\,\langle1_{\{x_{i}\in[y_{i},y_{i}+\epsilon_{i}]\}}\rangle_{x}\\
 & =\langle\delta(x-y)\rangle_{x},
\end{align*}
where the symbol $\langle\rangle$ denotes the average over many repetitions
of the experiment, over realizations for a stochastic model, or over
time, the indicator function $1_{x\in S}$ is $1$ if $x\in S$ and
zero otherwise, and the Dirac $\delta$-distribution acting on a vector
is understood as $\delta(x)=\Pi_{i=1}^{N}\delta(x_{i})$. The symbol
$p(x)$ can be regarded as a probability density, but we will here
use it in a more general sense, also applied to deterministic systems,
for example where the values of $x$ follow a deterministic equation
of motion. It holds that $p$ is normalized in the sense
\begin{align}
1 & =\int\,p(x)\,dx.\label{eq:normalization_p}
\end{align}
Evaluating for the observable function $f$ the expectation value
$\langle f(x)\rangle$, we may use the Taylor representation of $f$
to write
\begin{align*}
\langle f(x)\rangle & :=\int\,p(x)\,f(x)\,dx\\
 & =\sum_{n_{1},\ldots,n_{N}=0}^{\infty}\frac{f^{(n_{1},\ldots,n_{N})}(0)}{n_{1}!\cdots n_{N}!}\,\langle x_{1}^{n_{1}}\cdots x_{N}^{n_{N}}\rangle\\
 & =\sum_{n=0}^{\infty}\sum_{i_{1},\ldots,i_{n}=1}^{N}\frac{f_{i_{1}\cdots i_{n}}^{(n)}(0)}{n!}\,\langle\prod_{l=1}^{n}x_{i_{l}}\rangle,
\end{align*}
where we denoted by $f^{(n_{1},\ldots,n_{N})}(x):=\big(\frac{\partial}{\partial x_{1}}\big)^{n_{i}}\cdots\big(\frac{\partial}{\partial x_{N}}\big)^{n_{N}}\,f(x)$
the $n_{1}$-th to $n_{N}$-th derivative of $f$ by its arguments;
the alternative notation for the Taylor expansion denotes the $n$-th
derivative by $n$ (possibly) different $x$ as $f_{i_{1}\cdots i_{n}}^{(n)}(x):=\prod_{l=1}^{n}\frac{\partial}{\partial x_{i_{l}}}f(x)$.
We here defined\textbf{ }the \textbf{moments} as 
\begin{align}
\langle x_{1}^{n_{1}}\cdots x_{N}^{n_{N}}\rangle & :=\int\,p(x)\,x_{1}^{n_{1}}\cdots x_{N}^{n_{N}}\,dx\label{eq:def_moments}
\end{align}
 of the system's state variables. Knowing only the latter, we are
hence able to evaluate the expectation value of arbitrary observables
that possess a Taylor expansion.

Alternatively, we may write our observable $f$ in its Fourier representation
$f(x)=\iFtr{\hat{f}}x=\frac{1}{\left(2\pi\right)^{N}}\,\int\,\hat{f}(\omega)\,e^{i\omega^{\T}x}\,d\omega$
so that we get for the expectation value
\begin{align}
\langle f(x)\rangle & =\frac{1}{\left(2\pi\right)^{N}}\int\,\hat{f}(\omega)\,\int\,p(x)\,e^{i\omega^{\T}x}\,dx\,d\omega\nonumber \\
 & =\frac{1}{\left(2\pi\right)^{N}}\int\,\hat{f}(\omega)\,\langle e^{i\omega^{\T}x}\rangle_{x}\,d\omega,\label{eq:exp_f_Fourier}
\end{align}
where $\omega^{\T}x=\sum_{i=1}^{N}\omega_{i}x_{i}$ denotes the Euclidean
scalar product.

We see that we may alternatively determine the functions $\langle e^{i\omega^{\T}x}\rangle_{x}$
for all $\omega$ to characterize the distribution of $x$, motivating
the definition
\begin{align}
Z(j) & :=\langle e^{j^{\T}x}\rangle_{x}\nonumber \\
 & =\int\,p(x)\,e^{j^{\T}x}\,dx.\label{eq:def_char_fctn}
\end{align}
Note that we can express $Z$ as the Fourier transform of $p$, so
it is clear that it contains the same information as $p$ (for distributions
$p$ for which a Fourier transform exists). The function $Z$ is called
the \textbf{characteristic function} or \textbf{moment generating
function} \citep[p. 32]{Gardiner85}. The argument $j$ of the function
is sometimes called the ``source'', because in the context of quantum
field theory, these variables correspond to particle currents. We
will adapt this customary name here, but without any physical implication.
The moment generating function $Z$ is identical to the partition
function $\mathcal{Z}$ in statistical physics, apart from the lacking
normalization of the latter. From the normalization (\ref{eq:normalization_p})
and the definition (\ref{eq:def_char_fctn}) follows that
\begin{align}
Z(0) & =1.\label{eq:normalization_Z}
\end{align}

We may wonder how the moments, defined in (\ref{eq:def_moments}),
relate to the characteristic function (\ref{eq:def_char_fctn}). We
see that we may obtain the moments by a simple differentiation of
$Z$ as
\begin{align}
\langle x_{1}^{n_{1}}\cdots x_{N}^{n_{N}}\rangle & =\left.\left\{ \prod_{i=1}^{N}\partial_{i}^{n_{i}}\right\} \,Z(j)\right|_{j=0},\label{eq:moments_generation}
\end{align}
where we introduced the short hand notation $\partial_{i}^{n_{i}}=\frac{\partial^{n_{i}}}{\partial j_{i}^{n_{i}}}$
and set $j=0$ after differentiation. Conversely, we may say that
the moments are the Taylor coefficients of $Z$, from which follows
the identity
\begin{align*}
Z(j) & =\sum_{n_{1},\ldots,n_{N}}\frac{\langle x_{1}^{n_{1}}\ldots x_{N}^{n_{N}}\rangle}{n_{1}!\ldots n_{N}!}\,j_{1}^{n_{1}}\ldots j_{N}^{n_{N}}.
\end{align*}

\subsection{Transformation of random variables\label{sec:Transformation-of-random}}

Often one knows the statistics of some random variable $x$ but would
like to know the statistics of $y$, a function of $x$
\begin{align*}
y & =f(x).
\end{align*}
The probability densities transform as
\begin{align*}
p_{y}(y) & =\int dx\,p_{x}(x)\,\delta(y-f(x)).
\end{align*}
It is obvious that the latter definition of $p_{y}$ is properly normalized:
integrating over all $y$, the Dirac distribution reduces to a unit
factor so that the normalization condition for $p_{x}$ remains. What
does the corresponding moment-generating function look like?

We obtain it directly from its definition (\ref{eq:def_char_fctn})
as
\begin{align*}
Z_{y}(j) & =\langle e^{j^{\T}y}\rangle_{y}\\
 & =\int dy\,p_{y}(y)\,e^{j^{\T}y}\\
 & =\int dy\,\int dx\,p_{x}(x)\,\delta(y-f(x))\,e^{j^{\T}y}\\
 & =\int dx\,p_{x}(x)\,e^{j^{\T}f(x)}\\
 & =\langle e^{j^{\T}f(x)}\rangle_{x},
\end{align*}
where we swapped the order of the integrals in the third line and
performed the integral over $y$ by employing the property of the
Dirac distribution. So we only need to replace the source term $j^{\T}x\to j^{\T}f(x)$
to obtain the transformed moment generating function.

\subsection{Cumulants\label{sub:Cumulants}}

For a set of independent variables the probability density factorizes
as $p^{\mathrm{indep.}}(x)=p_{1}(x_{1})\cdots p_{N}(x_{N})$. The
characteristic function, defined by (\ref{eq:def_char_fctn}), then
factorizes as well $Z^{\mathrm{indep.}}(j)=Z_{1}(j{}_{1})\cdots Z_{N}(j_{N})$.
Hence the $k$-th ($k\le N$) moment $\langle x_{1}\ldots x_{k}\rangle=\langle x_{1}\rangle\ldots\langle x_{k}\rangle$
decomposes into a product of $n$ first moments of the respective
variable. We see in this example that the higher order moments contain
information which is already contained in the lower order moments.

One can therefore ask if it is possible to define an object that only
contains the dependence at a certain order and removes all dependencies
that are already contained in lower orders. The observation that the
moment-generating function in the independent case decomposes into
a product, leads to the idea to consider its logarithm
\begin{align}
W(j) & :=\ln\,Z(j),\label{eq:def_W}
\end{align}
because for independent variables it consequently decomposes into
a sum $W^{\mathrm{indep.}}(j)=\sum_{i}\ln\,Z_{i}(j_{i})$. The Taylor
coefficients of $W^{\mathrm{indep.}}$ therefore do not contain any
mixed terms, because $\partial_{k}\partial_{l}W^{\mathrm{indep.}}\big|_{j=0}=0\quad\forall k\neq l$.
The same it obviously true for higher derivatives. This observation
motivates the definition of the \textbf{cumulants} as the Taylor coefficients
of $W$

\begin{align}
\llangle x_{1}^{n_{1}}\ldots x_{N}^{n_{N}}\rrangle & :=\left.\left\{ \prod_{i=1}^{N}\partial_{i}^{n_{i}}\right\} W(j)\right|_{j=0},\label{eq:def_cumulant}
\end{align}
which we here denote by double angular brackets $\llangle\circ\rrangle$.
For independent variables, as argued above, we have $\llangle x_{1}\ldots x_{N}\rrangle^{\mathrm{indep.}}=0$. 

The function $W$ defined by (\ref{eq:def_W}) is called the \textbf{cumulant
generating function}. We may conversely express it as a Taylor series
\begin{align}
W(j) & =\ln\,Z(j)=\sum_{n_{1},\ldots,n_{N}}\frac{\llangle x_{1}^{n_{1}}\ldots x_{N}^{n_{N}}\rrangle}{n_{1}!\ldots n_{N}!}\,j_{1}^{n_{1}}\ldots j_{N}^{n_{N}}\label{eq:cumulant_Taylor}\\
 & =:\sum_{n_{1},\ldots,n_{N}}\frac{W^{(n_{1},\ldots,n_{N})}(0)}{n_{1}!\ldots n_{N}!}\,j_{1}^{n_{1}}\ldots j_{N}^{n_{N}},\nonumber 
\end{align}
where we introduced the notation $f^{(n)}$ for the $n$-th derivative
of the function. The cumulants are hence the Taylor coefficients of
the cumulant-generating function. The normalization (\ref{eq:normalization_Z})
of $Z(0)=1$ implies 
\begin{align*}
W(0) & =0.
\end{align*}
For the cumulants this particular normalization is, however, not crucial,
because a different normalization $\tilde{Z}(j)=C\,Z(j)$ would give
an inconsequential additive constant $\tilde{W}(j)=\ln(C)+W(j)$,
which therefore does not affect the cumulants, which contain at least
one derivative. The definition $W(j):=\ln\,\mathcal{Z}(j)$ for a
partition function $\mathcal{Z}$ would hence lead to the same cumulants.
In statistical physics, this latter definition of $W$ corresponds
to the free energy \citep{NegeleOrland98}.

\subsection{Connection between moments and cumulants\label{sec:Connection-moments-cumulants}}

Since both, moments and cumulants, characterize a distribution one
may wonder if and how these objects are related. The situation up
to this point is this:
\begin{center}
\includegraphics[bb=0 0 235 87]{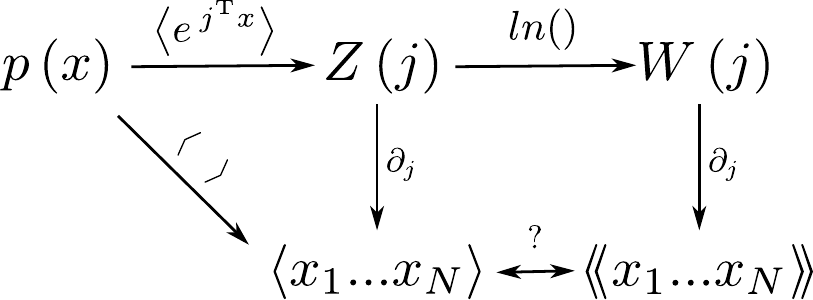}
\par\end{center}

We know how to obtain the moment generating function $Z$ from the
probability $p$, and the cumulant generating function from $Z$ by
the logarithm. The moments and cumulants then follow as Taylor coefficients
from their respective generating functions. Moreover, the moments
can also directly be obtained by the definition of the expectation
value. What is missing is a direct link between moments and cumulants.
This link is what we want to find now.

To this end we here consider the case of $N$ random variables $x_{1},\ldots,x_{N}$.
At first we restrict ourselves to the special case of the $k$-point
moment ($1\le k\le N$)
\begin{eqnarray}
\langle x_{1}\cdots x_{k}\rangle & = & \left.\partial_{1}\cdots\partial_{k}\,Z(j)\right|_{j=0},\label{eq:raw_corr_generated}
\end{eqnarray}
where individual variables only appear in single power.

It is sufficient to study this special case, because a power of $x^{n}$
with $n>1$ can be regarded by the left hand side of (\ref{eq:raw_corr_generated})
as the $n$-fold repeated occurrence of the same index. We therefore
obtain the expressions for repeated indices by first deriving the
results for all indices assumed different and setting indices indentical
in the final result. We will come back to this procedure at the end
of the section. 

Without loss of generality, we are here only interested in $k$-point
correlation functions with consecutive indices from $1$ to $k$,
which can always be achieved by renaming the components $x_{i}$.
We express the moment generating function using (\ref{eq:def_W})
as
\begin{align*}
Z(j) & =\exp(W(j)).
\end{align*}
Taking derivative by $j$ as in (\ref{eq:raw_corr_generated}), we
anticipate due to the exponential function that the term $\exp(W(j))$
will be reproduced, but certain pre-factors will be generated. We
therefore define the function $f_{k}(j)$ as the prefactor appearing
in the $k$-fold derivative of $Z(j)$ as
\begin{align*}
\partial_{1}\cdots\partial_{k}\,Z(j) & =\partial_{1}\cdots\partial_{k}\,\exp(W(j))\\
 & =:f_{k}(j)\,\exp(W(j)).
\end{align*}
Obviously due to (\ref{eq:raw_corr_generated}) and $\exp(W(0))=1$,
the function evaluated at zero is the $k$-th moment
\begin{align*}
f_{k}(0) & =\langle x_{1}\cdots x_{k}\rangle.
\end{align*}
We now want to obtain a recursion formula for $f_{k}$ by applying
the product rule as
\begin{align*}
\partial_{k}\underbrace{\big(f_{k-1}(j)\,\exp(W(j))\Big)}_{\partial_{1}\cdots\partial_{k-1}\,Z(j)} & \stackrel{\text{product rule}}{=}\underbrace{\left(\partial_{k}f_{k-1}+f_{k-1}\,\partial_{k}W\right)}_{f_{k}}\exp(W(j)),
\end{align*}
from which we obtain
\begin{align}
f_{k} & =\partial_{k}f_{k-1}+f_{k-1}\,\partial_{k}W.\label{eq:recursion_moments_from_cumulants_k_point}
\end{align}
The explicit first three steps lead to (starting from $f_{1}(j)\equiv\partial_{1}W(j)$)
\begin{align}
f_{1} & =\partial_{1}W\label{eq:correlators_moments_epxlicit_1_2_3}\\
f_{2} & =\partial_{1}\partial_{2}W+\left(\partial_{1}W\right)\left(\partial_{2}W\right)\nonumber \\
f_{3} & =\partial_{1}\partial_{2}\partial_{3}W+\left(\partial_{1}\partial_{3}W\right)\left(\partial_{2}W\right)+\left(\partial_{1}W\right)\left(\partial_{2}\partial_{3}W\right)\nonumber \\
 & +\left(\partial_{1}\partial_{2}W+\left(\partial_{1}W\right)\left(\partial_{2}W\right)\right)\,\partial_{3}W\nonumber \\
 & =\partial_{1}\partial_{2}\partial_{3}W+\left(\partial_{1}W\right)\left(\partial_{2}\partial_{3}W\right)+\left(\partial_{2}W\right)\left(\partial_{1}\partial_{3}W\right)+\left(\partial_{3}W\right)\left(\partial_{1}\partial_{2}W\right)+\left(\partial_{1}W\right)\left(\partial_{2}W\right)\left(\partial_{3}W\right).\nonumber 
\end{align}
The structure shows that the moments are composed of all combinations
of cumulants of all lower orders. More specifically, we see that
\begin{itemize}
\item the number of derivatives in each term is the same, here three
\item the three derivatives are partitioned in all possible ways to act
on $W$, from all derivatives acting on the same $W$ (left most term
in last line) to each acting on a separate $W$ (right most term).
\end{itemize}
Figuratively, we can imagine these combinations to be created by having
$k$ places and counting all ways of forming $n$ subgroups of sizes
$l_{1},\ldots,l_{n}$ each, so that $l_{1}+\ldots+l_{n}=k$. On the
example $k=3$ we would have
\begin{align*}
\langle1\,2\,3\rangle & =\underbrace{\llangle1\,2\,3\rrangle}_{n=1\;l_{1}=3}\\
 & +\underbrace{\llangle1\rrangle\llangle2\,3\rrangle+\llangle2\rrangle\llangle3\,1\rrangle+\llangle3\rrangle\llangle1\,2\rrangle}_{n=2;\,l_{1}=1\le l_{2}=2}\\
 & +\underbrace{\llangle1\rrangle\llangle2\rrangle\llangle3\rrangle}_{n=3;\,l_{1}=l_{2}=l_{3}=1}.
\end{align*}
We therefore suspect that the general form can be written as

\begin{align}
f_{k} & =\sum_{n=1}^{k}\sum_{\begin{array}[t]{c}
\{1\le l_{1}\le\ldots,\le l_{n}\le k\}\\
\sum_{i}l_{i}=k
\end{array}}\quad\sum_{\sigma\in P(\{l_{i}\},k)}\left(\partial_{\sigma(1)}\cdots\partial_{\sigma(l_{1})}W\right)\ldots\left(\partial_{\sigma(k-l_{n}+1)}\cdots\partial_{\sigma(k)}W\right),\label{eq:f_k_combinatorial_mom_corr}
\end{align}
where the sum over $n$ goes over all numbers of subsets of the partition,
the sum 
\begin{eqnarray*}
\sum_{\begin{array}[t]{c}
\{1\le l_{1}\le\ldots,\le l_{n}\le k\}\\
\sum_{i}l_{i}=k
\end{array}}
\end{eqnarray*}
goes over all sizes $l_{1},\ldots,l_{n}$ of each subgroup, which
we can assume to be ordered by the size $l_{i}$, and $P(\{l_{i}\},k)$
is the set of all permutations of the numbers $1,\ldots,k$ that,
for a given partition $\{1\le l_{1}\le\ldots\le l_{n}\le k\}$, lead
to a different term: Obviously, the exchange of two indices within
a subset does not cause a new term, because the differentiation may
be performed in arbitrary order.\ifthenelse{\boolean{lecture}}{ We
will prove this assertion in the exercises in \subref{exercises_cumulants_moments}c).}

Setting all sources to zero $j_{1}=\ldots=j_{k}=0$ leads to the expression
for the $k$-th moment by the $1,\ldots,k$-point cumulants
\begin{align}
\label{eq:k_point_corr_from_cum}\\
\langle x_{1}\cdots x_{k}\rangle & =\sum_{n=1}^{k}\sum_{\begin{array}[t]{c}
\{1\le l_{1}\le\ldots,\le l_{n}\le k\}\\
\sum_{i}l_{i}=k
\end{array}}\quad\sum_{\sigma\in P(\{l_{i}\},k)}\llangle x_{\sigma(1)}\cdots x_{\sigma(l_{1})}\rrangle\cdots\llangle x_{\sigma(k-l_{n}+1)}\cdots x_{\sigma(k)}\rrangle.\nonumber 
\end{align}

\begin{itemize}
\item So the recipe to determine the $k$-th moment is: Draw a set of $k$
points, partition them in all possible ways into disjoint subgroups
(using every point only once). Now assign, in all possible ways that
lead to a different composition of the subgroups one variable to each
of the dots in each of these combinations. The $i$-th subgroup of
size $l_{i}$ corresponds to a cumulant of order $l_{i}$. The sum
over all such partitions and all permutations yields the $k$-th moment
expressed in terms of cumulants of order $\le k$.
\end{itemize}
We can now return to the case of higher powers in the moments, the
case that $m\ge2$ of the $x_{i}$ are identical. Since the appearance
of two differentiations by the same variable in (\ref{eq:raw_corr_generated})
is handled in exactly the same way as for $k$ different variables,
we see that the entire procedure remains the same: In the final result
(\ref{eq:k_point_corr_from_cum}) we just have $m$ identical variables
to assign to different places. All different assignments of these
variables to positions need to be counted separately.

\section{Gaussian distribution and Wick's theorem}

We will now study a special case of a distribution that plays an essential
role in all further development, the Gaussian distribution. In a way,
field theory boils down to a clever reorganization of Gaussian integrals.
In this section we will therefore derive fundamental properties of
this distribution.

\subsection{Gaussian distribution\label{sec:Gaussian-distribution}}

A Gaussian distribution of $N$ centered (mean value zero) variables
$x$ is defined for a positive definite symmetric matrix $A$ as

\begin{eqnarray}
p(x) & \propto & \exp\Big(-\frac{1}{2}x^{\T}Ax\Big).\label{eq:Gauss_unnormalized}
\end{eqnarray}
A more general formulation for symmetry is that $A$ is self-adjoint
with respect to the Euclidean scalar product (see \secref{Appendix-Self-adjoint-matrices}).
As usual, positive definite means that the bilinear form $x^{\T}\,A\,x>0\quad\forall x\neq0$.
Positivity equivalently means that all eigenvalues $\lambda_{i}$
of $A$ are positive. The properly normalized distribution is
\begin{align}
p(x) & =\frac{\det(A)^{\frac{1}{2}}}{(2\pi)^{\frac{N}{2}}}\,\exp\left(-\frac{1}{2}x^{\T}Ax\right);\label{eq:N_dim_Gauss}
\end{align}
this normalization factor is derived in \secref{Appendix_Normalization_Gaussian}.

\subsection{Moment and cumulant generating function of a Gaussian\label{sub:Completion_of_square_Gaussian}}

The moment generating function $Z(j)$ follows from the definition
(\ref{eq:def_char_fctn}) for the Gaussian distribution (\ref{eq:N_dim_Gauss})
by the substitution $y=x-A^{-1}j$, which is the $N$-dimensional
version of the ``completion of the square''. With the normalization
$C=\frac{\det(A)^{\frac{1}{2}}}{(2\pi)^{\frac{N}{2}}}$ we get

\begin{align}
Z(j) & =\langle e^{j^{\T}x}\rangle_{x}\label{eq:Z_Gauss}\\
 & =C\,\int\,\Pi_{i}dx_{i}\exp\Big(-\frac{1}{2}x^{\T}Ax+\underbrace{j^{\T}x}_{\frac{1}{2}\,\big(A^{-1}\,j\big)^{\T}A\,x+\frac{1}{2}\,x^{\T}A\,\big(A^{-1}\,j\big)}\Big)\nonumber \\
 & =C\,\int\,\Pi_{i}dx_{i}\,\exp\Big(-\frac{1}{2}\underbrace{\left(x-A^{-1}j\right)^{\T}}_{y^{\T}}\,A\,\underbrace{\left(x-A^{-1}j\right)}_{y}+\frac{1}{2}\,j^{\T}\,A^{-1}\,j\Big)\nonumber \\
 & =\underbrace{C\,\int\,\Pi_{i}dy_{i}\,\exp\Big(-\frac{1}{2}y^{\T}A\,y\Big)}_{=1}\,\exp\Big(\frac{1}{2}\,j^{\T}\,A^{-1}\,j\Big)\nonumber \\
 & =\exp\,\Big(\frac{1}{2}\,j^{\T}\,A^{-1}\,j\Big).\nonumber 
\end{align}
 The integral measures do not change form the third to the fourth
line, because we only shifted the integration variables. We used from
the fourth to the fifth line that $p$ is normalized, which is not
affected by the shift, because the boundaries of the integral are
infinite. The cumulant generating function $W(j)$ defined by \eqref{def_W}
then is
\begin{align}
W(j) & =\ln\,Z(j)\nonumber \\
 & =\frac{1}{2}\,j^{\T}A^{-1}\,j.\label{eq:W_Gauss}
\end{align}
Hence the second order cumulants are 
\begin{align}
\llangle x_{i}x_{j}\rrangle & =\left.\partial_{i}\partial_{j}W\right|_{j=0}\label{eq:cumulants_Gauss}\\
 & =A_{ij}^{-1},\nonumber 
\end{align}
where the factor $\frac{1}{2}$ is canceled, because, by the product
rule, the derivative first acts on the first and then on the second
$j$ in (\ref{eq:W_Gauss}), both of which yield the same term due
to the symmetry of $A^{-1T}=A^{-1}$ (The symmetry of $A^{-1}$ follows
from the symmetry of $A$, because $\mathbf{1}=A^{-1}A=A^{\T}A^{-1\T}=A\,A^{-1\T}$;
because the inverse of $A$ is unique it follows that $A^{-1\T}=A^{-1}$).

All cumulants other than the second order (\ref{eq:cumulants_Gauss})
vanish, because (\ref{eq:W_Gauss}) is already the Taylor expansion
of $W$, containing only second order terms and the Taylor expansion
is unique. This property of the Gaussian distribution will give rise
to the useful theorem by Wick in the following subsection.

Eq. (\ref{eq:cumulants_Gauss}) is or course the covariance matrix,
the matrix of second cumulants. We therefore also write the Gaussian
distribution as
\begin{eqnarray*}
x & \sim & \N(0,A^{-1}),
\end{eqnarray*}
where the first argument $0$ refers to the vanishing mean value. 

\subsection{Wick's theorem\label{sec:Wicks-theorem}}

For the Gaussian distribution introduced in \secref{Gaussian-distribution},
all moments can be expressed in terms of products of only second cumulants
of the Gaussian distribution. This relation is known as \textbf{Wick's
theorem} \citep{ZinnJustin96,Kleinert89}.

Formally this result is a special case of the general relation between
moments and cumulants (\ref{eq:k_point_corr_from_cum}): In the Gaussian
case only second cumulants (\ref{eq:cumulants_Gauss}) are different
from zero. The only term that remains in (\ref{eq:k_point_corr_from_cum})
is hence a single partition in which all subgroups have size two,
i.e. $l_{1}=\ldots=l_{n}=2$; each such sub-group corresponds to a
second cumulant. In particular it follows that all moments with odd
power $k$ of $x$ vanish. For a given even $k$, the sum over all
$\sigma\in P[\{2,\ldots,2\}](k)$ includes only those permutations
$\sigma$ that lead to different terms

\begin{align}
\langle x_{1}\cdots x_{k}\rangle_{x\sim\N(0,A^{-1})} & =\sum_{\sigma\in P(\{2,\ldots,2\},k)}\quad\llangle x_{\sigma(1)}x_{\sigma(2)}\rrangle\cdots\llangle x_{\sigma(k-1)}x_{\sigma(k)}\rrangle\nonumber \\
 & \stackrel{(\ref{eq:cumulants_Gauss})}{=}\sum_{\sigma\in P(\{2,\ldots,2\},k)}A_{\sigma(1)\sigma(2)}^{-1}\cdots A_{\sigma(k-1)\sigma(k)}^{-1}.\label{eq:Wick_moments}
\end{align}

We can interpret the latter equation in a simple way: To calculate
the $k$-th moment of a Gaussian distribution, we need to combine
the $k$ variables in all possible, distinct pairs and replace each
pair $(i,j)$ by the corresponding second cumulant $\llangle x_{i}x_{j}\rrangle=A_{ij}^{-1}$.
Here ``distinct pairs'' means that we treat all $k$ variables as
different, even if they may in fact be the same variable, in accordance
to the note at the end of \secref{Connection-moments-cumulants}.
In the case that a subset of $n$ variables of the $k$ are identical,
this gives rise to a \textbf{combinatorial factor}. Figuratively,
we may imagine the computation of the $k$-th moment as composed out
of so called \textbf{contractions}: Each pair of variables is contracted
by one Gaussian integral. This is often indicated by an angular bracket
that connects the two elements that are contracted. In this graphical
notation, the fourth moment $\langle x_{1}x_{2}x_{3}x_{4}\rangle$
of an $N$ dimensional Gaussian can be written as

\begin{align}
\langle x_{1}x_{2}x_{3}x_{4}\rangle_{x\sim\N(0,A^{-1})}= & \begC1{x_{1}}\endC1{x_{2}}\begC2{x_{3}}\endC2{x_{4}}+\begC1{x_{1}}\begC2{x_{2}}\endC1{x_{3}}\endC2{x_{4}}+\begC1{x_{1}}\begC2{x_{2}}\endC2{x_{3}}\endC1{x_{4}}\nonumber \\
= & \llangle x_{1}x_{2}\rrangle\llangle x_{3}x_{4}\rrangle+\llangle x_{1}x_{3}\rrangle\llangle x_{2}x_{4}\rrangle+\llangle x_{1}x_{4}\rrangle\llangle x_{2}x_{3}\rrangle\nonumber \\
= & A_{12}^{-1}\,A_{34}^{-1}+A_{13}^{-1}\,A_{24}^{-1}+A_{14}^{-1}\,A_{23}^{-1}.\label{eq:fourth_moment_example}
\end{align}
To illustrate the appearance of a combinatorial factor, we may imagine
the example that all $x_{1}=x_{2}=x_{3}=x_{4}=x$ in the previous
example are identical. We see from \eqref{fourth_moment_example}
by setting all indices to the same value that we get the same term
three times in this case, namely
\begin{eqnarray*}
\langle x^{4}\rangle & = & 3\,\llangle x^{2}\rrangle^{2}.
\end{eqnarray*}

\subsection{Graphical representation: Feynman diagrams\label{sub:Graphical-representation:ContractionsFeynman}}

An effective language to express contractions, such as \eqref{fourth_moment_example}
is the use of Feynman diagrams. The idea is simple: Each contraction
of a centered Gaussian variable is denoted by a straight line that
we define as
\begin{align*}
\langle x_{i}x_{j}\rangle_{x\sim\N(0,A^{-1})}=\llangle x_{i}x_{j}\rrangle=A_{ij}^{-1} & =\begC1{x_{i}}\endC1{x_{j}}=:\quad\Diagram{\vertexlabel^{i}f\vertexlabel^{j}}
\quad,
\end{align*}
in field theory also called the \textbf{bare propagator} between $i$
and $j$. In the simple example of a multinomial Gaussian studied
here, we do not need to assign any direction to the connection.

A fourth moment in this notation would read
\begin{eqnarray*}
\langle x_{1}x_{2}x_{3}x_{4}\rangle_{x\sim\N(0,A^{-1})} & = & \Diagram{\vertexlabel^{1}f\vertexlabel^{2}\\
\phantom{a}\\
\phantom{a}\\
\vertexlabel_{3}f\vertexlabel_{4}
}
\quad+\quad\Diagram{\vertexlabel^{1}f\vertexlabel^{3}\\
\phantom{a}\\
\phantom{a}\\
\vertexlabel_{2}f\vertexlabel_{4}
}
\quad+\quad\Diagram{\vertexlabel^{1}f\vertexlabel^{4}\\
\phantom{a}\\
\phantom{a}\\
\vertexlabel_{2}f\vertexlabel_{3}
}
\\
\\
 & = & A_{12}^{-1}\,A_{34}^{-1}+A_{13}^{-1}\,A_{24}^{-1}+A_{14}^{-1}\,A_{23}^{-1}.
\end{eqnarray*}

If all $x$ are identical, we can derive this combinatorial factor
again in an intuitive manner: We fix one ``leg'' of the first contraction
at one of the four available $x$. The second leg can then choose
from the three different remaining $x$ to be contracted. For the
remaining two $x$ there is only a single possibility left. So in
total we have three different pairings. The choice of the initial
leg among the four $x$ does not count as an additional factor, because
for any of these four initial choices, the remaining choices would
lead to the same set of pairings, so that we would count the same
contractions four times. These four initial choices hence do not lead
to different partitions of the set in \eqref{Wick_moments}. The factor
three from this graphical method of course agrees to the factor three
we get by setting all indices $1,\ldots,4$ equal in \eqref{fourth_moment_example}.
Hence, we have just calculated the forth moment of a one-dimensional
Gaussian with the result
\begin{align*}
\langle x^{4}\rangle_{x\sim\N} & =3\llangle x^{2}\rrangle.
\end{align*}

\subsection{Appendix: Self-adjoint operators\label{sec:Appendix-Self-adjoint-matrices}}

We denote as $(x,y)$ a scalar product. We may think of the Euclidean
scalar product $(x,y)=\sum_{i=1}^{N}x_{i}y_{i}$ as a concrete example.
The condition for symmetry of $A$ can more accurately be stated as
the operator $A$ being self-adjoint with respect to the scalar product,
meaning that
\begin{align*}
(x,A\,y) & \stackrel{\text{def. adjoint}}{=:}(A^{\T}\,x,y)\,\quad\forall x,y\\
A & ^{\T}=A.
\end{align*}

If a matrix $A$ is self-adjoint with respect to the Euclidean scalar
product $(\cdot,\cdot)$, its diagonalizing matrix $U$ has orthogonal
column vectors with respect to the same scalar product, because from
the general form of a basis change into the eigenbasis $\diag(\{\lambda_{i}\})=U^{-1}\,A\,U$
follows that $(U^{-1\T},A\,U)\stackrel{\text{def. of adjoint}}{=}(A^{\T}\,U^{-1\T},U)\stackrel{\text{symm. of }(\cdot,\cdot)}{=}(U,A^{\T}\,U^{-1\T})\stackrel{A\text{ self. adj.}}{=}(U,A\,U^{-1\T})$.
So the column vectors of $U^{-1T}$ need to be parallel to the eigenvectors
of $A$, which are the column vectors of $U$, because eigenvectors
are unique up to normalization. If we assume them normalized we hence
have $U^{-1\T}=U$ or $U^{-1}=U^{\T}$. It follows that $(Uv,Uw)=(v,U^{\T}U\,w)=(v,w)$,
the condition for the matrix $U$ to be unitary with respect to $(\cdot,\cdot)$,
meaning its transformation conserves the scalar product.

\subsection{Appendix: Normalization of a Gaussian\label{sec:Appendix_Normalization_Gaussian}}

The equivalence between positivity and all eigenvalues being positive
follows from diagonalizing $A$ by an orthogonal transform $U$
\begin{align*}
\diag(\{\lambda_{i}\}) & =U^{\T}\,A\,U,
\end{align*}
where the columns of $U$ are the eigenvectors of $A$ (see \secref{Appendix-Self-adjoint-matrices}
for details). The determinant of the orthogonal transform, due to
$U^{-1}=U^{\T}$ is $|\det(U)|=1$, because $1=\det(\mathbf{1})=\det(U^{\T}U)=\det(U)^{2}$.
The orthogonal transform therefore does not affect the integration
measure. In the coordinates system of eigenvectors $v$ we can then
rewrite the normalization integral as

\begin{eqnarray*}
 &  & \int_{-\infty}^{\infty}\Pi_{i}dx_{i}\exp\Big(-\frac{1}{2}x^{\T}Ax\Big)\\
 & \stackrel{x=U\,v}{=} & \int_{-\infty}^{\infty}\Pi_{k}dv_{k}\exp\Big(-\frac{1}{2}v^{\T}U^{\T}AUv\Big)\\
 & = & \int_{-\infty}^{\infty}\Pi_{k}dv_{k}\exp\Big(-\frac{1}{2}\sum_{i}\lambda_{i}v_{i}^{2}\Big)\\
 & = & \Pi_{k}\sqrt{\frac{2\pi}{\lambda_{k}}}=(2\pi)^{\frac{N}{2}}\det(A)^{-\frac{1}{2}},
\end{eqnarray*}
where we used in the last step that the determinant of a matrix equals
the product of its eigenvalues.

\section{Perturbation expansion\label{sec:Perturbation-expansion}}

\subsection{General case}

Only few problems can be solved exactly. We therefore rely on perturbative
methods to evaluate the quantities of physical interest. One such
method follows the known avenue of a perturbation expansion: If a
part of the problem is solvable exactly, we can try to obtain corrections
in a perturbative manner, if the additional parts of the theory are
small compared to the solvable part.

First, we introduce a new concept, which we call the \textbf{action
$S(x)$}. It is just another way to express the probability distribution.
The main difference is that the notation using the action typically
does not care about the proper normalization of $p(x)$, because the
two are related by
\begin{align*}
p(x) & \propto\exp(S(x)).
\end{align*}
We will see in the sequel, that the normalization can be taken care
of diagrammatically. We saw an example of an action in the last section
in (\ref{eq:Gauss_unnormalized}): The action of the Gaussian is $S(x)=-\frac{1}{2}x^{\T}Ax$.

Replacing $p(x)$ by $\exp(S(x))$ in the definition of the moment
generating function (\ref{eq:def_char_fctn}), we will call the latter
$\Z(j)$. We therefore obtain the normalized moment generating function
as

\begin{align}
Z(j) & =\frac{\Z(j)}{\Z(0)},\label{eq:def_Z_partition}\\
\Z(j) & =\int\,dx\,\exp\left(S(x)+j^{\T}x\right).\nonumber 
\end{align}
We here denote as $\Z$ the unnormalized partition function, for which
in general $\Z(0)\neq1$ and $Z$ is the properly normalized moment
generating function that obeys $Z(0)=1$.

As initially motivated, let us assume that the problem can be decomposed
into a part $S_{0}(x)$, of which we are able to evaluate the partition
function $\Z_{0}(j)$ exactly, and a perturbing part $\epsilon V(x)$
as

\begin{align*}
S(x) & =S_{0}(x)+\epsilon V(x).
\end{align*}
We here introduced the small parameter $\epsilon$ that will serve
us to organize the perturbation expansion. Concretely, we assume that
we are able to compute the integral 
\begin{align}
\Z_{0}(j) & =\int\,dx\,\exp\left(S_{0}(x)+j^{\T}x\right).\label{eq:Z_0_perturbation}
\end{align}
As an example we may think of $S_{0}(x)=-\frac{1}{2}x^{\T}Ax$, a
Gaussian distribution (\ref{eq:Gauss_unnormalized}). We are, however,
not restricted to perturbations around a Gaussian theory, although
this will be the prominent application of the method presented here
and in fact in most applications of field theory. Let us further assume
that the entire partition function can be written as

\begin{align}
\Z(j) & =\int\,dx\,\exp\left(S_{0}(x)+\epsilon V(x)+j^{\T}x\right)\nonumber \\
 & =\int\,dx\,\exp\left(\epsilon V(x)+j^{\T}x\right)\,\exp\left(S_{0}(x)\right),\label{eq:Z_perturbation}
\end{align}
where all terms of the action that are not part of the solvable theory
are contained in the \textbf{potential} $V(x)$, multiplied by a prefactor
$\epsilon$ that is assumed to be small. The name ``potential''
is here chosen in reminiscence of the origin of the term in interacting
systems, where the pairwise potential, mediating the interaction between
the individual particles, is often treated as a perturbation. For
us, the $V$ is just an arbitrary smooth function of the $N$-dimensional
vector $x$ of which we will assume that a Taylor expansion exists.

The form of \eqref{Z_perturbation} shows that we may interpret the
moment generating function as the ratio of expectation values 
\begin{align}
Z(j) & =\frac{\langle\exp\left(\epsilon V(x)+j^{\T}x\right)\rangle_{0}}{\langle\exp\left(\epsilon V(x)\right)\rangle_{0}},\label{eq:pert-exp-V}
\end{align}
where $\langle\ldots\rangle_{0}=\int dx\ldots\exp(S_{0}(x))$ is the
expectation value with respect to our solvable theory \eqref{Z_0_perturbation}
at $j=0$ \citep[see also Peirl's method in ref.][p. 164]{Binney92};
note that, do to the lack of normalization, the latter is not a proper
expectation value, though. Since we assumed that (\ref{eq:Z_0_perturbation})
can be computed, we may obtain all expectation values from $\Z_{0}$
as
\begin{align*}
\langle x_{1}\cdots x_{k}\rangle_{0} & =\partial_{1}\cdots\partial_{k}\Z_{0}(j)\Big|_{j=0}.
\end{align*}

Recalling our initial motivation to introduce moments in \secref{Probabilities},
we immediately see that the problem reduces to the calculation of
all moments $\langle\cdots\rangle_{0}$ appearing as a result of a
Taylor expansion of the terms $\exp(\epsilon V(x))$ and $\exp(\epsilon V(x)+j^{\T}x)$.

We also note that if we are after the cumulants obtained from the
cumulant generating function $W$, we may omit the normalization factor
$\langle\exp\left(\epsilon V(x)\right)\rangle_{0}$, because
\begin{align*}
W(j) & =\ln\,Z(j)=\ln\Z(j)-\ln\Z(0).
\end{align*}
Since the cumulants, by \eqref{cumulant_Taylor}, are derivatives
of $W$, the additive constant term $-\ln\Z(0)$ does not affect their
value.

\subsection{Special case of a Gaussian solvable theory\label{sec:perturbation_Gaussian}}

Now we will specifically study the Gaussian theory as an example,
so we assume that $\Z_{0}=Z_{0}$ in \eqref{Z_0_perturbation} is
of Gaussian form \eqref{Z_Gauss}
\begin{align*}
Z_{0}(j) & =\exp\left(\frac{1}{2}\,j^{\T}\,A^{-1}\,j\right),
\end{align*}
because this special case is fundamental for the further developments.
In calculating the moments that contribute to \eqref{pert-exp-V},
we may hence employ Wick's theorem (\ref{eq:Wick_moments}). Let us
first study the expression we get for the normalization factor $\Z(0)$.

We get with the series representation $\exp(\epsilon V(x))=1+\epsilon V(x)+\frac{\epsilon^{2}}{2!}V^{2}(x)+O(\epsilon^{3}$)
the lowest order approximation $Z_{0}(j)$ and correction terms $Z_{V}(j)$
from \eqref{pert-exp-V} as 
\begin{align}
\Z(0) & =\Z_{0}(0)+\Z_{V}(0)\nonumber \\
\Z_{V}(0) & :=\langle\epsilon V(x)+\frac{\epsilon^{2}}{2!}V^{2}(x)+\ldots\rangle_{0}.\label{eq:Z_V_expectation}
\end{align}
In deriving the formal expressions, our aim is to obtain graphical
rules to perform the expansion. We therefore write the Taylor expansion
of the potential $V$ as
\begin{align}
V(x) & =\sum_{n_{1},\ldots,n_{N}}\,\frac{V^{(n_{1},\ldots n_{N})}}{n_{1}!\cdots n_{N}!}\;x_{1}^{n_{1}}\cdots x_{N}^{n_{N}}\label{eq:Taylor_v-1}\\
 & =\sum_{n=0}^{\infty}\,\sum_{i_{1},\ldots,i_{n}=1}^{N}\,\frac{1}{n!}\,V_{i_{1}\cdots i_{n}}^{(n)}\prod_{k=1}^{n}x_{i_{k}},\nonumber 
\end{align}
 where $V^{(n_{1},\ldots n_{N})}=\frac{\partial^{n_{1}+\ldots+n_{N}}V(0)}{\partial_{1}^{n_{1}}\cdots\partial_{N}^{n_{N}}}$
are the derivatives of $V$ evaluated at $x=0$ and $V_{i_{1}\cdots i_{n}}^{(n)}=\frac{\partial^{n}V(0)}{\partial x_{i_{1}}\cdots\partial x_{i_{n}}}$
is the derivative by $n$ arbitrary arguments. We see that the two
representations are identical, because each of the indizes $i_{1},\ldots,i_{n}$
takes on any of the values $1,\ldots,N$. Hence there are $\left(\begin{array}{c}
n\\
n_{k}
\end{array}\right)$ combinations that yield a term $x_{k}^{n_{k}}$, because this is
the number of ways by which any of the $n$ indizes $i_{l}$ may take
on the particular value $i_{l}=k$. So we get a combinatorial factor
$\frac{1}{n!}\left(\begin{array}{c}
n\\
n_{k}
\end{array}\right)=\frac{1}{(n-n_{k})!n_{k}!}$. Performing the same consideration for the remaining $N-1$ coordinates
brings the second line of (\ref{eq:Taylor_v-1}) into the first.

We now extend the graphical notation in terms of Feynman diagrams
to denote the Taylor coefficients of the potential by \textbf{interaction
vertices}
\begin{eqnarray}
\epsilon\,\frac{V_{i_{1}\cdots i_{n}}^{(n)}}{n!}\,\prod_{k=1}^{n}x_{i_{k}} & =: & \begin{array}{ccc}
i_{1} &  & i_{n}\\
 & \Diagram{x}
\\
i_{2} &  & \ldots
\end{array}.\label{eq:def_interaction_Feynman}
\end{eqnarray}

The corrections $\Z_{V}(0)$ require, to first order in $\epsilon$,
the calculation of the moments 
\begin{align}
 & \langle x_{i_{1}}\cdots x_{i_{n}}\rangle_{0},\label{eq:moment_gaussian_expansion}
\end{align}
So with Wick's theorem the first order correction terms are

\begin{align}
 & \epsilon\,\sum_{n=0}^{\infty}\,\sum_{i_{1},\ldots,i_{n}=1}^{N}\,\frac{1}{n!}\,V_{i_{1}\cdots i_{n}}^{(n)}\,\sum_{\sigma\in P(\{2,\ldots,2\},n)}A_{\sigma(1)\sigma(2)}^{-1}\cdots A_{\sigma(n-1)\sigma(n)}^{-1},\label{eq:explicit_first_order}
\end{align}
where $\sigma$ are all permutations that lead to distinct pairings
of the labels $i_{1},\ldots,i_{n}$.

Continuing the expansion up to second order in $\epsilon$ we insert
(\ref{eq:Taylor_v-1}) into $\exp(\epsilon V(x))$ and using $\exp\sum=\prod\exp$
to get
\begin{align}
\exp(\epsilon V(x)) & =\exp\,\big(\epsilon\sum_{n=0}^{\infty}\sum_{i_{1},\ldots,i_{n}=1}^{N}\epsilon\,\frac{V_{i_{1}\cdots i_{n}}^{(n)}}{n!}\;\prod_{k=1}^{n}x_{i_{k}}\big)\label{eq:expansion_V_second_kind}\\
\nonumber \\
 & =1+\epsilon\,\sum_{n=0}^{\infty}\sum_{i_{1},\ldots,i_{n}=1}^{N}\frac{V_{i_{1}\cdots i_{n}}^{(n)}}{n!}\;\prod_{k=1}^{n}x_{i_{k}}+\frac{\epsilon^{2}}{2!}\,\sum_{n,m=0}^{\infty}\sum_{\{i_{k},j_{l}\}}\frac{V_{i_{1}\cdots i_{n}}^{(n)}}{n!}\,\frac{V_{j_{1}\cdots j_{m}}^{(m)}}{m!}\,\prod_{k=1}^{n}x_{i_{k}}\prod_{l=1}^{m}x_{j_{l}}+\ldots\nonumber 
\end{align}

The last line shows that we get a sum over each index $i_{k}$. We
see, analogous to the factor $\epsilon^{2}/2!$, that a contribution
with $k$ vertices has an overall factor $\epsilon^{k}/k!$. If the
$k$ vertices are all different, we get the same term multiple times
due to the sums over the index tuples $i_{1},\ldots i_{n}$. The additional
factor corresponds to the number of ways to assign the $k$ vertices
to $k$ places. So if the $k$ vertices in total are made up of groups
of $r_{i}$ identical vertices each, with $k=\sum_{i=1}^{n}r_{i}$,
we get another factor $\frac{k!}{r_{1}!\cdots r_{n}!}$.

We need to compute the expectation value $\langle\ldots\rangle_{0}$
of the latter expression, according to (\ref{eq:Z_V_expectation}).
For a Gaussian solvable part this task boils down to the application
of Wick's theorem. These expressions soon become unwieldy, but we
can make use of the graphical language introduced above and derive
the so called \textbf{Feynman rules} to compute the corrections in
$\Z_{V}(0)$ at order $k$ in $\epsilon$:
\begin{itemize}
\item At order $k$, which equals the number of interaction vertices, each
term comes with a factor $\frac{\epsilon^{k}}{k!}$.
\item If vertices repeat $r_{i}$ times the factor is $\frac{\epsilon^{k}}{r_{1}!\cdots r_{n}!}$.
\item A graph representing this correction consists of $k$ interaction
vertices (factor $\frac{V_{i_{1}\cdots i_{n}}^{(n)}}{n!}$); in each
such vertex $n$ lines cross.
\item We need to consider all possible combinations of $k$ such vertices
that are generated by (\ref{eq:expansion_V_second_kind}).
\item The legs of the interaction vertices are joined in all possible ways
into pairs; this is because we take the expectation value with regard
to the Gaussian in (\ref{eq:Z_V_expectation}) (due to the permutations
$\sum_{\sigma\in P[\{2,\ldots,2\}](q)}$); every pair of joined legs
is denoted by a connecting line from $x_{i}$ to $x_{j}$, which end
on the corresponding legs of the interaction vertices; each such connection
yields to a factor $A_{ij}^{-1}$
\item We get a sum over each index $i_{k}$.
\end{itemize}
We will exemplify these rules in the following example on a toy model.

\subsection{Example: Example: ``$\phi^{3}+\phi^{4}$'' theory\label{sub:Example_phi3_phi4_pert-1}}

As an example let us study the system described by the action
\begin{align}
S(x) & =S_{0}(x)+\epsilon\,V(x)\label{eq:def_S_phi34}\\
V(x) & =\frac{\alpha}{3!}x^{3}+\frac{\beta}{4!}x^{4}\nonumber \\
S_{0}(x) & =-\frac{1}{2}Kx^{2}+\frac{1}{2}\,\ln\,\frac{K}{2\pi},\nonumber 
\end{align}
with $K>0$. We note that the action is already in the form to extract
the Taylor coefficients $V^{(3)}=\alpha$ and $V^{(4)}=\beta$. Here
the solvable part of the theory, $S_{0}$, is a one-dimensional Gaussian.
The constant term $\frac{1}{2}\ln\frac{K}{2\pi}$ is the normalization,
which we could drop as well, since we will ultimately calculate the
ratio (\ref{eq:pert-exp-V}), where this factor drops out. With this
normalization, a contraction therefore corresponds to the Gaussian
integral 
\begin{eqnarray*}
\Diagram{\vertexlabel_{x}f\vertexlabel_{x}}
\quad & =\quad & \langle xx\rangle_{0}\\
 & = & \sqrt{\frac{K}{2\pi}}\,\int\,x^{2}\,\exp\left(-\frac{1}{2}Kx^{2}\right)\,dx\\
 & = & \llangle x^{2}\rrangle_{0}=K^{-1},
\end{eqnarray*}
the variance of the unperturbed distribution, following from the general
form (\ref{eq:cumulants_Gauss}) for the second cumulants of a Gaussian
distribution for the one-dimensional case considered here. Alternatively,
two-fold integration by parts yields the same result.

The first order correction to the denominator $\Z(0)$ is therefore
\begin{align*}
\Z_{V,1}(0) & =\epsilon\,\Big\langle\big(\frac{\alpha}{3!}x^{3}+\frac{\beta}{4!}x^{4}\big)\Big\rangle_{0}\\
 & =\epsilon\,\frac{\alpha}{3!}\,\big\langle x^{3}\big\rangle_{0}+\epsilon\,\frac{\beta}{4!}\langle x^{4}\rangle_{0}\\
 & =0+3\cdot\Feyn{f0flfluf0f0flflu}\\
 & =0+\epsilon\,\frac{\beta}{4!}\,3\,K^{-2},
\end{align*}
where the first term $\propto x^{3}$ vanishes, because the Gaussian
is centered (has zero mean). We here use the notation of the interaction
vertex $\Feyn{f0xf0}$ as implying the prefactor $\epsilon$, as defined
in (\ref{eq:def_interaction_Feynman}) and the factor $\frac{\beta}{4!}$,
which is the Taylor coefficient of the potential. We have two connecting
lines, hence the factor $(K^{-1})^{2}$. The factor $3$ appearing
in the third line can be seen in two ways: 1. By the combinations
to contract the four lines of the vertex: We choose one of the four
legs arbitrarily; we then have three choices (factor $3$) to connect
this leg to one of the three remaining ones; the remaining two legs
can be combined in a single manner then (factor $1$). Choosing any
other of the four legs to begin with leads to the same combinations,
so there is no additional factor $4$ (would double-count combinations).
2. By the result $\langle x^{4}\rangle_{0}=3!!=3\cdot1=3$, valid
for a unit variance Gaussian.

At second order we get
\begin{align*}
\Z_{V,2}(0) & =\frac{\epsilon^{2}}{2!}\,\big\langle\frac{\alpha}{3!}x^{3}\,\frac{\alpha}{3!}x^{3}\big\rangle_{0}+\frac{2!}{1!1!}\,\frac{\epsilon^{2}}{2!}\,\underbrace{\big\langle\frac{\alpha}{3!}x^{3}\,\frac{\beta}{4!}x^{4}\big\rangle_{0}}_{=0}+\frac{\epsilon^{2}}{2!}\,\big\langle\frac{\beta}{4!}x^{4}\,\frac{\beta}{4!}x^{4}\big\rangle_{0}\\
 & =3\cdot2\cdot\Feyn{fflfluf}+3\cdot3\cdot\Feyn{f0flfluf0ff0flflu}\\
 & +4\cdot3\cdot2\cdot\Feyn{mflflum}+\big(\begin{array}{c}
4\\
2
\end{array}\big)^{2}\cdot2\cdot\Feyn{f0flfluf0f0flfluf0f0flfluf0}+3\cdot\Feyn{f0flfluf0f0flfluf0}\cdot3\cdot\Feyn{f0flfluf0f0flfluf0}\\
 & =\frac{\epsilon^{2}}{2!}\,\big(\frac{\alpha}{3!}\big)^{2}\,K^{-3}\,\underbrace{(3\cdot2+3\cdot3)}_{=15=(6-1)!!}\\
 & +\frac{\epsilon^{2}}{2!}\,\big(\frac{\beta}{4!}\big)^{2}\,K^{-4}\,\underbrace{(4\cdot3\cdot2+\big(\begin{array}{c}
4\\
2
\end{array}\big)^{2}\cdot2+3\cdot3)}_{=105=(8-1)!!}.
\end{align*}
We dropped from the first to the second line the term with an uneven
power in $x$, because we have a centered Gaussian. The combinatorial
factors, such as $3\cdot2$ for the first diagram, correspond to the
number of combinations by which the legs of the two vertices can be
contracted in the given topology. The factor $\frac{2!}{1!1!}$ is
due to the number of ways in which the sum in (\ref{eq:expansion_V_second_kind})
produces the same term. The expressions in the underbraces show that
alternatively, we can obtain the results from the expression of the
$k$-th moment of a Gaussian with variance $K^{-1}$, which is $(k-1)!!\,K^{-\frac{k}{2}}$
\ifthenelse{\boolean{lecture}}{(see exercises in \secref{exercises_feynman_wick}).}

\subsection{External sources}

Now let us extend this reasoning to $\Z(j)$, which is a function
of $j$. Analogously as for the potential, we may expand the source
term $j^{\T}x$ into its Taylor series
\begin{align}
\exp(j^{\T}x) & =\exp(\sum_{l=1}^{N}j_{l}x_{l})\nonumber \\
 & =\sum_{m=0}^{\infty}\frac{1}{m!}\sum_{l_{1}\cdots l_{m}=1}^{N}\,\prod_{k=1}^{m}j_{l_{k}}x_{l_{k}.}\label{eq:expansion_j}
\end{align}
So for $\Z(j)$, instead of \eqref{moment_gaussian_expansion}, we
need to evaluate the moments
\begin{align}
 & \langle\underbrace{x_{i_{1}}\cdots x_{i_{n}}}_{n\text{ single factors }x}\,\underbrace{x_{l_{1}}\cdots x_{l_{m}}}_{m\text{ factors }x}\rangle_{0}.\label{eq:moment_incl_source}
\end{align}
So in addition to the $n$ single factors $x$ from the interaction
vertices, we get $m$ additional factors due to the source terms $j_{l}x_{l}$.
By Wick's theorem, we need to pair all these $x_{i}$ in all possible
ways into pairs (expressed by sum over all distinct pairings $\sigma\in P(\{2,\ldots,2\},q+m$),
so the generalization of \eqref{explicit_first_order} at first order
in $\epsilon$ (higher orders in $\epsilon$ are analogous to \eqref{expansion_V_second_kind})
reads 
\begin{align}
 & \sum_{m=0}^{\infty}\frac{1}{m!}\sum_{l_{1}\cdots l_{m}=1}^{N}\,\sum_{n=0}^{\infty}\sum_{i_{1},\ldots,i_{n}=1}^{N}\epsilon\,\frac{V_{i_{1}\cdots i_{n}}^{(n)}}{n!}\,\sum_{\sigma\in P(\{2,\ldots,2\},m+n)}A_{\sigma(1)\sigma(2)}^{-1}\cdots A_{\sigma(n+m-1)\sigma(n+m)}^{-1}.\label{eq:perturb_first_order}\\
 & .\nonumber 
\end{align}
So the additional graphical rules are:
\begin{itemize}
\item In a way, the source term $j_{i}x_{i}$ act like a monopole interaction
vertex; these terms are represented by a line ending in an \textbf{external
leg} to which we assign the name $j_{i}$: $\feyn{\vertexlabel_{j_{i}}f}$
\item We need to construct all graphs including those where lines end on
an arbitrary number of external points $j_{i}$.
\item A graph with $r$ external lines contributes to the $l$-th moment,
because after differentiating $\Z(j)$ $l$-times and setting $j=0$
in the end, this is the only remaining term.
\item For a graph with $l$ external lines, we have an additional factor
$\frac{1}{l!}$ in much the same way as interaction vertices. By Wick's
theorem and \eqref{explicit_first_order}, we need to treat each of
these $l_{i}$ factors $j_{i}$ as distinct external legs to arrive
at the right combinatorial factor. Each external leg $j_{i}$ comes
with a sum $\sum_{i=1}^{N}$.
\end{itemize}
These rules are summarized in \tabref{Diagrammatic-rules-scriptZ}.
We will exemplify these rules in the example in \subref{Example_phi3_phi4_pert},
but first reconsider the normalization factor appearing in \eqref{pert-exp-V}
in the following section.
\begin{table}[h]
\begin{centering}
\begin{tabular}{cc|ccc|cc}
meaning & \hspace{1cm} & \hspace{1cm} & algebraic term & \hspace{1cm} & \hspace{1cm} & graphical representation\tabularnewline
\hline 
perturbation order $k$ &  &  & $\frac{\epsilon^{k}}{k!}$ &  &  & number of interaction vertices\tabularnewline
\hline 
each internal index is summed over &  &  & $\sum_{i_{k}=1}^{N}$ &  &  & \tabularnewline
\hline 
interaction vertex with $n$ legs &  &  & $\epsilon\,\frac{V_{i_{1}\cdots i_{n}}^{(n)}}{n!}$ &  &  & $\Diagram{\vertexlabel^{i_{1}}fd & fu\vertexlabel^{i_{n}}\\
\vertexlabel^{i_{2}}fu & fd\vertexlabel^{\ldots}
}
$\tabularnewline
\hline 
internal line &  &  & $A_{ik}^{-1}$ &  &  & $\vertexlabel_{x_{i}}\Diagram{f}
\vertexlabel_{x_{k}}$\tabularnewline
contraction of two internal $x_{i},x_{k}$ &  &  &  &  &  & \tabularnewline
\hline 
external line &  &  & $\sum_{k}A_{ik}^{-1}\,j_{k}$ &  &  & $\vertexlabel_{x_{i}}\Diagram{f}
\vertexlabel_{j_{k}}$\tabularnewline
contraction of internal or external $x_{i}$ and external $x_{k}$ &  &  &  &  &  & \tabularnewline
\end{tabular}
\par\end{centering}
\caption{Diagrammatic rules for the perturbative expansion of $\protect\Z_{V}(j)$.\label{tab:Diagrammatic-rules-scriptZ}}
\end{table}

\subsection{Cancellation of vacuum diagrams\label{sec:Cancellation-of-vacuum}}

To arrive at an expression for the perturbation expansion (\ref{eq:perturbation_general})
of the normalized moment generating function $Z(j)$ (\ref{eq:def_Z_partition}),
whose derivatives yield all moments, we need to divide by $\Z(0)$,
the partition function at source value $j=0$. By the rules derived
in the previous section, we see that the diagrams contributing to
$\Z(0)$ are so called \textbf{vacuum diagrams}: Diagrams without
external lines. An example appearing at first order in $\epsilon$
in a theory with a four point interaction vertex is:
\begin{align*}
 & \Diagram{flfluf0f0flfluf0}
\\
\end{align*}
But applying the same set of rules to the calculation of $\Z(j)$,
we see that the expansion also generates exactly the same vacuum diagrams.
This can be seen from \eqref{perturb_first_order}: At given order
$k$, among the pairings $\sigma$ there are in particular those that
decompose into two disjoint sets, such that all external lines are
contracted with only a subset of $k^{\prime}$ interaction vertices.
We could formally write these as
\begin{align}
\sum_{\sigma\in P(\{2,\ldots,2\},q)\times P(\{2,\ldots,2\},r)} & =\sum_{\sigma_{a}\in P(\{2,\ldots,2\},q)}\,\sum_{\sigma_{b}\in P(\{2,\ldots,2\},r)}.\label{eq:decomposition_permutation}
\end{align}
The remaining $k-k^{\prime}$ vertices are contracted only among one
another, without any connection to the first cluster. An example at
first order and with two external lines is:
\begin{align*}
 & \Diagram{flfluf0f0flfluf0}
\times\Diagram{\vertexlabel_{j_{i}}f\vertexlabel_{j_{k}}}
.\\
\end{align*}

Let us now fix the latter part of the diagram, namely those vertices
that are connected to external legs and let us assume it is composed
of $k^{\prime}$ vertices. We want to investigate, to all orders in
$k$, by which vacuum diagrams such a contribution is multiplied.
At order $k=k^{\prime}$ there cannot be any additional vertices in
the left vacuum part; we get our diagram times $1$ at this order;
the factor $1$ stems from (\ref{eq:expansion_V_second_kind}). At
order $k=k^{\prime}+1$, we get a multiplication with all vacuum diagrams
that have a single vertex. At order $k=k^{\prime}+k^{\prime\prime}$,
we hence get a multiplicative factor of all vacuum diagrams with $k^{\prime\prime}$
vertices. So we see that our particular contribution is multiplied
with all possible vacuum diagrams. To see that they exactly cancel
with those from the denominator $\Z(0)$, we are left to check that
they arise with the same combinatorial factor in both terms. The number
of permutations in (\ref{eq:decomposition_permutation}) is obviously
the same as those in the computation of the vacuum part in the denominator,
as explained in \secref{perturbation_Gaussian}
\begin{align*}
\frac{\Z(j)}{\Z(0)}= & \frac{\big(1+\Diagram{f0flfluf0f0flfluf0\\
\phantom{x}\\
\phantom{x}\\
\phantom{x}
}
+\ldots\big)\times\Diagram{\vertexlabel_{j_{i}}f\vertexlabel_{j_{k}}}
+\ldots}{\big(1+\Diagram{f0flfluf0f0flfluf0\\
\\
\\
\\
\\
\\
}
+\ldots\big)}.\\
\end{align*}
 Also the powers $\epsilon^{k^{\prime}}\cdot\epsilon^{k^{\prime\prime}}=\epsilon^{k}$
obviously add up to the right number. We still need to check the factor
that takes care of multiple occurrences of vertices. In total we have
$k$ vertices. Let us assume a single type of vertex for simplicity.
We have $k$ such vertices in total (in the left and in the right
part together). If $k^{\prime}\le k$ of these appear in the right
part of the diagram, we have $\big(\begin{array}{c}
k\\
k^{\prime}
\end{array}\big)=\frac{k!}{(k-k^{\prime})!\,k^{\prime}!}$ ways of choosing this subset from all of them. Each of these choices
will appear and will yield the same algebraic expression. So we get
a combinatorial factor
\begin{align*}
\frac{1}{k!}\,\frac{k!}{(k-k^{\prime})!k^{\prime}!} & =\frac{1}{(k-k^{\prime})!}\cdot\frac{1}{k^{\prime}!}.
\end{align*}
The first factor on the right hand side is just the factor that appears
in the corresponding vacuum diagram in the denominator. The second
factor is the one that appear in the part that is connected to the
external lines.

We therefore conclude that each diagram with external legs is multiplied
by all vacuum diagrams with precisely the same combinatorial factors
as they appear in the normalization $\Z(0)$. So all vacuum diagrams
are canceled and what remains in $Z_{V}$ are only diagrams that are
connected to external lines:

\begin{align*}
Z(j) & =\frac{\Z(j)}{\Z(0)}=Z_{0}(j)+Z_{V}(j)\\
Z_{V}(j) & =\sum\,\text{graphs(}\Delta=A^{-1},\epsilon V)\text{ with external legs ending on }j\\
\left.Z_{V}^{(l_{1},\ldots,l_{N})}(j)\right|_{j=0} & =\langle x^{l_{1}}\cdots x^{l_{N}}\rangle\\
 & =\sum\,\text{graphs(}\Delta=A^{-1},\epsilon V)\text{ with }l_{1}+\ldots+l_{N}\text{ external legs replaced by }j_{i}^{l_{i}}\to l_{i}!.
\end{align*}
where the rules summarized in the table \tabref{Diagrammatic-rules-scriptZ}
above apply to translate diagrams into their algebraic counterpart
and the latter term $l_{i}!$ arises from the $l_{i}$ derivatives
acting on the external source $j_{i}$ coming in the given power $l_{i}$.

\subsection{Equivalence of graphical rules for $n$-point correlation and $n$-th
moment}

We here want to see that the graphical rules for computing the $n$-th
moment of a single variable $\langle x^{n}\rangle$ are the same as
those for the $n$-point correlation function $\langle x_{1}\cdots x_{n}\rangle$
with $n$ different variables. To see this, we express the moment
generating function for the single variable $Z(j)$ as $Z(j_{1},\ldots,j_{n})=\int dx\,p(x)\,e^{x\,\sum_{i}j_{i}}$
so that the $n$-th moment can alternatively be expressed as
\begin{align*}
\partial_{j_{1}}\cdots\partial_{j_{n}}Z(j_{1},\ldots,j_{n})\big|_{j_{i}=0} & =\langle x^{n}\rangle\\
 & =\partial_{j}^{n}Z(j)\big|_{j=0}.
\end{align*}
This definition formally has $n$ different sources, all coupling
to the same $x$. The combinatorial factors constructed by the diagrams
are the same as those obtained by the $n$-fold derivative: We have
$n(n-1)\cdots1=n!$ ways of assigning the $n$ different $j_{i}$
to the external legs, all of which in this case yield the same result.

\subsection{Example: ``$\phi^{3}+\phi^{4}$'' theory\label{sub:Example_phi3_phi4_pert}}

As an example let us study the system described by the action (\ref{eq:def_S_phi34}).
At zeroth order, the moment generating function (\ref{eq:moment_generating_expansion})
therefore is
\begin{align}
Z_{0}(j) & =\exp\left(\frac{1}{2}K^{-1}j^{2}\right).\label{eq:Z0_phi_34}
\end{align}

At first order in $\epsilon$ we need all contributions with a single
interaction vertex. If it is the three-point vertex, we only get a
contribution that has a single external leg $j$ that contributes,
which corresponds to a so-called \textbf{tadpole diagram}, a diagram
with a single external leg and the two remaining legs connected to
a loop
\begin{align}
\Diagram{\vertexlabel_{j}ff0flfluf0}
= & \epsilon\,\frac{\alpha}{3!}\,j\,\underbrace{\langle\underbrace{x^{3}}_{q=3}\cdot\underbrace{x}_{r=1}\rangle_{0}}_{3(K^{-1})^{2}}\label{eq:first_way_correction}\\
= & \epsilon\frac{\alpha}{3!}\,3\,(K^{-1})^{2}\,j=\epsilon\frac{\alpha}{2}\,(K^{-1})^{2}\,j.\label{eq:second_way_correction}
\end{align}
We may obtain the value of this contribution in two ways:
\begin{enumerate}
\item In the first way, corresponding to (\ref{eq:first_way_correction}),
we directly use the expansions coefficients of (\ref{eq:Taylor_v-1})
at the desired order in $\epsilon$, here $\epsilon^{1}$, and the
coefficients of (\ref{eq:expansion_j}) at the desired order, here
$j^{1}$, collect all factors $x$ of the product (here $x^{4}$)
and obtain their value under the Gaussian distribution by Wick's theorem
(\ref{eq:Wick_moments}), corresponding to a direct evaluation of
(\ref{eq:perturb_first_order}). So we here get $\langle x^{4}\rangle_{0}=3(K^{-1})^{2}$,
because there are $3$ distinct pairings of the first $x$ with the
three remaining ones and then only one combination is left and we
have one propagator line.
\item Alternatively, corresponding to (\ref{eq:second_way_correction}),
we may use the graphical rules derived in the previous section to
get the same result: We have a factor $\frac{\epsilon^{1}}{1!}$,
because we are at first order (one interaction vertex). The three-point
vertex comes with a factor $\frac{\alpha}{3!}$. There is one external
leg, so $\frac{j}{1!}$. The combinatorial factor $3$ arises from
the three choices of attaching the external source $j$ to one of
the three legs of the three point vertex. The remaining two legs of
the three point vertex can then be contracted in only a single way.
\end{enumerate}
Because the diagram has a single leg it contributes to the first moment.
We see that the four point vertex does not contribute to the mean
at this order, because it would give a contribution $\propto\langle x^{5}\rangle_{0}=0$,
which vanishes by Wick's theorem.

Calculating corrections to the mean at second order in $\epsilon$,
we get four different non-vanishing contributions with one external
leg\ifthenelse{\boolean{lecture}}{, calculated in the exercises}.
One of them is

\begin{align*}
\Diagram{\vertexlabel_{j}ff0flfluf0f0flfluf0}
= & \frac{\epsilon^{2}}{1!1!}\,\frac{\alpha}{3!}\,\frac{\beta}{4!}\,3\cdot4\cdot3\cdot\,j\,K^{-4}=\epsilon^{2}\,\frac{\alpha\beta}{4}\,K^{-4}\,j.\\
\end{align*}

The combinatorial factor arises as follows: The external leg $j$
is connected to the three point vertex ($3$ possibilities). The remaining
two legs of the three point vertex need to be connected to two of
the legs of the four point vertex. We may choose one of the legs of
the three point vertex arbitrarily and connect it to one of the four
legs of the four point vertex ($4$ possibilities). The other leg
then has $3$ possibilities left. Had we chosen the other leg of the
three point vertex, we would have gotten the same combinations, so
no additional factor two. Since we have two different interaction
vertices, we get a factor $\frac{\epsilon^{2}}{1!1!}$ form the exponential
function of the interaction potential $V$. 

Diagrams with two external legs that contribute to the second moment
are
\begin{align}
\Diagram{\vertexlabel_{j}f\vertexlabel_{j}}
 & =\frac{j^{2}}{2!}K^{-1},\label{eq:zeroth_order_second_moment}
\end{align}
where the combinatorial factor is one, because there is a unique way
to contract the pair of factors $x$ attached to each $j$. This can
also be seen from the explicit calculation as in point 1. above, as
$\frac{\epsilon^{0}}{0!}\,\frac{j^{2}}{2!}\,\langle x^{2}\rangle_{0}=\frac{j^{2}}{2}K^{-1}$.
The only contribution with one interaction vertex is
\begin{align}
\Diagram{\vertexlabel^{j}fd & f0flfluf0\\
\vertexlabel^{j}fu
}
 & =4\cdot3\cdot\epsilon\frac{\beta}{4!}\,K^{-3}\,\frac{j^{2}}{2!}=\epsilon\frac{\beta}{4}\,K^{-3}\,j^{2}.\label{eq:first_order_second_moment}
\end{align}
At moments higher than one, having two or more external legs, we may
also get unconnected contributions that factorize. For example a second
order contribution to the second moment is
\begin{align}
\Diagram{\vertexlabel_{j}ff0flfluf0}
\times\Diagram{f0flfluf0f\vertexlabel_{j}}
 & =2\cdot3\cdot3\cdot\frac{\epsilon^{2}}{2}\,\left(\frac{\alpha}{3!}\right)^{2}K^{-4}\frac{j^{2}}{2!}=\left(\epsilon\frac{\alpha}{2}\,\right)^{2}K^{-4}\,\frac{j^{2}}{2!},\label{eq:unconnected_second_moment}\\
\nonumber 
\end{align}
being one-half the square of (\ref{eq:second_way_correction}) (Combinatorial
factor: Two vertices to choose to attach the first leg times three
legs to choose from and three legs to choose for attaching the other
external leg). We recognize that this term is a contribution to the
second moment stemming from the product of two contributions from
the first moment. If we calculate the variance, the second cumulant,
we know that exactly these terms will be subtracted. In a way, they
do not carry any new information. We will see in the next section
how these redundant terms are removed in the graphical language.

\section{Linked cluster theorem\label{sec:Linked-cluster-theorem}}

The relations of the different generating functions, the action, the
moments, and cumulants up to this point can be summarized as follows:
\begin{center}
\includegraphics{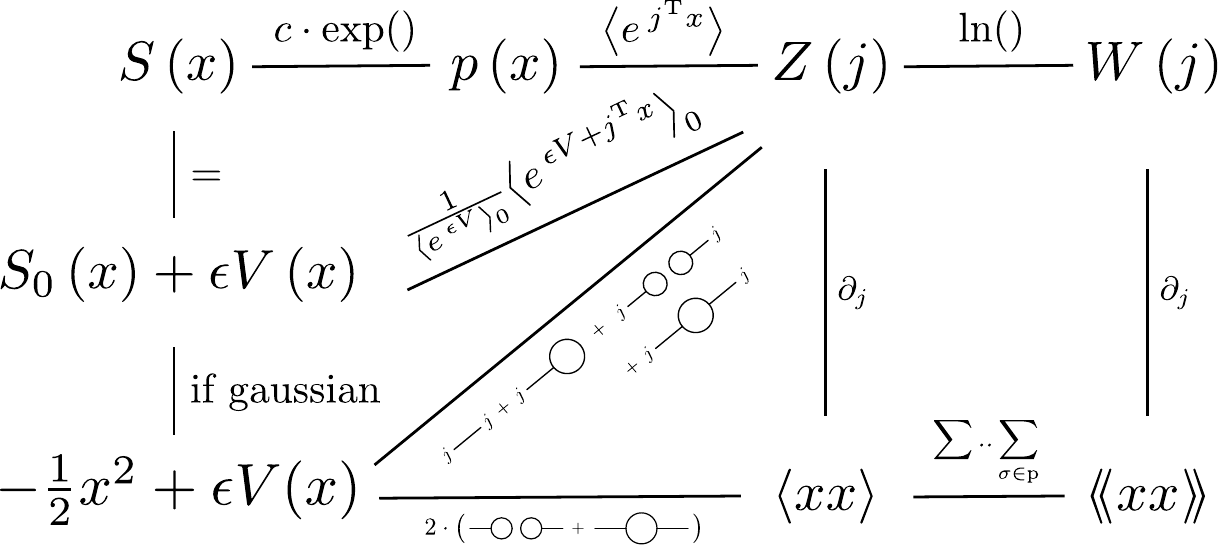}
\par\end{center}

We saw in the last section (in \secref{Cancellation-of-vacuum}) that
the topology of certain graphs allowed us to exclude them from the
expansion of $Z$: the absence of external lines in the vacuum graphs
lead to their cancellation by the normalization. In the coming section
we will derive a diagrammatic expansion of $W$ and the cumulants
and will investigate the topological features of the contributing
graphs. Stated differently, we want to find direct links from the
action $S$ to the cumulant generating function $S$ and to the cumulants.

In the preceding example we noticed that we obtained in step four
a diagram combined of two unconnected diagrams, the first part of
which already appeared at the lower order $\epsilon$. Similarly,
determining corrections $\propto j^{2}$, we would find that similar
diagrams decompose into unconnected components that already appeared
at linear order in $j$. It would be more efficient to only calculate
each diagram exactly once.

We have already faced a similar problem in \subref{Cumulants}, when
we determined the moment generating function of a factorizing density,
a density of independent variables. The idea there was to obtain the
Taylor expansion of $\ln\,Z$ instead of $Z$, because the logarithm
converts a product into a sum. A Taylor expansion can therefore only
contain mixed terms in different $j_{i}$ and $j_{k}$ if these are
part of the same connected component, also called a \textbf{linked
cluster}. We will explore the same idea here to see how this result
comes out more formally in a moment.

\subsection{General proof of the linked cluster theorem\label{sec:General-proof-linked-cluster}}

The \textbf{linked cluster theorem} that we will derive here is fundamental
to organize the perturbative treatment, because it drastically reduces
the number of diagrams that need to be computed. 

To proceed, we again assume we want to obtain a perturbation expansion
of $W(j)=\ln\,Z(j)$ around a theory $W_{0}(j)=\ln\,Z_{0}(j)$. We
here follow loosely the derivation of \citet[p. 120ff]{ZinnJustin96}.

We consider here the general (not necessarily Gaussian) case, where
we know all cumulants of $\Z_{0}$, we may expand the exponential
function in its Taylor series and employ \eqref{k_point_corr_from_cum}
to determine all appearing moments of $x$ as products of cumulants.
Using our result, \eqref{moments_generation}, from \secref{Probabilities},
we see that instead of writing the moments $\langle x_{1}\cdots x_{k}\rangle_{0}$
as expectation values with respect to $\Z_{0}$, we may as well write
them as derivatives of the moment generating function: Each term $\cdots x_{i}\cdots$
will hence be replaced by $\cdots\partial_{i}\cdots$, so that in
total we may write \eqref{Z_perturbation} as 
\begin{align}
\Z(j) & =\int\,dx\,\exp(\epsilon V(\underbrace{x}_{\to\nabla_{j}}))\,\exp\left(S_{0}(x)+j^{\T}x\right)\label{eq:moment_generating_expansion}\\
 & =\exp(\epsilon V(\nabla_{j}))\,\underbrace{\int\,dx\,\exp\left(S_{0}(x)+j^{\T}x\right)}_{=\Z_{0}(j)}\nonumber \\
 & =\exp(\epsilon V(\nabla_{j}))\,\Z_{0}(j)\nonumber \\
 & =\exp(\epsilon V(\nabla_{j}))\,\exp(W_{0}(j))\,\Z_{0}(0)\nonumber 
\end{align}
where $\nabla_{j}=(\partial_{1},\ldots,\partial_{N})^{\T}$ is the
nabla operator, a vector containing the derivative by $j_{k}$, denoted
as $\partial_{k}$, in the $k-$th entry. The latter expression is
defined by the Taylor expansion of the exponential function.

The following proof of connectedness of all contributions, unlike
the results presented in \secref{perturbation_Gaussian}, does not
rely on $\Z_{0}$ being Gaussian. We here start from the general expression
\eqref{moment_generating_expansion} to derive an expansion of $W(j)$,
using the definition \eqref{def_W} to write
\begin{align}
\exp(W(j))=Z(j)=\frac{\Z(j)}{\Z(0)} & =\exp\left(\epsilon V(\nabla_{j})\right)\,\exp\left(W_{0}(j)\right)\,\frac{\Z_{0}(0)}{\Z(0)}\label{eq:perturbation_general}\\
W_{V}(j):=W(j)-W_{0}(j) & =\ln\Big(\,\exp\left(-W_{0}(j)\right)\,\exp\left(\epsilon V(\nabla_{j})\right)\,\exp\left(W_{0}(j)\right)\Big)+\underbrace{\ln\,\frac{\Z_{0}(0)}{\Z(0)}}_{\text{const.}}\nonumber 
\end{align}
where in the second step we multiplied by $\exp(-W_{0}(j))$ and then
took the $\ln$. The latter term $\ln\,\frac{\Z_{0}(0)}{\Z(0)}$ is
just a constant making sure that $W(0)=0$. Since we are ultimately
interested in the derivatives of $W$, namely the cumulants, we may
drop the constant and ensuring $W(0)=0$ dropping the zeroth order
Taylor coefficient in the final result. The last expression shows
that we obtain the full cumulant generating function as $W_{0}$ plus
an additive correction $W_{V}$, which depends on $V$. The aim is
to derive diagrammatic rules to compute $W_{V}$.

The idea is now to prove \textbf{connectedness} of all contributions
by induction, dissecting the operator $\exp\left(\epsilon V(\nabla_{j})\right)$
into infinitesimal operators of slices $\frac{1}{L}$ as
\begin{align}
\exp\left(\epsilon V(\nabla_{j})\right) & =\lim_{L\to\infty}(1+\frac{\epsilon}{L}\,V(\nabla_{j}))^{L}.\label{eq:exponential_expansion}
\end{align}
Each operator of the form $1+\frac{\epsilon}{L}\,V(\nabla_{j})$ only
causes an infinitesimal perturbation provided that $\frac{\epsilon}{L}\ll1$.
We formally keep the $\epsilon$-dependence here for later comparison
with the results obtained in \secref{perturbation_Gaussian}.

We start the induction by noting that at order $\epsilon^{0}$ we
have $W_{V}=0$, so it contains no diagrams. In particular, there
are no disconnected components. 

To make the induction step, we assume that, for large $L$ given and
fixed, the assumption is true until some $0\le l\le L$, which is
that $W_{l}(j)$ is composed of only connected components, where
\begin{align*}
\exp\left(W_{l}(j)\right):= & (1+\frac{\epsilon}{L}\,V(\nabla_{j}))^{l}\,\exp\left(W_{0}(j)\right).
\end{align*}
We then get
\begin{eqnarray*}
W & = & \lim_{L\to\infty}\,W_{L}.
\end{eqnarray*}
Hence we need to show that 
\begin{align*}
\exp(W_{l+1}(j)) & =(1+\frac{\epsilon}{L}\,V(\nabla_{j}))\,\exp(W_{l}(j))
\end{align*}
is still composed only out of connected components. To this end we
again multiply by $\exp\left(-W_{l}(j)\right)$, take the logarithm
and expand $\ln(1+\frac{\epsilon}{L}x)=\frac{\epsilon}{L}\,x+O((\frac{\epsilon}{L})^{2})$
to get
\begin{align}
W_{l+1}(j)-W_{l}(j) & =\frac{\epsilon}{L}\,\left(\exp\left(-W_{l}(j)\right)\,V(\nabla_{j})\,\exp\left(W_{l}(j)\right)\right)+O\big(\big(\frac{\epsilon}{L}\big)^{2}\big).\label{eq:expansion_W_iteration}
\end{align}
Expanding the potential into its Taylor representation (\ref{eq:Taylor_v-1}),
we need to treat individual terms of the form
\begin{align}
 & \frac{V_{i_{1}\cdots i_{n}}^{(n)}}{n!}\,\exp\left(-W_{l}(j)\right)\,\partial_{i_{1}}\cdots\partial_{i_{n}}\,\exp\left(W_{l}(j)\right).\label{eq:single_step-1}
\end{align}
Since the differential operator is multiplied by the respective Taylor
coefficient $\frac{V^{(n)}}{n!}$ from (\ref{eq:Taylor_v-1}), and
noting that the two exponential factors cancel each other after application
of the differential operator to the latter one, what remains is a
set of connected components of $W_{l}(j)$ tied together by the vertex
$\frac{V^{(n)}}{n!}$. The $j$-dependence is ultimately within $W_{0}$,
so that the operation of the differential operator generates all kind
of derivatives of $W_{0}$. We see that disconnected components cannot
appear, because in each iteration step of the form (\ref{eq:single_step-1})
there is only a single interaction vertex. Each leg of such a vertex
corresponds to the appearance of one $\partial_{i_{k}}$, which, by
acting on $W_{l}(j)$ attaches to one one such component.

As an example, consider a one-dimensional theory with the interaction
$\epsilon V(x)=\epsilon\,\frac{x^{4}}{4!}$. We use the symbol with
superscript $l(j)$ to denote $W_{l}(j)$ as a function of $j$ and
the number of legs $n$ as the number of derivatives taken
\begin{align*}
W_{l}^{(n)}(j) & =:\Diagram{\vertexlabel_{n} &  &  & \vertexlabel_{\ldots}\\
fd & !c{l(j)} & fu\\
\vertexlabel^{1}fu &  & fd\vertexlabel^{2}
}
\\
\end{align*}
In this notation, a single step produces the new diagrams

\begin{align}
 & W_{l+1}(j)-W_{l}(j)\nonumber \\
= & \exp(-W_{l}(j))\,\quad\Diagram{\vertexlabel_{\partial_{j}} &  & \vertexlabel_{\partial_{j}}\\
fd & fu\\
\vertexlabel^{\partial_{j}}fu & fd\vertexlabel^{\partial_{j}}
}
\quad\exp(W_{l}(j))\\
= & \exp(-W_{l}(j))\,\Bigg[\Diagram{!c{l(j)} &  &  & !c{l(j)}\\
 & fd & fu\\
!c{l(j)} & fu & fd & !c{l(j)}
}
\quad+\quad\left(\begin{array}{c}
4\\
2
\end{array}\right)\cdot\Diagram{!c{l(j)} & f0flfluf0f0flfluf0 & !c{l(j)}}
\quad\label{eq:one_step_W_graphical}\\
 & \phantom{\exp(-W_{l}(j))}+\quad\Diagram{mflflum!c{l(j)}}
+\quad4\cdot\Diagram{!c{l(j)}ffflfluf!c{l(j)}}
+\quad\left(\begin{array}{c}
4\\
2
\end{array}\right)\cdot\Diagram{ &  & !c{l(j)}\\
!c{l(j)}f0flfluf0 & fu\\
 & fd & !c{l(j)}
}
\Bigg]\,\exp(W_{l}(j))\quad.
\end{align}
By construction, because every differential operator is attached to
one leg of the interaction vertex, we do not produce any unconnected
components. The combinatorial factors are the same as usual but can
also be derived from the rules of differentiation: For the first term,
each of the four differential operators needs to act on $W_{l}(j)$,
so a factor $1$; For the second term: There are $\left(\begin{array}{c}
4\\
2
\end{array}\right)$ ways of choosing two of the four derivatives that should act on the
same $W_{l}$ and two which remain to act on a new $W_{l}$ from the
exponential function. The other factors follow by analogous arguments.
We see that only sums of connected components are produced, proving
the assumption of connectedness by induction.

What remains to be shown is that the connected diagrams produced by
the iterative application of \eqref{expansion_W_iteration} come with
the same factor as those that are produced by the direct perturbation
expansion in \secref{perturbation_Gaussian}, for the example of a
Gaussian theory as the underlying exactly solvable model. We therefore
rewrite the recursion step as
\begin{align}
W_{l+1}(j) & =\underline{1}\cdot W_{l}(j)+\underline{\frac{\epsilon}{L}}\cdot\sum_{n=1}^{\infty}\sum_{i_{1}\cdots i_{n}=1}^{N}\frac{V^{(n)}}{n!}\,\exp\left(-W_{l}(j)\right)\,\partial_{i_{1}}\cdots\partial_{i_{n}}\,\exp\left(W_{l}(j)\right).\label{eq:iterative_W}
\end{align}
Here we can ommit the constant term of $V$, so starting at $n=1$,
because the constant may be absorbed into the normalization constant.
The latter expression shows that each step adds to $W_{l}(j)$ the
set of diagrams from the term (\ref{eq:single_step-1}) on the right
hand side to obtain $W_{l+1}(j)$. The additional diagrams, as argued
above, combine the connected elements already contained in $W_{l}$
with vertices from \eqref{Taylor_v-1}.

We now want to show that we only need to include those new diagrams
that add exactly one vertex to each diagram already contained in $W_{l}$
and that we do not need to consider situations where the additional
vertex ties together two components that each have one or more interaction
vertices. Stated differently, only one leg of the interaction vertex
shown in (\ref{eq:one_step_W_graphical}) must attach to a component
in $W_{l}$, while all remaining legs must be attached to $W_{0}$.
To understand why this is, we need to consider the overall factor
in front of a resulting diagram with $k$ interaction vertices after
$L$ iterations of (\ref{eq:iterative_W}). Each step of (\ref{eq:iterative_W}),
by the first term, copies all diagrams as they are and, by the second
term, adds those formed by help of an additional interaction vertex.
Following the modification of one component through the iteration,
in each step we hence have the binary choice to either leave it as
it is or to combine it with other components by help of an additional
vertex.

We first consider the case that each of the $k$ vertices is picked
up in a different step (at different $l$) in the iteration. To formalize
this idea, we need to distinguish the terms in $W_{l}$
\begin{align*}
W_{l}(j) & =W_{0}(j)+W_{V,l}(j)=\Diagram{!c{0(j)}}
+\Diagram{!c{V,l(j)}}
\end{align*}
into those of the solvable theory $W_{0}$, which are independent
of $\epsilon$, and the corrections in $W_{V,l}$ that each contain
at least one interaction vertex and hence at least one factor of $\epsilon$.
For the example shown in (\ref{eq:one_step_W_graphical}), this means
that we need to insert $W_{0}+W_{V,l}$ at each ``leave'', multiply
out and only keep those graphs that contain at most one contribution
from $W_{V,l}$; all other contributions would add a diagram with
more than one additional vertex and we would hence need less than
$k$ steps to arrive at a diagram of order $k$.

Each such step comes with a factor $\frac{\epsilon}{L}$ and there
are $\left(\begin{array}{c}
L\\
k
\end{array}\right)$ ways to select $k$ steps out of the $L$ in which the second term
rather than the first term of (\ref{eq:iterative_W}) acted on the
component in question. So in total we get a factor
\begin{align}
\left(\frac{\epsilon}{L}\right)^{k}\left(\begin{array}{c}
L\\
k
\end{array}\right) & =\frac{\epsilon^{k}}{k!}\,\frac{L(L-1)\cdots(L-k+1)}{L^{k}}\stackrel{L\to\infty}{\to}\frac{\epsilon^{k}}{k!},\label{eq:factor_pick_up_one}
\end{align}
which is independent of $L$.

Now consider the case that we pick up the $k$ vertices along the
iteration (\ref{eq:iterative_W}) such that in one step we combined
two sub-components with each one or more vertices; a diagram where
the vertex combines two or more components from $W_{V,l}$. Consequently,
to arrive at $k$ vertices in the end, we only need $k^{\prime}<k$
iteration steps in which the latter rather than the first term of
(\ref{eq:iterative_W}) acted on the component. The overall factor
therefore is
\begin{align}
\left(\frac{\epsilon}{L}\right)^{k}\left(\begin{array}{c}
L\\
k^{\prime}
\end{array}\right) & =\frac{\epsilon^{k}}{k^{\prime}!}\,\frac{L(L-1)\cdots(L-k^{\prime}+1)}{L^{k}}\stackrel{L\gg k^{\prime}}{=}\frac{\epsilon^{k}}{k^{\prime}!}\,\frac{1}{L^{k-k^{\prime}}}\stackrel{L\to\infty}{=}0.\label{eq:factor_pick_up_many}
\end{align}
In the limit that we are interested in we can hence neglect the latter
option and conclude that we only need to consider in each step the
addition of a single vertex to any previously existing component.
The very same argument shows why the neglected terms of $O(\frac{\epsilon}{L})^{2}$
that we dropped when expanding $\ln(1+\frac{\epsilon}{L})$, do not
contribute to the final result in the limit $L\to\infty$: Such terms
would increse the order of the term in a single iteration step by
two or more - consequently we would need $k^{\prime}<k$ steps to
arrive at an order $k$ contribution - the combinatorial factor would
hence be $\propto L^{k-k^{\prime}}$, as shown above, so these terms
do not contribute.

We see that after $L$ steps all possible diagrams are produced, starting
from those with $k=0$ interaction vertices and ending with those
that have $k=L$ interaction verticesand the overall factor for each
diagram is as in the perturbation expansion derived in \secref{perturbation_Gaussian}:
the connected diagrams come with the same factor as in $Z_{V}$. The
factor (\ref{eq:factor_pick_up_one}) also obviously follows from
the series representation of the exponential function in \eqref{exponential_expansion}.
All other constituents of the diagram are, by construction, identical
as well.

So to summarize, we have found the simple rule to calculate $W(j)$:
\begin{align}
W(j) & =\ln\,Z(j)\label{eq:linked_cluster}\\
 & =W_{0}(j)+\sum_{\text{connected diagrams}}\in Z_{V}(j),\nonumber 
\end{align}
where the same rules of construction apply for $Z_{V}$ that are outlined
in \secref{perturbation_Gaussian}.

\subsection{Dependence on $j$ - external sources - two complimentary views}

There are two different ways how one may interpret the iterative construction
(\ref{eq:iterative_W}): We may either consider the $W_{l}(j)$ appearing
on the right hand side as a function of $j$, or we may expand this
function in powers of $j$. In the graphical representation above
(\ref{eq:one_step_W_graphical}), we used the former view.

In the following, instead, we want to follow the latter view, exhibiting
explicitly the $j$-dependence on the external legs. The two representations
are, of course, equivalent.

Each element
\begin{eqnarray*}
W_{l}^{(1)}(j) & = & \Diagram{f!c{l(j)}}
\end{eqnarray*}
that appears in the first term of (\ref{eq:one_step_W_graphical})
is a function of $j$. Note that the diagrams produced in (\ref{eq:one_step_W_graphical})
look like a vacuum diagrams. We will reconcile this apparent discrepancy
now. Let us for concreteness imagine the first step of the iteration,
so $l=1$: Then all cumulants appearing in the last expression belong
to the unperturbed theory, hence the first term of (\ref{eq:one_step_W_graphical})
takes the form
\begin{align}
W_{1}(j) & =\Diagram{!c{0(j)} &  &  & !c{0(j)}\\
 & fd & fu\\
!c{0(j)} & fu & fd & !c{0(j)}
}
+\ldots.\label{eq:W_1_j_dep}
\end{align}
Now imagine we have the unperturbed theory represented in its cumulants
and let us assume that only the first three cumulants are non-vanishing

\begin{align}
W_{0}(j) & =\sum_{n=1}^{3}\frac{1}{n!}\,W_{0}^{(n)}(0)\,j^{n}\nonumber \\
 & =\feyn{\vertexlabel^{j}f!c{0(0)}}+\frac{1}{2!}\quad\feyn{\vertexlabel^{j}f!c{0(0)}f\vertexlabel^{j}}+\frac{1}{3!}\quad\Diagram{\vertexlabel^{j}fd & !c{0(0)} & f\vertexlabel^{j}\\
\vertexlabel^{j}fu
}
\quad,\label{eq:W_0_Taylor}
\end{align}
where the superscipt $0(0)$ is meant to indicate that the cumulants
of the solvable theory are just numbers that are independent of $j$
and the entire $j$-dependence of $W_{0}(j)$ is explicit on the factors
$j$ on the legs in (\ref{eq:W_0_Taylor}).

We may therefore make the $j$-dependence in (\ref{eq:W_1_j_dep})
explicit by inserting the latter representation for each $W_{0}^{(1)}(j)$,
which we obtain by differentiating (\ref{eq:W_0_Taylor}) once
\begin{align}
W_{0}^{(1)}(j) & =\feyn{f!c{0(0)}}+\quad\feyn{f!c{0(0)}f\vertexlabel^{j}}+\frac{1}{2!}\quad\Diagram{fd & !c{0(0)} & f\vertexlabel^{j}\\
\vertexlabel^{j}fu
}
\quad,\label{eq:W_0_1_Taylor}
\end{align}
removing one $j$ from each term and using the product rule. The tadpole
diagram $\feyn{f!c{0(0)}}$ signifies the mean value of the solvable
theory $\langle x\rangle=W_{0}^{(1)}(0)$. These diagrams would of
course vanish if $W_{0}$ was a centered Gaussian.

Making this replacement for each of the symbols for $W_{0}^{(1)}(j)$
in (\ref{eq:W_1_j_dep}) produces all diagrams, where all possible
combinations of the above terms appear on the legs of the interaction
vertex

\begin{align*}
\Diagram{!c{0(j)} &  &  & !c{0(j)}\\
 & fd & fu\\
!c{0(j)} & fu & fd & !c{0(j)}
}
 & \quad=\quad\Diagram{!c{0(0)} &  &  & !c{0(0)}\\
 & fd & fu\\
!c{0(0)} & fu & fd & !c{0(0)}
}
\quad+\quad4\cdot\quad\,\Diagram{ & !c{0(0)} &  &  & !c{0(0)}\\
 &  & fd & fu\\
\vertexlabel_{j}f & !c{0(0)} & fu & fd & !c{0(0)}
}
\quad+\ldots+\quad4\cdot\frac{1}{2!}\cdot\quad\Diagram{ & !c{0(0)} &  &  & !c{0(0)}\\
 &  & fd & fu\\
 & !c{0(0)} & fu & fd & !c{0(0)} & fu\vertexlabel^{j}\\
 &  &  &  &  & fd\vertexlabel^{j}
}
\quad+\ldots\quad,\\
\end{align*}
where the factor $4$ in the second term comes from the four possible
legs of the interaction vertex to attach the $j$-dependence and the
factor $4$ in the third term comes for the same reason. The $\frac{1}{2!}$
is the left-over of the factor $\frac{1}{3!}$ of the third cumulant
and a factor $3$ due to the product rule from the application of
the $\partial_{j}$ to either of the three legs of the third cumulant.

This explicit view, having the $j$-dependence on the external legs,
allows us to understand (\ref{eq:linked_cluster}) as a rule to construct
the cumulants of the theory directly, because differentiating amounts
to the removal of the $j$ on the corresponding leg. We can therefore
directly construct the cumulents from all connected diagrams with
a given number of external legs corresponding to the order of the
cumulant

\begin{align*}
\left.W^{(l_{1},\ldots,l_{N})}(j)\right|_{j=0} & =\llangle x^{l_{1}}\cdots x^{l_{N}}\rrangle\\
 & =W_{0}^{(l_{1},\ldots,l_{N})}(j)\big|_{j=0}+\sum_{\text{connected diagrams}}\in Z_{V}(j)\,\text{with }l_{1}+\ldots+l_{N}\text{ external legs replaced by }j_{i}^{l_{i}}\to l_{i}!.
\end{align*}
We saw a similar example in \subref{Example_phi3_phi4_pert} in the
calculation of the expectation value, derived from diagrams with a
single external leg.

\subsection{Example: Connected diagrams of the ``$\phi^{3}+\phi^{4}$'' theory\label{sec:Example_Connecteddiagrams_phi34}}

As an example let us study the system described by the action (\ref{eq:def_S_phi34}).
We want to determine the cumulant generating function until second
order in the vertices. To lowest order we have with (\ref{eq:Z0_phi_34})
$W_{0}(j)=\frac{1}{2}Kj^{2}$. To first order, we need to consider
with (\ref{eq:linked_cluster}) all connected diagrams with one vertex.
We get one first order correction with one external leg
\begin{align}
\Diagram{\vertexlabel_{j}{f}{}f0{fl}{}{flu}{}f0}
= & 3\cdot K^{-2}\,j\,\epsilon\frac{\alpha}{3!}=\epsilon\frac{\alpha}{2}\,K^{-2}\,j.\label{eq:first_order_phi43_1}
\end{align}
The correction to the second cumulant is
\begin{align}
\Diagram{\vertexlabel^{j}fd & f0{fl}{}{flu}{}f0\\
\vertexlabel^{j}fu
}
 & =4\cdot3\cdot\epsilon\frac{\beta}{4!}\,K^{-3}\,\frac{j^{2}}{2!}=\epsilon\frac{\beta}{4}\,K^{-3}\,j^{2}.\label{eq:first_order_phi34_2}
\end{align}
In addition, we of course have the bare interaction vertices connected
to external sources, i.e. a contribution to the third and fourth cumulants
\begin{align*}
\Diagram{\vertexlabel^{j}fdf\vertexlabel_{j}\\
\vertexlabel^{j}fu
}
 & =3\cdot2\cdot1\cdot\epsilon\frac{\alpha}{3!}\,K^{-3}\,\frac{j^{3}}{3!}=\epsilon\frac{\alpha}{3!}\,K^{-3}\,j^{3}\\
\\
\Diagram{\vertexlabel^{j}fd & fu\vertexlabel^{j}\\
\vertexlabel^{j}fu & fd\vertexlabel^{j}
}
 & =\epsilon\frac{\beta}{4!}\,K^{-4}\,j^{4}.
\end{align*}
These are all corrections at first order.

At second order we have the contributions to the first cumulant
\begin{align}
\Diagram{\vertexlabel_{j}ffflfluf}
 & =4\cdot3\cdot2\cdot\,\frac{\epsilon^{2}}{1!1!}\,\frac{\alpha}{3!}\,\frac{\beta}{4!}\,K^{-4}\,j\label{eq:second_order_phi34_mean_1}\\
\nonumber \\
\Diagram{\vertexlabel_{j}ff0flfluf0f0flfluf0}
 & =3\cdot4\cdot3\,\frac{\epsilon^{2}}{1!1!}\,\frac{\alpha}{3!}\,\frac{\beta}{4!}\,K^{-4}\,j\label{eq:second_order_phi34_mean_2}
\end{align}

\begin{fmffile}{Two_loop_FeynMF}%
	\begin{eqnarray}
		\parbox{30mm}{
		\begin{fmfgraph*}(80,60) 
			\fmfleft{i}
			\fmflabel{j}{i}
			\fmfright{o}
			\fmf{plain}{i,v1,v1,v2}
			\fmf{plain,left}{v2,o,v2}
		\end{fmfgraph*}
		} &= & 4 \cdot 3 \cdot 3 \cdot \frac{\epsilon^2}{1!1!}\frac{\alpha}{3!}\frac{\beta}{4!} K^{-4} \; j.
	\end{eqnarray}	
\end{fmffile}The corrections to the second cumulant are
\begin{align}
\Diagram{\vertexlabel_{j}ff0flfluf0f\vertexlabel_{j}}
 & =2\cdot3\cdot3\cdot2\cdot\frac{\epsilon^{2}}{2!}\,\left(\frac{\alpha}{3!}\right)^{2}\,K^{-4}\,\frac{j^{2}}{2!}\label{eq:second_order_phi34_var_1}\\
\nonumber \\
\Diagram{\vertexlabel_{j}ffflfluff\vertexlabel_{j}}
 & =2\cdot4\cdot4\cdot3\cdot2\cdot\frac{\epsilon^{2}}{2!}\,\left(\frac{\beta}{4!}\right)^{2}\,K^{-5}\,\frac{j^{2}}{2!}\label{eq:second_order_phi34_var_2}\\
\nonumber \\
\Diagram{\vertexlabel^{j}fd & ff0flfluf0\\
\vertexlabel^{j}fu
}
 & =2\cdot3\cdot2\cdot3\cdot\frac{\epsilon^{2}}{2!}\,\left(\frac{\alpha}{3!}\right)^{2}\,K^{-4}\,\frac{j^{2}}{2!}\label{eq:second_order_phi34_var3}\\
\nonumber \\
\Diagram{\vertexlabel^{j}fd & f0flfluf0f0flfluf0\\
\vertexlabel^{j}fu
}
 & =2\cdot4\cdot3\cdot4\cdot3\cdot\frac{\epsilon^{2}}{2!}\,\left(\frac{\beta}{4!}\right)^{2}\,K^{-5}\,\frac{j^{2}}{2!}.\label{eq:second_order_phi34_var4}\\
\nonumber \\
\Diagram{fs0c & fs0c\\
\vertexlabel^{j}ff & ff\vertexlabel^{j}
}
 & =2\cdot4\cdot4\cdot3\cdot3\cdot\frac{\epsilon^{2}}{2!}\,\left(\frac{\beta}{4!}\right)^{2}\,K^{-5}\,\frac{j^{2}}{2!}
\end{align}

Here the first factor $2$ comes from the two identical vertices to
choose from to attach the external legs. We could go on to the third
and fourth cumulants, but stop here. We notice that there are some
elements repeating, such as in \eqref{second_order_phi34_var3}, which
is composed of a bare three-point interaction vertex and \eqref{first_order_phi43_1}.
Remembering the proof of the linked cluster theorem, this is what
we should expect: Each order combines the bare interaction vertices
with all diagrams that have already be generated up to this order.
In \secref{Vertex-generating-function} we will see how we can constrain
this proliferation of diagrams.

\section{Functional preliminaries\label{sec:funcional_preliminaries}}

In this section we collect some basic rules of functional calculus
that will be needed in the subsequent sections. In this section we
assume that $f:\mathcal{C}\mapsto\mathbb{R}$ is a functional that
maps from the space of smooth functions $\mathcal{C}$ to the real
numbers.

\subsection{Functional derivative}

The derivative of a functional in the point $x$ is defined as
\begin{align}
\frac{\delta f[x]}{\delta x(t)} & :=\lim_{\epsilon\to0}\frac{1}{\epsilon}\,f[x+\epsilon\,\delta(\circ-t)]-f[x]\label{eq:def_func_deriv}\\
 & =\frac{d}{d\epsilon}F(\epsilon)\Big|_{\epsilon=0}\qquad F(\epsilon):=f[x+\epsilon\,\delta(\circ-t)],\nonumber 
\end{align}
where the second equal sign only holds if the limit exists. Linearity
of the definition of the derivative is obvious. Note that one always
differentiates with respect to one particular time $t$. The functional
derivative by $x(t)$ therefore measures how sensitive the functional
depends on the argument in the point $x(t)$.

\subsubsection{Product rule\label{sub:func_Product-rule}}

Since the functional derivative can be traced back to the ordinary
derivative, all known rules carry over. In particular, the product
rule reads
\begin{align}
\frac{\delta}{\delta x(t)}(f[x]g[x]) & =\frac{d}{d\epsilon}\,(F(\epsilon)G(\epsilon))\label{eq:func_product_rule}\\
 & =F^{\prime}(0)G(0)+F(0)G^{\prime}(0)=\frac{\delta f[x]}{\delta x(t)}g[x]+f[x]\frac{\delta g[x]}{\delta x(t)}.\nonumber 
\end{align}

\subsubsection{Chain rule}

With $g:\mathcal{C}\mapsto\mathcal{C}$, the chain rule follows from
the $n$-dimensional chain rule by discretizing the $t$-axis in $N$
bins of width $h$ and applying the chain rule in $\mathbb{R}^{N}$
and then taking the limit of the infinitesimal discretization

\begin{align}
\frac{\delta}{\delta x(t)}f[g[x]] & =\frac{d}{d\epsilon}\,f[g[x+\epsilon\delta(\circ-t)]\nonumber \\
 & =\lim_{h\to0}\,\frac{d}{d\epsilon}\,f(g[x+\epsilon\delta(\circ-t)](h),\ldots,g[x+\epsilon\delta(\circ-t)](Nh))\nonumber \\
 & \stackrel{N-\text{dim chain rule}}{=}\lim_{h\to0}\sum_{i=1}^{N}\,\frac{\partial f}{\partial y_{i}}\,\frac{\partial g[x+\epsilon\delta(\circ-t)](ih)}{\partial\epsilon}\label{eq:func_chain}\\
 & =\lim_{h\to0}\underbrace{\sum_{i=1}^{N}h}_{\to\int\,ds}\,\frac{1}{h}\frac{\partial f}{\partial y_{i}}\,\frac{\delta g[x](ih)}{\delta x(t)}\nonumber \\
 & =\int\,ds\,\frac{\delta f[g]}{\delta y(s)}\,\frac{\delta g[x](s)}{\delta x(t)}.\nonumber 
\end{align}

\subsubsection{Special case of the chain rule: Fourier transform}

In the case of a Fourier transform $x(t)=\frac{1}{2\pi}\int\,e^{i\omega t}X(\omega)\,d\omega$,
we may apply the chain rule to obtain the derivative of the functional
$\hat{f}$ defined on the Fourier transform $X$ by

\begin{align*}
\hat{f}[X] & :=f[\underbrace{\frac{1}{2\pi}\int\,e^{i\omega\circ}X(\omega)\,d\omega}_{\equiv x(\circ)}],
\end{align*}
where $\circ$ is the argument of the function $x(\circ)$ on which
the functional $f$ depends. We obtain by using 

\begin{align*}
\frac{\delta\hat{f}[X]}{\delta X(\omega)}=\frac{\delta}{\delta X(\omega)}\,f\left[\frac{1}{2\pi}\int\,e^{i\omega\circ}X(\omega)\,d\omega\right] & \stackrel{}{=}\int\underbrace{\frac{e^{i\omega s}}{2\pi}}_{\frac{\delta x(s)}{\delta X(\omega)}}\,\frac{\delta f[x]}{\delta x(s)}\,ds.
\end{align*}
So the relationship between a functional and the functional of the
Fourier transform has the inverse transformation properties than a
function, indicated by the opposite sign of $\omega$ and the appearance
of the factor $1/2\pi$.

We will frequently encounter expressions of the form
\begin{align}
 & \int\,\frac{\delta f[x]}{\delta x(s)}\,y(s)\,ds\label{eq:derivative_Fourier_transform}\\
= & \int\frac{1}{2\pi}\int\frac{\delta f[x]}{\delta x(s)}\,e^{i\omega s}Y(\omega)\,d\omega\,ds\nonumber \\
= & \int\underbrace{\frac{1}{2\pi}\int e^{i\omega s}\frac{\delta f[x]}{\delta x(s)}\,ds}_{=\frac{\delta\hat{f}[X]}{\delta X(\omega)}}\,Y(\omega)\,d\omega\nonumber \\
= & \int\,\frac{\delta\hat{f}}{\delta X(\omega)}\,Y(\omega)\,d\omega,\nonumber 
\end{align}
which are hence invariant under Fourier transform. We will make use
of this property when evaluating Feynman diagrams in Fourier domain.

\subsection{Functional Taylor series}

The perturbative methods we have met so far often require the form
of the action to be an algebraic functional of the fields. We obtain
such a form by functional Taylor expansion. Assume we have a functional
$f[x]$ of a field $x(t)$. We seek the analogue to the usual Taylor
transform, which is a representation of the functional as the series
\begin{align*}
f[x] & =\sum_{n=0}^{\infty}\int dt_{1}\cdots\int dt_{n}\,a_{n}(t_{1},\ldots,t_{n})\,\prod_{i=1}^{n}x(t_{i}),
\end{align*}
where we assume $a_{n}$ to be symmetric with respect to permutations
of its arguments. Taking the $k$-th functional derivative $\frac{\delta}{\delta x(t)}$
we get by the product rule
\begin{align*}
\frac{\delta^{k}}{\delta x(s_{1})\cdots\delta x(s_{k})}f[x]\Big|_{x=0} & =\sum_{\left(i_{1},...,i_{k}\right)\in S\left(1,..,k\right)}a_{k}\left(s_{i_{1}},...,s_{i_{k}}\right)\overbrace{=}^{a_{k}\text{ symm.}}k!\,a_{k}(s_{1},\ldots,s_{k}),
\end{align*}
as only the term with $n=k$ remains after setting $x=0$ ($S\left(1,..,k\right)$
indicates the symmetric group, i.e. all permutations of $1,..,k$).
The application of the first derivative yields, by product rule, the
factor $k$ by applying the differentiation to any of the $k$ factors,
the second application yields $k-1$ and so on. We therefore need
to identify $k!\,a_{k}=\delta^{k}f/\delta x^{k}$ and obtain the form
reminiscent of the usual $n$-dimensional Taylor expansion
\begin{align}
f[x] & =\sum_{n=0}^{\infty}\int dt_{1}\cdots\int dt_{n}\,\frac{1}{n!}\,\frac{\delta^{n}f}{\delta x(t_{1})\cdots\delta x(t_{n})}_{n}\,\prod_{i=1}^{n}x(t_{i}).\label{eq:func_Taylor}
\end{align}
The generalization to an expansion around another point than $x\equiv0$
follows by replacing $x\to x-x^{0}$. The generalization to functional
that depend on several fields follows by application of the functional
Taylor expansion for each dependence.

\section{Functional formulation of stochastic differential equations\label{sec:Martin-Siggia-Rose-De-Dominicis}}

We here follow \citet{Chow10_1009} to derive the Martin-Siggia-Rose-DeDominicis-Janssen
\citep{Martin73,janssen1976_377,dedominicis1976_247,DeDomincis78_353,Altland01,Chow15}
path integral representation of a stochastic differential equation
and \citet{Wio89_7312} to obtain the Onsager-Machlup path integral
\citep{Onsager53}. We generalize the notation to also include the
Stratonovich convention as in \citep{Wio89_7312}. \citet{hertz2016_arxiv}
also provide a pedagogical survey of the Martin-Siggia-Rose path integral
formalism for the dynamics of stochastic and disordered systems. The
material of this section has previously been made publicly available
as \citep{Schuecker16b_arxiv}.

The presented functional formulation of dynamics is advantageous in
several respects. First, it recasts the dynamical equations into a
path-integral, where the dynamic equations give rise to the definition
of an ``action''. In this way, the known tools from theoretical
physics, such as perturbation expansions with the help of Feynman
diagrams or the loopwise expansions to obtain a systematic treatment
of fluctuations \citep{ZinnJustin96}, can be applied. Within neuroscience,
the recent review \citep{Chow15} illustrates the first, the work
by \citep{Buice07_051919} the latter approach. Moreover, this formulation
will be essential for the treatment of disordered systems in \prettyref{sec:Sompolinsky-Crisanti-Sommers-theory},
following the spirit of the work by \citet{DeDomincis78_353} to obtain
a generating functional that describes an average system belonging
to an ensemble of systems with random parameters.

Many dynamic phenomena can be described by differential equations.
Often, the presence of fluctuations is represented by an additional
stochastic forcing. We therefore consider the \textbf{stochastic differential
equation} (SDE) 
\begin{eqnarray}
dx(t) & = & f(x)\,dt+g(x)\,dW(t)\label{eq:SDE}\\
x(0+) & = & a,\nonumber 
\end{eqnarray}
where $a$ is the initial value and $dW$ a stochastic increment.
Stochastic differential equations are defined as the limit $h\to0$
of a dynamics on a discrete time lattice of spacing $h$. For discrete
time $t_{l}=lh$, $l=0,\ldots,M$, the solution of the SDE consists
of the discrete set of points $x_{l}=x(t_{l})$. For the discretization
there are mainly two conventions used, the Ito and the Stratonovich
convention \citep{Gardiner09}. In case of additive noise ($g(x)=\mathrm{const.}$),
where the stochastic increment in \eqref{SDE} does not depend on
the state $x$, the two conventions yield the same continuous-time
limit \citep{Gardiner09}. However, as we will see, different discretization
conventions of the drift term lead to different path integral representations.
The Ito convention defines the symbolic notation of \eqref{SDE} to
be interpreted as 
\begin{eqnarray*}
x_{i}-x_{i-1} & = & f(x_{i-1})\,h+a\delta_{i1}+g(x_{i-1})\,\xi_{i},
\end{eqnarray*}
where $\xi_{i}$ is a stochastic increment that follows a probabilistic
law. A common choice for $\xi_{i}$ is a normal distribution $\rho(\xi_{i})=\N(0,\,hD)$,
called a Wiener increment. Here the parameter $D$ controls the variance
of the noise. The term $a\delta_{i1}$ ensures that, in the absence
of noise $\xi_{1}=0$ and assuming that $x_{i\le0}=0$, the solution
obeys the stated initial condition $x_{1}=a$. If the variance of
the increment is proportional to the time step $h$, this amounts
to a $\delta$-distribution in the autocorrelation of the noise $\xi=\frac{dW}{dt}dt$.
The Stratonovich convention, also called mid-point rule, instead interprets
the SDE as
\begin{eqnarray}
x_{i}-x_{i-1} & = & f\left(\frac{x_{i}+x_{i-1}}{2}\right)\,h+a\delta_{i1}+g(\frac{x_{i}+x_{i-1}}{2})\,\xi_{i}.\label{eq:SDE_stratonovich}
\end{eqnarray}
Both conventions can be treated simultaneously by defining
\begin{eqnarray}
x_{i}-x_{i-1} & = & f(\mbox{\ensuremath{\alpha}}x_{i}+(1-\alpha)x_{i-1})\,h+a\delta_{i1}+g(\mbox{\ensuremath{\alpha}}x_{i}+(1-\alpha)x_{i-1})\,\xi_{i}\label{eq:discrete_sde}\\
\alpha & \in & [0,1].\nonumber 
\end{eqnarray}
Here $\alpha=0$ corresponds to the Ito convention and $\alpha=\frac{1}{2}$
to Stratonovich. 

In the following we will limit the treatment to so-called additive
noise, where the function $g(x)=1$ is the identity. The two conventions,
Ito and Stratonovich then converge to the same limit, but their representation
still bears some differences. Both conventions appear in the literature.
For this reason, we here keep the derivation general, keeping the
value $\alpha\in[0,1]$ arbitrary.

If the noise is drawn independently for each time step, which is the
definition of the noise being white, the probability density of the
points $x_{1},\ldots,x_{M}$ along the path $x(t)$ can be written
as

\begin{eqnarray}
p(x_{1},\ldots,x_{M}|a) & \equiv & \int\Pi_{i=1}^{M}d\xi_{i}\,\rho(\xi_{i})\,\delta(x_{i}-y_{i}(\xi_{i},x_{i-1})),\label{eq:p_of_x}
\end{eqnarray}
where, by \eqref{discrete_sde}, $y_{i}(\xi_{i},x_{i-1})$ is understood
as the solution of \eqref{discrete_sde} at time point $i$ given
the noise realization $\xi_{i}$ and the solution until the previous
time point $x_{i-1}$: The solution of the SDE starts at $i=0$ with
$x_{0}=0$ so that $\xi_{1}$ and $a$ together determine $x_{1}$.
In the next time step, $\xi_{2}$ and $x_{1}$ together determine
$x_{2}$, and so on. In the Ito-convention ($\alpha=0$) we have an
explicit solution $y_{i}(\xi_{i},x_{i-1})=x_{i-1}+f(x_{i-1})\,h+a\delta_{i1}+\xi_{i}$,
while the Stratonovich convention yields an implicit equation, since
$x_{i}$ appears as an argument of $f$ in \eqref{SDE_stratonovich}.
We will see in \eqref{normalization_strato} that the latter gives
rise to a non-trivial normalization factor for $p$, while for the
former this factor is unity.

The notation $y_{i}(\xi_{i},x_{i-1})$ indicates that the solution
only depends on the last time point $x_{i}$, but not on the history
longer ago. This property is called the \textbf{Markov property} of
the process. The form of \eqref{p_of_x} also shows that the density
is correctly normalized, because integrating over all paths
\begin{align}
 & \int dx_{1}\cdots\int dx_{M}\,p(x_{1},\ldots,x_{M}|a)=\int\Pi_{i=1}^{M}d\xi_{i}\rho(\xi_{i})\,\underbrace{\int dx_{i}\,\delta(x_{i}-y_{i}(\xi_{i},x_{i-1}))}_{=1}\label{eq:normalization_p_path}\\
= & \Pi_{i=1}^{M}\int d\xi_{i}\rho(\xi_{i})=1\nonumber 
\end{align}
yields the normalization condition of $\rho(\xi_{i})$, $i=1,\ldots,M$,
the distribution of the stochastic increments.

In section \prettyref{sec:Onsager-Machlup-path-integral} we will
look at the special case of Gaussian noise and derive the so called
Onsager-Machlup path integral \citep{Onsager53}. This path integral
has a square in the action, originating from the Gaussian noise. For
many applications, this square complicates the analysis of the system.
The formulation presented in \prettyref{subsec:Martin-Siggia-Rose-De-Dominicis}
removes this square on the expense of the introduction of an additional
field, the so called response field. This formulation has the additional
advantage that responses of the system to perturbations can be calculated
in compact form, as we will see below.

\subsection{Onsager-Machlup path integral{*}\label{sec:Onsager-Machlup-path-integral}}

Using \eqref{p_of_x} and the substitution $\delta(y)\,dy=\delta(\phi(x_{i+1}))\phi^{\prime}dx_{i+1}$
with $y=\phi(x_{i})=\xi_{i}(x_{i})$ obtained by solving \eqref{discrete_sde}
for $W_{i}$

\begin{eqnarray}
W_{i}(x_{i}) & = & x_{i}-x_{i-1}-f(\mbox{\ensuremath{\alpha}}x_{i}+(1-\alpha)x_{i-1})\,h-a\delta_{i-1,0}\nonumber \\
\frac{\partial W_{i}}{\partial x_{i}} & =\phi^{\prime}= & 1-\alpha f^{\prime}h\label{eq:normalization_strato}
\end{eqnarray}
we obtain

\begin{eqnarray}
p(x_{1},\ldots,x_{M}|a) & = & \int\Pi_{i=1}^{M}d\xi_{i}\,\rho(\xi_{i})\,\times\label{eq:density}\\
 & \times & \delta(\xi_{i}-\underbrace{x_{i}-x_{i-1}-f(\mbox{\ensuremath{\alpha}}x_{i}+(1-\alpha)x_{i-1})\,h-a\delta_{i-1,0}}_{\equiv\xi_{i}(x_{i})})\,(1-\alpha f^{\prime}h).\nonumber \\
 & = & \Pi_{i=1}^{M}\rho(x_{i}-x_{i-1}-f(\mbox{\ensuremath{\alpha}}x_{i}+(1-\alpha)x_{i-1})\,h-a\delta_{i-1,0})\,\left(1-\alpha h\,f^{\prime}(\mbox{\ensuremath{\alpha}}x_{i}+(1-\alpha)x_{i-1})\right).\nonumber 
\end{eqnarray}
For the case of a Gaussian noise $\rho(\xi_{i})=\N(0,\,Dh)=\frac{1}{\sqrt{2\pi Dh}}\,e^{-\frac{\xi_{i}^{2}}{2Dh}}$
the variance of the increment is 
\begin{eqnarray}
\langle\xi_{i}\xi_{j}\rangle & = & \begin{cases}
Dh & \quad i=j\\
0 & \quad i\neq j
\end{cases}\label{eq:GWN_OM}\\
 & = & \delta_{ij}\,Dh.\nonumber 
\end{eqnarray}
Using the Gaussian noise and then taking the limit $M\to\infty$ of
eq. \eqref{density} with $1-\alpha f^{\prime}h\to\exp(-\alpha f^{\prime}h)$
we obtain
\begin{eqnarray*}
p(x_{1},\ldots,x_{M}|a) & = & \Pi_{i=1}^{M}\rho(x_{i}-x_{i-1}-f(\mbox{\ensuremath{\alpha}}x_{i}+(1-\alpha)x_{i-1})\,h-a\delta_{i-1,0})\,(1-\alpha f^{\prime}h)+O(h^{2})\\
 & = & \Pi_{i=1}^{M}\frac{1}{\sqrt{2\pi Dh}}\,\exp\left[-\frac{1}{2Dh}(x_{i}-x_{i-1}-f(\mbox{\ensuremath{\alpha}}x_{i}+(1-\alpha)x_{i-1})\,h-a\delta_{i-1,0})^{2}-\alpha f^{\prime}h\right]+O(h^{2})\\
 & = & \left(\frac{1}{\sqrt{2\pi Dh}}\right)^{M}\,\exp\left[-\frac{1}{2D}\sum_{i=1}^{M}\left[(\frac{x_{i}-x_{i-1}}{h}-f(\mbox{\ensuremath{\alpha}}x_{i}+(1-\alpha)x_{i-1})-a\frac{\delta_{i-1,0}}{h})^{2}-\alpha f^{\prime})\right]h\right]+O(h^{2}).
\end{eqnarray*}
We will now define a symbolic notation by recognizing $\lim_{h\to0}\frac{x_{i}-x_{i-1}}{h}=\partial_{t}x(t)$
as well as $\lim_{h\to0}\frac{\delta_{i0}}{h}=\delta(t)$ and $\lim_{h\to0}\sum_{i}f(hi)\,h=\int f(t)\,dt$
\begin{align}
p[x|x(0+)=a]\,\mathcal{D}_{\sqrt{2\pi Dh}}x & =\exp\left(-\frac{1}{2D}\int_{0}^{T}(\partial_{t}x-f(x)-a\delta(t))^{2}-\alpha f^{\prime}\,dt\right)\D_{\sqrt{2\pi Dh}}x\label{eq:OM_pathint}\\
 & :=\lim_{M\to\infty}p(x_{1},\ldots,x_{M}|a)\frac{dx_{1}}{\sqrt{2\pi Dh}}\ldots\frac{dx_{M}}{\sqrt{2\pi Dh}},\nonumber 
\end{align}
where we defined the integral measure $\mathcal{D}_{\sqrt{2\pi Dh}}x:=\Pi_{i=1}^{M}\frac{dx_{i}}{\sqrt{2\pi Dh}}$
to obtain a normalized density $1=\int\mathcal{D}_{\sqrt{2\pi Dh}}x\,p[x|x(0+)=a]$.

\subsection{Martin-Siggia-Rose-De Dominicis-Janssen (MSRDJ) path integral\label{subsec:Martin-Siggia-Rose-De-Dominicis}}

The square in the action \eqref{OM_pathint} sometimes has disadvantages
for analytical reasons, for example if quenched averages are to be
calculated, as we will do in \prettyref{sec:Sompolinsky-Crisanti-Sommers-theory}.
To avoid the square we will here introduce an auxiliary field, the
\textbf{response field} $\tx$ (the name will become clear in \prettyref{sec:Response-function-in}).
This field enters the probability functional \eqref{p_of_x} by representing
the $\delta$-distribution by its Fourier integral
\begin{eqnarray}
\delta(x) & = & \frac{1}{2\pi i}\int_{-i\infty}^{i\infty}\,d\tilde{x}\,e^{\tilde{x}x}.\label{eq:Fourier_delta}
\end{eqnarray}
Replacing the $\delta$-distribution at each time slice by an integral
over $\tx_{i}$ at the corresponding slice, \eqref{p_of_x} takes
the form 
\begin{align}
p(x_{1},\ldots,x_{M}|a) & =\prod_{i=1}^{M}\left\{ \int d\xi_{i}\rho(\xi_{i})\,\int_{-i\infty}^{i\infty}\frac{d\tilde{x}_{i}}{2\pi i}\,\exp\left(\tilde{x}_{i}(x_{i}-x_{i-1}-f(\alpha x_{i}+(1-\alpha)x_{i-1})h-\xi_{i}-a\delta_{i-1,0})-\alpha f^{\prime}h\right)\right\} \nonumber \\
 & =\prod_{i=1}^{M}\left\{ \int_{-i\infty}^{i\infty}\frac{d\tilde{x}_{i}}{2\pi i}\,\exp\left(\tilde{x}_{i}(x_{i}-x_{i-1}-f(\alpha x_{i}+(1-\alpha)x_{i-1})h-a\delta_{i-1,0})-\alpha f^{\prime}h+W_{\xi}(-\tilde{x}_{i})\right)\right\} \label{eq:general_path_prob}\\
\nonumber \\
W_{\xi}(-\tilde{x}) & \equiv\ln\int d\xi_{i}\rho(\xi_{i})\,e^{-\tilde{x}\xi_{i}}=\langle e^{-\tilde{x}\xi_{i}}\rangle_{\xi_{i}}.\nonumber 
\end{align}
Here $W_{\xi}(-\tilde{x})$ is the cumulant generating function of
the noise process (see \prettyref{sec:Probabilities}) evaluated at
$-\tilde{x}$. Note that the index $i$ of the field $\tilde{x}_{i}$
is the same as the index of the noise variable $\xi_{i}$, which allows
the identification of the definition of the cumulant generating function.
The distribution of the noise therefore only appears in the probability
density in the form of $W_{\xi}(-\tx)$. For Gaussian noise \prettyref{eq:GWN_OM}
the cumulant generating function is
\begin{align}
W_{\xi}(-\tilde{x}) & =\frac{Dh}{2}\tilde{x}^{2}.\label{eq:Gaussian_generator_discrete}
\end{align}

\subsection{Moment generating functional}

The probability distribution \eqref{general_path_prob} is a distribution
for the random variables $x_{1},\ldots,x_{M}$. We can alternatively
describe the probability distribution by the moment-generating functional
(see \prettyref{sec:Probabilities}) by adding the terms $\sum_{l=1}^{M}j_{l}x_{l}h$
to the action and integrating over all paths
\begin{align}
Z(j_{1},\ldots,j_{M}) & :=\Pi_{l=1}^{M}\left\{ \int_{-\infty}^{\infty}dx_{l}\,\exp\left(j_{l}x_{l}h\right)\right\} \,p(x_{1},\ldots,x_{M}|a).\label{eq:generating_functional_def}
\end{align}
Moments of the path can be obtained by taking derivatives (writing
$\bl=(j_{1},\ldots,j_{M})$)

\begin{align}
\left.\frac{\partial}{\partial(h\,j_{k})}Z(\bl)\right|_{\bl=0} & =\Pi_{l=1}^{M}\left\{ \int_{-\infty}^{\infty}dx_{l}\right\} \,p(x_{1},\ldots,x_{M}|a)\,x_{k}\nonumber \\
 & \equiv\langle x_{k}\rangle.\label{eq:moments_func_deriv}
\end{align}
The generating functional takes the explicit form
\begin{eqnarray}
Z(\bl) & = & \Pi_{l=1}^{M}\left\{ \int_{-\infty}^{\infty}dx_{l}\exp\left(j_{l}x_{l}h\right)\,\int_{-i\infty}^{i\infty}\frac{d\tilde{x}_{l}}{2\pi i}\right\} \times\label{eq:generating_functional_discrete}\\
 &  & \times\exp\left(\sum_{l=1}^{M}\tilde{x}_{l}(x_{l}-x_{l-1}-f(\alpha x_{l}+(1-\alpha)x_{l-1})h-a\delta_{l-1,0})-\alpha f^{\prime}h+W_{\xi}(-\tilde{x}_{l})\right),\nonumber 
\end{eqnarray}
where we used $\prod_{l=1}^{M}\exp(W_{\xi}(-\tx_{l}))=\exp(\sum_{l=1}^{M}W_{\xi}(-\tx_{l}))$.

Letting $h\to0$ we now define the path integral as the generating
functional \eqref{generating_functional_discrete} and introduce the
notations $\Pi_{l=1}^{M}\int_{-\infty}^{\infty}dx_{l}\stackrel{h\to0}{\to}\int\mathcal{D}x$
as well as $\Pi_{l=1}^{M}\int_{-i\infty}^{i\infty}\frac{d\tilde{x}_{l}}{2\pi i}\stackrel{h\to0}{\to}\int\mathcal{D}_{2\pi i}\tilde{x}$.
Note that the different integral boundaries are implicit in this notation,
depending on whether we integrate over $x(t)$ or $\tilde{x}(t)$. 

Introducing in addition the cumulant generating functional of the
noise process as
\begin{align*}
W_{\xi}[-\tilde{x}]=\ln\,Z_{\xi}[-\tx] & =\ln\left\langle \exp\left(-\int_{-\infty}^{\infty}\tilde{x}(t)\,dW(t)\right)\right\rangle {}_{dW}\\
 & :=\lim_{h\to0}\ln\langle\exp(\sum_{l=1}^{M}-\tx_{l}\xi_{l})\rangle_{\xi}\\
 & =\lim_{h\to0}\sum_{l=1}^{M}\ln\langle\exp(-\tx_{l}\xi_{l})\rangle_{\xi_{l}}
\end{align*}
we may write symbolically for the probability distribution \eqref{general_path_prob}
\begin{eqnarray}
p[x|x(0+)=a] & = & \int\mathcal{D}_{2\pi i}\tilde{x}\,\exp\left(\int_{-\infty}^{\infty}\tilde{x}(t)(\partial_{t}x-f(x)-a\delta(t))-\alpha f^{\prime}\,dt+W_{\xi}[-\tx]\right)\label{eq:martin_siggia_rose_general}\\
 & = & \int\mathcal{D}_{2\pi i}\tilde{x}\,\exp\left(\tilde{x}^{\T}(\partial_{t}x-f(x)-a\delta(t))-\int_{-\infty}^{\infty}\alpha f^{\prime}\,dt+W_{\xi}[-\tx]\right).\nonumber 
\end{eqnarray}
In the the second line we use the definition of the inner product
on the space of functions 
\begin{align}
x^{\T}y & :=\int_{-\infty}^{\infty}x(t)y(t)\,dt.\label{eq:inner_product}
\end{align}
This vectorial notation also reminds us of the discrete origin of
the path integral. Note that the lattice derivative appearing in \eqref{martin_siggia_rose_general}
follows the definition $\partial_{t}x=\lim_{h\to0}\frac{1}{h}\left(x_{t/h}-x_{t/h-1}\right)$.
The convention is crucial for the moment-generating function to be
properly normalized, as shown in \eqref{normalization_p_path}: Only
the appearance of $x_{t/h}$ alone within the Dirac $\delta$ allows
the path integral $\int\D x$ to be performed to yield unity.

We compactly denote the generating functional \eqref{generating_functional_discrete}
as

\begin{align}
Z[j] & =\int\D x\,\int\D_{2\pi i}\tilde{x}\,\exp\left(\tilde{x}^{\T}(\partial_{t}x-f(x)-a\delta(t))+j^{\T}x-\alpha1^{\T}f^{\prime}(x)+W_{\xi}[-\tx]\right).\label{eq:MSR_Z_tilde_J}
\end{align}
For Gaussian white noise we have with \eqref{Gaussian_generator_discrete}
the moment generating functional $W_{\xi}[-\tilde{x}]=\frac{D}{2}\,\tilde{x}^{\T}\tx$.
If in addition, we adopt the Ito convention, i.e. setting $\alpha=0$,
we get
\begin{align}
Z[j] & =\int\D x\,\int\D_{2\pi i}\tilde{x}\,\exp\left(\tilde{x}^{\T}(\partial_{t}x-f(x)-a\delta(t))+\frac{D}{2}\tx^{\T}\tx+j^{\T}x\right).\label{eq:msr_gwn}
\end{align}
For $M\to\infty$ and $h\to0$ the source term is $\exp\left(\sum_{l=1}^{M}j_{l}\,x_{l}h\right)\stackrel{h\to0}{\rightarrow}\exp\left(\int j(t)x(t)\,dt\right)\equiv\exp(j^{\T}x)$.
So the derivative on the left hand side of \eqref{moments_func_deriv}
turns into the functional derivative 
\begin{align*}
\frac{\partial}{\partial(hj_{k})}Z(\bl) & \equiv\lim_{\epsilon\to0}\frac{1}{\epsilon}\left(Z(j_{1},\ldots,j_{k}+\frac{\epsilon}{h},\,j_{k+1},\ldots,j_{M}]-Z(j_{1},\ldots,j_{k},\ldots,j_{M})\right)\stackrel{h\to0}{\rightarrow}\frac{\delta}{\delta j(t)}Z[j],
\end{align*}
and the moment becomes $\langle x(t)\rangle$ at time point $t=hk$.

We can therefore express the $n$-th moment of the process by formally
performing an $n$-fold functional derivative
\begin{align*}
\langle\underbrace{x(t)\cdots x(s)}_{n}\rangle & =\frac{\delta^{n}}{\delta j(t)\cdots\delta j(s)}\,Z[j]\Big|_{j=0}.
\end{align*}

\subsection{Response function in the MSRDJ formalism\label{sec:Response-function-in}}

The path integral \eqref{general_path_prob} can be used to determine
the response of the system to an external perturbation. To this end
we consider the stochastic differential equation \eqref{SDE} that
is perturbed by a time-dependent drive $-\tilde{j}(t)$

\begin{eqnarray*}
dx(t) & = & (f(x(t))-\tilde{j}(t))\,dt+dW(t)\\
x(0+) & = & a.
\end{eqnarray*}

In the following we will only consider the Ito convention and set
$\alpha=0$. We perform the analogous calculation that leads from
\eqref{SDE} to \eqref{generating_functional_discrete} with the additional
term $-\tilde{j}(t)$ due to the perturbation. In the sequel we will
see that, instead of treating the perturbation explicitly, it can
be expressed with the help of a second source term. The generating
functional including the perturbation is

\begin{align}
Z(\bl,\tbj) & =\Pi_{l=1}^{M}\left\{ \int_{-\infty}^{\infty}dx_{l}\int_{-i\infty}^{i\infty}\frac{d\tilde{x}_{l}}{2\pi i}\right\} \times\nonumber \\
 & \times\exp\left(\sum_{l=1}^{M}\tilde{x}_{l}(x_{l}-x_{l-1}-f(x_{l-1})h-a\delta_{l-1,0})+j_{l}x_{l}h+\tilde{x}_{l}\tilde{j}_{l-1}h+W_{\xi}(-\tx_{l})\right)\label{eq:msr_z}\\
 & =\int\D x\,\int\D_{2\pi i}\tilde{x}\,\exp\left(\int_{-\infty}^{\infty}\tilde{x}(t)(\partial_{t}x-f(x)-a\delta(t))+j(t)x(t)+\tilde{j}(t-)\tilde{x}(t)\,dt+W_{\xi}[-\tx]\right),\nonumber 
\end{align}
where we moved the $\tilde{j}-$dependent term out of the parenthesis.

Note that the external field $\tj_{l-1}$ couples to the field $\tx_{l}$,
because $\tj(t)$ must be treated along the same lines as $f(x(t))$;
in particular both terms' time argument must be delayed by a single
time slice. As before, the moments of the process follow as functional
derivatives \eqref{moments_func_deriv} $\left.\frac{\delta}{\delta j(t)}Z[j,\tilde{j}]\right|_{j=\tilde{j}=0}=\langle x(t)\rangle$.
Higher order moments follow as higher derivatives, in complete analogy
to \eqref{moments_generation}.

The additional dependence on $\tilde{j}$ allows us to investigate
the response of arbitrary moments to a small perturbation localized
in time, i.e. $\tilde{j}(t)=-\epsilon\delta(t-s)$. In particular,
we characterize the average response of the first moment with respect
to the unperturbed system by the \textbf{response function} $\chi(t,s)$

\begin{align}
\chi(t,s) & :=\lim_{\epsilon\to0}\frac{1}{\epsilon}\left(\langle x(t)\rangle_{\tilde{j}=-\epsilon\delta(\cdot-s)}-\langle x(t)\rangle_{\tilde{j}=0}\right)\label{eq:MSR_response}\\
 & =\lim_{\epsilon\to0}\frac{1}{\epsilon}\left.\frac{\delta}{\delta j(t)}\left(Z[j,\tilde{j}-\epsilon\delta(t-s)]-Z[j,\tilde{j}]\right)\right|_{j=\tilde{j}=0}\nonumber \\
 & =\left.-\frac{\delta}{\delta j(t)}\frac{\delta}{\delta\tilde{j}(s)}Z[j,\tilde{j}]\right|_{j=\tilde{j}=0}\nonumber \\
 & =-\langle x(t)\,\tilde{x}(s)\rangle,\nonumber 
\end{align}
where we used the definition of the functional derivative from the
third to the fourth line.

So instead of treating a small perturbation explicitly, the response
of the system to a perturbation can be obtained by a functional derivative
with respect to $\tilde{j}$: $\tilde{j}$ couples to $\tilde{x}$,
$\tilde{j}$ contains perturbations, therefore $\tilde{x}$ measures
the response and is the so called response field. The response function
$\chi(t,s)$ can then be used as a kernel to obtain the mean response
of the system to a small external perturbation of arbitrary temporal
shape.

There is an important difference for the response function between
the Ito and Stratonovich formulation, that is exposed in the time-discrete
formulation. For the perturbation $\tilde{j}(t)=-\epsilon\delta(t-s)$,
we obtain the perturbed equation, where $\frac{s}{h}$ denotes the
discretized time point at which the perturbation is applied. The perturbing
term must be treated analogously to $f$, so 
\begin{eqnarray*}
x_{i}-x_{i-1} & = & f(\mbox{\ensuremath{\alpha}}x_{i}+(1-\alpha)x_{i-1})\,h+\epsilon\left(\alpha\delta_{i,\frac{s}{h}}+(1-\alpha)\delta_{i-1,\frac{s}{h}}\right)+\xi_{i}\\
\alpha & \in & [0,1].
\end{eqnarray*}
Consequently, the value of the response function $\chi(s,s)$ at the
time of the perturbation depends on the choice of $\alpha$. We denote
as $x_{j}^{\epsilon}$ the solution after application of the perturbation,
as $x_{j}^{0}$ the solution without; for $i<j$ the two are identical
and the equal-time response is
\begin{align}
\chi(s,s) & =\lim_{\epsilon\to0}\frac{1}{\epsilon}\left(x_{\frac{s}{h}}^{\epsilon}-x_{\frac{s}{h}}^{0}\right)\label{eq:equal_time_response}\\
 & =\lim_{\epsilon\to0}\frac{1}{\epsilon}\left(f(\mbox{\ensuremath{\alpha}}x_{\frac{s}{h}}^{\epsilon}+(1-\alpha)x_{\frac{s}{h}-1})-f(\mbox{\ensuremath{\alpha}}x_{\frac{s}{h}}^{0}+(1-\alpha)x_{\frac{s}{h}-1})\right)\,h+\alpha\delta_{\frac{s}{h},\frac{s}{h}}+(1-\alpha)\delta_{\frac{s}{h}-1,\frac{s}{h}}\nonumber \\
 & \stackrel{h\to0}{=}\alpha,\nonumber 
\end{align}
because the contribution of the deterministic evolution vanishes due
to the factor $h$. So for $\alpha=0$ (Ito convention) we have $\chi(s,s)=0$,
for $\alpha=\frac{1}{2}$ (Stratonovich) we have $\chi(s,s)=\frac{1}{2}$.
The Ito-convention is advantageous in this respect, because it leads
to vanishing contributions in Feynman diagrams (see \prettyref{sec:Perturbation-theory-for-MSR})
with response functions at equal time points \citep{Chow15}. In \eqref{msr_z}
this property is reflected by the displacement of the indices in the
term $\tilde{x}_{l}\tilde{j}_{l-1}h$.

By the same argument follows that 
\begin{align}
\langle x(t)\tx(s)\rangle & \equiv\begin{cases}
0\quad\forall t\le s & \text{Ito}\\
0\quad\forall t<s & \text{Stratonovich}\\
\frac{1}{2}\quad t=s & \text{Stratonovich}
\end{cases}.\label{eq:Ito_Strato}
\end{align}

We also observe that the initial condition contributes a term $-a\delta_{l,0}$.
Consequently, the initial condition can alternatively be included
by setting $a=0$ and instead calculate all moments from the generating
functional $Z[j,\tilde{j}-a\delta]$ instead of $Z[j,\tilde{j}]$.
In the following we will therefore skip the explicit term ensuring
the proper initial condition as it can be inserted by choosing the
proper value for the source $\tilde{j}$. See also \citep[Sec. 5.5]{hertz2016_arxiv}.

For the important special case of Gaussian white noise \eqref{GWN_OM},
the generating functional, including the source field $\tilde{j}$
coupling to the response field, takes the form

\begin{align}
Z[j,\tilde{j}] & =\int\D x\,\int\D_{2\pi i}\tilde{x}\,\exp\left(\tilde{x}^{\T}(\partial_{t}x-f(x))+\frac{D}{2}\tx^{\T}\tx+j^{\T}x+\tilde{j}^{\T}\tilde{x}\right),\label{eq:MSR_GWN}
\end{align}
where we again used the definition of the inner product \eqref{inner_product}.

\section{Ornstein-Uhlenbeck process: The free Gaussian theory\label{sec:Ornstein-Uhlenbeck}}

\subsection{Definition}

We will here study a first example of application of the MSRDJ formalism
to a linear stochastic differential equation, the Ornstein-Uhlenbeck
process \citep{Risken96}. This example is fundamental to all further
development, as it is the free Gaussian part of the theory, the dynamic
counterpart of the Gaussian studied in \secref{Gaussian-distribution}.
The stochastic differential equation (\ref{eq:SDE}) in this case
is

\begin{align}
dx & =m\,x\,dt+dW,\label{eq:SDE_n_dim}\\
x & \in\mathbb{R}^{N},\nonumber \\
m & \in\mathbb{R}^{N\times N},\nonumber \\
\langle dW{}_{i}(t)dW_{j}(s)\rangle & =D_{ij}\,\delta_{t,s}\,dt,\nonumber 
\end{align}
where $dW_{i}$ are Wiener increments that may be correlated with
covariance matrix $D_{ij}$. The generalization of the \secref{Martin-Siggia-Rose-De-Dominicis}
to this set of $N$ coupled stochastic differential equations is straight
forward and left as an exercise. The result is the action 
\begin{align}
S[x,\tilde{x}] & =\int\,\tx^{\T}(t)\,\left(\partial_{t}-m\right)\,x(t)+\tx(t)^{\T}\frac{D}{2}\tx(t)\,dt\label{eq:action_OUP}\\
 & =\tx^{\T}\,\left(\partial_{t}-m\right)\,x+\tx^{\T}\frac{D}{2}\tx,\nonumber 
\end{align}
where the transposed $^{\T}$ in the first line is mean with respect
to the $N$ different components and in the second line in addition
for the time argument; as a consequence, we need to think about the
matrix $D$ in the second line as containing an additional $\delta(t-s)$.
We see that this notation considers different time points on the same
footing as different components of $x$.

We may write the action in a more symmetric form by introducing the
compound field $y(t)=\left(\begin{array}{c}
x(t)\\
\tx(t)
\end{array}\right)$ as

\begin{align}
S[y]=S[x,\tx] & =-\frac{1}{2}\,y^{\T}\,A\,y\nonumber \\
 & =-\frac{1}{2}\iint\,y^{\T}(t)\,A(t,s)\,y(s)\,dt\,ds,\nonumber \\
A(t,s) & =\left(\begin{array}{cc}
0 & \partial_{t}+m^{\T}\\
-\partial_{t}+m & -D
\end{array}\right)\,\delta(t-s),\label{eq:def_A_OUP}
\end{align}
where the transposed in the first line is meant as referring to the
field index (i.e. distinguishing between $x$ and $\tx$) as well
as to the time argument. The minus sign in the upper right entry follows
from integration by parts as $\int\tx(t)\,\left(\partial_{t}-m\right)x(t)\,dt=\int x(t)\,\left(-\partial_{t}-m\right)\tx(t)\,dt$,
assuming that the boundary terms vanish.

\subsection{Propagators in time domain\label{sub:Propagators}}

The moment generating functional $Z[j,\tj]$, corresponding to \eqref{def_A_OUP}
is
\begin{align}
Z[j,\tj] & =\int\D x\int\D\tx\,\exp\left(S[x,\tx]+j^{\T}x+\tj^{\T}\tx\right)\nonumber \\
Z[\bar{j}] & =\int\D y\,\exp\Big(-\frac{1}{2}\,y^{\T}A\,y+\bar{j}^{\T}\,y\Big),\label{eq:Z_j_OUP}
\end{align}
where we introduced $\bar{j}=\left(\begin{array}{c}
j\\
\tj
\end{array}\right)$. Following the derivation in \secref{Gaussian-distribution}, we
need to determine the propagators $\Delta$ in the sense
\begin{align}
\Delta= & A^{-1}\label{eq:def_propagator}\\
\int\,A(s,t)\,\Delta(t,u)\,dt & =\diag(\delta(s-u)),\nonumber 
\end{align}
which is the time-continuous analogue of \eqref{Z_Gauss}. The diagonal
matrix of Dirac $\delta$ is the continuous version of the identity
matrix with respect to the matrix multiplication $\int\,f(t)\,g(t)\,dt$,
the inner product on our function space.

The latter form also explains the name propagator of Green's function:
By its definition \eqref{def_propagator}, $\Delta$ is the fundamental
solution of the linear differential operator $A$. This means given
we want to solve the inhomogeneous problem
\begin{align}
\int A(t,s)\,y(s)\,ds & =f(t).\label{eq:lin_diffeq}
\end{align}
We see that the application of $\int\,du\,A(t,u)\circ$ from left
on $y(u)$, defined as
\begin{align*}
y(u) & =\int\Delta(u,s)\,f(s)\,ds,
\end{align*}
reproduces with the property \eqref{def_propagator} the right hand
side $f(t)$ of \eqref{lin_diffeq}. So $\Delta$ is indeed the Green's
function or fundamental solution to $A$.

An analogous calculation as the completion of square (see exercises)
then leads to
\begin{align}
Z[\bar{j}] & =\exp\left(\frac{1}{2}\,\bar{j}^{\T}\Delta\,\bar{j}\right).\label{eq:Z_j_OUP_integrated}
\end{align}
So we need to determine the four entries of the two-by-two matrix
\begin{align*}
\Delta(t,u) & =\left(\begin{array}{cc}
\Delta_{xx}(t,u) & \Delta_{x\tx}(t,u)\\
\Delta_{\tx x}(t,u) & \Delta_{\tx\tx}(t,u)
\end{array}\right)=\left(\begin{array}{cc}
\langle x(t)x(u)\rangle & \langle x(t)\tx(u)\rangle\\
\langle\tx(t)x(u)\rangle & \langle\tx(t)\tx(u)\rangle
\end{array}\right),
\end{align*}
where the latter equality follows from comparing the second derivatives
of (\ref{eq:Z_j_OUP}) to those of (\ref{eq:Z_j_OUP_integrated}),
setting $\bar{j}=0$ in the end. The factor $\frac{1}{2}$ in (\ref{eq:Z_j_OUP_integrated})
drops out, because the first differentiation, by product rule, needs
to act on each of the two occurrences of $\bar{j}$ in (\ref{eq:Z_j_OUP_integrated})
in turn for the diagonal element, and acting on each of the off-diagonal
elements, producing two identical terms in either case. The elements
are hence the correlation and response functions of the fields $x$
and $\tx$.

\subsection{Propagators in Fourier domain\label{sub:Propagators-in-Fourier}}

The inversion of (\ref{eq:def_propagator}) can easiest be done in
frequency domain. The Fourier transforms as $y(t)=\iFtr Yt=\frac{1}{2\pi}\int\,e^{i\omega t}\,Y(\omega)\,d\omega$
is a unitary transform, hence does not affect the integration measures
and moreover transforms scalar products
\begin{align}
x^{\T}y & :=\int\,x(t)\,y(t)\,dt\label{eq:scalar_Fourier}\\
 & =\iint\frac{d\omega}{2\pi}\,\frac{d\omega^{\prime}}{2\pi}\,X(\omega)\,Y(\omega^{\prime})\,\underbrace{\int\,e^{i\left(\omega+\omega^{\prime}\right)t}dt}_{2\pi\,\delta(\omega+\omega^{\prime})}\nonumber \\
 & =\int\,\frac{d\omega}{2\pi}\,X(-\omega)Y(\omega)=:X{}^{\T}Y,\nonumber 
\end{align}
If we use the convention that every $\int_{\omega}=\int\frac{d\omega}{2\pi}$
comes with a factor $(2\pi)^{-1}$ we get for a linear differential
operator $A[\partial_{t}]$ that $y^{\T}A[\partial_{t}]y\to Y^{\T}A[i\omega]Y$.
We therefore obtain (\ref{eq:def_A_OUP}) in Fourier domain with $Y=\left(\begin{array}{c}
X\\
\tilde{X}
\end{array}\right)$ as
\begin{align}
S[X,\tilde{X}] & =-\frac{1}{2}Y^{\T}AY\nonumber \\
A(\omega^{\prime},\omega) & =2\pi\,\delta(\omega^{\prime}-\omega)\,\left(\begin{array}{cc}
0 & i\omega+m^{\T}\\
-i\omega+m & -D
\end{array}\right).\label{eq:inv_prop_OUP_Fourier}
\end{align}
We see that the form of $A$ is self-adjoint with respect to the scalar
product (\ref{eq:scalar_Fourier}), because bringing $A$ to the left
hand side, we need to transpose and transform $\omega\to-\omega$,
which leaves $A$ invariant. Hence with the Fourier transformed sources
$\bar{J}$, we have a well-defined Gaussian integral
\begin{align}
Z[\bar{J}] & =\exp\left(-\frac{1}{2}\,Y^{\T}AY+\bar{J}^{\T}Y\right).\label{eq:Z_OUP_Fourier}
\end{align}
Since (\ref{eq:inv_prop_OUP_Fourier}) is diagonal in frequency domain,
we invert the two-by-two matrix separately at each frequency. The
moment generating function in frequency domain (\ref{eq:Z_j_OUP_integrated})
therefore follows by determining the inverse of $A$ in the sense
\begin{align*}
\int\frac{d\omega^{\prime}}{2\pi}\,A(\omega,\omega^{\prime})\Delta(\omega^{\prime},\omega^{\prime\prime}) & =2\pi\,\delta(\omega-\omega^{\prime\prime}),
\end{align*}
because $2\pi\delta$ is the identity with regard to our scalar product
$\int\frac{d\omega}{2\pi}$. So we obtain
\begin{align}
Z[\bar{J}] & =\exp\left(\frac{1}{2}\iint_{\omega^{\prime}\omega}\,\bar{J}^{\T}(-\omega)\,\Delta(\omega,\omega^{\prime})\,\bar{J}(\omega^{\prime})\right)=\exp\left(\frac{1}{2}\,\bar{J}^{\T}\Delta\,\bar{J}\right),\label{eq:Z_J_OUP}\\
\Delta(\omega,\omega^{\prime}) & \stackrel{(\ref{eq:Z_J_OUP}),(\ref{eq:inv_prop_OUP_Fourier})}{=}2\pi\,\delta(\omega-\omega^{\prime})\,\left(\begin{array}{cc}
\left(-i\omega+m\right)^{-1}D\left(i\omega+m^{\T}\right)^{-1} & \left(-i\omega+m\right)^{-1}\\
\left(i\omega+m^{\T}\right)^{-1} & 0
\end{array}\right),\nonumber \\
 & =(2\pi)^{2}\,\left(\begin{array}{cc}
\frac{\delta^{2}Z}{\delta J(-\omega)\delta J(\omega^{\prime})} & \frac{\delta^{2}Z}{\delta J(-\omega)\delta\tJ(\omega^{\prime})}\\
\frac{\delta^{2}Z}{\delta\tJ(-\omega)\delta J(\omega^{\prime})} & \frac{\delta^{2}Z}{\delta\tJ(-\omega)\delta\tJ(\omega^{\prime})}
\end{array}\right)\nonumber \\
 & \stackrel{(\ref{eq:Z_OUP_Fourier})}{=}\left(\begin{array}{cc}
\langle X(\omega)X(-\omega^{\prime})\rangle & \langle X(\omega)\tilde{X}(-\omega^{\prime})\rangle\\
\langle\tilde{X}(\omega)X(-\omega^{\prime})\rangle & 0
\end{array}\right)\nonumber 
\end{align}
where the signs of the frequency arguments in the second last line
are flipped with respect to the signs of the frequencies in $J(\omega)$,
because the source term is $\bar{J}^{\T}Y$, involving the inverse
of the sign. The additional factor $(2\pi)^{-2}$ in the forth line
comes from the source terms $\bar{J}^{\T}Y=\int\frac{d\omega}{2\pi}\bar{J}^{\T}(-\omega)Y(\omega)$,
which yield a factor $(2\pi)^{-1}$ upon each differentiation. Overall,
we see that for each contraction of a pair of $X^{\alpha},X^{\beta}\in\{X,\tilde{X}\}$
we get a term
\begin{eqnarray*}
\langle X^{\alpha}(\omega^{\prime})X^{\beta}(\omega)\rangle & = & (2\pi)^{2}\,\frac{\delta^{2}Z}{\delta J^{\alpha}(-\omega^{\prime})\,\delta J^{\beta}(\omega)}\\
 & = & \Delta_{\alpha\beta}(\omega^{\prime},\omega)\\
 & \propto & 2\pi\,\delta(\omega-\omega^{\prime}).
\end{eqnarray*}
The Fourier transform $\Ftr f{\omega}$ is a linear functional of
a function $f$, so that the functional derivative follows as
\begin{eqnarray*}
\frac{\delta}{\delta f(s)}\Ftr f{\omega} & = & \frac{\delta}{\delta f(s)}\int\,e^{-i\omega t}\,f(t)\,dt=e^{-i\omega s}.
\end{eqnarray*}
Assuming a one-dimensional process in the following, $m<0\in\mathbb{R}$,
we can apply the chain rule (\ref{eq:func_chain}) to calculate the
covariance function in time domain as

\begin{align}
\Delta_{xx}(t,s) & \equiv\langle x(t)x(s)\rangle\label{eq:free_covariance-1-1}\\
 & =\left.\frac{\delta^{2}}{\delta j(t)\delta j(s)}\,Z[j,\tj]\right|_{j=\tj=0}\nonumber \\
 & =\int d\omega^{\prime}\,d\omega\,\underbrace{e^{-i\omega^{\prime}t}\,e^{-i\omega s}}_{=\frac{\delta J(\omega^{\prime})}{\delta j(t)}\,\frac{\delta J(\omega)}{\delta j(s)}}\,\underbrace{\left.\frac{\delta^{2}}{\delta J(\omega^{\prime})\delta J(\omega)}\,Z[J,\tJ]\right|_{J=\tJ=0}}_{(2\pi)^{-2}\Delta_{xx}(-\omega^{\prime},\omega)\propto(2\pi)^{-1}\delta(\omega+\omega^{\prime})}\nonumber \\
 & \stackrel{\omega^{\prime}=-\omega}{=}\int\frac{d\omega}{2\pi}\,e^{i\omega(t-s)}\,\left(-i\omega+m\right)^{-1}D\left(i\omega+m\right)^{-1}\nonumber \\
 & =\frac{1}{2\pi i}\,\int_{-i\infty}^{i\infty}dz\,e^{z(t-s)}\,\left(-z+m\right)^{-1}D\left(z+m\right)^{-1}\nonumber \\
 & \stackrel{t>s}{=}\frac{-D}{2m}\,e^{m(t-s)},\nonumber 
\end{align}
where we used the functional chain rule (\ref{eq:func_chain}) in
the third step, got a factor $2$ two derivatives acting in the two
possible orders of the $J$ (canceled by $\frac{1}{2}$ from the \eqref{Z_J_OUP}),
and used the residue theorem in the last, closing the contour in the
half plane with $\Re(z)<0$ to ensure convergence. Note that $m<0$
to ensure stability of (\ref{eq:SDE_n_dim}), so that the covariance
is positive as it should be. The minus sign arises from the winding
number due to the form $(-z+m)^{-1}=-(z-m)^{-1}$ of the pole. For
$t<s$ it follows by symmetry that $\Delta_{xx}(t,s)=\frac{-D}{2m}\,e^{m|t-s|}.$
In the last step we assumed a one-dimensional dynamics, the penultimate
line also holds for $N$ dimensions. For $N$ dimensions, we would
need to transform into the space of eigenvectors of the matrix and
apply the residue theorem for each of these directions separately.

The response functions are
\begin{align}
\Delta_{x\tx}(t,s)= & \langle x(t)\tx(s)\rangle=\Delta_{\tx x}(s-t)\label{eq:free_response}\\
= & \int d\omega^{\prime}\,d\omega\,\underbrace{e^{-i\omega^{\prime}t}\,e^{-i\omega s}}_{=\frac{\delta J(\omega^{\prime})}{\delta j(t)}\,\frac{\delta\tJ(\omega)}{\delta j(s)}}\,\underbrace{\left.\frac{\delta^{2}}{\delta J(\omega^{\prime})\delta\tJ(\omega)}\,Z[J,\tJ]\right|_{J=\tJ=0}}_{(2\pi)^{-2}\Delta_{x\tx}(-\omega^{\prime},\omega)\propto(2\pi)^{-1}\,\delta(\omega^{\prime}+\omega)}\nonumber \\
= & \int\,\frac{d\omega}{2\pi}e^{i\omega(t-s)}\,(-i\omega+m)^{-1}\nonumber \\
= & -\frac{1}{2\pi i}\int_{-i\infty}^{i\infty}\,e^{z(t-s)}\,(z-m)^{-1}\,dz\nonumber \\
= & -H(t-s)\,e^{m\,(t-s)},\nonumber 
\end{align}
which is consistent with the interpretation of the response to a Dirac-$\delta$
perturbation considered in \secref{Response-function-in}. We assumed
a one-dimensional dynamics in the last step. The Heaviside function
arises if $t<s$: One needs to close the integration contour in the
right half plane to get a vanishing contribution along the arc, but
no pole is encircled, because $m<0$ for stability.

For the diagrammatic formulation, we follow the convention proposed
in \citep[p.136ff, Fig. 4.2]{Fischer91}: We represent the response
function function by a straight line with an arrow pointing in the
direction of time propagation, a correlation function as a line with
two incoming arrows
\begin{align}
\Delta(t,s) & =\left(\begin{array}{cc}
\langle x(t)x(s)\rangle & \langle x(t)\tx(s)\rangle\\
\langle\tx(t)x(s)\rangle & \langle\tx(t)\tx(s)\rangle
\end{array}\right)=\left(\begin{array}{cc}
\quad\vertexlabel^{x(t)}\Diagram{fVfA}
\vertexlabel^{x(s)}\quad & \quad\vertexlabel^{x(t)}\Diagram{fV}
\vertexlabel^{\tilde{x}(s)}\quad\\
\quad\vertexlabel^{\tilde{x}(t)}\Diagram{fA}
\vertexlabel^{x(s)}\quad & 0
\end{array}\right).\label{eq:propagator_matrix_OUP}
\end{align}
The propagators of the linear and hence Gaussian theory are also often
called \textbf{bare propagators}. In contrast, propagators including
perturbative corrections are called \textbf{full propagators}. The
arrows are chosen such that they are consistent with the flow of time,
reflected by the properties:
\begin{itemize}
\item Response functions are causal, i.e. $\langle x(t)\tilde{x}(s)\rangle=0$
is $t\le s$. For $t=s$ the vanishing response relies on the Ito-convention
(see \secref{Response-function-in}).
\item As a consequence, all loops formed by propagators $\Feyn{fA}$ connecting
to a vertex at which $x(t)$ and $\tilde{x}(s)$ interact at identical
time points (see also coming section) or in a causal fashion, i.e.
$s\ge t$, vanish.
\item Correlations between pairs of response fields vanish $\langle\tilde{x}(t)\tilde{x}(s)\rangle$.
\item For zero external sources $j=\tj=0$, the expectation values of the
fields vanish $\langle x(t)\rangle=0$, as well as for the response
field $\langle\tilde{x}(t)\rangle=0$, because the action \eqref{Z_j_OUP}
is a centered Gaussian.
\end{itemize}

\section{Perturbation theory for stochastic differential equations\label{sec:Perturbation-theory-for-MSR}}

We now want to combine the perturbative method developed in \secref{Perturbation-expansion}
with the functional representation of stochastic differential equations
introduced in \secref{Martin-Siggia-Rose-De-Dominicis}. The Orsntein-Uhlenbeck
process studied as a special case in \secref{Ornstein-Uhlenbeck}
in this context plays the role of the solvable, Gaussian part of the
theory. We here want to show how to calculate perturbative corrections
that arise from non-linearites in the stochastic differential equation,
corresponding to the non-Gaussian part of the action.

\subsection{Vanishing moments of response fields\label{sec:Vanishing-response-field}}

We now would like to extend the system from the previous section to
the existence of a non-linearity in the stochastic differential equation
(\ref{eq:SDE_n_dim}) of the form
\begin{align}
dx & =f(x)\,dt-\tj\,dt+d\xi,\label{eq:nonlin_diffeq}
\end{align}
where $f(x)$ is some non-linear function of $x$. We first want to
show that, given the value of the source $j=0$, all moments of the
response field vanish. In the derivation of the path-integral representation
of $Z$ in \secref{Martin-Siggia-Rose-De-Dominicis}, we saw that
$Z$ belongs to a properly normalized density, as demonstrated by
(\ref{eq:normalization_p_path}), so $Z[j=0]=1$. The same normalization
of course holds in the presence of an arbitrary value of $\tj$ in
\eqref{nonlin_diffeq}, because $\tj$ corresponds to an additional
term on the right hand side of the stochastic differential equation
and our derivation of $Z$ holds for any right hand side. As a consequence
we must have
\begin{align}
Z[0,\tj] & \equiv1\qquad\forall\,\tj.\label{eq:Z_const_j_tilde}
\end{align}
We hence conclude that any derivative by $\tj$ of (\ref{eq:Z_const_j_tilde})
must vanish, so that all moments of $\tx$ vanish 
\begin{align*}
\frac{\delta^{n}}{\delta\tj(t_{1})\cdots\delta\tj(t_{n})}Z[0,\tj]=\langle\tx(t_{1})\cdots\tx_{n}(t_{n})\rangle & \equiv0\qquad\forall\,n>0.
\end{align*}
We note that the latter condition holds irrespective of the value
of $\tj$; we may also evaluate the moments of $\tx$ at some non-zero
$\tj$, corresponding to a particular value of the inhomogeneity on
the right hand side of (\ref{eq:nonlin_diffeq}).

\subsection{Vanishing response loops\label{sec:Vanishing-response-loops}}

We would like to treat the non-linear function $f(x)$ in (\ref{eq:nonlin_diffeq})
perturbatively, so we consider its Taylor expansion $f(x(t))=f^{(1)}(0)\,x(t)+\sum_{n=2}^{\infty}\frac{f^{(n)}(0)}{n!}\,x(t)^{n}$.
We here restrict the choice of $f$ to functions with $f(0)=0$, because
an offset can be absorbed into a non-vanishing external source field
$\tilde{j}\neq0$. For clarity of notation we here treat the one-dimensional
case, but the extension to $N$ dimensions is straight forward. We
may absorb the linear term in the propagator, setting $m:=f^{(1)}(0)$
as the linear case (\ref{eq:action_OUP}). The remaining terms yield
interaction vertices in the action
\begin{align*}
S[x,\tx] & =\underbrace{\tx^{\T}(\partial_{t}-f^{(1)}(0))\,x+\tx^{\T}\frac{D}{2}\tx}_{S_{0}[x,\tx]}\quad\underbrace{-\sum_{n=2}^{\infty}\frac{f^{(n)}(0)}{n!}\,\tx^{\T}x^{n}}_{V[x,\tx]},
\end{align*}
which are of the form
\begin{align}
V[x,\tx]= & \sum_{n=2}^{\infty}\frac{f^{(n)}(0)}{n!}\underbrace{\tilde{x}^{\T}\,x^{n}}_{\int\tilde{x}(t)\,x^{n}(t)\,dt}\,=\Diagram{ & fuV\\
f0\vertexlabel^{\tilde{x}(t)}fV & fV\vertexlabel^{x(t)}\quad,\\
 & fdV\\
 & \ldots
}
\label{eq:interaction_vertex_Langevin}
\end{align}
where the ellipses indicates the remaining legs attached to an $x$,
one leg for each power in $x$. 

We saw in the previous section that the response functions in the
Gaussian case are causal, i.e. $\langle x(t)\tilde{x}(s)\rangle=0$
for $t\le s$ and also that $\langle\tilde{x}(t)\tilde{x}(s)\rangle=0\quad\forall t,s$.
We will now show that this property is conserved in presence of an
arbitrary non-linearity that mediates a causal coupling. To this end
consider a perturbative correction to the response function with a
single interaction vertex. Since the interaction vertices are of the
form (\ref{eq:interaction_vertex_Langevin}), they couple only equal
time arguments (see underbrace in \eqref{interaction_vertex_Langevin}).
A contribution to a response function $\langle x(t)\tilde{x}(s)\rangle$
requires a bare propagator $\Feyn{fA}$ from $\tilde{x}(s)$ to one
of the three right legs of the vertex (\ref{eq:interaction_vertex_Langevin})
and one additional propagator $\Feyn{fA}$ from the left leg of the
vertex to one of the external $x(t)$. The remaining $x$-legs of
the vertex need to be contracted by the propagator $\Feyn{fVfA}$.
Since both propagators to the external legs mediate a causal interaction
and the vertex forces the intermediate time points of both propagators
to be identical, it implies that the correction is unequal zero only
for $t_{i}>s\quad\forall i$. We also see from this argument, that
a generalization of this argument to causal interactions is straight
forward.

By the inductive nature of the proof of connectedness in \secref{General-proof-linked-cluster},
this argument holds for arbitrary orders in perturbation theory, since
the connected diagrams with $i+1$ vertices are formed from those
with $i$: If causality holds for response functions with $i$ vertices,
this property obviously transcends to order $i+1$ by the above argument,
hence it holds at arbitrary order.

The same line of arguments shows that all correlators of the form
$\langle\tilde{x}(t)\cdots\tilde{x}(s)\rangle=0$ vanish. We know
this property already from the general derivation in \secref{Vanishing-response-field},
which only required the normalization condition and of course holds
for arbitrary non-linearities $f$. Often one finds in the literature
diagrammatic arguments for the vanishing moments, which we will show
here for completeness.

Indeed, at lowest order, the form of (\ref{eq:propagator_matrix_OUP})
shows that second moments of $\tilde{x}$ vanish. The first moment
of $\tilde{x}$, by differentiating (\ref{eq:Z_j_OUP_integrated})
by $\delta Z/\delta\tilde{j}(t)\big|_{\tilde{j}=0}=\langle\tilde{x}(j)\rangle=\int\Delta_{\tilde{x}(t)x(s)}j(s)\,ds$
as well vanishes for $j=0$, which even holds for $\tilde{j}\neq0$
due to the absence of $\Delta_{\tilde{x}\tilde{x}}=0$. The independence
of $\tilde{j}$ it is consistent with the possibility to absorb the
source term $\tilde{j}$ in the inhomogeneity of the differential
equation.

In the non-linear case, corrections to the mean value would come from
graphs with one external $\tilde{j}$ leg. Such a leg must be connected
by the response function $\Feyn{fA}$ to one of the $x$-legs of the
vertex, so that again a free $\tilde{x}(t)$ leg of the vertex remains.
Due to the vanishing mean $\tilde{x}$, we only have the option to
connect this free leg to one of the $x(t)$-legs of the vertex by
another response function. We still get a vanishing contribution,
because response functions (in the here considered Ito-convention)
vanish at equal time points, $\langle x(t)\tilde{x}(t)\rangle=0$
(see \secref{Response-function-in}), and all time points of fields
on the interaction vertex are identical. The generalization to general
causal relationships, i.e. $\tilde{x}(t)$, $x(s)$ with $t\ge s$
on the vertex, holds analogously. The same property holds in the Stratonovich
convention, as outlined below.

The same argument holds for all higher moments of $\tilde{x}$, where
for each external line $\tilde{j}$ one propagator $\Feyn{fA}$ attaches
to the corresponding $x$-legs of the vertex. The remaining single
$\tilde{x}$-leg of the vertex again cannot be connected in a way
that would lead to a non-vanishing contribution. The argument generalizes
to higher order corrections, by replacing the bare propagators by
the full propagators, which, by the argument given above, have the
same causality properties.

Comparing this result to the literature \citep[see p. 4914 after eq. (9)]{DeDomincis78_353}
and \citep[see eq. (7)]{janssen1976_377}, a difference is that these
works considered the Stratonovich convention. An additional term $-\frac{1}{2}f^{\prime}(x)$
is present in the action (\ref{eq:MSR_Z_tilde_J}), because the Stratonovich
convention amounts to $\alpha=\frac{1}{2}$. The response function
at zero time lag then is $\langle x(t)\tilde{x}(t)\rangle=\frac{1}{2}$
(see \secref{Response-function-in}). The contributions of loops
closed by response functions $\langle\tilde{x}(t)x(t)\rangle$ that
end on the same vertex of the form (\ref{eq:interaction_vertex_Langevin})
are
\begin{align*}
\Diagram{fV & flAfV\\
 & gdV\\
 & \ldots
}
 & =n\,\underbrace{\langle x(t)\tilde{x}(t)\rangle}_{=\frac{1}{2}}\frac{f^{(n)}}{n!}x(t)^{n-1}\\
 & =\frac{1}{2}\partial_{x}\frac{f^{(n)}}{\left(n-1\right)!}x(t)^{n-1},
\end{align*}
where two of the $n$ $x$-legs are shown explicitly. The combinatorial
factor $n$ in the first line stems from the $n$ possible ways to
attach the propagator to one of the $n$ factors $x$ of the vertex.
The last line shows that the remaining term is the opposite of the
contribution $-\frac{1}{2}f^{\prime}(x)$ that comes from the functional
determinant in (\ref{eq:MSR_Z_tilde_J}). In conclusion, all contributions
of closed loops of response functions on the same vertex are canceled
in the Stratonovich convention in the same way as in the Ito convention.

\subsection{Feynman rules for SDEs in time domain and frequency domain\label{sec:Feynman-rules-Fourier}}

An arbitrary given action first needs to be converted into algebraic
form in the fields, typically by Taylor expansion. We then have a
stochastic differential equation with a linear part on the left and
some nonlinearity on the right, for example 
\begin{align}
dx(t)+x(t)\,dt-\frac{\alpha}{2!}x^{2}(t)\,dt & =dW(t).\label{eq:nonlin_sde}
\end{align}
The action is therefore $S[x,\tx]=S_{0}[x,\tx]-\frac{\alpha}{2!}\,\tx^{\T}x^{2}$,
where $S_{0}[x,\tx]=\tx^{\T}(\partial_{t}+1)\,x+\frac{D}{2}\tx^{\T}\tx$
is the Gaussian part. After having determined the propagators corresponding
to the Gaussian part $S_{0}$ of the action, given in frequency domain
by (\ref{eq:Z_J_OUP}) with $m=-1$,
\begin{eqnarray}
\Delta(t,s) & = & \left(\begin{array}{cc}
\langle x(t)x(s)\rangle & \langle x(t)\tx(s)\rangle\\
\langle\tx(t)x(s)\rangle & \langle\tx(t)\tx(s)\rangle
\end{array}\right)=\left(\begin{array}{cc}
\quad\vertexlabel^{x(t)}\Diagram{fVfA}
\vertexlabel^{x(s)}\quad & \quad\vertexlabel^{x(t)}\Diagram{fV}
\vertexlabel^{\tilde{x}(s)}\quad\\
\quad\vertexlabel^{\tilde{x}(t)}\Diagram{fA}
\vertexlabel^{x(s)}\quad & 0
\end{array}\right)\nonumber \\
 & = & \left(\begin{array}{cc}
\frac{1}{2D}\,e^{-|t-s|} & -H(t-s)\,e^{-(t-s)}\\
-H(s-t)\,e^{-(s-t)} & 0
\end{array}\right)\label{eq:propagator_MSRDJ_example}
\end{eqnarray}
 we need to evaluate the Feynman diagrams of corrections that contain
the interaction vertex in (\ref{eq:nonlin_sde})
\begin{align*}
\Diagram{\Diagram{ &  & \vertexlabel^{x(t)}\\
 & fuV\\
f0\vertexlabel^{\tilde{x}(t)}fV\\
 & fdV\\
 &  & \vertexlabel_{x(t)}
}
}
=-\frac{\alpha}{2!}\,\tilde{x}^{\T}x^{2} & =-\frac{\alpha}{2!}\int\,dt\,\tilde{x}(t)x^{2}(t).
\end{align*}
A perturbation correction to the mean value at first order (one interaction
vertex) is hence caused by the diagram
\begin{align*}
\Diagram{\vertexlabel^{j(t)}f!{fV}{\Delta_{x\tx}}f0!{flV}{\Delta_{xx}}fluVf0}
 & =-1\cdot\frac{\alpha}{2!}\,\int dt^{\prime}\Delta_{x\tx}(t,t^{\prime})\Delta_{xx}(t^{\prime},t^{\prime})\\
 & =\frac{\alpha}{2!}\,\int dt^{\prime}H(t-t^{\prime})\,e^{-(t-t^{\prime})}\frac{D}{2}\\
 & =\frac{\alpha}{2!}\,\int_{0}^{\infty}d\tau\,e^{-\tau}\frac{D}{2}\\
 & =\frac{\alpha D}{4}.
\end{align*}

For problems that are time-translation invariant, often a formulation
in Fourier domain leads to simpler expressions. By help of \subref{Propagators-in-Fourier},
we transfer the Feynman rules from time to frequency domain. We first
express the interaction vertex in terms of the Fourier transforms
of the fields to get

\begin{align*}
\stackrel{\cF}{\to}\Diagram{\Diagram{ &  & \vertexlabel^{X(\omega_{2})}\\
 & fuV\\
f0\vertexlabel^{\tilde{X}(\omega_{1})}f0fV\\
 & fdV\\
\\
 &  & \vertexlabel_{X(\omega_{3})}
}
\quad=}
 & -\frac{\alpha}{2!}\,\iiint\,\frac{d\omega_{1}}{2\pi}\,\frac{d\omega_{2}}{2\pi}\,\frac{d\omega_{3}}{2\pi}\,\underbrace{\int\,dt\,e^{i(\omega_{1}+\omega_{2}+\omega_{3})t}}_{2\pi\,\delta(\omega_{1}+\omega_{2}+\omega_{3})}\,\tilde{X}(\omega_{1})X(\omega_{2})X(\omega_{3})\\
= & -\frac{\alpha}{2!}\,\iint\,\frac{d\omega_{1}}{2\pi}\,\frac{d\omega_{2}}{2\pi}\,\tilde{X}(\omega_{1})X(\omega_{2})X(-\omega_{1}-\omega_{2}).
\end{align*}
 So we get from the Dirac-$\delta$ that the frequencies at each
vertex need to sum up to zero. We may thereofore think of the fequency
``flowing'' through the vertex and obeying a conservation equation
- the frequencies flowing into a vertex also must flow out. We note
that we get one factor $(2\pi)^{-1}$ less than the number of legs
of the vertex. The number of factors $(2\pi)^{-1}$ therefore equals
the number of remaining momentum integrals.

Moreover, we see that every external leg comes with a factor $(2\pi)^{-1}$
from the integration over $\omega$ and a factor $2\pi$ from the
connecting propagator, so that the overall number of such factors
is not affected. Each propagator connecting an internal pair of $X$
or $\tilde{X}$ comes, by (\ref{eq:Z_J_OUP}), with a factor $2\pi$.
Due to the consevation of the frequencies also at each propagator
by the Dirac $\delta$, in the final expression there are hence as
many frequency integrals left as we have factors $(2\pi)^{-1}$. We
may therefore also only keep a single frequency dependence of the
propagator and write the term $2\pi\delta$ explicitly, hence defining
\begin{align}
2\pi\delta(\omega+\omega^{\prime})\,\Delta(\omega) & :=\Delta(\omega,-\omega^{\prime}),\label{eq:def_single_freq_prop}
\end{align}
to get the matrix of propagators
\begin{align}
2\pi\delta(\omega+\omega^{\prime})\,\Delta(\omega) & =\left(\begin{array}{cc}
\langle X(\omega)X(\omega^{\prime})\rangle & \langle X(\omega)\tilde{X}(\omega^{\prime})\rangle\\
\langle\tilde{X}(\omega)X(\omega^{\prime})\rangle & 0
\end{array}\right)\label{eq:propagator_simple}\\
 & =2\pi\delta(\omega+\omega^{\prime})\,\left(\begin{array}{cc}
\left(-i\omega-1\right)^{-1}D\left(i\omega-1\right)^{-1} & \left(-i\omega-1\right)^{-1}\\
\left(i\omega-1\right)^{-1} & 0
\end{array}\right).\nonumber 
\end{align}
As an example, the first order correction to the first moment then
has the form
\begin{align}
\Diagram{\vertexlabel^{J(-\omega)}fVf0flVfluVf0}
 & =\int\,\frac{d\omega}{2\pi}J(-\omega)\,\int\frac{d\omega^{\prime}}{2\pi}2\pi\delta(\omega+\omega^{\prime})\,\Delta_{x\tx}(\omega)\,\iint\frac{d\omega_{1}}{2\pi}\frac{d\omega_{2}}{2\pi}\,\frac{-\alpha}{2!}\,2\pi\delta(\omega^{\prime}+\omega_{1}+\omega_{2})\,2\pi\delta(\omega_{1}+\omega_{2})\,\Delta_{xx}(\omega_{1})\label{eq:correction_mean}\\
 & =\int\,\frac{d\omega}{2\pi}J(-\omega)\,2\pi\,\delta(\omega)\,\Delta_{x\tx}(\omega)\,\frac{-\alpha}{2!}\,\iint\frac{d\omega_{1}}{2\pi}\,\Delta_{xx}(\omega_{1})\nonumber \\
 & =\int\,\frac{d\omega}{2\pi}J(-\omega)\,2\pi\,\delta(\omega)\,\left(-i\omega-1\right)^{-1}\,\frac{-\alpha}{2!}\,\int\,\frac{d\omega_{1}}{2\pi}\,\left(-i\omega_{1}+1\right)^{-1}D\left(i\omega_{1}+1\right)^{-1}\nonumber \\
 & =J(0)\,\frac{-\alpha}{2!}\,\int\,\frac{d\omega_{1}}{2\pi}\,\left(-i\omega_{1}+1\right)^{-1}D\left(i\omega_{1}+1\right)^{-1},
\end{align}
where the connecting external line $\Feyn{fV}=\left(-i\omega+1\right)^{-1}$
has the shown sign, because $J(-\omega)$ couples to $X(\omega)$,
so we need to take the upper right element in (\ref{eq:propagator_simple}).
The last factor $\Feyn{fVfA}=\Delta_{xx}(\omega_{1})=\left(-i\omega_{1}-1\right)^{-1}D\left(i\omega_{1}-1\right)^{-1}$
is the covariance function connecting the two $X(\omega_{1})$ and
$X(\omega_{2})$ legs of the vertex.

Since originally each integral over $\omega_{i}$ comes with $(2\pi)^{-1}$
and each conservation of sums of $\omega$ in either a propagator
or a vertex comes with $2\pi\delta(\sum_{i}\omega_{i})$, we have
as many factors $(2\pi)^{-1}$ as we have independent momentum integrals.
We summarize the rules as follows:
\begin{itemize}
\item An external leg ending on $J(\omega)$ attaches to a variable $X(-\omega)$
within the diagram and analogous for $\tJ(\omega)$ and $\tilde{X}(-\omega)$.
\item At each vertex, the sum of all $\omega$ flowing into the vertex must
sum up to zero, since we get a term $\propto\delta(\sum_{i=1}^{n}\omega_{i})$.
\item The frequencies that enter a propagator line must also exit, since
we get a term $\propto\delta(\omega+\omega^{\prime})$.
\item We have as many factors $(2\pi)^{-1}$ as we have independent $\omega$
integrals left after all constraints of $\omega$-conservation have
been taken into account.
\item The number of $\omega$ integrals hence must correspond to the number
of loops: all other frequencies are fixed by the external legs.
\end{itemize}
So we may infer the frequencies on each propagator line by rules analogous
to Kirchhoff's law: Treating the frequencies as if they were conserved
currents.

Using these rules we could have written down the fourth line in \eqref{correction_mean}
directly.

The above integral by
\begin{align}
 & \frac{1}{2\pi}\int\,d\omega_{1}\,\left(-i\omega_{1}+1\right)^{-1}D\left(i\omega_{1}+1\right)^{-1}\label{eq:int_mean}\\
 & =\frac{1}{2\pi i}\int_{-i\infty}^{i\infty}\,dz\,\left(-z+1\right)^{-1}D\left(z+1\right)^{-1}\nonumber \\
 & =\frac{1}{2\pi i}\int_{\gamma}\,dz\,\left(-z+1\right)^{-1}D\left(z+1\right)^{-1}\nonumber \\
 & =\frac{D}{2}.\nonumber 
\end{align}
hence evaluates to $\frac{\alpha}{2}\,\frac{D}{2}\,J(0).$ We here
closed the path $\gamma$ in the positive direction, which is the
left half-plane (with $\Re(z)<0$), we get a $+1$ from the winding
number. We encircle the pole $z=-1$ from the right factor and need
to replace $z=-1$ in the left term.

The result, being proportional to $J(0)$, shows that the correction
only affects the stationary expectation value at $\omega=0$, which
therefore is (by the functional chain rule)
\begin{align*}
\langle x(t)\rangle & =\frac{\delta W}{\delta j(t)}\Big|_{j=0}=\int\,\underbrace{e^{-i\omega t}}_{\frac{\delta J(\omega)}{\delta j(t)}}\frac{\delta\hat{W}}{\delta J(\omega)}\Big|_{J=0}d\omega=\int e^{-i\omega t}\delta(-\omega)\,\frac{\alpha D}{4}\,d\omega=\frac{\alpha D}{4},
\end{align*}
which is valid to first order in $\alpha$. We here used that due
to the source being of the form $J^{\dagger}X=\int\frac{d\omega}{2\pi}\,J(-\omega)X(\omega)$
that $\frac{\delta\hat{W}}{\delta J(\omega)}=\frac{1}{2\pi}\,2\pi\frac{\alpha D}{4}\delta(-\omega)$.
This value is also naively expected, by noting that the variance in
the unperturbed system is $\langle x^{2}\rangle=\frac{D}{2}$, so
the expectation value of the non-linear term on the right hand side
of (\ref{eq:nonlin_sde}) is $\alpha\,\frac{D}{4}$.

\subsection{Diagrams with more than a single external leg}

In calculating diagrams with more that a single external leg, we remember
that the $n$-fold repetition of an external leg must come from the
factor
\begin{eqnarray*}
 &  & \exp(j^{\T}x)=\sum_{n}\frac{(j^{\T}x)^{n}}{n!}.
\end{eqnarray*}
So a diagram with $n$-legs of identical type $j$ comes with $n$
time-integrals and a factor $n!^{-1}$. This is completely analogous
to the case of the $n$-fold repetition of an interaction vertex.

It is instructive to first derive the correction to $W$ - hence we
compute the $j$-dependent contribution - and only in a second step
differentiate the result by $j$ to obtain the correction to the cumulants.

For example, a diagram contributing to the correction of the variance
of the process would come with a factor $\frac{1}{2!}\iint dt\,ds\,j(t)\,j(s)$
prior to taking the second derivative by $j$. Concretely, let us
consider the diagram

\begin{align*}
\Diagram{2\cdot2\cdot2\cdot\vertexlabel_{j(t)}!{f}{\Delta_{xx}}f0!{fl}{\Delta_{xx}}!{fluA}{\Delta_{\tx x}}f0!{fA}{\Delta_{\tx x}}\vertexlabel_{j(s)}}
 & =\frac{1}{2!}\,\iint dt\,ds\,j(t)j(s)\,\underbrace{2\cdot2\cdot2\cdot\frac{1}{2!}\big(\frac{\alpha}{2!}\big)^{2}\,\iint\,dt^{\prime}ds^{\prime}\,\Delta_{xx}(t,t^{\prime})\,\Delta_{xx}(t^{\prime},s^{\prime})\Delta_{\tx x}(t^{\prime},s^{\prime})\,\Delta_{\tx x}(s^{\prime},s)}_{=:f(t,s)}\\
 & =:\frac{1}{2!}\,\iint dt\,ds\,j(t)\,j(s)\,f(t,s).
\end{align*}
The combinatorial factor arises from two possibilities of connecting
the $\Delta_{xx}$ propagator of the left external leg to either of
the vertices and the two possibilities of choosing the incoming $x$-leg
of the vertex to which we connect this external leg. Another factor
two arises from the two possibilities of connecting the $\Delta_{xx}$
propagator to either of the two $x$-legs of the right vertex. All
other contractions are uniquely determined then; so in total we have
a factor $2\cdot2\cdot2$.

In calculating the contribution to the covariance function $\llangle x(t)x(s)\rrangle$,
the second cumulant of the process, we need to take the second functional
derivative. Because the factor $j$ appears twice, we obtain by the
application of the functional product rule the correction to $\delta^{2}W/\delta j(t)\,\delta j(s)$
\begin{eqnarray*}
 &  & \frac{1}{2!}\big(f(t,s)+f(s,t)\big),
\end{eqnarray*}
which is a manifestly symmetric contribution as it has to be for a
covariance function. A single term $f(t,s)$ is not necessarily symmetric,
as seen from the appearance of the non-symmetric functions $\Delta_{\tx x}$.

We may calculate the same contribution in frequency domain. To assign
the frequencies to the legs we use that at each line the frequencies
must have opposite sign on either end and the sums of all frequencies
at a vertex must sum up to zero; the frequency of the left field of
the propagator is the argument of the corresponding function $\Delta(\omega)$,
according to (\ref{eq:def_single_freq_prop}). So we get
\begin{eqnarray}
\nonumber \\
I & := & 2\cdot2\cdot2\cdot\qquad\Diagram{\vertexlabel_{J(\omega)}!{f}{-\omega\;\Delta_{xx}\;\omega}ff0!{fl}{(-\omega+\omega^{\prime})\,\Delta_{xx}\,(\omega-\omega^{\prime})}!{fluA}{-\omega^{\prime}\;\Delta_{\tx x}\;\omega^{\prime}}f0f!{fA}{-\omega\;\Delta_{\tx x}\;\omega}\vertexlabel_{J(-\omega)}}
\nonumber \\
\nonumber \\
\nonumber \\
 & = & \frac{1}{2!}\,\int\frac{d\omega}{2\pi}\,J(\omega)\,\underbrace{\int\frac{d\omega^{\prime}}{2\pi}\,2\cdot2\cdot2\cdot\frac{1}{2!}\,\big(\frac{\alpha}{2!}\big)^{2}\,\Delta_{xx}(-\omega)\,\Delta_{xx}(-\omega+\omega^{\prime})\,\Delta_{\tx x}(-\omega^{\prime})\,\Delta_{\tx x}(-\omega)}_{=:F(\omega)}\,J(-\omega)\nonumber \\
 & =: & \frac{1}{2!}\,\int\frac{d\omega}{2\pi}\,J(\omega)\,F(\omega)\,J(-\omega).\label{eq:two_leg_diagram}
\end{eqnarray}
We observe that the contribution can be written as an integral over
one frequency, the frequency within the loop $\omega^{\prime}$. Each
of the sources is attached by a propagator to this loop integral.
We will see in the following that the inner integral is an effective
1PI vertex.

The contribution to the variance therefore becomes with the functional
chain rule and $\delta J(\omega)/\delta j(t)=e^{-i\omega t}$
\begin{align*}
\frac{\delta^{2}W}{\delta j(t)\delta j(s)} & =\iint\,d\omega\,d\omega^{\prime}\,e^{-i\omega t}\,e^{-i\omega^{\prime}s}\,\frac{\delta^{2}W}{\delta J(\omega)\delta J(\omega^{\prime})}.
\end{align*}
By the last line in (\ref{eq:two_leg_diagram}) and the application
of the product rule we see that $\frac{\delta^{2}I}{\delta J(\omega)\delta J(\omega^{\prime})}=\frac{1}{2!}\,\frac{1}{2\pi}\,\delta(\omega+\omega^{\prime})\,\big(F(\omega)+F(-\omega)\big)$
so that
\begin{align}
\frac{\delta^{2}I}{\delta j(t)\delta j(s)} & =\int\,\frac{d\omega}{2\pi}\,e^{i\omega(s-t)}\,\frac{1}{2!}\big(F(\omega)+F(-\omega)\big).\label{eq:F_contrib_I}
\end{align}
Again, the product rule causes a symmetric contribution of the diagram.
The back transform can be calculated with the help of the residue
theorem. Multiple poles of order $n$ can be treated by Cauchy's differential
formula

\begin{eqnarray*}
f^{(n)}(a) & = & \frac{n!}{2\pi i}\oint\,\frac{f(z)}{(z-a)^{n+1}}\,dz.
\end{eqnarray*}

\subsection{Appendix: Unitary Fourier transform\label{sub:Unitary-transform-Fourier}}

A unitary transform is defined as an isomorphism that preserves the
inner product. In our example the space is the vector space of all
functions and the inner (scalar) product is
\begin{eqnarray}
(f,g) & = & \int_{-\infty}^{\infty}f^{\ast}(t)\,g(t)\,dt.\label{eq:scalar_prod_C}
\end{eqnarray}
The Fourier transform is a linear mapping of a function $f(t)$ to
$F(\omega)$, which can be understood as the projection onto the orthogonal
basis vectors $u_{\omega}(t):=\frac{1}{2\pi}e^{i\omega t}$. The basis
is orthogonal because
\begin{eqnarray}
(u_{\omega},u_{\omega^{\prime}}) & = & \int_{-\infty}^{\infty}\frac{e^{i(\omega^{\prime}-\omega)t}}{(2\pi)^{2}}\,dt=\frac{\delta(\omega^{\prime}-\omega)}{2\pi}.\label{eq:ortho_fourier_modes}
\end{eqnarray}
The Fourier transform is a unitary transformation, because it preserves
the form of the scalar product on the two spaces
\begin{eqnarray}
(f,g):=\int\,f^{\ast}(t)\,g(t)\,dt & = & \int d\omega\,\int d\omega^{\prime}\underbrace{\int dt\,\frac{e^{i(-\omega+\omega^{\prime})t}}{(2\pi)^{2}}}_{\equiv\delta(-\omega+\omega^{\prime})}\,F^{\ast}(\omega)G(\omega^{\prime})\nonumber \\
 & = & \int\frac{d\omega}{2\pi}\,F^{\ast}(\omega)\,G(\omega)=:(F,G).\label{eq:ortho_fourier}
\end{eqnarray}
So the scalar products in the two spaces have the same form.

Changing the path integral from $\int\D x(t)$ to $\int\D X(\omega)$,
each individual time integral can be expressed by all frequency integrals
as 
\begin{eqnarray}
\int dx(t) & = & \int d\omega\,\frac{e^{i\,t\omega}}{2\pi}\,\int dX(\omega).\label{eq:transform_X_to_x}
\end{eqnarray}
The transform (\ref{eq:transform_X_to_x}) is a multiplication with
the (infinite dimensional) matrix $U_{t\omega}=\frac{e^{i\,t\omega}}{2\pi}$.
This matrix $U\equiv(u_{-\infty},\ldots,u_{\infty})$ has the property
\begin{eqnarray*}
(U^{T\ast}U)_{\omega\omega^{\prime}} & \stackrel{(\ref{eq:ortho_fourier_modes})}{=} & \frac{\delta(\omega-\omega^{\prime})}{2\pi},
\end{eqnarray*}
which is the infinite dimensional unit matrix, from which follows
in particular that $|\det(U)|=\mathrm{const}$. Hence changing the
path integral $\int\D x(t)$ to $\int\D X(\omega)$ we only get a
constant from the determinant. Since we are only interested in derivatives
of generating functionals, this constant has no consequence. However,
the integration boundaries change. The integral $\int\D x(t)$ goes
over all real-valued functions $x(t)$. Hence the corresponding Fourier
transforms $X(\omega)$ have the property $X(-\omega)=X(\omega)^{\ast}$.

The action in (\ref{eq:def_A_OUP}) instead of the standard scalar
product on $\mathbb{C}$ (\ref{eq:scalar_prod_C}) employs the Euclidean
scalar product between functions $x$ and $y$ of the form
\begin{align*}
x^{\T}y & =\int\,dt\,x(t)\,y(t).
\end{align*}
As a consequence, in frequency domain we get
\begin{align}
\int\,\frac{d\omega}{2\pi}\,X(-\omega)\,Y(\omega) & .\label{eq:def_inner_prod_dagger}
\end{align}

\section{Dynamic mean-field theory for random networks\label{sec:Sompolinsky-Crisanti-Sommers-theory}}

Systems with many interacting degrees of freedom present a central
quest in physics. While disordered equilibrium systems show fascinating
properties such as the spin-glass transition \citep{Parisi80_1101,Sompolinsky81},
new collective phenomena arise in non-equilibrium systems: Large random
networks of neuron-like units can exhibit chaotic dynamics \citep{Sompolinsky88_259,Vreeswijk96,Monteforte10_268104}
with important functional consequences. In particular, information
processing capabilities show optimal performance close to the onset
of chaos \citep{Legenstein07_323,Sussillo09_544,Toyoizumi11_051908}.

Until today, the seminal work by \citet{Sompolinsky88_259} has a
lasting influence on the research field of random recurrent neural
networks, presenting a solvable random network model with deterministic
continuous-time dynamics that admits a calculation of the transition
to a chaotic regime and a characterization of chaos by means of Lyapunov
exponents. Many subsequent studies have built on top of this work
\citep{rajan10_011903,Hermann12_018702,wainrib13_118101,Aljadeff15_088101,Kadmon15_041030,Goedeke16_arxiv}.

The presentation in the original work \citep{Sompolinsky88_259},
published in \emph{Physical Review Letters}, summarizes the main steps
of the derivations and the most important results. In this chapter
we would like to show the formal calculations that reproduce the most
important results. After communication with A Crisanti we could confirm
that the calculations by the original authors are indeed to large
extent identical to the presentation here. The original authors recently
published an extended version of their work \citep{Crisanti18_062120}.

Possible errors in this document should not be attributed to the original
authors, but to the authors of this manuscript. In deriving the theory,
we also present a recent extension of the model to stochastic dynamics
due to additive uncorrelated Gaussian white noise \citep{Schuecker18_041029}.
The original results of \citep{Sompolinsky88_259} are obtained by
setting the noise amplitude $D=0$ in all expressions. The here presented
material has previously been made publicly available as \citep{Schuecker16b_arxiv}.

\subsection{Definition of the model and generating functional\label{sec:generating_functional}}

We study the coupled set of first order stochastic differential equations
\begin{align}
d\bx(t)+\bx(t)\,dt & =\bJ\phi(\bx(t))\,dt+d\boldsymbol{\xi}(t),\label{eq:diffeq_motion}
\end{align}
where 
\begin{align}
J_{ij} & \sim\begin{cases}
\N(0,\frac{g^{2}}{N})\,\text{i.i.d.} & \text{for }i\neq j\\
0 & \text{for }i=j
\end{cases}\label{eq:connectivity_distribution}
\end{align}
are i.i.d. Gaussian random couplings, $\phi$ is a non-linear gain
function applied element-wise, the $dW_{i}$ are pairwise uncorrelated
Wiener processes with $\langle d\xi{}_{i}(t)d\xi_{j}(s)\rangle=D\,\delta_{ij}\delta_{st}\,dt$.
For concreteness we will use 
\begin{align}
\phi(x) & =\tanh(x),\label{eq:gain_function}
\end{align}
as in the original work \citep{Sompolinsky88_259}.

We formulate the problem in terms of a generating functional from
which we can derive all moments of the activity as well as response
functions. Introducing the notation $\tbx^{\T}\bx=\sum_{i}\int\,\tx_{i}(t)x_{i}(t)\,dt$,
we obtain the moment-generating functional as derived in \prettyref{sec:Martin-Siggia-Rose-De-Dominicis}

\begin{align}
Z[\bl,\tbj](\bJ) & =\int\D\bx\int\D\tbx\,\exp\Big(S_{0}[\bx,\tbx]-\tbx^{\T}\bJ\phi\left(\bx\right)+\bl^{\T}\bx+\tbj^{\T}\tbx\Big)\nonumber \\
\text{with }S_{0}[\bx,\tbx] & =\tbx^{\T}\left(\partial_{t}+1\right)\bx+\frac{D}{2}\tbx^{\T}\tbx,\label{eq:def_S0}
\end{align}
where the measures are defined as $\int\D\bx=\lim_{M\to\infty}\Pi_{j=1}^{N}\Pi_{l=1}^{M}\int_{-\infty}^{\infty}dx_{j}^{l}$
and $\int\D\tbx=\lim_{M\to\infty}\Pi_{j=1}^{N}\Pi_{l=1}^{M}\int_{-i\infty}^{i\infty}\frac{d\tilde{x}_{j}^{l}}{2\pi i}$.
Here the superscript $k$ denotes the $k$-th time slice and we skip
the subscript $\D_{2\pi i}$, as introduced in \prettyref{eq:OM_pathint}
in \prettyref{sec:Onsager-Machlup-path-integral}, in the measure
of $\D\tbx$. The action $S_{0}$ is defined to contain all single
unit properties, therefore excluding the coupling term $-\tbx^{\T}\bJ\phi\left(\bx\right)$,
which is written explicitly.

\subsection{Property of self-averaging\label{sub:Property-of-self-averaging}}

We see from \prettyref{eq:def_S0} that the term that couples the
different neurons has a special form, namely
\begin{align}
h_{i}(t) & :=[\bJ\phi\left(\bx\right)]_{i}\nonumber \\
 & =\sum_{j}J_{ij}\phi(x_{j}(t)),\label{eq:def_h}
\end{align}
which is the sum of many contributions. In the first exercises (see
\prettyref{sec:Probabilities-moments-cumulant}), we have calculated
the distribution of the sum of independent random numbers. We found
that the sum approaches a Gaussian if the terms are weakly correlated,
given the number of constituents is sufficiently large. In general,
such results are called \textbf{concentration of measure} \citep[i. p. section VII]{Touchette09},
because its probability distribution, in mathematics often called
a \textbf{measure}, becomes very peaked around its mean value.

In the following derivation we are going to find a similar behavior
for $h_{i}$ due to the large number of synaptic inputs summed up
in \prettyref{eq:def_h}. The latter statement is about the temporal
statistics of $h_{i}\sim\N(\mu_{i},\sigma_{i}^{2})$. We can try to
make a conceptually analogous, but different statement about the statistics
of $h_{i}$ with respect to the randomness of $J_{ij}$: The couplings
$J_{ij}$ are constant in time; they are therefore often referred
to as \textbf{frozen} or \textbf{quenched disorder}. Observing that
each $h_{i}$ approaches a Gaussian the better the larger the $N$,
we may ask how much the parameters $\mu_{i}$ and $\sigma_{i}$ of
this Gaussian vary from one realization of $J_{ij}$ to another. If
this variability becomes small, because $i$ was chosen arbitrary,
this implies that also the variability from one neuron $i$ to another
neuron $k$ at one given, fixed $J_{ij}$ must be small - this property
is called \textbf{self-averaging}: The average over the disorder,
over an ensemble of systems, is similar to the average over many units
$i$ in a single realization from the ensemble. As a result, we may
hope to obtain a low-dimensional description of the statistics for
one typical unit. This is what we will see in the following.

As a consequence of the statistics of $h_{i}$ to converge to a well-defined
distribution in the $N\to\infty$ limit, we may hope that the entire
moment generating functional $Z[\bl](\bJ)$, which, due to $\bJ$
is a random object, shows a concentration of measure as well. The
latter must be understood in the sense that for most of the realizations
of $\bJ$ the generating functional $Z$ is close to its average $\langle Z[\bl](\bJ)\rangle_{\bJ}$.
We would expect such a behavior, because mean and variance of the
$h_{i}$ approach certain, fixed values, the more precise the larger
the network size is. Such a statement makes an assertion about an
ensemble of networks. In this case it is sufficient to calculate the
latter. It follows that all quantities that can be calculated from
$Z[\bl](\bJ)$ can then also be - approximately - obtained from $\langle Z[\bl](\bJ)\rangle_{\bJ}$.
Each network is obtained as one realization of the couplings $J_{ij}$
following the given probabilistic law \prettyref{eq:connectivity_distribution}.
The goal of the mean-field description derived in the following is
to find such constant behavior independent of the actual realization
of the frozen disorder.

The assumption that quantities of interest are self-averaging is implicit
in modeling approaches that approximate neuronal networks by networks
with random connectivity; we expect to find that observables of interest,
such as the rates, correlations, peaks in power spectra, are independent
of the particular realization of the randomness.

To see the concept of self-averaging more clearly, we may call the
distribution of the activity in the network $p[\bx](\bJ)$ for one
particular realization $\bJ$ of the connectivity. Equivalently, we
may express it as its Fourier transform, the moment-generating functional
$Z[\bl](\bJ)$. Typically we are interested in some experimental observables
$O[\bx]$. We may for example think of the population-averaged autocorrelation
function
\begin{align*}
\langle O_{\tau}[\bx]\rangle_{\bx(\bJ)} & =\frac{1}{N}\sum_{i=1}^{N}\langle x_{i}(t+\tau)x_{i}(t)\rangle_{\bx(\bJ)},
\end{align*}
where the expectation value $\langle\rangle_{\bx(\bJ)}$ is over realizations
of $\bx$ for one given realization of $\bJ$. It is convenient to
express the observable in its Fourier transform $O[\bx]=\int\D\bl\,\hat{O}[\bl]\,\exp(\bl^{\T}\bx)$
(with suitably defined $\hat{O}$ and measure $\D$) using \prettyref{eq:exp_f_Fourier}
\begin{align*}
\langle O[\bx]\rangle_{\bx}(\bJ) & =\int\D\bl\,\hat{O}[\bl]\,Z[\bl](\bJ),
\end{align*}
where naturally the moment generating functional appears as $Z[\bl](\bJ)=\langle\exp(\bl^{\T}\bx)\rangle_{x}(\bJ)$.
The mean observable averaged over all realizations of $\bJ$ can therefore
be expressed as
\begin{align*}
\langle\langle O[\bx]\rangle_{\bx(\bJ)}\rangle_{\bJ} & =\int\D\bl\,\hat{O}[\bl]\,\langle Z[\bl](\bJ)\rangle_{\bJ},
\end{align*}
in terms of the generating functional that is averaged over the frozen
disorder, as anticipated above.

\begin{figure}
\begin{centering}
\includegraphics[scale=0.7]{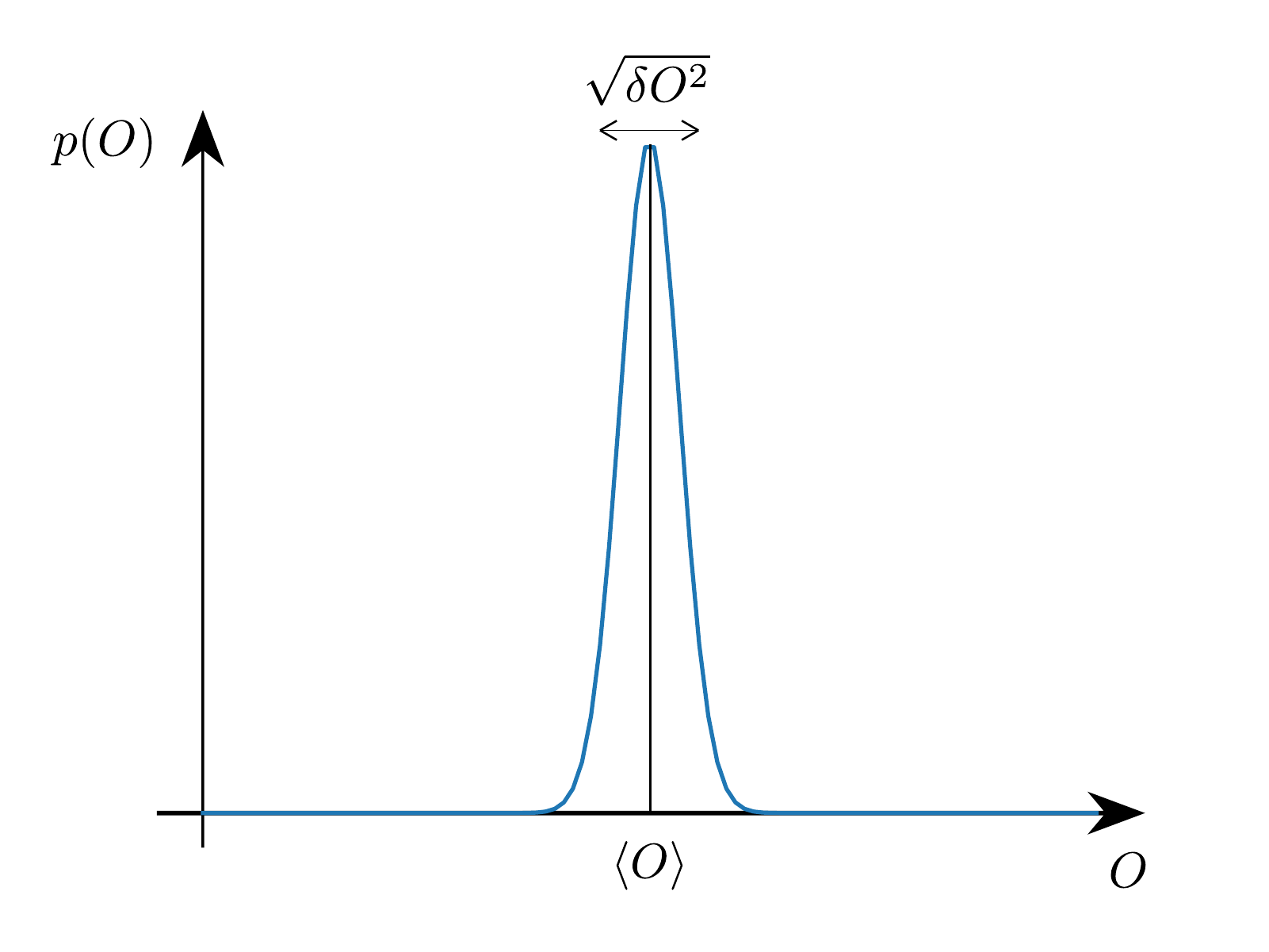}
\par\end{centering}
\caption{\textbf{Self-averaging observable $O$. }The variability $\delta O$
over different realizations of the random disorder is small, so that
with high probability, the measured value in one realization is close
to the expectation value $\langle O\rangle$ over realizations.\label{fig:Self-averaging-observable}}
\end{figure}

We call a quantity self-averaging, if its variability with respect
to the realization of $\bJ$ is small compared to a given bound $\epsilon$.
Here $\epsilon$ may for example be determined by the measurement
accuracy of an experiment. With the short hand $\delta O[\bx]:=O[\bx]-\langle\langle O[\bx]\rangle_{\bx}(\bJ)\rangle_{\bJ}$
we would like to have

\begin{align}
\langle[\langle\delta O[x]\rangle_{x}(\bJ)]^{2}\rangle_{\bJ}=\langle\left(\int\D\bx\,p[\bx](\bJ)\,\delta O[\bx]\right)^{2}\rangle_{\bJ} & \ll\epsilon,\label{eq:expectation_observable}
\end{align}
a situation illustrated in \prettyref{fig:Self-averaging-observable}.
Analogously to the mean, the variance of the observable can be expressed
in terms of the average of the product of a pair of generating functionals
\begin{align}
Z_{2}[\bl,\bl^{\prime}] & :=\langle Z[\bl](\bJ)\,Z[\bl^{\prime}](\bJ)\rangle_{\bJ}\label{eq:replica_Z}
\end{align}
as
\begin{align}
\langle\delta Q^{2}(\bJ)\rangle_{\bJ} & =\iint\,\D\bl\D\bl^{\prime}\,\delta\hat{O}[\bl]\,\delta\hat{O}[\bl^{\prime}]\,Z_{2}[\bl,\bl^{\prime}].\label{eq:variance_disorder}
\end{align}
Taking the average over products of generating functional is called
the \textbf{replica method}: we replicate a system with identical
parameters and average the product.

In the particular case that $Z_{2}[\bl,\bl^{\prime}]$ factorizes
into a product of two functionals that individually depend on $\bl$
and $\bl^{\prime}$, the variance of any observable vanishes. We will
see in the following that to leading order in $N$, the number of
neurons, this will be indeed the case for the model studied here.

\subsection{Average over the quenched disorder\label{sub:Disorder-average}}

We now assume that the system \prettyref{eq:diffeq_motion} shows
self-averaging behavior, independent of the particular realization
of the couplings, as explained above. To capture these properties
that are generic to the ensemble of the models, we introduce the averaged
functional
\begin{eqnarray}
\bar{Z}[\bl,\tbj] & := & \langle Z[\bl,\tbj](\bJ)\rangle_{\bJ}\label{eq:disorder_averaged_Z}\\
 & = & \int\Pi_{ij}dJ_{ij}\,\mathcal{N}(0,\frac{g^{2}}{N},J_{ij})\,Z[\bl,\tbj](\bJ).\nonumber 
\end{eqnarray}
We use that the coupling term $\exp(-\sum_{i\neq j}J_{ij}\int\tilde{x}_{i}(t)\phi(x_{j}(t))\,dt)$
in \eqref{def_S0} factorizes into $\Pi_{i\neq j}\exp(-J_{ij}\int\tilde{x}_{i}(t)\phi(x_{j}(t))\,dt)$
as does the distribution over the couplings (due to $J_{ij}$ being
independently distributed). We make use of the couplings appearing
linearly in the action so that we may rewrite the term depending on
the connectivity $J_{ij}$ for $i\neq j$
\begin{align}
 & \int dJ_{ij}\mathcal{N}(0,\frac{g^{2}}{N},J_{ij})\,\exp\left(-J_{ij}y_{ij}\right)=\left\langle \exp(-J_{ij}y_{ij})\right\rangle _{J_{ij}\sim\N(0,\frac{g^{2}}{N})}\label{eq:completion_of_square}\\
\text{with }y_{ij} & :=\int\,\tilde{x}_{i}(t)\phi(x_{j}(t))\,dt.\nonumber 
\end{align}
The form in the first line is that of the moment generating function
(\eqref{def_char_fctn}) of the distribution of the $J_{ij}$ evaluated
at the point $-y_{ij}$. For a general distribution of i.i.d. variables
$J_{ij}$ with the $n$-th cumulant $\kappa_{n}$, we hence get with
\prettyref{eq:cumulant_Taylor}
\begin{align*}
\left\langle \exp(-J_{ij}y_{ij})\right\rangle _{J_{ij}} & =\exp(\sum_{n}\frac{\kappa_{n}}{n!}\,(-y_{ij})^{n}).
\end{align*}
For the Gaussian case studied here, where the only non-zero cumulant
is $\kappa_{2}=\sigma^{2}$, we hence get
\begin{align*}
\left\langle \exp\left(-J_{ij}y_{ij}\right)\right\rangle _{J_{ij}\sim\N(0,\frac{g^{2}}{N})} & =\exp\left(\frac{g^{2}}{2N}y_{ij}^{2}\right)\\
 & =\exp\left(\frac{g^{2}}{2N}\left(\int\tilde{x}_{i}(t)\phi(x_{j}(t))\,dt\right)^{2}\right).
\end{align*}
We reorganize the last term including the sum $\sum_{i\neq j}$ as
\begin{eqnarray*}
 &  & \frac{g^{2}}{2N}\sum_{i\neq j}\left(\int\tilde{x}_{i}(t)\phi(x_{j}(t))\,dt\right)^{2}\\
 & = & \frac{g^{2}}{2N}\sum_{i\neq j}\int\int\tilde{x}_{i}(t)\phi(x_{j}(t))\,\tilde{x}_{i}(t^{\prime})\phi(x_{j}(t^{\prime}))\,dt\,dt^{\prime}\\
 & = & \frac{1}{2}\int\int\left(\sum_{i}\tilde{x}_{i}(t)\tilde{x}_{i}(t^{\prime})\right)\,\left(\frac{g^{2}}{N}\sum_{j}\phi(x_{j}(t))\phi(x_{j}(t^{\prime}))\right)\,dt\,dt^{\prime}\\
 &  & -\frac{g^{2}}{2N}\,\int\int\sum_{i}\tilde{x}_{i}(t)\tilde{x}_{i}(t^{\prime})\phi(x_{i}(t))\phi(x_{i}(t^{\prime}))\,dt\,dt^{\prime},
\end{eqnarray*}
where we used $\left(\int f(t)dt\right)^{2}=\int\int f(t)f(t^{\prime})\,dt\,dt^{\prime}$
in the first step and $\sum_{ij}x_{i}y_{j}=\sum_{i}x_{i}\sum_{j}y_{j}$
in the second. The last term is the diagonal element that is to be
taken out of the double sum. It is a correction of order $N^{-1}$
and will be neglected in the following. The disorder-averaged generating
functional \prettyref{eq:disorder_averaged_Z} therefore takes the
form
\begin{eqnarray}
\bar{Z}[\bl,\tbj] & = & \int\D\bx\int\D\tbx\,\exp\Big(S_{0}[\bx,\tbx]+\bl^{\T}\bx+\tbj^{\T}\tbx\Big)\times\label{eq:Zbar_pre}\\
 &  & \times\exp\Big(\frac{1}{2}\int_{-\infty}^{\infty}\int_{-\infty}^{\infty}\left(\sum_{i}\tilde{x}_{i}(t)\tilde{x}_{i}(t^{\prime})\right)\,\underbrace{\left(\frac{g^{2}}{N}\sum_{j}\phi(x_{j}(t))\phi(x_{j}(t^{\prime}))\right)}_{=:Q_{1}(t,t^{\prime})}\,dt\,dt^{\prime}\Big).\nonumber 
\end{eqnarray}
The coupling term in the last line shows that both sums go over all
indices, so the system has been reduced to a set of $N$ identical
systems coupled to one another in an identical manner. The coupling
term contains quantities that depend on four fields. We now aim to
decouple these terms into terms of products of pairs of fields. The
aim is to make use of the central limit theorem, namely that the quantity
$Q_{1}$ indicated by the curly braces in \prettyref{eq:Zbar_pre}
is a superposition of a large ($N$) number of (weakly correlated)
contributions, which will hence approach a Gaussian distribution.
Introducing $Q_{1}$ as a new variable is therefore advantageous,
because we know that the systematic fluctuation expansion is an expansion
for the statistics close to a Gaussian. To lowest order, fluctuations
are neglected alltogether. The outcome of the saddle point or tree
level approximation to this order is the replacement of $Q_{1}$ by
its expectation value. To see this, let us define

\begin{align}
Q_{1}(t,s):= & \frac{g^{2}}{N}\sum_{j}\phi(x_{j}(t))\phi(x_{j}(s))\label{eq:def_Q1}
\end{align}
and enforce this condition by inserting the Dirac-$\delta$ functional
\begin{align}
 & \delta[-\frac{N}{g^{2}}Q_{1}(s,t)+\sum_{j}\phi(x_{j}(s))\,\phi(x_{j}(t))]\label{eq:Hubbard_Stratonovich}\\
= & \int\D Q_{2}\,\exp\left(\iint\,Q_{2}(s,t)\left[-\frac{N}{g^{2}}Q_{1}(s,t)+\sum_{j}\,\phi(x_{j}(s))\,\phi(x_{j}(t))\right]\,ds\,dt\right).\nonumber 
\end{align}
We here note that as for the response field, the field $Q_{2}\in i\mathbb{R}$
is purely imaginary due to the Fourier representation \prettyref{eq:Fourier_delta}
of the $\delta$. The enforcement of a constraint by such a conjugate
auxiliary field is a common practice in large $N$ field theory \citep{Moshe03}.

We aim at a set of self-consistent equations for the auxiliary fields.
We treat the theory as a field theory in the $Q_{1}$ and $Q_{2}$
in their own right. We therefore introduce one source $k$, $\tilde{k}$
for each of the fields to be determined and drop the source terms
for $x$ and $\tilde{x}$; this just corresponds to a transformation
of the random variables of interest (see \prettyref{sec:Transformation-of-random}).
Extending our notation by defining $Q_{1}^{\T}Q_{2}:=\iint\,Q_{1}(s,t)\,Q_{2}(s,t)\,ds\,dt$
and $\tx^{\T}Q_{1}\tx:=\iint\,\tx(s)\,Q_{1}(s,t)\,\tx(t)\,ds\,dt$
we hence rewrite \prettyref{eq:Zbar_pre} as

\begin{eqnarray}
\bar{Z}_{Q}[k,\tilde{k}] & := & \int\D Q_{1}\int\D Q_{2}\,\exp\left(-\frac{N}{g^{2}}Q_{1}^{\T}Q_{2}+N\,\ln\,Z[Q_{1},Q_{2}]+k^{\T}Q_{1}+\tilde{k}^{\T}Q_{2}\right)\label{eq:Zbar}\\
Z[Q_{1},Q_{2}] & = & \int\D x\int\D\tx\,\exp\Big(S_{0}[x,\tx]+\frac{1}{2}\tx^{\T}Q_{1}\tx+\phi(x)^{\T}Q_{2}\phi(x)\Big),\nonumber 
\end{eqnarray}
where the integral measures $\D Q_{1,2}$ must be defined suitably.
In writing $N\,\ln\,Z[Q_{1},Q_{2}]$ we have used that the auxiliary
fields couple only to sums of fields $\sum_{i}\phi^{2}(x_{i})$ and
$\sum_{i}\tx_{i}^{2}$, so that the generating functional for the
fields $\bx$ and $\tbx$ factorizes into a product of $N$ factors
$Z[Q_{1},Q_{2}]$. The latter only contains functional integrals over
the two scalar fields $x$, $\tx$. This shows that we have reduced
the problem of $N$ interacting units to that of a single unit exposed
to a set of external fields $Q_{1}$ and $Q_{2}$.

The remaining problem can be considered a field theory for the auxiliary
fields $Q_{1}$ and $Q_{2}$. The form \prettyref{eq:Zbar} clearly
exposes the $N$ dependence of the action for these latter fields
in \prettyref{eq:Zbar}: It is of the form $\int dQ\exp(Nf(Q))\,dQ$,
which, for large $N$, suggests a saddle point approximation.

\begin{figure}
\begin{centering}
\includegraphics[scale=0.5]{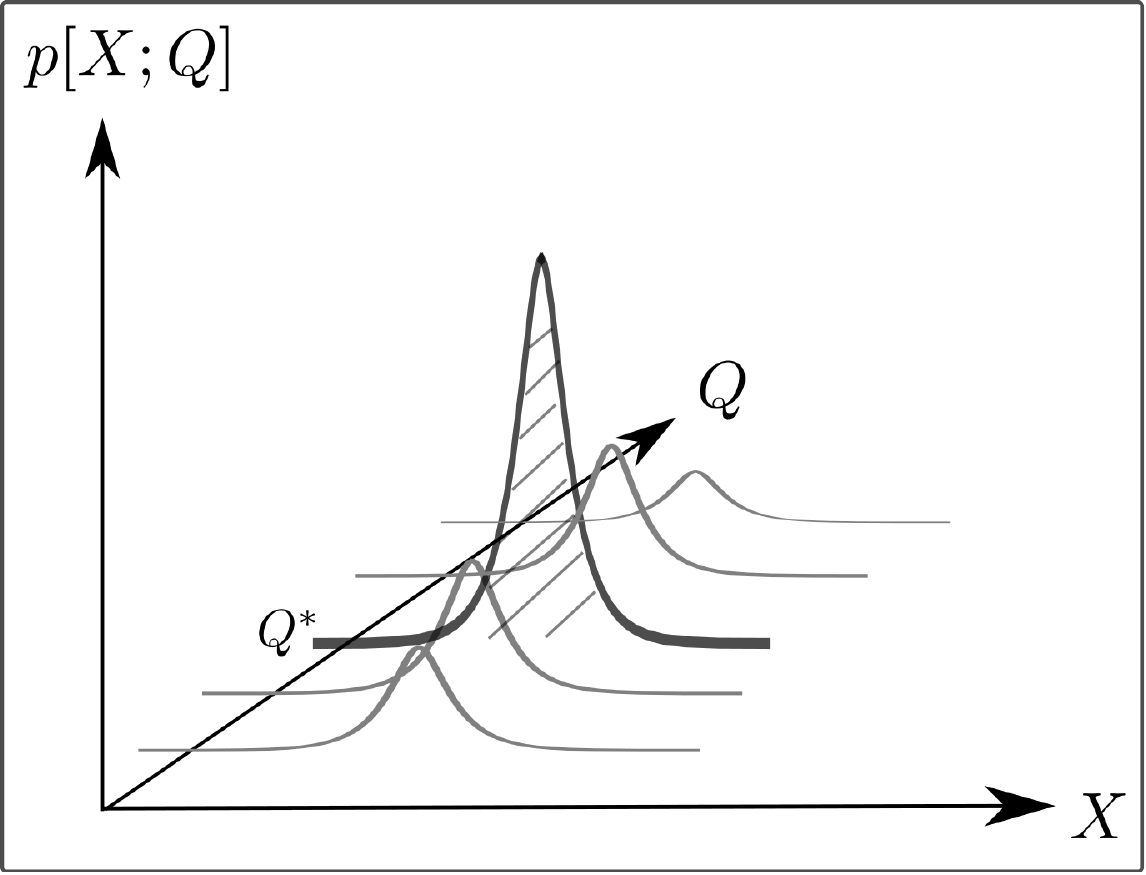}
\par\end{centering}
\caption{\textbf{Finding saddle point by maximizing contribution to probability:}
The contribution to the overall probability mass depends on the value
of the parameter $Q$, i.e. we seek to maximize $b[Q]:=\int\protect\D x\,p[\protect\bx;Q]$
\eqref{def_b}. The point at which the maximum is attained is denoted
as $Q^{\ast}$, the value $b[Q^{\ast}]$ is indicated by the hatched
area.\label{fig:intuition_saddle_point_max}}
\end{figure}

In the saddle point approximation \citep{Sompolinsky82_6860} we seek
the stationary point of the action determined by 
\begin{eqnarray}
0=\frac{\delta S[Q_{1},Q_{2}]}{\delta Q_{\{1,2\}}}=\frac{\delta}{\delta Q_{\{1,2\}}}\left(-\frac{N}{g^{2}}Q_{1}^{\T}Q_{2}+N\,\ln Z[Q_{1},Q_{2}]\right) & = & 0.\label{eq:def_saddle_Q1_Q2}
\end{eqnarray}
This procedure corresponds to finding the point in the space $(Q_{1},Q_{2})$
which provides the dominant contribution to the probability mass.
This can be seen by writing the probability functional as $p[\bx]=\iint\D Q_{1}\D Q_{2}\,p[\bx;Q_{1},Q_{2}]$
with 
\begin{align}
p[\bx;Q_{1},Q_{2}] & =\exp\left(-\frac{N}{g^{2}}Q_{1}^{\T}Q_{2}+\sum_{i}\ln\int\D\tx\,\exp\Big(S_{0}[x_{i},\tx]+\frac{1}{2}\tx^{\T}Q_{1}\tx+\phi(x_{i})^{\T}Q_{2}\phi(x_{i})\Big)\right)\nonumber \\
b[Q_{1},Q_{2}] & :=\int\D\bx\,p[\bx;Q_{1},Q_{2}],\label{eq:def_b}
\end{align}
where we defined $b[Q_{1},Q_{2}]$ as the contribution to the entire
probability mass for a given value of the auxiliary fields $Q_{1},Q_{2}$.
Maximizing $b$ therefore amounts to the condition \prettyref{eq:def_saddle_Q1_Q2},
illustrated in \prettyref{fig:intuition_saddle_point_max}. We here
used the convexity of the exponential function. 

A more formal argument to obtain \prettyref{eq:def_saddle_Q1_Q2}
proceeds by introducing the Legendre-Fenchel transform of $\ln\bar{Z}$
as 
\begin{eqnarray*}
\Gamma[Q_{1}^{\ast},Q_{2}^{\ast}] & := & \sup_{k,\tilde{k}}\,k^{\T}Q_{1}^{\ast}+\tilde{k}^{\T}Q_{2}^{\ast}-\ln\bar{Z}_{Q}[k,\tilde{k}],
\end{eqnarray*}
the vertex generating functional or effective action (see \prettyref{sec:Vertex-generating-function}
and \citep{ZinnJustin96,NegeleOrland98}). It holds that $\frac{\delta\Gamma}{\delta q_{1}}=k$
and $\frac{\delta\Gamma}{\delta q_{2}}=\tilde{k}$, the equations
of state \prettyref{eq:equation_of_state}. The tree-level or mean-field
approximation amounts to the approximation $\Gamma[Q_{1}^{\ast},Q_{2}^{\ast}]\simeq-S[q_{1},q_{2}]$,
as derived in \prettyref{eq:Gamma0}. The equations of state, for
vanishing sources $k=\tilde{k}=0$, therefore yield the saddle point
equations 
\begin{align*}
0=k=\frac{\delta\Gamma}{\delta Q_{1}^{\ast}} & =-\frac{\delta S}{\delta Q_{1}^{\ast}}\\
0=\tilde{k}=\frac{\delta\Gamma}{\delta Q_{2}^{\ast}} & =-\frac{\delta S}{\delta Q_{2}^{\ast}},
\end{align*}
identical to \prettyref{eq:def_saddle_Q1_Q2}. This more formal view
has the advantage of being straight forwardly extendable to loopwise
corrections (see \prettyref{sec:Loopwise-gamma}).

The functional derivative in the stationarity condition \prettyref{eq:def_saddle_Q1_Q2}
applied to $\ln Z[Q_{1},Q_{2}]$ produces an expectation value with
respect to the distribution \prettyref{eq:def_b}: the fields $Q_{1}$
and $Q_{2}$ here act as sources. This yields the set of two equations
\begin{eqnarray}
0=-\frac{N}{g^{2}}\,Q_{1}^{\ast}(s,t)+\frac{N}{Z}\,\left.\frac{\delta Z[Q_{1},Q_{2}]}{\delta Q_{2}(s,t)}\right|_{Q^{\ast}} & \leftrightarrow & Q_{1}^{\ast}(s,t)=g^{2}\left\langle \phi(x(s))\phi(x(t))\right\rangle _{Q^{\ast}}=:g^{2}C_{\phi(x)\phi(x)}(s,t)\label{eq:saddle_Q1_Q2}\\
0=-\frac{N}{g^{2}}\,Q_{2}^{\ast}(s,t)+\frac{N}{Z}\,\left.\frac{\delta Z[Q_{1},Q_{2}]}{\delta Q_{1}(s,t)}\right|_{Q^{\ast}} & \leftrightarrow & Q_{2}^{\ast}(s,t)=\frac{g^{2}}{2}\langle\tilde{x}(s)\tilde{x}(t)\rangle_{Q^{\ast}}=0,\nonumber 
\end{eqnarray}
where we defined the average autocorrelation function $C_{\phi(x)\phi(x)}(s,t)$
of the non-linearly transformed activity of the units. The second
saddle point $Q_{2}^{\ast}=0$ vanishes. This is because all expectation
values of only $\tx$ fields vanish, as shown in \prettyref{sec:Vanishing-response-field}.
This is true in the system that is not averaged over the disorder
and remains true in the averaged system, since the average is a linear
operation, so expectation values become averages of their counterparts
in the non-averaged system. If $Q_{2}$ was non-zero, it would alter
the normalization of the generating functional through mixing of retarded
and non-retarded time derivatives which then yield acausal response
functions \citep{Sompolinsky82_6860}. 

The expectation values $\langle\rangle_{Q^{\ast}}$ appearing in \eqref{saddle_Q1_Q2}
must be computed self-consistently, since the values of the saddle
points, by \prettyref{eq:Zbar}, influence the statistics of the fields
$\bx$ and $\tbx$, which in turn determines the functions $Q_{1}^{\ast}$
and $Q_{2}^{\ast}$ by \eqref{saddle_Q1_Q2}.

Inserting the saddle point solution into the generating functional
\eqref{Zbar} we get
\begin{align}
\bar{Z}^{\ast} & \propto\int\D x\int\D\tx\,\exp\,\Big(S_{0}[x,\tx]+\frac{g^{2}}{2}\tx^{\T}C_{\phi(x)\phi(x)}\tx\Big).\label{eq:Z_bar_star}
\end{align}
As the saddle points only couple to the sums of fields, the action
has the important property that it decomposes into a sum of actions
for individual, non-interacting units that feel a common field with
self-consistently determined statistics, characterized by its second
cumulant $C_{\phi(x)\phi(x)}$. Hence the saddle-point approximation
reduces the network to $N$ non-interacting units, or, equivalently,
a single unit system. The step from \prettyref{eq:Zbar_pre} to \prettyref{eq:Z_bar_star}
is therefore the replacement of the term $Q_{1}$, which depends on
the very realization of the $x$ by $Q_{1}^{\ast}$, which is a given
function, the form of which depends only on the statistics of the
$x$. This step allows the decoupling of the equations and again shows
the self-averaging nature of the problem: the particular realization
of the $x$ is not important; it suffices to know their statistics
that determines $Q_{1}^{\ast}$ to get the dominant contribution to
$Z$.

\begin{figure}
\begin{centering}
\includegraphics[scale=0.5]{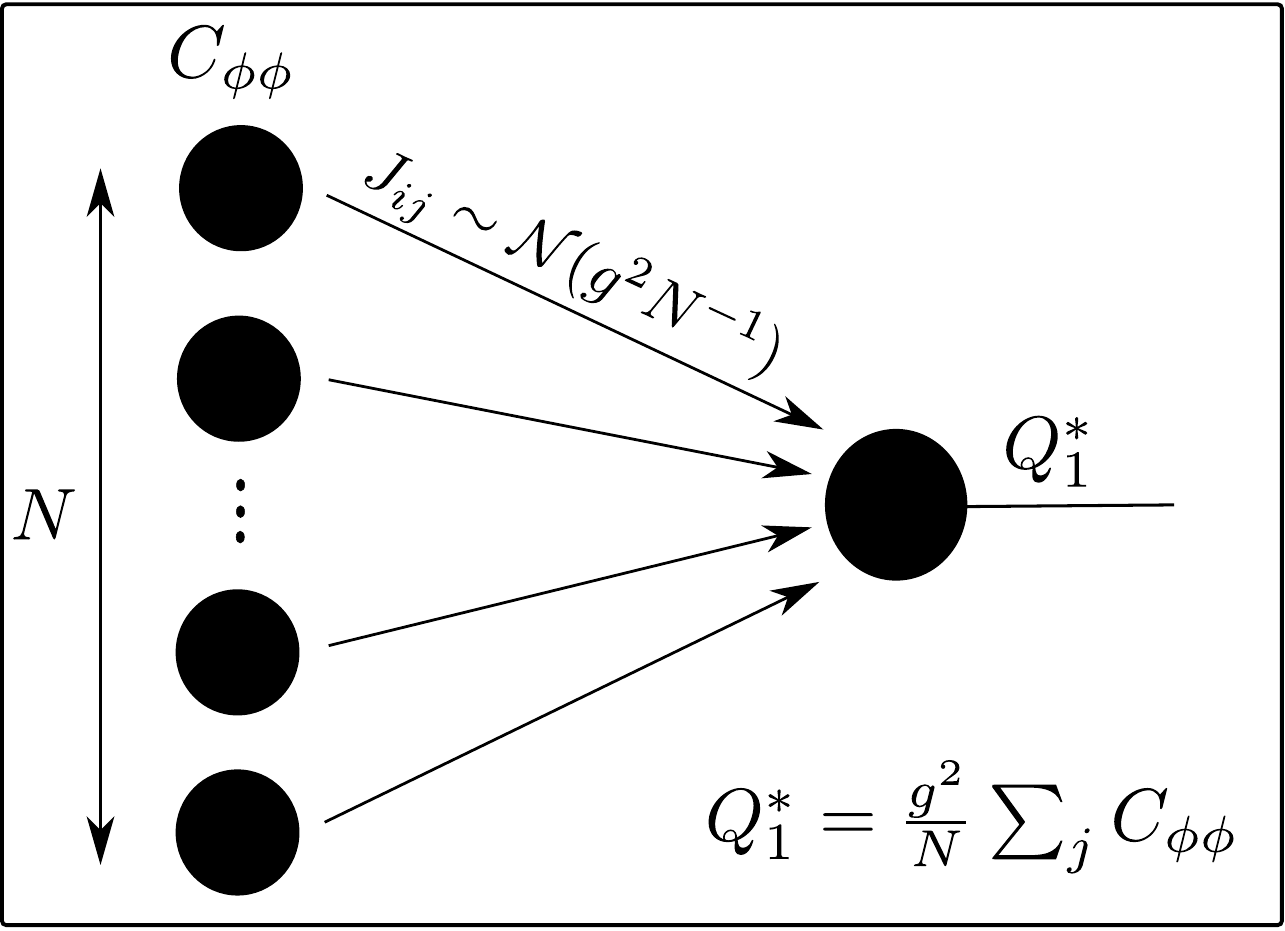}
\par\end{centering}
\caption{\textbf{Interpretation of the saddle point value $Q_{1}^{\ast}$ given
by eq. \prettyref{eq:saddle_Q1_Q2}:} The summed covariances $C_{\phi\phi}$
received by a neuron in the network, weighted by the synaptic couplings
$J_{ij}$, which have Gaussian statistics with variance $g^{2}N^{-1}$.\label{fig:intuition_saddle_point}}
\end{figure}

The second term in \prettyref{eq:Z_bar_star} is a Gaussian noise
with a two point correlation function $C_{\phi(x)\phi(x)}(s,t)$.
The physical interpretation is the noisy signal each unit receives
due to the input from the other $N$ units. Its autocorrelation function
is given by the summed autocorrelation functions of the output activities
$\phi(x_{i}(t))$ weighted by $g^{2}N^{-1}$, which incorporates
the Gaussian statistics of the couplings. This intuitive picture is
shown in \prettyref{fig:intuition_saddle_point}.

The interpretation of the noise can be appreciated by explicitly considering
the moment generating functional of a Gaussian noise with a given
autocorrelation function $C(t,t^{\prime})$, which leads to the cumulant
generating functional $\ln Z_{\zeta}[\tilde{x}]$ that appears in
the exponent of \eqref{Z_bar_star} and has the form
\begin{eqnarray*}
\ln\,Z_{\zeta}[\tilde{x}] & = & \ln\langle\exp\left(\int\tx(t)\,\zeta(t)\,dt\right)\rangle\\
 & = & \frac{1}{2}\int_{-\infty}^{\infty}\int_{-\infty}^{\infty}\tilde{x}(t)\,C(t,t^{\prime})\,\tilde{x}(t^{\prime})\,dt\,dt^{\prime}\\
 & = & \frac{1}{2}\tx^{\T}\,C\,\tx.
\end{eqnarray*}
Note that the effective noise term only has a non-vanishing second
cumulant. This means the effective noise is Gaussian, as the cumulant
generating function is quadratic. It couples pairs of time points
that are correlated.

This is the starting point in \citep[ eq. (3)]{Sompolinsky88_259},
stating that the effective mean-field dynamics of the network is given
by that of a single unit
\begin{eqnarray}
(\partial_{t}+1)\,x(t) & = & \eta(t)\label{eq:effective_mf_Langevin}
\end{eqnarray}
driven by a Gaussian noise $\eta=\zeta+\frac{d\xi}{dt}$ with autocorrelation
$\langle\eta(t)\eta(s)\rangle=g^{2}\,C_{\phi(x)\phi(x)}(t,s)+D\delta(t-s)$.
In the cited paper the white noise term $\propto D$ is absent, though.

We may either formally invert the operator $-S^{(2)}$ corresponding
to the action \prettyref{eq:Z_bar_star} to obtain the propagators
of the system as in the case of the Ornstein-Uhlenbeck processes \prettyref{sec:Ornstein-Uhlenbeck}.
Since we only need the propagator $\Delta_{xx}(t,s)=\langle x(t)x(s)\rangle=:C_{xx}(t,s)$
here, we may alternatively multiply the equation \prettyref{eq:effective_mf_Langevin}
for time points $t$ and $s$ and take the expectation value with
respect to the noise $\eta$ on both sides, which leads to
\begin{align}
\left(\partial_{t}+1\right)\left(\partial_{s}+1\right)C_{xx}(t,s) & =g^{2}\,C_{\phi(x)\phi(x)}(t,s)+D\delta(t-s),\label{eq:cov_xx_diffeq}
\end{align}
where we defined the covariance function of the activities $C_{xx}(t,s):=\langle x(t)x(s)\rangle$.
In the next section we will rewrite this equation into an equation
of a particle in a potential.\textbf{}

\subsection{Stationary statistics: Self-consistent autocorrelation of as motion
of a particle in a potential\label{sub:particle_motion}}

We are now interested in the stationary statistics of the system,
i.e. $C_{xx}(t,s)=:c(t-s)$. The inhomogeneity in \eqref{cov_xx_diffeq}
is then also time-translation invariant, $C_{\phi(x)\phi(x)}(t+\tau,t)$
is only a function of $\tau$. Therefore the differential operator
$\left(\partial_{t}+1\right)\left(\partial_{s}+1\right)c(t-s)$, with
$\tau=t-s$, simplifies to $(-\partial_{\tau}^{2}+1)\,c(\tau)$ so
we get
\begin{eqnarray}
(-\partial_{\tau}^{2}+1)\,c(\tau) & = & g^{2}\,C_{\phi(x)\phi(x)}(t+\tau,t)+D\,\delta(\tau).\label{eq:diffeq_auto}
\end{eqnarray}
Once \prettyref{eq:diffeq_auto} is solved, we know the covariance
function $c(\tau)$ between two time points $\tau$ apart as well
as the variance $c(0)=:c_{0}$. Since by the saddle point approximation
in \prettyref{sub:Disorder-average} the expression \prettyref{eq:Z_bar_star}
is the generating functional of a Gaussian theory, the $x$ are zero
mean Gaussian random variables. We might call the field $x(t)=:x_{1}$
and $x(t+\tau)=:x_{2}$, which follow the distribution 
\begin{align*}
(x_{1},x_{2}) & \sim\N\Big(0,\left(\begin{array}{cc}
c(0) & c(\tau)\\
c(\tau) & c(0)
\end{array}\right)\Big).
\end{align*}
 Consequently the second moment completely determines the distribution.
We can therefore obtain $C_{\phi(x)\phi(x)}(t,s)=g^{2}f_{\phi}(c(\tau),c(0))$
with 
\begin{align}
f_{u}(c,c_{0}) & =\langle u(x_{1})u(x_{2})\rangle_{(x_{1},x_{2})\sim\N\Big(0,\left(\begin{array}{cc}
c_{0} & c\\
c & c_{0}
\end{array}\right)\Big)}\label{eq:def_f}\\
 & =\iint\,u\Bigg(\sqrt{c_{0}-\frac{c^{2}}{c_{0}}}\,z_{1}+\tfrac{c}{\sqrt{c_{0}}}\,z_{2}\Bigg)u\Bigg(\sqrt{c_{0}}\,z_{2}\Bigg)\,Dz_{1}Dz_{2}
\end{align}
with the Gaussian integration measure $Dz=\exp(-z^{2}/2)/\sqrt{2\pi}\,dz$
and for a function $u(x)$. Here, the two different arguments of $u(x)$
are by construction Gaussian with zero mean, variance $c(0)=c_{0}$,
and covariance $c(\tau)$. Note that \prettyref{eq:def_f} reduces
to one-dimensional integrals for $f_{u}(c_{0},c_{0})=\langle u(x)^{2}\rangle$
and $f_{u}(0,c_{0})=\langle u(x)\rangle^{2}$, where $x$ has zero
mean and variance $c_{0}$.

We note that $f_{u}(c(\tau),c_{0})$ in \prettyref{eq:def_f} only
depends on $\tau$ through $c(\tau)$. We can therefore obtain it
from the ``potential'' $g^{2}f_{\Phi}(c(\tau),c_{0})$ by 
\begin{eqnarray}
C_{\phi(x)\phi(x)}(t+\tau,t) & =: & \frac{\partial}{\partial c}\,g^{2}f_{\Phi}(c(\tau),c_{0})\label{eq:def_potential_v}
\end{eqnarray}
where $\Phi$ is the integral of $\phi$, i.e. $\Phi(x)=\int_{0}^{x}\phi(x)\,dx=\ln\,\cosh(x)$.
The property $\frac{\partial}{\partial c}\,f_{\Phi}(c,c_{0})=f_{\Phi^{\prime}}(c(\tau),c_{0})$
is known as Price's theorem \citep{PapoulisProb}. \ifthenelse{\boolean{lecture}}{It
is shown in the exercises in \prettyref{sub:exercises_disorder_SCS}.}
Note that the representation in \prettyref{eq:def_f} differs from
the one used in \citep[eq. (7)]{Sompolinsky88_259}. The expression
used here is also valid for negative $c(\tau)$ in contrast to the
original formulation. We can therefore express the differential equation
for the autocorrelation with the definition of the potential $V$
\begin{eqnarray}
V(c;c_{0}) & := & -\frac{1}{2}c^{2}+g^{2}f_{\Phi}(c(\tau),c_{0})-g^{2}f_{\Phi}(0,c_{0}),\label{eq:def_potential_V}
\end{eqnarray}
where the subtraction of the last constant term is an arbitrary choice
that ensures that $V(0;c_{0})=0$. The equation of motion \prettyref{eq:diffeq_auto}
therefore takes the form

\begin{eqnarray}
\partial_{\tau}^{2}\,c(\tau) & = & -V^{\prime}(c(\tau);c_{0})-D\,\delta(\tau),\label{eq:eq_motion_cxx}
\end{eqnarray}
so it describes the motion of a particle in a (self-consistent) potential
$V$ with derivative $V^{\prime}=\frac{\partial}{\partial c}V$. The
$\delta$-distribution on the right hand side causes a jump in the
velocity that changes from $\frac{D}{2}$ to $-\frac{D}{2}$ at $\tau=0$,
because $c$ is symmetric ($c(\tau)=c(-\tau)$) and hence $\dot{c}(\tau)=-\dot{c}(-\tau)$
and moreover the term $-V^{\prime}(c(\tau);c_{0})$ does not contribute
to the kink. The equation must be solved self-consistently, as the
initial value $c_{0}$ determines the effective potential $V(\cdot,c_{0})$
via \eqref{def_potential_V}. The second argument $c_{0}$ indicates
this dependence.

\begin{figure}
\begin{centering}
\includegraphics[scale=2]{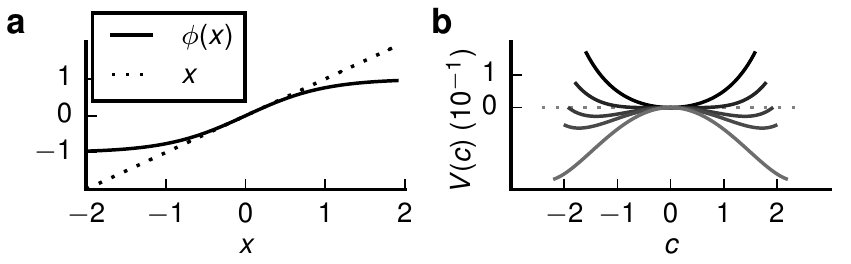}
\par\end{centering}
\caption{\textbf{Effective potential for the noise-less case $D=0$. a }The
gain function $\phi(x)=\tanh(x)$ close to the origin has unit slope.
Consequently, the integral of the gain function $\Phi(x)=\ln\cosh(x)$
close to origin has the same curvature as the parabola $\frac{1}{2}x^{2}$.
\textbf{b }Self-consistent potential for $g=2$ and different values
of $c_{0}=1.6,1.8,1.924,2,2.2$ (from black to light gray). The horizontal
gray dotted line indicates the identical levels of initial and finial
potential energy for the self-consistent solution $V(c_{0};c_{0})=0$,
corresponding to the initial value that leads to a monotonously decreasing
autocovariance function that vanishes for $\tau\to\infty$.\label{fig:effective_potential}}
\end{figure}

The gain function $\phi(x)=\tanh(x)$ is shown in \prettyref{fig:effective_potential}a,
while \prettyref{fig:effective_potential}b shows the self-consistent
potential for the noiseless case $D=0$.

The potential is formed by the interplay of two opposing terms. The
downward bend is due to $-\frac{1}{2}c^{2}$. The term $g^{2}f_{\Phi}(c;c_{0})$
is bent upwards. We get an estimate of this term from its derivative
$g^{2}f_{\phi}(c,c_{0})$: Since $\phi(x)$ has unit slope at $x=0$
(see \prettyref{fig:effective_potential}a), for small amplitudes
$c_{0}$ the fluctuations are in the linear part of $\phi$, so $g^{2}f_{\phi}(c,c_{0})\simeq g^{2}c$
for all $c\le c_{0}$. Consequently, the potential $g^{2}f_{\Phi}(c,c_{0})=\int_{0}^{c}g^{2}f_{\phi}(c^{\prime},c_{0})\,dc^{\prime}\stackrel{c<c_{0}\ll1}{\simeq}g^{2}\frac{1}{2}c^{2}$
has a positive curvature at $c=0$.

For $g<1$, the parabolic part dominates for all $c_{0}$, so that
the potential is bent downwards and the only bounded solution in the
noiseless case $D=0$ of \prettyref{eq:eq_motion_cxx} is the vanishing
solution $c(t)\equiv0$.

For $D>0$, the particle may start at some point $c_{0}>0$ and, due
to its initial velocity, reach the point $c(\infty)=0$. Any physically
reasonable solution must be bounded. In this setting, the only possibility
is a solution that starts at a position $c_{0}>0$ with the same initial
energy $V(c_{0};c_{0})+E_{\mathrm{kin}}^{0}$ as the final potential
energy $V(0;c_{0})=0$ at $c=0$. The initial kinetic energy is given
by the initial velocity $\dot{c}(0+)=-\frac{D}{2}$ as $E_{\mathrm{kin}}^{(0)}=\frac{1}{2}\dot{c}(0+)^{2}=\frac{D^{2}}{8}$.
This condition ensures that the particle, starting at $\tau=0$ at
the value $c_{0}$ for $\tau\to\infty$ reaches the local maximum
of the potential at $c=0$; the covariance function decays from $c_{0}$
to zero.

For $g>1$, the term $g^{2}f_{\Phi}(c;c_{0})$ can start to dominate
the curvature close to $c\simeq0$: the potential in \prettyref{fig:effective_potential}b
is bent upwards for small $c_{0}$. For increasing $c_{0}$, the fluctuations
successively reach the shallower parts of $\phi$, hence the slope
of $g^{2}f_{\phi}(c,c_{0})$ diminishes, as does the curvature of
its integral, $g^{2}f_{\Phi}(c;c_{0})$. With increasing $c_{0}$,
the curvature of the potential at $c=0$ therefore changes from positive
to negative.

In the intermediate regime, the potential assumes a double well shape.
Several solutions exist in this case. One can show that the only stable
solution is the one that decays to $0$ for $\tau\to\infty$ \citep{Sompolinsky88_259,Crisanti18_062120}.
In the presence of noise $D>0$ this assertion is clear due to the
decorrelating effect of the noise, but it remains true also in the
noiseless case.

By the argument of energy conservation, the corresponding value $c_{0}$
can be found numerically as the root of 
\begin{eqnarray}
V(c_{0};c_{0})+E_{\mathrm{kin}}^{(0)} & \stackrel{!}{=} & 0\label{eq:initial_c0}\\
E_{\mathrm{kin}}^{(0)} & = & \frac{D^{2}}{8},\nonumber 
\end{eqnarray}
for example with a simple bisectioning algorithm.

\begin{figure}
\begin{centering}
\includegraphics[scale=2]{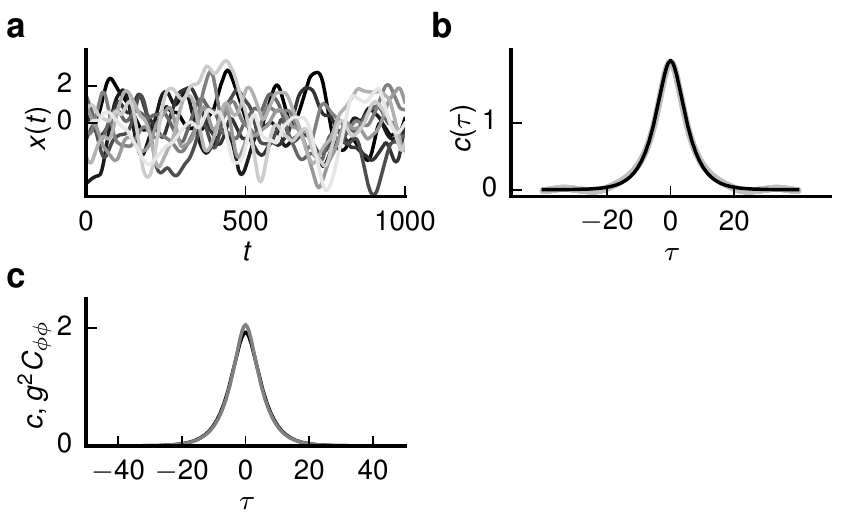}
\par\end{centering}
\caption{\textbf{Self-consistent autocovariance function from dynamic mean-field
theory in the noise-less case.} Random network of $5000$ Gaussian
coupled units with with $g=2$ and vanishing noise $D=0$. \textbf{a}
Activity of the first $10$ units as function of time. \textbf{b}
Self-consistent solution of covariance $c(\tau)$ (black) and result
from simulation (gray). The theoretical result is obtained by first
solving \prettyref{eq:initial_c0} for the initial value $c_{0}$
and then integrating \prettyref{eq:acf_system}. \textbf{c }Self-consistent
solution (black) as in b and $C_{\phi\phi}(\tau)=g^{2}f_{\phi}(c(\tau),c_{0})$
given by \prettyref{eq:def_potential_v} (gray). Duration of simulation
$T=1000$ time steps with resolution $h=0.1$ each. Integration of
\prettyref{eq:diffeq_motion} by forward Euler method.\label{fig:Self-consistent-solution-corr}}
\end{figure}

The corresponding shape of the autocovariance function then follows
a straight forward integration of the differential equation \prettyref{eq:eq_motion_cxx}.
Rewriting the second order differential equation into a coupled set
of first order equations, introducing $\partial_{\tau}c=:y$, we get
for $\tau>0$
\begin{eqnarray}
\partial_{\tau}\left(\begin{array}{c}
y(\tau)\\
c(\tau)
\end{array}\right) & = & \left(\begin{array}{c}
c-g^{2}f_{\phi}(c,c_{0})\\
y(\tau)
\end{array}\right)\label{eq:acf_system}\\
\text{with initial condition}\nonumber \\
\left(\begin{array}{c}
y(0)\\
c(0)
\end{array}\right) & = & \left(\begin{array}{c}
-\frac{D}{2}\\
c_{0}
\end{array}\right).\nonumber 
\end{eqnarray}
The solution of this equation in comparison to direct simulation is
shown in \prettyref{fig:Self-consistent-solution-corr}. Note that
the covariance function of the input to a unit, $C_{\phi\phi}(\tau)=g^{2}f_{\phi}(c(\tau),c_{0})$,
bares strong similarities to the autocorrelation $c$, shown in \prettyref{fig:Self-consistent-solution-corr}c:
The suppressive effect of the non-linear, saturating gain function
is compensated by the variance of the connectivity $g^{2}>1$, so
that a self-consistent solution is achieved.

\subsection{Transition to chaos}

In this section, we will derive the largest Lyapunov exponent of the
system that allows us to assess the conditions under which the system
undergoes a transition into the chaotic regime. We will see that we
can also conclude from this calculation that the system, to leading
order in $N$ in the large $N$ limit, is indeed self-averaging. We
will here see that the dominant contribution to the moment-generating
function of the replicated system \prettyref{eq:replica_Z} indeed
factorizes.

\subsection{Assessing chaos by a pair of identical systems\label{sub:pair_of_systems}}

We now aim to study whether the dynamics is chaotic or not. To this
end, we consider a pair of identically prepared systems, in particular
with identical coupling matrix $\bJ$ and, for $D>0$, also the same
realization of the Gaussian noise. We distinguish the dynamical variables
$x^{\alpha}$ of the two systems by superscripts $\alpha\in\{1,2\}$.

\begin{figure}
\begin{centering}
\includegraphics[scale=2]{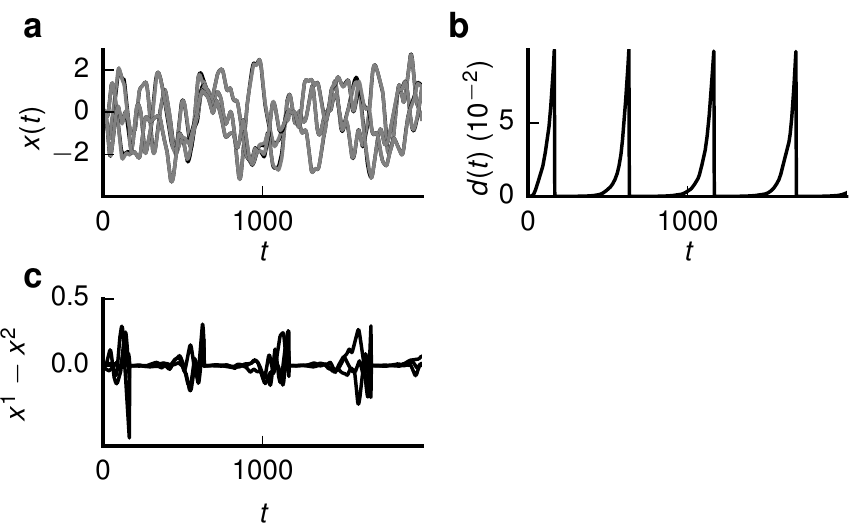}
\par\end{centering}
\caption{\textbf{Chaotic evolution. a }Dynamics of two systems starting at
similar initial conditions for chaotic case with $g=2$, $N=5000,$
$D=0.01$. Trajectories of three units shown for the unperturbed (black)
and the perturbed system (gray). \textbf{b }Absolute average squared
distance $d(t)$ given by \prettyref{eq:def_d_empirical} of the two
systems. \textbf{c }Difference $x_{1}-x_{2}$ for the first three
units. The second system is reset to the state of the first system
plus a small random displacement as soon as $d(t)>0.1$. Other parameters
as in \prettyref{fig:Self-consistent-solution-corr}. \label{fig:Chaotic-evolution-pair}}
\end{figure}

Let us briefly recall that the dynamical mean-field theory describes
empirical population-averaged quantities for a single network realization
(due to self-averaging). Hence, for large $N$ we expect that 
\begin{align*}
\frac{1}{N}\sum_{i=1}^{N}x_{i}^{\alpha}(t)x_{i}^{\beta}(s) & \simeq c^{\alpha\beta}(t,s)
\end{align*}
holds for most network realizations. To study the stability of the
dynamics with respect to perturbations of the initial conditions we
consider the population-averaged (mean-)squared distance between the
trajectories of the two copies of the network:
\begin{align}
\frac{1}{N}||x^{1}(t)-x^{2}(t)||^{2} & =\frac{1}{N}\sum_{i=1}^{N}\left(x_{i}^{1}(t)-x_{i}^{2}(t)\right)^{2}\label{eq:def_d_empirical}\\
 & =\frac{1}{N}\sum_{i=1}^{N}\left(x_{i}^{1}(t)\right)^{2}+\frac{1}{N}\sum_{i=1}^{N}\left(x_{i}^{2}(t)\right)^{2}-\frac{2}{N}\sum_{i=1}^{N}x_{i}^{1}(t)x_{i}^{2}(t)\nonumber \\
 & \simeq c^{11}(t,t)+c^{22}(t,t)-2c^{12}(t,t)\,.\nonumber 
\end{align}
This idea has also been employed in \citep{Derrida87_721}. Therefore,
we define the mean-field mean-squared distance between the two copies:
\begin{align}
d(t,s) & :=c^{11}(t,s)+c^{22}(t,s)-c^{12}(t,s)-c^{21}(t,s)\,,\label{eq:mean-squared-distance-1}
\end{align}
which gives for equal time arguments the actual mean-squared distance
$d(t):=d(t,t)\,$. Our goal is to find the temporal evolution of $d(t,s)\,$.
The time evolution of a pair of systems in the chaotic regime with
slightly different initial conditions is shown in \prettyref{fig:Chaotic-evolution-pair}.
Although the initial displacement between the two systems is drawn
independently for each of the four shown trials, the divergence of
$d(t)$ has a stereotypical form, which seems to be dominated by one
largest Lyapunov exponent. The aim of the remainder of this section
is to find this rate of divergence.

To derive an equation of motion for $d(t,s)$ it is again convenient
to define a generating functional that captures the joint statistics
of two systems and in addition allows averaging over the quenched
disorder \citep[see also ][Appendix 23, last remark]{ZinnJustin96}.

The generating functional is defined in analogy to the single system
\eqref{def_S0}
\begin{align}
Z[\{\bl^{\alpha},\tbj^{\alpha}\}_{\alpha\in\{1,2\}}](\bJ) & =\Pi_{\alpha=1}^{2}\Big\{\int\D\bx^{\alpha}\int\D\tbx^{\alpha}\,\exp\Big(\tbx^{\alpha\T}\big((\partial_{t}+1)\,\bx^{\alpha}-\sum_{j}\bJ\phi(\bx^{\alpha})\big)+\bl^{\alpha\T}\bx^{\alpha}+\tbj^{\alpha\T}\tbx^{\alpha}\Big)\Big\}\times\nonumber \\
 & \times\exp\Big(\frac{D}{2}\,(\tbx^{1}+\tbx^{2})^{\T}(\tbx^{1}+\tbx^{2})\Big)\Big\},\label{eq:Z_pair_pre}
\end{align}
where the last term is the moment generating functional due to the
white noise that is common to both subsystems. We note that the coupling
matrix $\bJ$ is the same in both subsystems as well. Using the notation
analogous to \eqref{def_S0} and collecting the terms that affect
each individual subsystem in the first, the common term in the second
line, we get

\begin{align}
Z[\{\bl^{\alpha},\tbj^{\alpha}\}_{\alpha\in\{1,2\}}](\bJ) & =\Pi_{\alpha=1}^{2}\Big\{\int\D\bx^{\alpha}\int\D\tbx^{\alpha}\,\exp\Big(S_{0}[\bx^{\alpha},\tbx^{\alpha}]-\tbx^{\alpha\T}\bJ\phi\left(\bx^{\alpha}\right)+\bl^{\alpha\T}\bx^{\alpha}+\tbj^{\alpha\T}\tbx^{\alpha}\Big)\Big\}\nonumber \\
 & \times\exp\left(D\tbx^{1\T}\tbx^{2}\right).\label{eq:Z_pair}
\end{align}
Here the term in the last line appears due to the mixed product of
the response fields in \eqref{Z_pair_pre}.

We will now perform the average over realizations in $\bJ$, as in
\prettyref{sub:Disorder-average} eq. \eqref{completion_of_square}.
We therefore need to evaluate the Gaussian integral
\begin{align}
 & \int dJ_{ij}\mathcal{N}(0,\frac{g^{2}}{N},J_{ij})\,\exp\left(-J_{ij}\sum_{\alpha=1}^{2}\tilde{x}_{i}^{\alpha\T}\phi(x_{j}^{\alpha})\right)\nonumber \\
 & =\exp\left(\frac{g^{2}}{2N}\sum_{\alpha=1}^{2}\left(\tilde{x}_{i}^{\alpha\T}\phi(x_{j}^{\alpha})\right)^{2}\right)\nonumber \\
 & \times\exp\left(\frac{g^{2}}{N}\,\tilde{x}_{i}^{1\T}\phi(x_{j}^{1})\,\tilde{x}_{i}^{2\T}\phi(x_{j}^{2})\right).\label{eq:quenched_avg_pair}
\end{align}
Similar as for the Gaussian integral over the common noises that gave
rise to the coupling term between the two systems in the second line
of \eqref{Z_pair}, we here obtain a coupling term between the two
systems, in addition to the terms that only include variables of a
single subsystem in the second last line. Note that the two coupling
terms are different in nature. The first, due to common noise, represents
common temporal fluctuations injected into both systems. The second
is static in its nature, as it arises from the two systems having
the same coupling $\bJ$ in each of their realizations that enter
the expectation value. The terms that only affect a single subsystem
are identical to those in \prettyref{eq:Zbar_pre}. We treat these
terms as before and here concentrate on the mixed terms, which we
rewrite (including the $\sum_{i\neq j}$ in \prettyref{eq:Z_pair}
and using our definition $\tilde{x}_{i}^{\alpha\T}\phi(x_{j}^{\alpha})=\int dt\,\tilde{x}_{i}^{\alpha}(t)\phi(x_{j}^{\alpha}(t))\,dt$)
as 
\begin{align}
 & \exp\Big(\frac{g^{2}}{N}\sum_{i\neq j}\,\tilde{x}_{i}^{1\T}\phi(x_{j}^{1})\,\tilde{x}_{i}^{2\T}\phi(x_{j}^{2})\Big)\label{eq:mixed_avg_pair}\\
= & \exp\Big(\iint\,\sum_{i}\tilde{x}_{i}^{1}(s)\tilde{x}_{i}^{2}(t)\underbrace{\frac{g^{2}}{N}\sum_{j}\phi(x_{j}^{1}(s))\,\phi(x_{j}^{2}(t))}_{=:T_{1}(s,t)}\,ds\,dt\Big)+O(N^{-1}),\nonumber 
\end{align}
where we included the self coupling term $i=j$, which is only a subleading
correction of order $N^{-1}$.

We now follow the steps in \prettyref{sub:Disorder-average} and introduce
three pairs of auxiliary variables. The pairs $Q_{1}^{\alpha},Q_{2}^{\alpha}$
are defined as before in \prettyref{eq:def_Q1} and \prettyref{eq:Hubbard_Stratonovich},
but for each subsystem, while the pair $T_{1},T_{2}$ decouples the
mixed term \prettyref{eq:mixed_avg_pair} by defining
\begin{align*}
T_{1}(s,t) & :=\frac{g^{2}}{N}\sum_{j}\phi(x_{j}^{1}(s))\,\phi(x_{j}^{2}(t)),
\end{align*}
as indicated by the curly brace in \prettyref{eq:mixed_avg_pair}.

Taken together, we can therefore rewrite the generating functional
\prettyref{eq:Z_pair} averaged over the couplings as
\begin{align}
\bar{Z}[\{\bl^{\alpha},\tbj^{\alpha}\}_{\alpha\in\{1,2\}}] & :=\langle Z[\{\bl^{\alpha},\tbj^{\alpha}\}_{\alpha\in\{1,2\}}](\bJ)\rangle_{\bJ}\label{eq:Zbar_pair_HS}\\
 & =\Pi_{\alpha=1}^{2}\left\{ \int\D Q_{1}^{\alpha}\int\D Q_{2}^{\alpha}\right\} \int\D T_{1}\int\D T_{2}\,\exp\Big(\Omega[\{Q_{1}^{\alpha},Q_{2}^{\alpha}\}_{\alpha\in\{1,2\}},T_{1},T_{2}]\Big)\nonumber \\
\Omega[\{Q_{1}^{\alpha},Q_{2}^{\alpha}\}_{\alpha\in\{1,2\}},T_{1},T_{2}] & :=-\sum_{\alpha=1}^{2}Q_{1}^{\alpha\T}Q_{2}^{\alpha}-T_{1}^{\T}T_{2}+\ln\,Z^{12}[\{Q_{1}^{\alpha},Q_{2}^{\alpha}\}_{\alpha\in\{1,2\}},T_{1},T_{2}]\nonumber \\
Z^{12}[\{Q_{1}^{\alpha},Q_{2}^{\alpha}\}_{\alpha\in\{1,2\}},T_{1},T_{2}] & =\Pi_{\alpha=1}^{2}\Big\{\int\D\bx^{\alpha}\int\D\tbx^{\alpha}\,\exp\Big(S_{0}[\bx^{\alpha},\tbx^{\alpha}]+\bl^{\alpha\T}\bx^{\alpha}+\tbj^{\alpha\T}\tbx^{\alpha}+\frac{1}{2}\tbx^{\alpha\T}Q_{1}^{\alpha}\tbx^{\alpha}+\frac{g^{2}}{N}\phi(\bx{}^{\alpha})^{\T}Q_{2}^{\alpha}\phi(\bx{}^{\alpha})\Big)\Big\}\nonumber \\
 & \times\exp\left(\tbx^{1\T}\left(T_{1}+D\right)\tbx^{2}+\frac{g^{2}}{N}\phi(\bx^{1})^{\T}T_{2}\phi(\bx^{2})\Big)\right).\nonumber 
\end{align}
We now determine, for vanishing sources, the fields $Q_{1}^{\alpha}$,
$Q_{2}^{\alpha}$, $T_{1}$, $T_{2}$ at which the contribution to
the integral is maximal by requesting $\frac{\delta\Omega}{\delta Q_{1,2}^{\alpha}}=\frac{\delta\Omega}{\delta T_{1,2}}\stackrel{!}{=}0$
for the exponent $\Omega$ of \eqref{Zbar_pair_HS}. Here again the
term $\ln\,Z^{12}$ plays the role of a cumulant generating function
and the fields $Q_{1}^{\alpha},Q_{2}^{\alpha},T_{1},T_{2}$ play the
role of sources, each bringing down the respective factor they multiply.
We denote the expectation value with respect to this functional as
$\langle\circ\rangle_{Q^{\ast},T^{\ast}}$ and obtain the self-consistency
equations
\begin{align}
Q_{1}^{\alpha\ast}(s,t) & =\frac{1}{Z^{12}}\,\frac{\delta Z^{12}}{\delta Q_{2}^{\alpha}(s,t)}=\frac{g^{2}}{N}\,\sum_{j}\langle\phi(x_{j}^{\alpha})\phi(x_{j}^{\alpha})\rangle_{Q^{\ast},T^{\ast}}\label{eq:saddle_pair}\\
Q_{2}^{\alpha\ast}(s,t) & =0\nonumber \\
T_{1}^{\ast}(s,t) & =\frac{1}{Z^{12}}\,\frac{\delta Z^{12}}{\delta T_{2}(s,t)}=\frac{g^{2}}{N}\,\sum_{j}\langle\phi(x_{j}^{1})\phi(x_{j}^{2})\rangle_{Q^{\ast},T^{\ast}}\nonumber \\
T_{2}^{\ast}(s,t) & =0.\nonumber 
\end{align}
The generating functional at the saddle point is therefore
\begin{align}
\bar{Z}^{\ast}[\{\bl^{\alpha},\tbj^{\alpha}\}_{\alpha\in\{1,2\}}] & =\iint\Pi_{\alpha=1}^{2}\D\bx^{\alpha}\D\tbx^{\alpha}\,\exp\Big(\sum_{\alpha=1}^{2}S_{0}[\bx^{\alpha},\tbx^{\alpha}]+\bl^{\alpha\T}\bx^{\alpha}+\tbj^{\alpha\T}\tbx^{\alpha}+\frac{1}{2}\tbx^{\alpha\T}Q_{1}^{\alpha\ast}\tbx^{\alpha}\Big)\times\nonumber \\
 & \times\exp\left(\tbx^{\alpha\T}\left(T_{1}^{\ast}+D\right)\tbx^{\beta}\right).\label{eq:Z_bar_pair_ast}
\end{align}
We make the following observations: 
\begin{enumerate}
\item The two subsystems $\alpha=1,2$ in the first line of \prettyref{eq:Z_bar_pair_ast}
have the same form as in \eqref{Z_bar_star}. This has been expected,
because the absence of any physical coupling between the two systems
implies that the marginal statistics of the activity in one system
cannot be affected by the mere presence of the second, hence also
their saddle points $Q_{1,2}^{\alpha}$ must be the same as in \eqref{Z_bar_star}.
\item The entire action is symmetric with respect to interchange of any
pair of unit indices. So we have reduced the system of $2N$ units
to a system of $2$ units.
\item If the term in the second line  of \eqref{Z_bar_pair_ast} was absent,
the statistics in the two systems would be independent. Two sources,
however, contribute to the correlations between the systems: The common
Gaussian white noise that gave rise to the term $\propto D$ and the
non-white Gaussian noise due to a non-zero value of the auxiliary
field $T_{1}^{\ast}(s,t)$.
\item Only products of pairs of fields appear in \eqref{Z_bar_pair_ast},
so that the statistics of the $x^{\alpha}$ is Gaussian.
\end{enumerate}
As for the single system, we can express the joint system by a pair
of dynamic equations
\begin{align}
\left(\partial_{t}+1\right)x^{\alpha}(t) & =\eta^{\alpha}(t)\quad\alpha\in\{1,2\}\label{eq:_effective_pair_eq}
\end{align}
together with a set of self-consistency equations for the statistics
of the noises $\eta^{\alpha}$ following from \eqref{saddle_pair}
\begin{align}
\langle\eta^{\alpha}(s)\,\eta^{\beta}(t)\rangle & =D\delta(t-s)+g^{2}\,\langle\phi(x^{\alpha}(s))\phi(x^{\beta}(t))\rangle.\label{eq:effective_pair_noise}
\end{align}
Obviously, this set of equations \eqref{_effective_pair_eq} and \eqref{effective_pair_noise}
marginally for each subsystem admits the same solution as determined
in \prettyref{sub:particle_motion}. Moreover, the joint system therefore
also possesses the fixed point $x^{1}(t)\equiv x^{2}(t)$, where the
activities in the two subsystems are identical, i.e. characterized
by $c^{12}(t,s)=c^{11}(t,s)=c^{22}(t,s)$ and consequently $d(t)\equiv0\,\forall t$
\prettyref{eq:mean-squared-distance-1}.

We will now investigate if this fixed point is stable. If it is, this
implies that any perturbation of the system will relax such that the
two subsystems are again perfectly correlated. If it is unstable,
the distance between the two systems may increase, indicating chaotic
dynamics.

We already know that the autocorrelation functions in the subsystems
are stable and each obey the equation of motion \eqref{eq_motion_cxx}.
We could use the formal approach, writing the Gaussian action as a
quadratic form and determine the correlation and response functions
as the inverse, or Green's function, of this bi-linear form. Here,
instead we employ a simpler approach: we multiply the equation \eqref{_effective_pair_eq}
for $\alpha=1$ and $\alpha=2$ and take the expectation value on
both sides, which leads to
\begin{align*}
\left(\partial_{t}+1\right)\left(\partial_{s}+1\right)\langle x^{\alpha}(t)x^{\beta}(s)\rangle & =\langle\eta^{\alpha}(t)\eta^{\beta}(s)\rangle,
\end{align*}
so we get for $\alpha,\beta\in\{1,2\}$
\begin{align}
\left(\partial_{t}+1\right)\left(\partial_{s}+1\right)c^{\alpha\beta}(t,s) & =D\delta(t-s)+g^{2}F_{\phi}\left(c^{\alpha\beta}(t,s),c^{\alpha\alpha}(t,t),c^{\beta\beta}(s,s)\right)\,,\label{eq:diffeq_cab}
\end{align}
where the function $F_{\phi}$ is defined as the Gaussian expectation
value 
\begin{align*}
F_{\phi}(c^{12},c^{1},c^{2}) & :=\E{\phi(x^{1})\phi(x^{2})}
\end{align*}
for the bi-variate Gaussian
\begin{align*}
\begin{pmatrix}x^{1}\\
x^{2}
\end{pmatrix} & \sim\mathcal{N}_{2}\left(0,\begin{pmatrix}c^{1} & c^{12}\\
c^{12} & c^{2}
\end{pmatrix}\right).
\end{align*}
First, we observe that the equations for the autocorrelation functions
$c^{\alpha\alpha}(t,s)$ decouple and can each be solved separately,
leading to the same equation \eqref{eq_motion_cxx} as before. As
noted earlier, this formal result could have been anticipated, because
the marginal statistics of each subsystem cannot be affected by the
mere presence of the respective other system. Their solutions
\begin{align*}
c^{11}(s,t)= & c^{22}(s,t)=c(t-s)
\end{align*}
then provide the ``background'', i.e., the second and third argument
of the function $F_{\phi}$ on the right-hand side, for the equation
for the crosscorrelation function between the two copies. Hence it
remains to determine the equation of motion for $c^{12}(t,s)$.

We first determine the stationary solution $c^{12}(t,s)=k(t-s)$.
We see immediately that $k(\tau)$ obeys the same equation of motion
as $c(\tau)$, so $k(\tau)=c(\tau)$. The distance \eqref{mean-squared-distance-1}
therefore vanishes. Let us now study the stability of this solution.
We hence need to expand $c^{12}$ around the stationary solution
\begin{align*}
c^{12}(t,s) & =c(t-s)+\epsilon\,k^{(1)}(t,s)\,,\:\epsilon\ll1\,.
\end{align*}
We expand the right hand side of \prettyref{eq:diffeq_cab} into a
Taylor series using Price's theorem and \eqref{def_f} 
\begin{align*}
F_{\phi}\left(c^{12}(t,s),c_{0},c_{0}\right) & =f_{\phi}\left(c^{12}(t,s),c_{0}\right)\\
 & =f_{\phi}\left(c(t-s),c_{0}\right)+\epsilon\,f_{\phi^{\prime}}\left(c(t-s),c_{0}\right)\,k^{(1)}(t,s)+O(\epsilon^{2}).
\end{align*}
Inserted into \eqref{diffeq_cab} and using that $c$ solves the lowest
order equation, we get the linear equation of motion for the first
order deflection
\begin{align}
\left(\partial_{t}+1\right)\left(\partial_{s}+1\right)\,k^{(1)}(t,s) & =g^{2}f_{\phi^{\prime}}\left(c(t-s),c_{0}\right)\,k^{(1)}(t,s).\label{eq:variational_equation}
\end{align}
In the next section we will determine the growth rate of $k^{(1)}$
and hence, by \eqref{mean-squared-distance-1}
\begin{align}
d(t) & =\underbrace{c^{11}(t,t)}_{c_{0}}+\underbrace{c^{22}(s,s)}_{c_{0}}\underbrace{-c^{12}(t,t)-c^{21}(t,t)}_{-2c_{0}-2\epsilon\,k^{(1)}(t,t)}\nonumber \\
 & =-2\epsilon\,k^{(1)}(t,t)\label{eq:relation_distance_c_12}
\end{align}
the growth rate of the distance between the two subsystems. The negative
sign makes sense, since we expect in the chaotic state that $c^{12}(t,s)\stackrel{t,s\to\infty}{=}0$,
so $k^{(1)}$ must be of opposite sign than $c>0$.

\subsection{Schr\"odinger equation for the maximum Lyapunov exponent}

We here want to reformulate the equation for the variation of the
cross-system correlation \prettyref{eq:variational_equation} into
a Schr\"odinger equation, as in the original work \citep[eq. 10]{Sompolinsky88_259}.

First, noting that $C_{\phi^{\prime}\phi^{\prime}}(t,s)=f_{\phi^{\prime}}\left(c(t-s),c_{0}\right)$
is time translation invariant, it is advantageous to introduce the
coordinates $T=t+s$ and $\tau=t-s$ and write the covariance $k^{(1)}(t,s)$
as $k(T,\tau)$ with $k^{(1)}(t,s)=k(t+s,t-s)$. The differential
operator $\left(\partial_{t}+1\right)\left(\partial_{s}+1\right)$
with the chain rule $\partial_{t}\to\partial_{T}+\partial_{\tau}$
and $\partial_{s}\to\partial_{T}-\partial_{\tau}$ in the new coordinates
is $(\partial_{T}+1)^{2}-\partial_{\tau}^{2}$. A separation ansatz
$k(T,\tau)=e^{\frac{1}{2}\kappa T}\,\psi(\tau)$ then yields the eigenvalue
equation
\begin{align*}
(\frac{\kappa}{2}+1)^{2}\psi(\tau)-\partial_{\tau}^{2}\psi(\tau) & =g^{2}f_{\phi^{\prime}}\left(c(\tau),c_{0}\right)\psi(\tau)
\end{align*}
for the growth rates $\kappa$ of $d(t)=-2k^{(1)}(t,t)=-2k(2t,0)$.
We can express the right hand side by the second derivative of the
potential \prettyref{eq:def_potential_V} $V(c(\tau);c_{0})$ so that
with 
\begin{eqnarray}
V^{\prime\prime}(c(\tau);c_{0}) & = & -1+g^{2}f_{\phi^{\prime}}\left(c(\tau),c_{0}\right)\label{eq:effective_potential}
\end{eqnarray}
we get the time-independent Schrödinger equation
\begin{eqnarray}
\left(-\partial_{\tau}^{2}-V^{\prime\prime}(c(\tau);c_{0})\right)\psi(\tau) & = & \underbrace{\left(1-\left(\frac{\kappa}{2}+1\right)^{2}\right)}_{=:E}\psi(\tau).\label{eq:Schroedinger}
\end{eqnarray}
The eigenvalues (``energies'') $E_{n}$ determine the exponential
growth rates $\kappa_{n}$ the solutions $k(2t,0)=e^{\kappa_{n}t}\,\psi_{n}(0)$
at $\tau=0$ with 
\begin{eqnarray}
\kappa_{n}^{\pm} & = & 2\left(-1\pm\sqrt{1-E_{n}}\right).\label{eq:roots_lambda}
\end{eqnarray}
We can therefore determine the growth rate of the mean-square distance
of the two subsystems in \prettyref{sub:pair_of_systems} by \prettyref{eq:relation_distance_c_12}.
The fastest growing mode of the distance is hence given by the ground
state energy $E_{0}$ and the plus in \prettyref{eq:roots_lambda}.
The deflection between the two subsystems therefore growth with the
rate
\begin{eqnarray}
\lambda_{\mathrm{max}} & = & \frac{1}{2}\kappa_{0}^{+}\label{eq:Lambda_max}\\
 & = & -1+\sqrt{1-E_{0}},\nonumber 
\end{eqnarray}
where the factor $1/2$ in the first line is due to $d$ being the
squared distance, hence the length $\sqrt{d}$ growth with half the
exponent as $d$.

\begin{figure}
\begin{centering}
\includegraphics[scale=2]{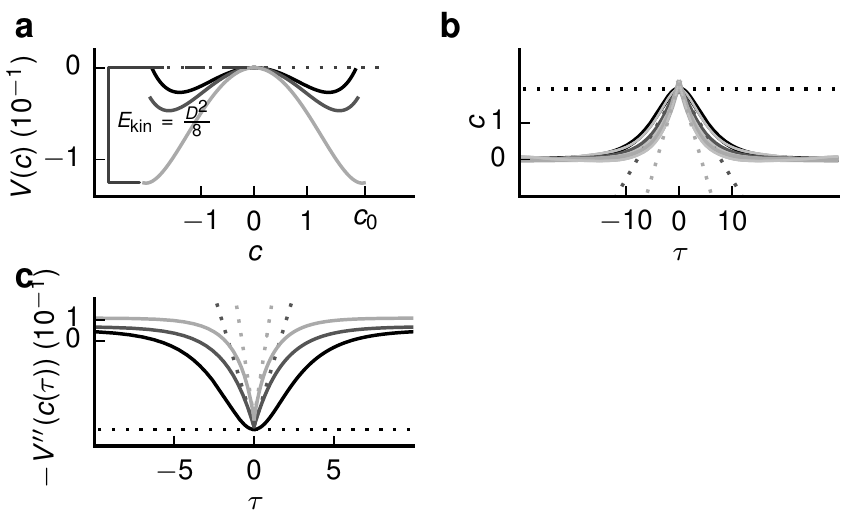}
\par\end{centering}
\caption{\textbf{Dependence of the self-consistent solution on the noise level
$D$.} \textbf{a} Potential that determines the self-consistent solution
of the autocorrelation function \eqref{def_potential_V}. Noise amplitude
$D>0$ corresponds to an initial kinetic energy $E_{\mathrm{kin}}=\frac{D^{2}}{8}$.
The initial value $c_{0}$ is determined by the condition $V(c_{0};c_{0})+E_{\mathrm{kin}}=0$,
so that the ``particle'' starting at $c(0)=c_{0}$ has just enough
energy to reach the peak of the potential at $c(\tau\to\infty)=0$.
In the noiseless case, the potential at the initial position $c(0)=c_{0}$
must be equal to the potential for $\tau\to\infty$, i.e. $V(c_{0};c_{0})=V(0)=0$,
indicated by horizontal dashed line and the corresponding potential
(black). \textbf{b} Resulting self-consistent autocorrelation functions
given by \prettyref{eq:acf_system}. The kink at zero time lag $\dot{c}(0-)-\dot{c}(0+)=\frac{D}{2}$
is indicated by the tangential dotted lines. In the noiseless case
the slope vanishes (horizontal dotted line). Simulation results shown
as light gray underlying curves.\textbf{ c} Quantum mechanical potential
appearing in the Schr\"odinger equation \prettyref{eq:Schroedinger}
with dotted tangential lines at $\tau=\pm0$. Horizontal dotted line
indicates the vanishing slope in the noiseless case. Other parameters
as in \prettyref{fig:effective_potential}.\label{fig:noise_dependence}}
\end{figure}
Energy conservation \prettyref{eq:initial_c0} determines $c_{0}$
also in the case of non-zero noise $D\neq0$, as shown in \prettyref{fig:noise_dependence}a.
The autocovariance function obtained from the solution of \prettyref{eq:acf_system}
agrees well to the direct simulation \prettyref{fig:noise_dependence}b.
The quantum potential appearing in \prettyref{eq:Schroedinger} is
shown in \prettyref{fig:noise_dependence}c.

\subsection{Condition for transition to chaos}

We can construct an eigensolution of \prettyref{eq:Schroedinger}
from \prettyref{eq:eq_motion_cxx}. First we note that for $D\neq0$,
$c$ has a kink at $\tau=0$. This can be seen by integrating \prettyref{eq:eq_motion_cxx}
from $-\epsilon$ to $\epsilon$, which yields
\begin{eqnarray*}
\lim_{\epsilon\to0}\int_{-\epsilon}^{\epsilon}\partial_{\tau}^{2}cd\tau & = & \dot{c}(0+)-\dot{c}(0-)\\
 & = & D.
\end{eqnarray*}
Since $c(\tau)=c(-\tau)$ is an even function it follows that $\dot{c}(0+)=-\dot{c}(0-)=-\frac{D}{2}$.
For $\tau\neq0$ we can differentiate \prettyref{eq:eq_motion_cxx}
with respect to time $\tau$ to obtain 
\begin{eqnarray*}
\partial_{\tau}\partial_{\tau}^{2}\,c(\tau) & = & \partial_{\tau}^{2}\,\dot{c}(\tau)\\
=-\partial_{\tau}V^{\prime}(c(\tau)) & = & -V^{\prime\prime}(c(\tau))\,\dot{c}(\tau).
\end{eqnarray*}
Comparing the right hand side expressions shows that $\left(\partial_{\tau}^{2}+V^{\prime\prime}(c(\tau))\right)\dot{c}(\tau)=0$,
so $\dot{c}$ is an eigensolution for eigenvalue $E_{n}=0$ of \prettyref{eq:Schroedinger}.

Let us first study the case of vanishing noise $D=0$ as in \citep{Sompolinsky88_259}.
The solution then $\dot{c}$ exists for all $\tau$. Since $c$ is
a symmetric function, $\Psi_{0}=\dot{c}$ has single node. The single
node of this solution implies there must be a state with zero nodes
that has even lower energy, i.e. $E_{0}<0$ . This, in turn, indicates
a positive Lyapunov exponent $\Lambda_{\mathrm{max}}$ according to
\prettyref{eq:Lambda_max}. This is the original argument in \citep{Sompolinsky88_259},
showing that at $g=1$ a transition from a silent to a chaotic state
takes place.

Our aim is to find the parameter values for which the transition to
the chaotic state takes place in the presence of noise. We know that
the transition takes place if the eigenvalue of the ground state of
the Schr\"odinger equation is zero. We can therefore explicitly try
to find a solution of \prettyref{eq:Schroedinger} for eigenenergy
$E_{n}=0$, i.e. we seek the homogeneous solution that satisfies all
boundary conditions, i.e. continuity of the solution as well as its
first and second derivative. We already know that $\dot{c}(\tau)$
is one homogeneous solution of \prettyref{eq:Schroedinger} for positive
and for negative $\tau$. For $D\neq0$, we can construct a continuous
solution from the two branches by defining

\begin{eqnarray}
y_{1}(\tau) & = & \begin{cases}
\dot{c}(\tau) & \tau\ge0\\
-\dot{c}(\tau) & \tau<0
\end{cases},\label{eq:def_y1}
\end{eqnarray}
which is symmetric, consistent with the search for the ground state.
In general, $y_{1}$  does not solve the Schr\"odinger equation,
because the derivative at $\tau=0$ is not necessarily continuous,
since by \eqref{diffeq_auto} $\partial_{\tau}y_{1}(0+)-\partial_{\tau}y_{1}(0-)=\ddot{c}(0+)+\ddot{c}(0-)=2(c_{0}-g^{2}f_{\phi}(c_{0};c_{0}))$.
Therefore $y_{1}$ is only an admissible solution, if the right hand
side vanishes. The criterion for the transition to the chaotic state
is hence

\begin{align}
0=\partial_{\tau}^{2}c(0\pm) & =c_{0}-g^{2}f_{\phi}\left(c_{0},c_{0}\right)\label{eq:ch_trans_noise}\\
 & =-V^{\prime}(c_{0};c_{0}).\nonumber 
\end{align}
The latter condition therefore shows that the curvature of the autocorrelation
function vanishes at the transition. In the picture of the motion
of the particle in the potential the vanishing acceleration at $\tau=0$
amounts to a potential with a flat tangent at $c_{0}$.

A necessary condition is the minimum of the potential 
\begin{align*}
V^{\prime\prime}(c_{0},c_{0}) & <0,
\end{align*}
because the ground state energy cannot be smaller than the potential,
as it is the sum of potential energy and kinetic energy. With \prettyref{eq:effective_potential}
the latter condition translates to

\begin{align*}
1 & \le g^{2}\langle\phi^{\prime}(x)^{2}\rangle
\end{align*}
It is equivalent to the spectral radius of the Jacobian $J_{ij}\phi^{\prime}$
of the dynamics \prettyref{eq:diffeq_motion} to be larger than one:
the point where linear stability is lost.

\begin{figure}
\centering{}\includegraphics{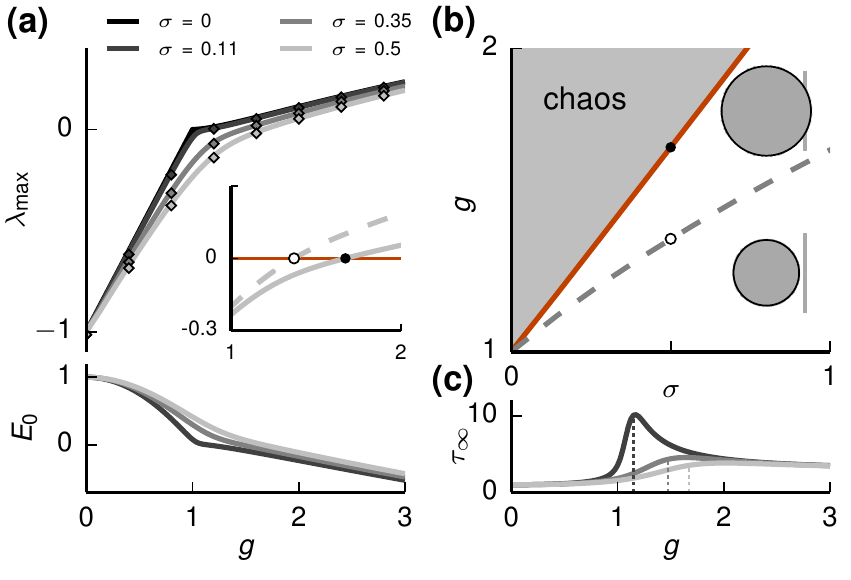}\caption{\textbf{Transition to chaos. (a)} Upper part of vertical axis: Maximum
Lyapunov exponent $\lambda_{\mathrm{max}}$ \prettyref{eq:Lambda_max}
as a function of the coupling strength $g$ for different input amplitude
levels. Mean-field prediction (solid curve) and simulation (diamonds).
Comparison to the upper bound $-1+g\sqrt{\langle\phi'(x){}^{2}\rangle}$
(dashed) for $\sqrt{D/2}=\sigma=0.5$ in inset. Zero crossings marked
with dots. Lower part of vertical axis: Ground state energy $E_{0}$
as a function of $g$.\textbf{ (b)} Phase diagram with transition
curve (solid red curve) obtained from \prettyref{eq:ch_trans_noise}
and necessary condition ($1=g^{2}\langle\phi^{\prime}(x)^{2}\rangle$,
gray dashed curve). Dots correspond to zero crossings in inset in
(a). Disk of eigenvalues of the Jacobian matrix for $\sqrt{D/2}=\sigma=0.8$
and $g=1.25$ (lower) and $g=2.0$ (upper) centered at $-1$ in the
complex plane (gray). Radius $\rho=g\sqrt{\langle\phi'(x){}^{2}\rangle}$
from random matrix theory (black). Vertical line at zero.\textbf{
(c)} Asymptotic decay time $\tau_{\infty}$ of autocorrelation function.
Vertical dashed lines mark the transition to chaos. Color code as
in (a). Network size of simulations $N=5000$. From \citep{Schuecker18_041029}.\label{fig:Transition-to-chaos}}
\end{figure}

The criterion for the transition can be understood intuitively. The
additive noise increases the peak of the autocorrelation at $\tau=0$.
In the large noise limit, the autocorrelation decays as $e^{-|\tau|}$,
so the curvature is positive. The decay of the autocorrelation is
a consequence of the uncorrelated external input. In contrast, in
the noiseless case, the autocorrelation has a flat tangent at $\tau=0$,
so the curvature is negative. The only reason for its decay is the
decorrelation due to the chaotic dynamics. The transition between
these two forces of decorrelation hence takes place at the point at
which the curvature changes sign, from dominance of the external sources
to dominance of the intrinsically generated fluctuations. The phase
diagram of the network is illustrated in \prettyref{fig:Transition-to-chaos}.
For a more detailed discussion please see \citep{Goedeke16_arxiv}.

A closely related calculation shows that condition for the the transition
to chaos in the absence of noise is identical to the condition for
a vanishing coupling between replicas. Therefore, in the chaotic regime
the system is, to leading order in $N$, also self-averaging. This
argument can be extended to the case with noise $D\neq0$. One finds
that also here the only physically admissible solution for the field
coupling the replicas is one that vanishes (see exercises).

\section{Vertex generating function\label{sec:Vertex-generating-function}}

We have seen in the previous sections that the statistics of a system
can be either described by the moment generating function $Z(j)$
or, more effectively, by the cumulant generating function $W(j)=\ln Z(j)$.
The decomposition of the action $S$ into a quadratic part $-\frac{1}{2}x^{\T}Ax$
and the remaining terms collected in $V(x)$ allowed us to derive
graphical rules in terms of Feynman diagrams to calculate the cumulants
or moments of the variables in an effective way (see \secref{Linked-cluster-theorem}).
We saw that the expansion of the cumulant generating function $W(j)$
in the general case $S_{0}(x)+\epsilon V(x)$ is composed of connected
components only. In the particular case of a decomposition as $S(x)=\frac{1}{2}x^{\T}Ax+\epsilon V(x)$,
we implicitly assume a quadratic approximation around the value $x=0$.
If the interacting part $V(x)$ and the external source $j$ is small
compared to the free theory, this is the natural choice. We will here
derive a method to systematically expand fluctuations around the true
mean value in the case that the interaction is strong, so that the
dominant point of activity is in general far away from zero. Let us
begin with an example to illustrate the situation.

\subsection{Motivating example for the expansion around a non-vanishing mean
value\label{sec:tanh_network}}

Let us study the fluctuating activity in a network of $N$ neurons
which obeys the set of coupled equations
\begin{align}
x_{i} & =\sum_{j}J_{ij}\phi(x_{j})+\mu_{i}+\xi_{i}\label{eq:mf_tanh_network}\\
\xi_{i} & \sim\N(0,D_{i})\qquad\langle\xi_{i}\xi_{j}\rangle=\delta_{ij}\,D_{i},\nonumber \\
\phi(x) & =\tanh(x-\theta).\nonumber 
\end{align}
Here the $N$ units are coupled by the synaptic weights $J_{ij}$
from unit $j$ to unit $i$. We may think about $x_{i}$ being the
membrane potential of the neuron and $\phi(x_{i})$ its firing rate,
which is a non-linear function $\phi$ of the membrane potential.
The non-linearity has to obey certain properties. For example, it
should typically saturate at high rates, mimicking the inability of
neurons to fire in rapid succession. The choice of $\phi(x)=\tanh(x)$
is common in the field of artificial neuronal networks. The term $\mu_{i}$
represents an additional input to the $i$-th neuron and $\xi_{i}$
is a centered Gaussian noise causing fluctuations within the network.

We may be interested in the statistics of the activity that arises
due to the interplay among the units. For illustrative purposes, let
us for the moment assume a completely homogeneous setting, where $J_{ij}=\frac{J_{0}}{N}\quad\forall\,i,j$
and $\mu_{i}=\mu$ as well as $D_{i}=D\quad\forall\,i$. Since the
Gaussian fluctuations are centered, we may obtain a rough approximation
by initially just ignoring its presence, leading us to a set of $N$
identical equations
\begin{align*}
x_{i} & =\frac{J_{0}}{N}\,\sum_{j=1}^{N}\phi(x_{j})+\mu.
\end{align*}
Due to the symmetry, we hence expect a homogeneous solution $x_{i}\equiv x\quad\forall\,i$,
which fulfills the equation
\begin{align}
x^{\ast} & =J_{0}\,\phi(x^{\ast})+\mu.\label{eq:self_consistent_mean_tanh}
\end{align}
There may, of course, also be asymmetric solutions to this equation,
those that break the symmetry of the problem. \ifthenelse{\boolean{lecture}}{We
will come back to these solutions in \secref{Spontaneous-symmetry-breaking}.}

\begin{figure}
\begin{centering}
\includegraphics{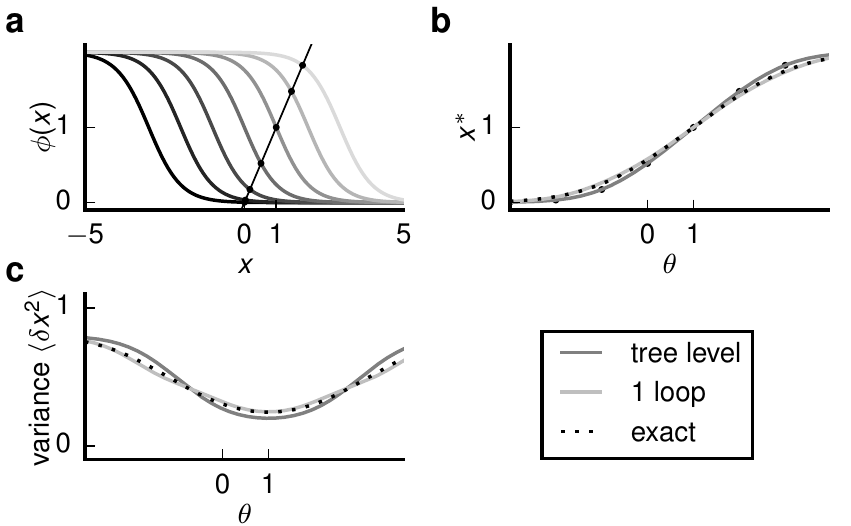}
\par\end{centering}
\caption{\textbf{Self-consistent solution of the mean activity in a single
population network.} \textbf{a} Tree level approximation of the self-consistent
activity given by intersection of left hand side of (\ref{eq:self_consistent_mean_tanh})
(thin black line with unit slope) and the right hand side; different
gray levels indicate thresholds $\theta\in[-3,4]$ from black to light
gray. Black dots mark the points of intersection, yielding the self-consistent
tree-level or mean-field approximation neglecting fluctuations. \textbf{b}
Self-consistent solution as function of the activation threshold $\theta$.
Black dotted curve: Exact numerical solution; mid gray: Tree level
(mean field) approximation neglecting fluctuations, as illustrated
in a and given by (\ref{eq:self_consistent_mean_tanh}); light gray:
one-loop correction given by (\ref{eq:one_loop_tanh}), including
fluctuation corrections. \textbf{c} Variance of $x$. Exact numerical
value (black) and mean-field approximation given by (\ref{eq:Gaussian_delta_x_tanh}).
Parameters: Mean input $\mu=1$, self-coupling $J_{0}=-1$, variance
of noise $D=1$.\label{fig:Illustration_self_cons_mean}}
\end{figure}

We note that even though we assumed the synaptic couplings to diminish
as $N^{-1}$, the input from the other units cannot be neglected compared
to the mean input $\mu$. So an approximation around the solution
with vanishing mean $\langle x\rangle=0$ seems inadaquate. Rather
we would like to approximate the statistics around the mean value
$x^{\ast}$ that is given by the self-consistent solution of (\ref{eq:self_consistent_mean_tanh}),
illustrated in \figref{Illustration_self_cons_mean}.

To take fluctuations into account, which we assume to be small, we
make the ansatz $x=x^{\ast}+\delta x$ and approximate 
\begin{align*}
\phi(x) & =\phi(x^{\ast})+\phi^{\prime}(x^{\ast})\,\delta x,
\end{align*}
which therefore satisfies the equation
\begin{align}
\delta x & =J_{0}\phi^{\prime}(x^{\ast})\,\delta x+\xi,\label{eq:linear_deviation}\\
\xi & \sim\N(0,\bar{D}).\nonumber 
\end{align}
Since \eqref{linear_deviation} is linearly related to the noise,
the statistics of $\delta x$ is 
\begin{eqnarray}
\delta x & \sim & \N(0,\underbrace{\frac{D}{|1-J_{0}\phi^{\prime}(x^{\ast})|^{2}}}_{=:\bar{D}}).\label{eq:Gaussian_delta_x_tanh}
\end{eqnarray}

We see that the denominator is only well-defined, if $J_{0}\phi^{\prime}(x^{\ast})\neq1$.
Otherwise the fluctuations will diverge and we have a critical point.
Moreover, we here assume that the graph of $J_{0}\phi$ cuts the identity
line with an angle smaller $45$ degrees, so that the fluctuations
of $\delta x$ are positively correlated to those of $\xi$ \textendash{}
they are related by a positive factor. If we had a time-dependent
dynamics, the other case would correspond to an unstable fixed point. 

Approximating the activity in this Gaussian manner, we see that we
get a correction to the mean activity as well: Taking the expectation
value on both sides of \eqref{mf_tanh_network}, and approximating
the fluctuations of $x$ by \eqref{Gaussian_delta_x_tanh} by expanding
the non-linearity to the next order we get

\begin{align}
x^{\ast}=\langle x^{\ast}+\delta x\rangle & =\mu+J_{0}\phi(x^{\ast})+J_{0}\phi^{\prime}(x^{\ast})\underbrace{\langle\delta x\rangle}_{=0}+J_{0}\,\frac{\phi^{\prime\prime}(x^{\ast})}{2!}\,\underbrace{\langle\delta x^{2}\rangle}_{=\bar{D}}+O(\delta x^{3}),\nonumber \\
x^{\ast}- & J_{0}\phi(x^{\ast})-\mu=J_{0}\,\frac{\phi^{\prime\prime}(x^{\ast})}{2!}\,\frac{D}{|1-J_{0}\phi^{\prime}(x^{\ast})|^{2}}+O(\delta x^{3}).\label{eq:one_loop_tanh}
\end{align}

So the left hand side does not vanish anymore, as it did at lowest
order; instead we get a fluctuation correction that depends on the
point $x^{\ast}$around which we expanded. So solving the latter equation
for $x^{\ast}$, we implicitly include the fluctuation corrections
of the chosen order: Note that the variance of the fluctuations, by
(\ref{eq:Gaussian_delta_x_tanh}), depends on the point $x^{\ast}$
around which we expand. We see from (\ref{eq:one_loop_tanh}) that
we get a correction to the mean with the same sign as the curvature
$\phi^{\prime\prime}$, as intuitively expected due to the asymmetric
``deformation'' of the fluctuations by $\phi$. The different approximations
(\ref{eq:self_consistent_mean_tanh}) and (\ref{eq:one_loop_tanh})
are illustrated in \figref{Illustration_self_cons_mean}. 

The analysis we performed here is ad hoc and limited to studying the
Gaussian fluctuations around the fixed point. In the following we
would like to generalize this approach to non-Gaussian corrections
and to a diagrammatic treatment of the correction terms.

\subsection{Legendre transform and definition of the vertex generating function
$\Gamma$}

\begin{figure}
\begin{centering}
\includegraphics{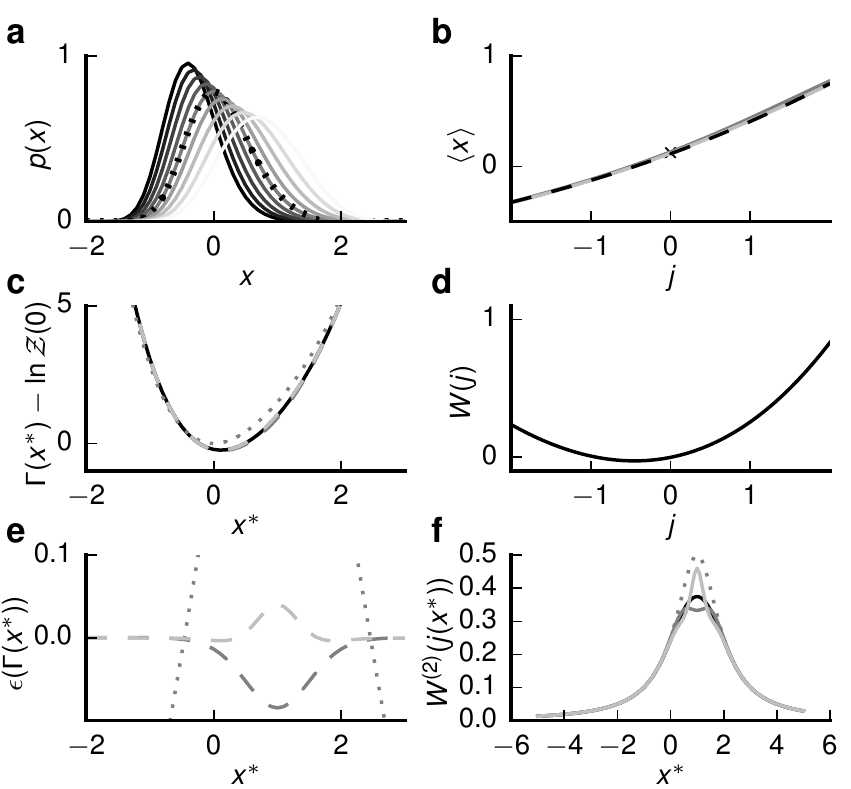}
\par\end{centering}
\begin{centering}
\par\end{centering}
\caption{\textbf{Loopwise expansion of $\Gamma$ for ``$\phi^{3}+\phi^{4}$''
theory.} \textbf{a} Probability density for action $S(x)=l\left(\frac{1}{2}x^{2}+\frac{\alpha}{3!}x^{3}+\frac{\beta}{4!}x^{4}\right)$
with $\alpha=1$, $\beta=-1$, $l=4$ for different values of the
source $j\in[-2,\ldots,2]$ (from black to light gray, black dotted
curve for $j=0$). The peak of the distribution shifts with $j$.
\textbf{b }The mean value $\langle x\rangle(j)$ as function of $j$.
One loop prediction of mean value at $j=0$ by tadpole diagram $\frac{l\alpha}{3!}\,\left(3\right)\frac{1}{l^{2}}=\frac{\alpha}{2l}$
is shown as black cross. Black dashed: exact; gray: solution of one-loop
approximation of equation of state, light gray: two-loop approximation.
\textbf{c }Effective action \textbf{$\Gamma(x^{\ast})$ }determined
numerically as $\Gamma(x^{\ast})-\ln\protect\Z(0)=\sup_{j}\,jx^{\ast}-W(j)-\ln\protect\Z(0)$
(black)\textbf{ }and by loopwise expansion (gray dotted: zero-loop,
gray dashed: one-loop, dark gray dashed: two-loop (see exercises).
\textbf{d} Cumulant generating function $W(j)$. \textbf{e} Error
$\epsilon=\Gamma^{x\,\text{loop}}-\Gamma$ of the loopwise expansions
of different orders $x$ (same symbol code as in c).\label{fig:Loopwise-expansion-phi4}}
\end{figure}

In the previous example in \secref{tanh_network} we aimed at a self-consistent
expansion around the true mean value $x^{\ast}$ to obtain corrections
to the expectation value due to fluctuations. The strategy was to
first perform an expansion around an arbitrarily chosen point $x^{\ast}$.
Then to calculate fluctuation corrections and only as a final step
we solve the resulting equation for the self-consistent value of $x^{\ast}$
that is equal to the mean. We will follow exactly the same line of
thoughts here, just formulating the problem with the help of an action,
because we ultimately aim at a diagrammatic formulation of the procedure.
Indeed, the problem from the last section can be formulated in terms
of an action, as will be shown in \secref{Loopwise-expansion-self-consistency}.

We will here follow the development pioneered in statistical physics
and field theory \citep{JonaLasinio64_1790,DeDominicis64_14} to define
the \textbf{effective action} or \textbf{vertex generating function}
(see also \citep[chapter 5]{Amit84}).

We write the cumulant generating function in its integral representation

\begin{align}
\exp\left(W(j)\right) & =Z(j)=\Z(0)^{-1}\int dx\,\exp\left(S(x)+j^{\T}x\right),\label{eq:W_as_Z-1}
\end{align}
to derive an equation that includes fluctuations. First we leave
this point $x^{\ast}$ arbitrary. This is sometimes called the \textbf{background
field method} \citep[Chapter 3.23.6]{Kleinert89}, allowing us to
treat fluctuations around some chosen reference field. We separate
the fluctuations of $\delta x=x-x^{\ast}$, insert this definition
into \eqref{W_as_Z-1} and bring the terms independent of $\delta x$
to the left hand side

\begin{align}
\exp\left(W(j)-j^{\T}x^{\ast}\right) & =\Z(0)^{-1}\int\,d\delta x\,\exp\left(S(x^{\ast}+\delta x)+j^{\T}\delta x\right).\label{eq:pre_Legendre}
\end{align}
We now make a special choice of $j$. For given $x^{\ast}$, we choose
$j$ so that $x^{\ast}=\langle x\rangle(j)$ becomes the mean. The
fluctuations of $\delta x$ then have vanishing mean value, because
$x^{\ast}\stackrel{!}{=}\langle x\rangle=\langle x^{\ast}+\delta x\rangle$.
Stated differently, we demand
\begin{align*}
0\stackrel{!}{=}\langle\delta x\rangle\equiv & \Z(0)^{-1}\int d\delta x\,\exp\left(S(x^{\ast}+\delta x)+j^{\T}\delta x\right)\,\delta x\\
\equiv & \Z(0)^{-1}\frac{d}{dj}\,\int d\delta x\,\exp\left(S(x^{\ast}+\delta x)+j^{\T}\delta x\right)\\
= & \frac{d}{dj}\,\exp\left(W(j)-j^{\T}x^{\ast}\right),
\end{align*}
where we used \eqref{pre_Legendre} in the last step. Since the exponential
function has the property $\exp(x)^{\prime}>0\quad\forall x$, the
latter expression vanishes at the point where the exponent is stationary
\begin{align}
\frac{d}{dj}\left(W(j)-j^{\T}x^{\ast}\right) & =0\label{eq:stationary_j}\\
\langle x\rangle(j)=\frac{\partial W(j)}{\partial j} & =x^{\ast}(j),\nonumber 
\end{align}
which shows again that $x^{\ast}(j)=\langle x\rangle(j)$ is the expectation
value of $x$ at a given value of the source $j$. 

The condition \eqref{stationary_j} has the form of a \textbf{Legendre
transform} from the function $W(j)$ to the new function, which we
call the \textbf{vertex generating function} or \textbf{effective
action}
\begin{align}
\Gamma(x^{\ast}) & :=\sup_{j}\,j^{\T}x^{\ast}-W(j).\label{eq:def_gamma}
\end{align}
The condition (\ref{eq:stationary_j}) implies that $j$ is chosen
such as to extremize $\Gamma(x^{\ast})$. We see that it must be the
supremum, because $W$ is a convex down function (see \secref{Convexity-W}).
It follows that $-W$ is convex up and hence the supremum of $j^{\T}x^{\ast}-W(j)$
at given $x^{\ast}$ is uniquely defined; the linear term does not
affect the convexity of the function, since its curvature is zero.

The Legendre transform has the property
\begin{align}
\frac{d\Gamma}{dx^{\ast}}(x^{\ast})= & j+\frac{\partial j^{\T}}{\partial x^{\ast}}x^{\ast}-\underbrace{\frac{\partial W^{\T}}{\partial j}}_{x^{\ast\T}}\frac{\partial j}{\partial x^{\ast}}\label{eq:equation_of_state}\\
= & j,\nonumber 
\end{align}
The latter equation is also called \textbf{equation of state}, as
its solution for $x^{\ast}$ allows us to determine the mean value
for a given source $j$, including all corrections due to fluctuations.
In statistical physics this mean value is typically an order parameter,
an observable that characterizes the state of the system.

The self-consistent solution if given by the equation of state \eqref{equation_of_state}.
The equation of state can be interpreted as a particle in a classical
potential $\Gamma(x^{\ast})$ and subject to a force $j$. The equilibrium
point of the particle, $x^{\ast}$, is then given by the equilibrium
of the two forces $j$ and $=-d\Gamma/dx^{\ast}\equiv-\Gamma^{(1)}(x^{\ast})$,
which is identical to the equation of state \eqref{equation_of_state}
\begin{eqnarray*}
0 & = & j-\Gamma^{(1)}(x^{\ast}).
\end{eqnarray*}

Comparing \eqref{stationary_j} and \eqref{equation_of_state} shows
that the functions $W^{(1)}$ and $\Gamma^{(1)}$ are inverse functions
of one another. It therefore follows by differentiation 
\begin{align}
\Gamma^{(1)}(W^{(1)}(j)) & =j\label{eq:inverse_W1_Gamma1}\\
\Gamma^{(2)}(W^{(1)}(j))\,W^{(2)}(j) & =1\nonumber 
\end{align}
that their Hessians are inverse matrices of each other
\begin{align}
\Gamma^{(2)} & =\left[W^{(2)}\right]^{-1}.\label{eq:WGamma_inv_Hessians}
\end{align}
From the convexity of $W$ therefore follows with the last expression
that also $\Gamma$ is a convex down function. The solutions of the
equation of state thus form convex regions. An example of the function
$\Gamma$ is shown in \figref{Loopwise-expansion-phi4}c.

One can see that the Legendre transform is involutive for convex functions:
applied twice it is the identity. Convexity is important here, because
the Legendre transform of any function is convex. In particular, applying
it twice, we arrive back at a convex function. So we only get an involution
for convex functions to start with. This given, we define
\begin{align*}
w(j):= & j^{\T}x^{\ast}-\Gamma(x^{\ast})\\
\text{with }\frac{d\Gamma(x^{\ast})}{dx^{\ast}} & =j
\end{align*}
it follows that
\begin{align}
\frac{dw(j)}{dj} & =x^{\ast}+j^{\T}\frac{\partial x^{\ast}}{\partial j}-\underbrace{\frac{\partial\Gamma}{\partial x^{\ast}}^{\T}}_{=j^{\T}}\frac{\partial x^{\ast}}{\partial j}=x^{\ast}(j)\label{eq:involution_Legendre}\\
 & =\langle x\rangle(j),\nonumber 
\end{align}
where the equal sign in the last line follows from our choice \eqref{stationary_j}
above. We hence conclude that $w(j)=W(j)+c$ with some inconsequential
constant $c$.

In the following we will investigate which effect the transition from
$W(j)$ to its Legendre transform $\Gamma(x^{\ast})$ has in terms
of Feynman diagrams. The relation between graphs contributing to $W$
and those that form $\Gamma$ will be exposed in \secref{Tree-level-expansion-Cumulants-Vertex}.

\subsection{Perturbation expansion of $\Gamma$\label{sec:Perturbation-expansion-of-Gamma}}

We have seen that we may obtain a self-consistency equation for the
mean value $x^{\ast}$ from the equation of state (\ref{eq:equation_of_state}).
The strategy therefore is to obtain an approximation of $\Gamma$
that includes fluctuation corrections and then use the equation of
state to get an approximation for the true mean value including these
very corrections. We will here obtain a perturbative procedure to
calculate approximations of $\Gamma$ and will find the graphical
rules for doing so. To solve a problem perturbatively we decompose
the action, as in \secref{Perturbation-expansion}, into $S(x)=S_{0}(x)+\epsilon V(x)$
with a part $S_{0}$ that can be solved exactly, i.e. for which we
know the cumulant generating function $W_{0}(j)$, and the remaining
terms collected in $\epsilon V(x)$. An example of a real world problem
applying this technique is given in \subref{Example-TAP-approximation}.
We here follow the presentation by \citep{Kuehn18_375004}.

To lowest order in perturbation theory, namely setting $\epsilon=0$,
we see that $W(j)=W_{0}(j)$; the corresponding leading order term
in $\Gamma$ is the Legendre transform
\begin{align}
\Gamma_{0}(x^{\ast}) & =\sup_{j}\,j^{\T}x^{\ast}-W_{0}(j).\label{eq:Gamma0_pert-1}
\end{align}
We now want to derive a recursive equation to obtain approximations
of the form
\begin{align}
\Gamma(x^{\ast}) & =:\Gamma_{0}(x^{\ast})+\Gamma_{V}(x^{\ast}),\label{eq:Gamma_pert_decomposition}
\end{align}
where we defined $\Gamma_{V}(x^{\ast})$ to contain all correction
terms due to the interaction potential $V$ to some order $\epsilon^{k}$
of perturbation theory.

Let us first see why the decomposition into a sum in \eqref{Gamma_pert_decomposition}
is useful. To this end, we first rewrite \eqref{pre_Legendre}, employing
\eqref{equation_of_state} to replace $j(x^{\ast})=\Gamma^{(1)}(x^{\ast})$
and by using $x=x^{\ast}+\delta x$ as
\begin{align}
\exp(-\Gamma(x^{\ast})) & =\Z^{-1}(0)\,\int dx\,\exp(S(x)+\Gamma^{(1)\T}(x^{\ast})(x-x^{\ast}))\label{eq:expGamma}\\
 & =\Z^{-1}(0)\,\int dx\,\exp(S_{0}(x)+\epsilon V(x)+\Gamma^{(1)\T}(x^{\ast})(x-x^{\ast})),\nonumber 
\end{align}
where we used in the second line the actual form of the perturbative
problem. Inserting the decomposition \eqref{Gamma_pert_decomposition}
of $\Gamma$ into the solvable and the perturbing part we can express
\eqref{expGamma} as
\begin{align*}
\exp(-\Gamma_{0}(x^{\ast})-\Gamma_{V}(x^{\ast})) & =\Z^{-1}(0)\,\int dx\,\exp\big(S_{0}(x)+\epsilon V(x)+\big(\Gamma_{0}^{(1)\T}(x^{\ast})+\Gamma_{V}^{(1)\T}(x^{\ast})\big)(x-x^{\ast})\big)\\
\exp(\underbrace{-\Gamma_{0}(x^{\ast})+\Gamma_{0}^{(1)\T}(x^{\ast})\,x^{\ast}}_{W_{0}(j)\big|_{j=\Gamma_{0}^{(1)}(x^{\ast})}}-\Gamma_{V}(x^{\ast})) & =\exp\big(\epsilon V(\partial_{j})+\Gamma_{V}^{(1)\T}(x^{\ast})(\partial_{j}-x^{\ast})\big)\,\underbrace{\Z^{-1}(0)\int dx\,\exp\big(S_{0}(x)+j^{\T}x)\big)\big|_{j=\Gamma_{0}^{(1)}(x^{\ast})}}_{W_{0}(j)\big|_{j=\Gamma_{0}^{(1)}(x^{\ast})}}\\
 & =\exp\big(\epsilon V(\partial{}_{j})+\Gamma_{V}^{(1)\T}(x^{\ast})(\partial_{j}-x^{\ast})\big)\,\exp\big(W_{0}(j)\big)\big|_{j=\Gamma_{0}^{(1)}(x^{\ast})},
\end{align*}
where we moved the perturbing part in front of the integral, making
the replacement $x\to\partial_{j}$ as in \eqref{perturbation_general}
and we identified the unperturbed cumulant generating function $\exp(W_{0}(j))\big|_{j=\Gamma_{0}(x^{\ast})}=\Z^{-1}(0)\,\int dx\,\exp\big(S_{0}(x)+\Gamma_{0}^{(1)\T}(x^{\ast})\,x\big)$
from the second to the third line. Bringing the term $\Gamma_{0}^{(1)\T}(x^{\ast})x^{\ast}$
to the left hand side, we get $-\Gamma_{0}(x^{\ast})+j^{\T}x^{\ast}=W_{0}(j)\big|_{j=\Gamma_{0}^{(1)}(x^{\ast})}$,
which follows from the definition (\ref{eq:Gamma0_pert-1}). Multiplying
with $\exp(-W_{0}(j))\big|_{j=\Gamma_{0}^{(1)}(x^{\ast})}$ from left
then leads to a recursive equation for $\Gamma_{V}$ 
\begin{align}
\exp(-\Gamma_{V}(x^{\ast})) & =\exp(-W_{0}(j))\,\exp\big(\epsilon V(\partial_{j})+\Gamma_{V}^{(1)\T}(x^{\ast})(\partial_{j}-x^{\ast})\big)\,\exp(W_{0}(j))\big|_{j=\Gamma_{0}^{(1)}(x^{\ast})},\label{eq:Gamma_V_recursion}
\end{align}
which shows that our ansatz \eqref{Gamma_pert_decomposition} was
indeed justified: we may determine $\Gamma_{V}$ recursively, since
$\Gamma_{V}$ appears again on the right hand side.

We want to solve the latter equation iteratively order by order in
the number of interaction vertices $k$. We know that to lowest order
\eqref{Gamma0_pert-1} holds, so $\Gamma_{V,0}=0$ in this case. The
form of the terms on the right hand side of \eqref{Gamma_V_recursion}
is then identical to \eqref{perturbation_general}, so we know that
the first order ($\epsilon^{1}$) contribution are all connected diagrams
with one vertex from $\epsilon V$ and connections formed by the cumulants
of $W_{0}(j)$, where finally we set $j=\Gamma_{0}^{(1)}(x^{\ast})$.
The latter step is crucial to be able to write down the terms explicitly.
Because $\Gamma^{(1)}$ and $W^{(1)}$ are inverse functions of one
another (following from \eqref{stationary_j} and \eqref{equation_of_state}),
this step expresses all cumulants in $W_{0}$ in terms of the first
cumulant:
\begin{align}
\llangle x^{n}\rrangle(x^{\ast}) & =W_{0}^{(n)}(\underbrace{\Gamma_{0}^{(1)}(x^{\ast})}_{\equiv j_{0}(x^{\ast})})\label{eq:replacement_j}\\
x^{\ast} & =W_{0}^{(1)}(j_{0})\quad\leftrightarrow\quad j_{0}=\Gamma_{0}^{(1)}(x^{\ast})\nonumber 
\end{align}
 The graphs then contain $n$-th cumulants of the unperturbed theory
$\llangle x^{n}\rrangle(x^{\ast})$: To evaluate them, we need to
determine $x^{\ast}=W_{0}^{(1)}(j_{0})$, invert this relation to
obtain $j_{0}(x^{\ast})$, and insert it into all higher derivatives
$W_{0}^{(n)}(j_{0}(x^{\ast}))$, giving us explicit functions of $x^{\ast}$.
The aforementioned graphs all come with a minus sign, due to the minus
on the left hand side of \eqref{Gamma_V_recursion}.

We want to solve (\ref{eq:Gamma_V_recursion}) iteratively order by
order in the number of vertices $k$, defining $\Gamma_{V,k}$. Analogous
to the proof of the linked cluster theorem, we arrive at a recursion
by writing the exponential of the differential operator in \eqref{Gamma_V_recursion}
as a limit

\begin{align}
\exp\big(\epsilon V(\partial_{j})+\Gamma_{V}^{(1)\T}(x^{\ast})(\partial_{j}-x^{\ast})\big) & =\lim_{L\to\infty}\left(1+\frac{1}{L}\left(\epsilon V(\partial_{j})+\Gamma_{V}^{(1)\T}(x^{\ast})(\partial_{j}-x^{\ast})\right)\right)^{L}.\label{eq:pert_term_factors}
\end{align}
Initially we assume $L$ to be fixed but large and choose some $0\le l\le L$.
We move the term $\exp(-W_{0}(j))$ to the left hand side of \eqref{Gamma_V_recursion}
and define $g_{l}(j)$ as the result after application of $l$ factors
of the right hand side as
\begin{align}
\exp(W_{0}(j)+g_{l}(j)):= & \left(1+\frac{1}{L}\left(\epsilon V(\partial_{j})+\Gamma_{V}^{(1)\T}(x^{\ast})\,(\partial_{j}-x^{\ast})\right)\right)^{l}\,\exp(W_{0}(j)),\label{eq:rec_G}
\end{align}
where, due to the unit factor in the bracket $\big(1+\ldots\big)^{l}$
we always get a factor $\exp(W_{0}(j))$, written explicitly. We obviously
have the initial condition
\begin{eqnarray}
g_{0} & \equiv & 0.\label{eq:G_0}
\end{eqnarray}
For $l=L\to\infty$ this expression collects all additional graphs
and we obtain the desired perturbative correction \eqref{Gamma_pert_decomposition}
of the effective action as the limit 
\begin{eqnarray}
-\Gamma_{V}(x^{\ast}) & = & \lim_{L\to\infty}\,g_{L}(j)\Big|_{j=\Gamma_{0}^{(1)}(x^{\ast})}.\label{eq:Gamma_V_final}
\end{eqnarray}
It holds the trivial recursion $\exp(W_{0}(j)+g_{l+1}(j))=\left(1+\frac{1}{L}\left(\epsilon V(\partial_{j})+\Gamma_{V}^{(1)\T}(x^{\ast})\,(\partial_{j}-x^{\ast})\right)\right)\,\exp(W_{0}(j)+g_{l}(j))$
from which we get a recursion for $g_{l}$

.

\begin{eqnarray}
 &  & g_{l+1}(j)-g_{l}(j)\label{eq:iteration_g}\\
 & = & \frac{\epsilon}{L}\,\exp(-W_{0}(j)-g_{l}(j))\,V(\partial_{j})\,\exp(W_{0}(j)+g_{l}(j))\label{eq:add_connected}\\
 & + & \frac{1}{L}\,\exp(-W_{0}(j)-g_{l}(j))\,\Gamma_{V}^{(1)}(x^{\ast})\left(\partial_{j}-x^{\ast}\right)\,\exp(W_{0}(j)+g_{l}(j))\label{eq:add_reducible}\\
 & + & \mathcal{O}(L^{-2}),\nonumber 
\end{eqnarray}
where we multiplied from left by $\exp(-W_{0}(j)-g_{l}(j))$, took
the logarithm and used $\ln(1+\frac{1}{L}x)=\frac{1}{L}x+\mathcal{O}(L^{-2})$.
To obtain the final result \eqref{Gamma_V_final}, we need to express
$j=\Gamma_{0}^{(1)}(x^{\ast})$ in $G_{l}(j)$.

\subsection{Generalized one-line irreducibility}

We now want to investigate what the iteration (\ref{eq:iteration_g})
implies in terms of diagrams. We therefore need an additional definition
of the topology of a particular class of graphs.

The term \textbf{one-line irredicibility} in the literature refers
to the absence of diagrams that can be disconnected by cutting a single
second order bare propagator (a line in the original language of Feynman
diagrams). In the slightly generalized graphical notation introduced
in \secref{Linked-cluster-theorem}, these graphs have the form
\begin{eqnarray*}
 &  & \Diagram{!c{k^{\prime}}f!c{0}f!c{k^{\prime\prime}}\quad,}
\end{eqnarray*}
where two sub-graphs of $k$ and $k^{\prime}$ vertices are joined
by a bare second order cumulant $\Feyn{f!c{0}f}$. We need to define
\textbf{irreducibility of a graph} in a more general sense here so
that we can extend the results also for perturbative expansions around
non-Gaussian theories. We will call a graph \textbf{reducible}, if
it can be decomposed into a pair of sub-graphs by disconnecting the
end point of a single vertex. In the Gaussian case, this definition
is identical to one-line reducibility, because all end points of vertices
necessarily connect to a second order propagator. This is not necessarily
the case if the bare theory has higher order cumulants. We may have
components of graphs, such as\begin{fmffile}{Exc3}
	\begin{eqnarray}	
		&\parbox{100mm}{
			\begin{fmfgraph*}(150,75)
				\fmfpen{.75thin}
				\fmftop{ou1,og1,ou2,og2,ou3,og3,ou4,og4,ou5,og5}
				\fmfbottom{uu1,ug1,uu2,ug2,uu3,ug3,uu4,ug4,uu5,ug5}
				\fmf{phantom}{ou1,g1,G2,ug2}
				\fmf{phantom}{ou2,v1,g2,ug3}
				\fmf{plain}{v1,g2}
				\fmf{phantom}{ou3,G1,v2,ug4}
				\fmf{phantom}{ou4,g3,G3,ug5}
				\fmf{plain, tension=1.25}{v2,ug4}
				\fmf{phantom}{ou3,v1,G2,ug1}
				\fmf{plain, tension = 1.25}{ou3,v1}
				\fmf{phantom}{v1,G2,ug1}
				\fmf{phantom}{ou4,G1,g2,ug2}
				\fmf{plain}{g2,ug2}
				\fmf{phantom}{ou5,g3,v2,ug3}
				\fmf{phantom,tension=1.}{ou5,g3}
				\fmf{plain,tension=1.}{g3,v2}
				\fmf{plain}{g2,v2}
				\fmf{plain}{g1,v1}
				\fmfv{decor.shape=circle,decor.filled=empty, decor.size=12.thin}{v1,v2}
			\end{fmfgraph*}
		}&
	\end{eqnarray}
\end{fmffile}

where the three-point interaction connects to two third (or higher)
order cumulants on either side. Disconnecting a single leg, either
to the left or to the right, decomposes the diagram into two parts.
We call such a diagram reducible and diagrams without this property
irreducible here.

We employ the following graphical notation: Since $g_{l}(j)=:\feyn{!c{g_{l}}}$
depends on $j$ only indirectly by the $j$-dependence of the contained
bare cumulants, we denote the derivative by attaching one leg, which
is effectively attached to one of the cumulants of $W_{0}$ contained
in $g_{l}$ 
\begin{eqnarray*}
\Diagram{\vertexlabel^{j}f!c{g_{l}}}
 & := & \partial_{j}\:\Diagram{!c{g_{l}}}
:=\partial_{j}g_{l}(j).
\end{eqnarray*}

We first note that \eqref{iteration_g} generates two kinds of contributions
to $g_{l+1}$, corresponding to the lines \eqref{add_connected} and
\eqref{add_reducible}, respectively. The first line causes contributions
that come from the vertices of $\epsilon V(\partial_{j})$ alone.
These are similar as in the linked cluster theorem \eqref{expansion_W_iteration}.
Determining the first order correction yields with $g_{0}=0$
\begin{eqnarray}
g_{1}(j) & = & \frac{\epsilon}{L}\,\exp(-W_{0}(j))\,V(\partial_{j})\,\exp(W_{0}(j))\label{eq:first_order_g1}\\
 & + & \mathcal{O}(L^{-2}),\nonumber 
\end{eqnarray}
which contains all graphs with a single vertex from $V$ and connections
formed by cumulants of $W_{0}$. These graphs are trivially irreducible,
because they only contain a single vertex.

The proof of the linked cluster theorem (see \secref{General-proof-linked-cluster})
shows how the construction proceeds recursively: correspondingly the
$l+1$-st step \eqref{add_connected} generates all connected graphs
from components already contained in $W_{0}+g_{l}$. These are tied
together with a single additional vertex from $\epsilon V(x)$. In
each step, we only need to keep those graphs where the new vertex
in \eqref{add_connected} joins at most one component from \textbf{$g_{l}$}
to an arbitrary number of components of $W_{0}$, hence we maximally
increase the number of vertices in each component by one. This is
so, because comparing the combinatorial factors \eqref{factor_pick_up_one}
and \eqref{factor_pick_up_many}, contributions formed by adding more
than one vertex (joining two or more components from $g_{l}$ by the
new vertex) in a single step are suppressed with at least $L^{-1}$,
so they vanish in the limit (\ref{eq:Gamma_V_final}).

The second term (\ref{eq:add_reducible}) is similar to (\ref{eq:add_connected})
with two important differences:

\begin{itemize}
\item The single appearance of the differential operator $\partial_{j}$
acts like a monopole vertex: the term therefore attaches an entire
sub-diagram contained in $\Gamma_{V}^{(1)}$ by a single link to any
diagram contained in $g_{l}$.
\item The differential operator appears in the form $\partial_{j}-x^{\ast}$.
As a consequence, when setting $j_{0}=\Gamma_{0}^{(1)}(x^{\ast})$
in the end in \eqref{Gamma_V_final}, all terms cancel where $\partial_{j}$
acts directly on $W_{0}(j)$, because $W_{0}^{(1)}(j_{0})=x^{\ast}$;
non-vanishing contributions only arise if the $\partial_{j}$ acts
on a component contained in $g_{l}$. Since vertices and cumulants
can be composed to a final graph in arbitrary order, the diagrams
produced by $\partial_{j}-x^{\ast}$ acting on $g_{l}$ are the same
as those in which $\partial_{j}-x^{\ast}$ first acts on $W_{0}$
and in a subsequent step of the iteration another $\partial_{j}$
acts on the produced $W_{0}^{(1)}$. So to construct the set of all
diagrams it is sufficient to think of $\partial_{j}$ as acting on
$g_{l}$ alone; the reversed order of construction, where $\partial_{j}$
first acts on $W_{0}$ and in subsequent steps of the iteration the
remainder of the diagram is attached to the resulting $W_{0}^{(1)}$,
is contained in the combinatorics.
\item These attached sub-diagrams from $\Gamma_{V}^{(1)}(x^{\ast})$ do
not depend on $j$; the $j$-dependence of all contained cumulants
is fixed to the value $j=\Gamma_{0}^{(1)}(x^{\ast})$, as seen from
\eqref{Gamma_V_recursion}. As a consequence, these sub-graphs cannot
form connections to vertices in subsequent steps of the iteration.
\end{itemize}
From the last point follows in addition, that the differentiation
in \eqref{add_reducible} with $\Gamma_{V}^{(1)}(x^{\ast})\equiv\partial_{x^{\ast}}\Gamma_{V}(x^{\ast})\stackrel{L\to\infty}{=}-\partial_{x^{\ast}}(g_{L}\circ\Gamma_{0}^{(1)}(x^{\ast}))$
produces an inner derivative $\Gamma_{0}^{(2)}$ attached to a single
leg of any component contained in $g_{L}$. Defining the additional
symbol
\begin{eqnarray*}
\Gamma_{0}^{(2)}(x^{\ast})=:\quad & \Diagram{g!p{0}g}
\end{eqnarray*}
allows us to write these contributions as
\begin{eqnarray}
\partial_{x^{\ast}}(g_{L}\circ\Gamma_{0}^{(1)}) & \equiv & (g_{L}^{(1)}\circ\Gamma_{0}^{(1)})\,\Gamma_{0}^{(2)}\Diagram{=\Diagram{!c{g_{L}}fg!p{0}g}
}
.\label{eq:subgraph_g_1}
\end{eqnarray}
So in total at step $l+1$, the line (\ref{eq:add_reducible}) contributes
graphs of the form
\begin{eqnarray}
g_{L}^{(1)}\,\Gamma_{0}^{(2)}\,g_{l}^{(1)} & =\Diagram{!c{g_{L}}fg!p{0}gf!c{g_{l}}}
 & \quad.\label{eq:reducible_general-1}
\end{eqnarray}
Since by their definition as a pair of Legendre transforms we have
\begin{eqnarray*}
1 & =\Gamma_{0}^{(2)}W_{0}^{(2)} & =\Diagram{g!p{0}gf!c{0}f}
\quad,
\end{eqnarray*}
we notice that the subtraction of the graphs (\ref{eq:reducible_general-1})
may cancel certain connected graphs produced by the line \eqref{add_connected}.
In the case of a Gaussian solvable theory $W_{0}$ this cancellation
is the reason why only one-line irreducible contributions remain.
We here obtain the general result, that these contributions cancel
all reducible components, according to the definition above.

To see the cancellation, we note that a reducible graph by our definition
has at least two components joined by a single leg of a vertex. Let
us first consider the case of a diagram consisting of exactly two
one-line irreducible sub-diagrams joined by a single leg. This leg
may either belong to the part $g_{L}^{(1)}$ or to $g_{l}^{(1)}$
in \eqref{reducible_general-1}, so either to the left or to the right
sub-diagram. In both cases, there is a second cumulant $W_{0}^{(2)}$
either left or right of $\Gamma_{0}^{(2)}$. This is because if the
two components are joined by a single leg, this particular leg must
have terminated on a $W_{0}^{(1)}$ prior to the formation of the
compound graph; in either case this term generates $W_{0}^{(1)}\stackrel{\partial_{j}}{\to}W_{0}^{(2)}$.

The second point to check is the combinatorial factor of graphs of
the form \eqref{reducible_general-1}. To construct a graph of order
$k$, where the left component has $k^{\prime}$ bare vertices and
the right has $k-k^{\prime}$, we can choose one of the $L$ steps
within the iteration in which we may pick up the left term by \eqref{add_reducible}.
The remaining $k-k^{\prime}$ vertices are picked up by \eqref{add_connected},
which are $\left(\begin{array}{c}
L-1\\
k-k^{\prime}
\end{array}\right)$ possibilities to choose $k-k^{\prime}$ steps from $L-1$ available
ones. Every addition of a component to the graph comes with $L^{-1}$.
Any graph in $\Gamma_{V}$ with $k^{\prime}$ vertices is $\propto\frac{\epsilon^{k^{\prime}}}{k^{\prime}!}$,
so together we get 
\begin{eqnarray}
\frac{L}{L}\,\frac{\epsilon^{k^{\prime}}}{k^{\prime}!}\,\left(\frac{\epsilon}{L}\right)^{k-k^{\prime}}\left(\begin{array}{c}
L-1\\
k-k^{\prime}
\end{array}\right) & \stackrel{L\to\infty}{\to} & \frac{\epsilon^{k}}{k^{\prime}!(k-k^{\prime})!}.\label{eq:comb_reducible}
\end{eqnarray}
The symmetry factors $s_{1},s_{2}$ of the two sub-graphs generated
by \eqref{reducible_general-1} enter the symmetry factor $s=s_{1}\cdot s_{2}\cdot c$
of the composed graph as a product, where $c$ is the number of ways
in which the two sub-graphs may be joined. But the factor $s$, by
construction, excludes those symmetries that interchange vertices
between the two sub-graphs. Assuming, without loss of generality,
a single sort of interaction vertex, there are $s^{\prime}=\left(\begin{array}{c}
k\\
k^{\prime}
\end{array}\right)$ ways of choosing $k^{\prime}$ of the $k$ vertices to belong to
the left part of the diagram. Therefore the symmetry factor $s$ is
smaller by the factor $s^{\prime}$ than the symmetry factor of the
corresponding reducible diagram constructed by \eqref{add_connected}
alone, because the latter exploits all symmetries, including those
that mix vertices among the sub-graphs. Combining the defect $s^{\prime}$
with the combinatorial factor \eqref{comb_reducible} yields $\frac{1}{k^{\prime}!(k-k^{\prime})!}/s^{\prime}=\frac{1}{k!}$,
which equals the combinatorial factor of the reducible graph.

Let us now study the general case of a diagram composed of an arbitrary
number of sub-diagrams of which $M$ are irreducible and connected
to the remainder of the diagram by exactly one link. The structure
of such a diagram is a (Cayley) tree and $M$ is the number of ``leaves''.
We assume furthermore that the whole diagram has $k$ vertices in
total and a symmetry factor $S$. We can replace $r=0,...,M$ of the
leaves by $\Gamma^{\left(1\right)}$-diagrams. We want to show that
the sum of these $M+1$ sub-diagrams vanishes. A diagram with $r$
replaced leaves yields the contribution
\begin{equation}
\frac{1}{k_{t}!\,\prod_{i=1}^{r}k_{i}!}\tilde{S}\cdot C,\label{eq:Prefactor_replaced_leaves}
\end{equation}
where $\tilde{S}$ is the symmetry factor of the diagram with replaced
leaves, $C$ is some constant equal for all diagrams under consideration
and $k_{t}$ and $k_{i}$ are the numbers of vertices in the ``trunk''
of the tree and in the $i$-th leaf, respectively, where $k_{t}+\sum_{i}^{r}\,k_{i}=k$.
Analogous to the case of two sub-diagrams, we can determine the relation
of $\tilde{S}$ to $S$: We have $\tilde{S}=S\left(\begin{array}{c}
k\\
k_{t},k_{1},...,k_{r}
\end{array}\right)^{-1}=S\frac{k!}{k_{t}!\,\prod_{i=1}^{r}k_{i}!}$, because in the diagram without explicit sub-diagrams, we have $\left(\begin{array}{c}
k\\
k_{t},k_{1},...,k_{r}
\end{array}\right)$ possibilities to distribute the vertices in the respective areas.
Therefore, the first two factors in \eqref{Prefactor_replaced_leaves}
just give $\frac{S}{k!}$, the prefactor of the original diagram.
Now, we have $\left(\begin{array}{c}
M\\
r
\end{array}\right)$ possibilities to choose $r$ leaves to be replaced and each of these
diagrams contributes with the sign $\left(-1\right)^{r}$. Summing
up all contributions leads to
\[
\frac{S\cdot C}{n!}\sum_{r=0}^{M}\left(\begin{array}{c}
M\\
r
\end{array}\right)\left(-1\right)^{r}=\frac{S\cdot C}{n!}\left(1-1\right)^{M}=0.
\]
In summary we conclude that all reducible graphs are canceled by \eqref{reducible_general-1}.

But there is a second sort of graphs produced by \eqref{reducible_general-1}
that does not exist in the Gaussian case: If the connection between
the two sub-components by $\Feyn{gpg}$ ends on a third or higher
order cumulant. These graphs cannot be produced by \eqref{add_connected},
so they remain with a minus sign. We show an example of such graphs
in the following \secref{Example_irreducible}. One may enumerate
all such diagrams by an expansion in terms of skeleton diagrams \citep{Kuehn18_375004}.

We now summarize the algorithmic rules derived from the above observations
to obtain $\Gamma$:
\begin{enumerate}
\item \label{enu:Calculate-Gamma0}Calculate $\Gamma_{0}(x^{\ast})=\sup_{j}\,j^{\T}x^{\ast}-W_{0}(j)$
explicitly by finding $j_{0}$ that extremizes the right hand side.
At this order $g_{0}=0$.
\item \label{enu:recursion}At order $k$ in the perturbation expansion:
\begin{enumerate}
\item \label{enu:connected}add all irreducible graphs in the sense of the
definition above that have $k$ vertices;
\item \label{enu:add-all-irreducible}add all graphs containing derivatives
$\Gamma_{0}^{(n)}$ as connecting elements that cannot be reduced
to the form of a graph contained in the expansion of $W_{V}(j_{0})$;
the graphs left out are the counterparts of the reducible ones in
$W_{V}(j_{0})$. The topology and combinatorial factors of these non-standard
contributions are generated iteratively by \eqref{iteration_g} from
the previous order in perturbation theory; this iteration, by construction,
only produces diagrams, where at least two legs of each $\Gamma_{0}^{(n)}$
connect to a third or higher order cumulant. We can also directly
leave out diagrams, in which a subdiagram contained in $W_{V}$ is
connected to the remainder of the diagram by a single leg of an interaction
vertex.  
\end{enumerate}
\item \label{enu:assign-the-factor-1-1}assign the factor $\frac{\epsilon^{k}}{r_{1}!\cdots r_{l+1}!}$
to each diagram with $r_{i}$-fold repeated occurrence of vertex $i$;
assign the combinatorial factor that arises from the possibilities
of joining the connecting elements as usual in Feynman diagrams (see
examples below).
\item \label{enu:j0_as_x_star-1-1}express the $j$-dependence of the $n$-th
cumulant $\llangle x^{n}\rrangle(x^{\ast})$ in all terms by the first
cumulant $x^{\ast}=\llangle x\rrangle=W_{0}^{(1)}(j_{0})$; this can
be done, for example, by inverting the last equation or directly by
using $j_{0}=\Gamma_{0}^{(1)}(x^{\ast})$; express the occurrence
of $\Gamma_{0}^{(2)}$ by its explicit expression.
\end{enumerate}

\subsection{Example\label{sec:Example_irreducible}}

As an example let us consider the case of a theory with up to third
order cumulants and a three point interaction vertex:

\begin{fmffile}{test7}
\fmfset{thin}{0.75pt}
\fmfset{decor_size}{4mm}
	\begin{eqnarray*}
	\epsilon V(x)=
	\parbox{30mm}{
		\begin{fmfgraph*}(40,40)
			\fmfsurroundn{i}{3}
			\fmf{plain}{i1,v1,i2}
			\fmf{plain}{i3,v1}
		\end{fmfgraph*}
	} \qquad \qquad W_{0}(j)= 
	\parbox{15mm}{
		\begin{fmfgraph}(30,30)
			\fmfsurroundn{i}{2}
			\fmf{phantom}{i1,v,i2}
			\fmfv{d.s=circle, d.filled=empty}{v}
		\end{fmfgraph}
	} \mkern-18mu = \; \quad
	\parbox{10mm}{
		\begin{fmfgraph*}(30,30)
			\fmfsurroundn{i}{2}	
			\fmf{plain}{i2,v}
			\fmf{phantom}{v,i1}
			\fmfv{d.s=circle, d.filled=empty}{v}
			\fmflabel{j}{i2}
		\end{fmfgraph*}		
	} + {1 \over 2} \; \quad
	\parbox{15mm}{
		\begin{fmfgraph*}(30,30)
			\fmfsurroundn{i}{2}	
			\fmf{plain}{i2,v}
			\fmf{plain}{v,i1}
			\fmfv{d.s=circle, d.filled=empty}{v}
			\fmflabel{j}{i2}
			\fmflabel{j}{i1}
		\end{fmfgraph*}		
	} + {1 \over 3!} \quad
	\parbox{15mm}{
		\begin{fmfgraph*}(30,30)
			\fmfsurroundn{i}{3}	
			\fmf{plain}{i2,v,i3}
			\fmf{plain}{v,i1}
			\fmfv{d.s=circle, d.filled=empty}{v}
			\fmflabel{j}{i1}
			\fmflabel{j}{i2}
			\fmflabel{j}{i3}
		\end{fmfgraph*}		
	}
	\end{eqnarray*}
\end{fmffile}

the first order of $g_{1}$ is then

\begin{fmffile}{test8}
\fmfset{thin}{0.75pt}
\fmfset{decor_size}{4mm}
\begin{align*}
	g_{1} \; =  \quad
	\parbox{15mm}{
		\begin{fmfgraph*}(30,30)
			\fmfsurroundn{i}{3}	
			\fmf{plain}{i2,v,i3}
			\fmf{plain}{v,i1}
			\fmfv{d.s=circle, d.filled=empty}{i1,i2,i3}
		\end{fmfgraph*}		
	} + \quad
	\parbox{15mm}{
		\begin{fmfgraph*}(40,30)
			\fmfsurroundn{i}{2}	
			\fmf{plain, tension=1.5}{i2,v}
			\fmf{plain, left=.7, tension=0.5}{v,i1,v}
			\fmfv{d.s=circle, d.filled=empty}{i1,i2}
		\end{fmfgraph*}		
	} \quad + \quad
	\parbox{15mm}{
		\begin{fmfgraph*}(30,30)
			\fmfsurroundn{i}{2}	
			\fmf{plain}{i2,i1}
			\fmf{plain, left=.7, tension=0.5}{i2,i1,i2}
			\fmf{plain, tension=0.2}{i1,i2}
			\fmfv{d.s=circle, d.filled=empty}{i2}
		\end{fmfgraph*}		
	}
\end{align*}
\end{fmffile}

and the second gives

\begin{fmffile}{test9}
\fmfset{thin}{0.75pt}
\fmfset{decor_size}{4mm}
\begin{align*}
	g_{2} - g_{1} = \quad
	\parbox{30mm}{
		\begin{fmfgraph*}(60,60)
			\fmfsurroundn{i}{4}
			\fmf{plain, tension=2}{i3,v1}	
			\fmf{plain}{v1,v2,v3,v4,v1}
			\fmf{plain, tension=2}{i1,v3}
			\fmf{phantom, tension=2.5}{i2,v2}
			\fmf{phantom, tension=2.5}{i4,v4}
			\fmfv{d.s=circle, d.filled=empty}{i3,i1,v2,v4}
		\end{fmfgraph*}		
	} +& \quad
	\parbox{30mm}{
		\begin{fmfgraph*}(60,30)
			\fmfleft{i1,i2}
			\fmfright{o1,o2}
			\fmf{plain, tension=1}{i1,v1,i2}	
			\fmf{plain}{v1,v2,v3}
			\fmf{plain, tension=1}{o1,v3,o2}
			\fmfv{d.s=circle, d.filled=empty}{i1,i2,v2,o1,o2}
		\end{fmfgraph*}		
	} + \; ... \; + \quad
	\parbox{30mm}{
		\begin{fmfgraph*}(100,30)
			\fmfleft{i1}
			\fmfright{o1}
			\fmf{plain, tension=2.5}{i1,v1}	
			\fmf{plain, left=.7}{v1,v2,v1}
			\fmf{plain, tension=2.5}{v2,v3}
			\fmf{plain, left=.7}{v3,o1,v3}
			\fmfv{d.s=circle, d.filled=empty}{i1,v2,o1}
		\end{fmfgraph*}		
	} \\[15pt]
	-& \; 
	\parbox{30mm}{
		\begin{fmfgraph*}(100,30)
			\fmfleft{i1,i2}
			\fmfright{o1,o2}
			\fmf{plain, tension=1}{i1,v1,i2}	
			\fmf{plain, tension=0.7}{v1,v2}
			\fmf{plain}{v2,n1}
			\fmf{wiggly}{n1,n2,n3}
			\fmfv{d.s=circle, d.filled=shaded}{n2}
			\fmfv{label=$\overbrace{\phantom{phantom}}^\text{$= 1$}$, label.angle=90, label.dist=0.7pt}{n3}
			\fmf{plain}{n3,n4}
			\fmf{plain, tension=0.7}{n4,v3}
			\fmf{plain, tension=1}{o1,v3,o2}
			\fmfv{d.s=circle, d.filled=empty}{i1,i2,v2,n4,o1,o2}
		\end{fmfgraph*}		
	} \quad - \; ... \; - \quad 
	\parbox{30mm}{
		\begin{fmfgraph*}(130,30)
			\fmfleft{i1}
			\fmfright{o1}
			\fmf{plain, tension=3.5}{i1,v1}	
			\fmf{plain, left=.7}{v1,v2,v1}
			\fmf{plain, tension=5}{v2,n1}
			\fmf{wiggly, tension=5}{n1,n2,n3}
			\fmfv{d.s=circle, d.filled=shaded}{n2}
			\fmfv{label=$\overbrace{\phantom{phantom}}^\text{$= 1$}$, label.angle=90, label.dist=0.7pt}{n3}
			\fmf{plain, tension=5}{n3,n4}
			\fmf{plain, tension=3.5}{n4,v3}
			\fmfv{d.s=circle, d.filled=empty}{n4}
			\fmf{plain, left=.7}{v3,o1,v3}
			\fmfv{d.s=circle, d.filled=empty}{i1,v2,o1}
		\end{fmfgraph*}		
	} \\[15pt]
	- & \quad
	\parbox{30mm}{
		\begin{fmfgraph*}(145,30)
			\fmfleft{i1}
			\fmfright{o1}
			\fmf{plain, tension=3.5}{i1,v1}	
			\fmf{plain, left=.7}{v1,v2,v1}
			\fmf{plain, tension=5}{v2,n1}
			\fmf{wiggly, tension=5}{n1,n2,n3}
			\fmfv{d.s=circle, d.filled=shaded}{n2}
			\fmfv{label=$\underbrace{\phantom{He sieh mal das ist Phantomas!}}_\text{additional non-cancelling diagram}$, label.angle=-90, label.dist=5pt}{n2}
			\fmf{plain, tension=5}{n3,v3}
			\fmfv{d.s=circle, d.filled=empty}{v3}
			\fmf{plain, left=.7}{v3,v4,v3}
			\fmf{plain,tension=3.5}{v4,o1}
			\fmfv{d.s=circle, d.filled=empty}{i1,v2,o1}
		\end{fmfgraph*}		
	} \\[10pt]
\end{align*}
\end{fmffile}

We see that the diagrams which can be composed out of two sub-diagrams
of lower order and are connected by a single line are cancelled. In
addition we get contributions from the term (\ref{eq:add_reducible}),
where $\Feyn{gpg}$ ties together two lower order components by attaching
to a cumulant of order three or higher on both sides. Such contributions
cannot arise from the term (\ref{eq:add_connected}) and are therefore
not canceled.

\subsection{Vertex functions in the Gaussian case\label{sec:Vertex-functions-Gaussian}}

When expanding around a Gaussian theory 
\begin{align*}
S_{0}(x) & =-\frac{1}{2}\,(x-x_{0})^{\T}A(x-x_{0}),
\end{align*}
the Legendre transform $\Gamma_{0}(x^{\ast})$ is identical to minus
this action, so we have (see \secref{Appendix-Legendre-transform-Gaussian}
for details)

\begin{align}
\Gamma_{0}(x^{\ast}) & =-S_{0}(x^{\ast})=\frac{1}{2}\,(x^{\ast}-x_{0})^{\T}A(x^{\ast}-x_{0}).\label{eq:Gamma_0_Gauss}
\end{align}
Hence writing the contributing diagrams to $\Gamma_{V}(x^{\ast})$,
given by \eqref{Gamma_V_recursion}, we see that (with the symmetry
of $A$ and the product rule)
\begin{align}
j_{0}(x^{\ast})=\Gamma_{0}^{(1)}(x^{\ast}) & =A(x^{\ast}-x_{0})\label{eq:Gamma_1_Gauss}
\end{align}
Here the step of expressing all cumulants by $x^{\ast}$ using (\ref{eq:Gamma_1_Gauss})
is trivial: The cumulant generating function is $W_{0}(j)=j\,x_{0}+\frac{1}{2}j^{\T}A^{-1}j$.
The first cumulant $W_{0}^{(1)}(j_{0})=x_{0}+A^{-1}j_{0}=x_{0}+A^{-1}A(x^{\ast}-x_{0})=x^{\ast}$
is, by construction, identical to $x^{\ast}$ and the second $\Delta=W^{(2)}(j)=A^{-1}$
is independent of $j$ and hence independent of $x^{\ast}$.

Applying the rules derived in \secref{Perturbation-expansion-of-Gamma},
we see that all connections are made by $\Delta=\Feyn{fcf}$. Hence,
all diagrams cancel which are connected of (at least) two components
connected by a single line, because each leg of a vertex necessarily
connects to a line. Also, there are no non-standard diagrams produced,
because there are only second cumulants in $W_{0}$ and because $\Gamma_{0}^{(2)}=[W_{0}^{(2)}]^{-1}=A$
is independent of $x^{\ast}$, so derivatives by $x^{\ast}$ cannot
produce non-standard terms with $\Gamma_{0}^{(>2)}$.

The cancelled diagrams are called \textbf{one-line reducible} or \textbf{one-particle-reducible}.
We therefore get the simple rule for the Gaussian case

\begin{align}
\Gamma_{V}(x^{\ast}) & =-\sum_{\text{1PI}}\in W_{V}(\Gamma_{0}^{(1)}(x^{\ast}))\label{eq:Gamma_V_1PI}\\
 & =-\sum_{\text{1PI}}\in W_{V}(j)\Big|_{j=A(x^{\ast}-x_{0})},\nonumber 
\end{align}
where the subscript $\text{1PI}$ stands for only including the \textbf{one
line irreducible diagrams}, those that cannot be disconnected by cutting
a single line.

Given we have all connected 1PI graphs of $W_{V}$, each external
leg $j$ is connected by a propagator $\Delta=A^{-1}$ to a source
$j$, canceling the factor $A$. Diagrammatically, we imagine that
we set $j=A(x^{\ast}-x_{0})$ in every external line of a graph, so
\begin{align*}
\ldots\feyn{ff0\vertexlabel_{j=A(x^{\ast}-x_{0})}f0} & =\ldots\underbrace{\Delta A}_{1}(x^{\ast}-x_{0}).
\end{align*}
 We therefore obtain the diagrammatic rules for obtaining the vertex
generating function for the perturbation expansion around a Gaussian:
\begin{itemize}
\item Determine all 1PI connected diagrams with any number of external legs.
\item Remove all external legs including the connecting propagator $\Delta=A^{-1}$.
\item Replace the resulting uncontracted $x$ on the vertex that was previously
connected to the leg by $x^{\ast}-x_{0}$.
\item For the expansion around a Gaussian theory, $W_{0}^{(2)}=A^{-1}$
is independent of $j_{0}$; so $x^{\ast}$ can only appear on an external
leg.
\end{itemize}
The fact that the external legs including the propagators are removed
is sometimes referred to as \textbf{amputation}. In the Gaussian case,
by the rules above, the equation of state (\ref{eq:equation_of_state})
amounts to calculating all 1PI diagrams with one external (amputated)
leg. We use the notation 
\begin{align*}
\frac{\partial\Gamma(x^{\ast})}{\partial x_{k}^{\ast}} & =\quad\Diagram{\vertexlabel^{x_{k}^{\ast}}gp}
\end{align*}
for such a derivative of $\Gamma$ by $x_{k}$, as introduced above.
We can also see the amputation for the correction terms directly:
Due to 
\begin{align*}
\frac{\partial\Gamma_{V}}{\partial x^{\ast}}=-\frac{\partial}{\partial x^{\ast}}\sum_{\mathrm{1PI}}\in W_{V}(\Gamma_{0}^{(1)}(x^{\ast})) & =-\sum_{\mathrm{1PI}}\in W_{V}^{(1)}(\Gamma_{0}^{(1)}(x^{\ast}))\Gamma_{0}^{(2)}(x^{\ast})
\end{align*}
each external leg is ``amputated'' by the inverse propagator $\Gamma_{0}^{(2)}=\left(W^{(2)}\right)^{-1}=A$
arising from the inner derivative.

\subsection{Example: Vertex functions of the ``$\phi^{3}+\phi^{4}$''-theory\label{sub:Example-Vertex-function-phi34}}

As an example let us study the action (\ref{eq:def_S_phi34}) with
$K=1$.  We have seen the connected diagrams that contribute to $W$
in \secref{Example_Connecteddiagrams_phi34}. With the results from
\secref{Perturbation-expansion-of-Gamma} we may now determine $\Gamma(x^{\ast})$.
To lowest order we have the Legendre transform of $W_{0}(j)=\frac{1}{2}j^{2}$,
which we determine explicitly as
\begin{align}
\Gamma_{0}(x^{\ast}) & =\sup_{j}\,x^{\ast}j-W_{0}(j),\nonumber \\
\frac{\partial}{\partial j}\left(x^{\ast}j-W_{0}(j)\right) & \stackrel{!}{=}0\leftrightarrow x^{\ast}=j,\nonumber \\
\Gamma_{0}(x^{\ast}) & =\left(x^{\ast}\right)^{2}-W_{0}(x^{\ast})=\frac{1}{2}\left(x^{\ast}\right)^{2},\label{eq:Gamma0_phi34}
\end{align}
So for a Gaussian theory, we have that $\Gamma_{0}(x^{\ast})=-S(x^{\ast})$.
The loopwise expansion studied in \secref{Loopwise-expansion-vertex-function}
will yield the same result. We will, however, see in \subref{Example-TAP-approximation}
that in the general case of a non-Gaussian solvable theory this is
not so.

The corrections of first order are hence the connected diagrams with
one interaction vertex (which are necessarily 1PI), where we need
to replace, according to (\ref{eq:Gamma_1_Gauss}), $j=\Gamma_{0}^{(1)}(x^{\ast})=x^{\ast}$
so we get from the diagrams with one external leg (compare \secref{Example_Connecteddiagrams_phi34})
\begin{align*}
\Diagram{\vertexlabel_{x^{\ast}}gf0flfluf0}
= & 3\cdot\,x^{\ast}\,\epsilon\frac{\alpha}{3!}=\epsilon\frac{\alpha}{2}\,x^{\ast}.
\end{align*}
We here used the notation $\Diagram{g}
$ for the amputated legs. From the correction with two external legs
we get
\begin{align*}
\Diagram{\vertexlabel^{x^{\ast}}\\
gd & f0flfluf0\\
\vertexlabel_{x^{\ast}}gu
}
 & =4\cdot3\cdot\frac{\left(x^{\ast}\right)^{2}}{2!}\,\epsilon\frac{\beta}{4!}=\epsilon\frac{\beta}{4}\,\left(x^{\ast}\right)^{2}.
\end{align*}
Finally we have the contributions from the bare interaction vertices
with three and four legs
\begin{align*}
3\cdot2\cdot\Diagram{\vertexlabel^{x^{\ast}}gdg\vertexlabel_{x^{\ast}}\\
\vertexlabel^{x^{\ast}}gu
}
 & =3\cdot2\cdot\frac{\left(x^{\ast}\right){}^{3}}{3!}\epsilon\frac{\alpha}{3!}\\
\\
\Diagram{\vertexlabel^{x^{\ast}}gd & gu\vertexlabel^{x^{\ast}}\\
\vertexlabel^{x^{\ast}}gu & gd\vertexlabel^{x^{\ast}}
}
 & =4\cdot3\cdot2\cdot\frac{\left(x^{\ast}\right)^{4}}{4!}\epsilon\frac{\beta}{4!}.
\end{align*}
The latter two terms show that the effective action contains, as a
subset, also the original vertices of the theory.

So in total we get the correction at first order to $\Gamma$

\begin{align}
\Gamma_{V,1}(x^{\ast}) & =-\epsilon\,\left(\frac{\alpha}{2}\,x^{\ast}+\frac{\beta}{4}\,\left(x^{\ast}\right)^{2}+\frac{\alpha}{3!}\,\left(x^{\ast}\right)^{3}+\frac{\beta}{4!}\,\left(x^{\ast}\right)^{4}\right).\label{eq:Gamma_1_phi34}
\end{align}
The expansion of $\Gamma$ including all corrections up to second
order in $\epsilon$ will be content of the exercises.

\subsection{Appendix: Explicit cancellation until second order\label{sub:Explicit-cancellation-reducible}}

Alternative to the general proof given above, we may see order by
order in $k$, that \eqref{Gamma_V_1PI} holds. At lowest order $W_{V}\equiv0$
and \eqref{Gamma0_pert-1} holds, so the assumption is true. Taking
into account the corrections that have one interaction vertex, we
get the additional term $W_{V,1}(\Gamma^{(1)}(x^{\ast}))=W_{V,1}(\Gamma_{0}^{(1)}(x^{\ast}))+O(k^{2})$.
We have replaced here the dependence on $\Gamma^{(1)}(x^{\ast})$
by the lowest order $\Gamma_{0}^{(1)}(x^{\ast})$, because $W_{V,1}$
already contains one interaction vertex, so the correction would already
be of second order. As there is only one interaction vertex, the contribution
is also 1PI. In addition, we get a correction to $j=j_{0}+j_{V,1}$,
inserted into 

\begin{align}
\Gamma(x^{\ast})) & =\left.j^{\T}x^{\ast}-W_{0}(j)-W_{V}(j)\right|_{j=\Gamma^{(1)}(x^{\ast})}\nonumber \\
W_{V}(j) & =\ln\,\exp(-W_{0}(j))\,\exp(\epsilon V(\nabla_{j}))\,\exp(W_{0}(j))\label{eq:Gamma_decomp-2}
\end{align}
leaves us with
\begin{align}
-\Gamma(x^{\ast}) & =\left.\underbrace{-j_{0}^{\T}x^{\ast}+W_{0}(j_{0})}_{-\Gamma_{0}(x^{\ast})}+j_{V,1}^{\T}\underbrace{\left(W_{0}^{(1)}(j_{0})-x^{\ast}\right)}_{=0}+W_{V,1}(j_{0})\right|_{j_{0}=\Gamma_{0}^{(1)}(x^{\ast})\quad j_{V,1}=\Gamma_{1}^{(1)}(x^{\ast})}+O(k^{2}),\nonumber \\
 & =\left.-\Gamma_{0}(x^{\ast})+W_{V,1}(j_{0})\right|_{j_{0}=\Gamma_{0}^{(1)}(x^{\ast})},\label{eq:first_order_Gamma_explicit-1}
\end{align}
where the shift of $j$ by $j_{V,1}$ in the two terms making up $\Gamma_{0}$
cancel each other. To first order, the assumption is hence true. At
second order we have
\begin{align*}
-\Gamma(x^{\ast}) & =\underbrace{-j_{0}^{\T}x^{\ast}+W_{0}(j_{0})}_{-\Gamma_{0}(x^{\ast})}+\\
 & +(j_{V,1}+j_{V,2})^{\T}\underbrace{\left(W_{0}^{(1)}(j_{0})-x^{\ast}\right)}_{=0}+\underline{\frac{1}{2}j_{1}^{\T}W_{0}^{(2)}j_{1}+j_{1}^{\T}W_{V,1}^{(1)}}+W_{V,1}(j_{0})+W_{V,2}(j_{0})\Big|_{j_{0}=\Gamma_{0}^{(1)}(x^{\ast})\quad j_{V,1}=\Gamma_{1}^{(1)}(x^{\ast})}+O(k^{3}).
\end{align*}
Using that $j_{1}=-W_{V,1}^{(1)}(\Gamma_{0}^{(1)}(x^{\ast}))\,\Gamma_{0}^{(2)}(x^{\ast})$,
following from (\ref{eq:first_order_Gamma_explicit-1}), we can combine
the two underlined terms by using $\Gamma_{0}^{(2)}(x^{\ast})W_{0}^{(2)}(j_{0})=1$
to obtain $W_{V,1}^{(1)\T}\,j_{1}=-W_{V,1}^{(1)\T}(j_{0})\,\Gamma_{0}^{(2)}(x^{\ast})\,W_{V,1}^{(1)\T}(j_{0})$.
We see that $\Gamma_{0}^{(2)}(x^{\ast})=\left(W_{0}^{(2)}(j_{0})\right)^{-1}$
amputates the propagator of the external legs of $W_{V,1}^{(1)}$.
The latter factor $W_{V,1}^{(1)}$ in any case has an external leg
connected to the remaining graph, also if the solvable theory has
non-vanishing mean $W_{0}^{(1)}(0)\neq0$, because $W$ by the linked
cluster theorem (see \secref{Linked-cluster-theorem}) only contains
connected diagrams whose end points are either $W_{0}^{(2)}(j)\,j$
or $W_{0}^{(1)}(j)$. In the first case, the derivative acting on
$W^{(2)}$ yields $0$ (by the assumption $W_{0}^{(\ge3)}=0$), acting
on $j$ yields $W^{(2)}(j)$. In the second case, the derivative acts
on the argument of $W^{(1)}(j)$ and hence also produces a factor
$W^{(2)}(j)$. In all cases, the term hence consists of two 1PI components
of first order connected by a single line. So in total we get 
\begin{align*}
-\Gamma(x^{\ast}) & =-\Gamma_{0}(x^{\ast})+W_{V,1}(j_{0})+\underbrace{W_{V,2}(j_{0})-\frac{1}{2}W_{V,1}^{(1)\T}(j_{0})\,\Gamma_{0}^{(2)}(x^{\ast})\,W_{V,1}^{(1)\T}(j_{0})}_{\sum_{\text{1PI}}\in W_{V,2}(j_{0})}.
\end{align*}
The last two terms together form the 1PI diagrams contained in $W_{V,2}(j_{0})$:
All diagrams of second order that are connected by a single link (coming
with a factor $1/2$, because they have two interaction vertices,
see \secref{Perturbation-expansion}) are canceled by the last term,
which produces all such contributions.

\subsection{Appendix: Convexity of $W$\label{sec:Convexity-W}}

We first show that the Legendre transform of any function $f(j)$
is convex. This is because for
\begin{align*}
g(x) & :=\sup_{j}\,j^{\T}x-f(j)
\end{align*}
we have with $\alpha+\beta=1$
\begin{align*}
g(\alpha x_{a}+\beta x_{b}) & =\sup_{j}\,j^{\T}(\alpha x_{a}+\beta x_{b})-(\alpha+\beta)\,f(j)\\
 & \le\sup_{j_{a}}\,\alpha\big(j_{a}^{\T}x_{a}-f(j_{a})\big)+\sup_{j_{b}}\,\beta\big(j_{b}^{\T}x_{b}-f(j_{b})\big)\\
 & =\alpha\,g(x_{a})+\beta\,g(x_{b}),
\end{align*}
which is the definition of a convex down function: the function is
always below the connecting chord. Hence we can only come back to
$W$ after two Legendre transforms if $W$ is convex to start with.

We now show that a differentiable $W$ is convex. This is because
$W^{(2)}$ is the covariance matrix, it is symmetric and therefore
has real eigenvalues. For covariance matrices these are in addition
always positive semi-definite \citep[p. 166]{ZinnJustin96}. This
can be seen from the following argument. Let us define the bilinear
form 
\begin{align*}
f(\eta) & :=\eta^{\T}W^{(2)}\eta.
\end{align*}
A positive semi-definite bilinear form has the property $f(\eta)\ge0\quad\forall\eta$.
Because $W^{(2)}$ is symmetric, the left and right eigenvectors are
identical. Therefore positive semi-definite also implies that all
eigenvalues must be non-negative. With $\delta x:=x-\langle x\rangle$
we can express $W_{kl}^{(2)}=\langle\delta x_{k}\delta x_{l}\rangle$,
because it is the covariance, so we may explicitly write $f(\eta)$
as 
\begin{align*}
f(\eta) & =\sum_{k,l}\eta_{k}W_{kl}^{(2)}\eta_{l}\\
 & =\Z^{-1}(j)\,\eta^{\T}\int\,dx\,\delta x\,\delta x^{\T}\,\exp\left(S(x)+j^{\T}x\right)\eta\\
 & =\Z^{-1}(j)\,\int\,dx\,\left(\eta^{\T}\delta x\right)^{2}\,\exp\left(S(x)+j^{\T}x)\right)\ge0,
\end{align*}
where $\Z^{-1}(j)=\int\,dx\,\exp\left(S(x)+j^{\T}x)\right)\ge0$.

Therefore even if $W(j)$ has vanishing Hessian on a particular segment
($W$ has a linear segment), $\sup_{j}j^{\T}x^{\ast}-W(j)$ has a
unique value for each given $x^{\ast}$ and hence $\Gamma(x^{\ast})$
is well defined.

\subsection{Appendix: Legendre transform of a Gaussian\label{sec:Appendix-Legendre-transform-Gaussian}}

For a Gaussian theory $S_{0}(x)=-\frac{1}{2}\,(x-x_{0})^{\T}A\,(x-x_{0})$
and $\Delta=A^{-1}$ we have

\begin{align*}
W_{0}(j) & =j^{\T}x_{0}+\frac{1}{2}j^{\T}\,\Delta\,j\\
\Gamma_{0}(x^{\ast}) & =\sup_{j}\,j^{\T}x^{\ast}-W_{0}(j).
\end{align*}
We find the extremum for $j$ as
\begin{align*}
0 & =\partial_{j}\left(j^{\T}x^{\ast}-W_{0}(j)\right)=\partial_{j}\left(j^{\T}(x^{\ast}-x_{0})-\frac{1}{2}j^{\T}\,\Delta\,j\right)=x^{\ast}-x_{0}-\Delta j\\
j & =\Delta^{-1}\left(x^{\ast}-x_{0}\right).
\end{align*}
Inserted into the definition of $\Gamma_{0}$ this yields (with $\Delta^{-1}=\Delta^{-1T}$)
\begin{align}
\Gamma_{0}(x^{\ast}) & =\left(x^{\ast}-x_{0}\right)^{\T}\Delta^{-1}(x^{\ast}-x_{0})-\frac{1}{2}\left(x^{\ast}-x_{0}\right)^{\T}\Delta^{-1}\Delta\,\Delta^{-1}\left(x^{\ast}-x_{0}\right).\nonumber \\
 & =\frac{1}{2}\left(x^{\ast}-x_{0}\right)^{\T}\,A\,\left(x^{\ast}-x_{0}\right)\label{eq:Gamma0_Gauss}\\
 & =-S_{0}(x^{\ast}).\nonumber 
\end{align}

\section{Application: TAP approximation\label{sub:Example-TAP-approximation}}

Suppose we are recording the activity of $N$ neurons. We bin the
spike trains with a small bin size $b$, so that the spike trains
are converted into a sequence of binary numbers $n_{i}\in[0,1]$ in
each time step for the neuron $i$. We would like to describe the
system by a joint probability distribution $p(n_{1},\ldots,n_{N})$
which we choose to maximize the entropy, while obeying the constraints
$\langle n_{i}\rangle=m_{i}$ and $\llangle n_{i}n_{j}\rrangle=c_{ij}$,
where the mean activity $m_{i}$ and the covariance $c_{ij}$ is measured
from data. The distribution is then of the Boltzmann type \citep{Jaynes57_620}\ifthenelse{\boolean{lecture}}{,
as shown in \prettyref{sec:maximum_entropy}} with an action

\begin{align}
S(n) & =\frac{\epsilon}{2}n^{\T}Kn+j^{\T}n\label{eq:pairwise_linked_cluster-1}\\
 & =\frac{\epsilon}{2}\sum_{k\neq l}n_{k}K_{kl}n_{l}+\underbrace{\sum_{k}j_{k}n_{k}}_{S_{0}},\nonumber 
\end{align}
We here want to illustrate the presented methods by deriving the Thouless-Anderson-Palmer
(TAP) \citep{Thouless77_593,Nakanishi97_8085,Tanaka98_2302} mean-field
theory of this pairwise model with with non-random couplings diagrammatically.
The TAP approximation plays an important role for spin glasses \citep{Fischer91},
but it is more recently also employed to efficiently train restricted
Boltzmann machines \citep{Gabrie15}. This expansion has an interesting
history. It has first been systematically derived by Vasiliev and
Radzhabov \citep{Vasiliev74} and was independently proposed by Thouless,
Anderson, and Palmer \citep{Thouless77_593}, but without proof. Later
Georges and Yedidia \citep{Georges91} found an iterative procedure
to compute also higher order corrections, but a direct diagrammatic
derivation has been sought for some time \citep[p. 28]{Opper01}.
The diagrammatic derivation here follows \citep{Kuehn18_375004}.

We want to treat the system perturbatively, where the part indicated
as $S_{0}$ is the solvable part of the theory, which is diagonal
in the index space of the units. Note that $K_{ij}$ only couples
units with different indices $i\neq j$, so we can treat $K_{ii}=0$.
We consider the part $\epsilon V(n)=\frac{\epsilon}{2}\sum_{k\neq l}n_{k}K_{kl}n_{l}$
perturbatively in $\epsilon$.

We use the double role of $j_{i}$, on the one hand being source
terms, on the other being parameters. We may separate these roles
by formally replacing $j_{i}\to j_{i}+h_{i}$ and setting the new
$j_{i}=0$ in the end. The calculation of the TAP mean-field theory
proceeds in a number of steps. We here follow the recipe given at
the end of \prettyref{sec:Perturbation-expansion-of-Gamma}.
\begin{enumerate}
\item Calculate \prettyref{eq:def_W} $W_{0}(j)=\ln\,\Z_{0}(j)-c$ (ignoring
the inconsequential constant c) of the solvable part.
\item Obtain the lowest order \prettyref{eq:Gamma0_pert-1} of the effective
action $\Gamma(m)$, introducing the notation $m_{i}=\langle n_{i}\rangle$,
which plays the role of $x^{\ast}$.
\item Then it holds that $\Gamma_{0}^{(1)}(m)=j_{0}$ and $W_{0}^{(1)}(j_{0})=m$,
as it should by the property \prettyref{eq:equation_of_state} of
the Legendre transform, i.e. $\Gamma_{0}^{(1)}$ and $W_{0}^{(1)}$
are inverse functions of one another.
\end{enumerate}
We treat the system perturbatively, considering the part indicated
as $S_{0}(n)$ in \prettyref{eq:pairwise_linked_cluster-1} as the
solvable part of the theory, in which the action decomposes into a
sum of single-spin problems
\begin{align*}
\Z_{0}(j) & =\prod_{i=1}^{N}(1+e^{j_{i}}).
\end{align*}
The cumulant generating function correspondingly becomes a sum
\begin{align}
W_{0}(j) & =\sum_{i=1}^{N}\ln(1+e^{j_{i}})+c,\label{eq:W0_binary}
\end{align}
where the constant $c$ arises from the normalization and is inconsequential.
To lowest order we therefore get the contribution $\Gamma_{0}(m)=\sup_{j}\,j^{\T}m-W_{0}(j)$
to the effective action. We determine the point of the supremum as
$\nabla_{j}\left(j^{\T}m-W_{0}(j)\right)\stackrel{!}{=}0$, so

\begin{align*}
m_{i}(j_{i}) & =\frac{e^{j_{i}}}{1+e^{j_{i}}}\qquad e^{j_{i}(m_{i})}=\frac{m_{i}}{1-m_{i}}.
\end{align*}
So the explicit form of the lowest order contribution is
\begin{align}
\Gamma_{0}(m) & =\sum_{i=1}^{N}j_{i}(m_{i})\,m_{i}-W_{0}(j(m))\label{eq:Gamma0_TAP}\\
 & =\sum_{i=1}^{N}\ln(m_{i})m_{i}+\ln(1-m_{i})(1-m_{i}),\nonumber 
\end{align}
which is the entropy of the independent distribution of binary variables
with mean $m_{i}$. The reason is that one can construct the distribution
$\exp(S_{0}(n))$ my maximizing the entropy for given mean value.
The condition for the constraint maximization of the entropy has the
form of a Legendre transform, which we are undoing here; the Legendre
transform is involutive.

We see that
\begin{align*}
j_{i}:=\Gamma_{0,i}^{(1)}(m) & =\ln(\frac{m_{i}}{1-m_{i}})
\end{align*}
and $W_{0}^{(1)}$ given by \prettyref{eq:propagators_binary} below
are indeed inverse functions of one another.
\begin{enumerate}
\item[4.] We now need to find the cumulants of the unperturbed system, required
to evaluate all corrections in $\Gamma_{V}$ up to second order in
$\epsilon$, i.e. $W_{0}^{(1)}(j)$ and $W_{0}^{(2)}(j)$; we will
use the diagrammatic notation here.
\item[5. ] According to the algorithm at the end of \prettyref{sec:Perturbation-expansion-of-Gamma},
we then need to express the cumulants in terms of $m$, by replacing
$j=j_{0}=\Gamma_{0}^{(1)}(m)$, using the insight from point 3 above.
\end{enumerate}
We obtain the cumulants of the free theory \eqref{W0_binary} as
\begin{align}
\partial_{i}W_{0} & =\frac{e^{j_{i}}}{1+e^{j_{i}}}=:m_{i}(j_{i}),\label{eq:propagators_binary}\\
\partial_{i}\partial_{j}W_{0} & =\delta_{ij}\,\Big(\frac{e^{j_{i}}}{1+e^{j_{i}}}-\frac{e^{2j_{i}}}{(1+e^{j_{i}})^{2}}\Big)=\delta_{ij}\,\frac{e^{j_{i}}}{(1+e^{j_{i}})^{2}}=\delta_{ij}\,\left.m_{i}\,(1-m_{i})\right|_{m_{i}=m_{i}(j_{i})},\nonumber 
\end{align}
The first line is the first cumulant, the second line the second cumulant
of the single binary variable.
\begin{enumerate}
\item[6. ] Now we need to determine all diagrams up to second order in $\epsilon$
that contribute to $\Gamma(m)$. Here we only need to compute the
diagrams with the features explained in \prettyref{sec:Perturbation-expansion-of-Gamma}.
This requires the knowledge of the cumulants $W^{(n)}(\Gamma_{0}^{(1)}(m))$
expressed in terms of $m$, as obtained under point 5 above. In the
perturbing part $\epsilon V(n)$, we only have a single interaction
vertex that is quadratic in the fields, namely $\epsilon\frac{V^{(2)}}{2!}=\frac{\epsilon}{2}\sum_{i\neq j}K_{ij}=\Feyn{fufd}$.
\end{enumerate}
If we truncate the perturbation expansion of $\Gamma(m)$ at second
order in $K_{ij}$, we need to consider all connected diagrams of
at most two such vertices, connected by the propagators \prettyref{eq:propagators_binary}.
The correction of first order in $K_{ij}$ yields the single term
\begin{align*}
-\Diagram{!c{0,i} & fufd & !c{0,j}}
= & -\frac{\epsilon}{2}\sum_{i\neq j}K_{ij}m_{i}m_{j}.
\end{align*}
We do not get any contribution from the second cumulant, because these
only join elements with identical indices and $K_{ii}=0$. The minus
sign appears here from \prettyref{eq:add_connected}. We obtain the
next order recursively from \prettyref{eq:add_connected}, dropping
all graphs which are connected out of two sub-graphs via a second
cumulant, such as $\Diagram{!c{0,i} & fufd & !c{0,j}fufd!c{0,k}}
=\frac{\epsilon^{2}}{2!}\,\sum_{i\neq j}\sum_{j\neq k}\,\frac{1}{2}K_{ij}\frac{1}{2}K_{jk}\,m_{i}\,m_{j}(1-m_{j})\,m_{k}$. The correction term of second order in $K$ is
\begin{align*}
\Diagram{-\Diagram{!c{0,i} & fufd & !c{0,j}\\
 & fdfu
}
=}
= & -\frac{1}{2!}\,\frac{\epsilon}{2}\sum_{i\neq j}\frac{\epsilon}{2}K_{ij}\sum_{k\neq l}K_{kl}\,\Big(\delta_{ik}\delta_{jl}m_{i}(1-m_{i})\,m_{j}(1-m_{j})+\delta_{il}\delta_{jk}m_{i}(1-m_{i})\,m_{j}(1-m_{j})\Big)\\
= & -\frac{\epsilon^{2}}{4}\,\sum_{i\neq j}K_{ij}^{2}\,m_{i}(1-m_{i})\,m_{j}(1-m_{j}),
\end{align*}
where we get the combinatorial factor $2$ from the two possible orientations
of attaching the second interaction vertex, as indicated by the Kronecker
$\delta$ expressions. The contribution $-\sum_{i\neq j}\sum_{k\neq l}K_{ij}K_{kl}m_{i}m_{j}m_{k}m_{l}$
does not appear, because the two interaction vertices are not connected
in this contribution and four terms of the form $-\sum_{i\neq j}\sum_{k\neq l}\,K_{ij}K_{kl}\delta_{ik}m_{i}(1-m_{i})\,m_{j}m_{l}$
disappear, because they are one particle reducible, since we may cut
the single line formed by the second cumulant connecting the two interaction
vertices.

\begin{figure}
\begin{centering}
\includegraphics{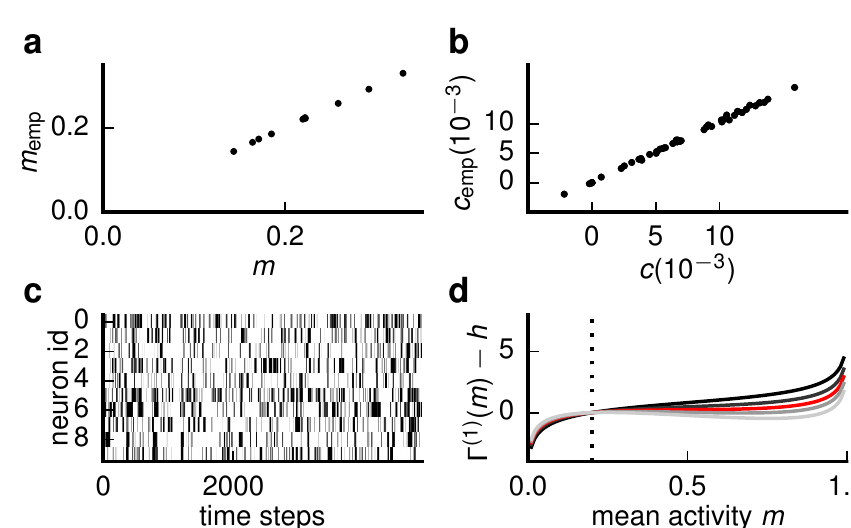}
\par\end{centering}
\caption{\textbf{Maximum entropy pairwise model and TAP mean-field theory.
}$N=10$ neurons. \textbf{(a)} Scatter plot of the given mean activity
of the neurons $m$ and the empirical estimate $m_{\mathrm{emp}}$
obtained by sampling of Glauber dynamics. Assigned mean activities
are normally distributed with mean $\langle m_{i}\rangle_{i}=0.2$
and standard deviation $\langle m_{i}^{2}\rangle_{i}-\langle m_{i}\rangle^{2}=0.05$,
clipped to $m_{i}\in[0.05,0.95]$. \textbf{(b)} Scatter plot of covariance.
Initial values chosen with correlation coefficients $k_{ij}\equiv\frac{c_{ij}}{\sqrt{m_{i}(1-m_{i})m_{j}(1-m_{j})}}$
randomly drawn from a normal distribution with mean $0.05$ and standard
deviation $0.03$, clipped to $k_{ij}\in[-1,1]$. \textbf{(c)} States
as a function of time step of the Glauber dynamics. Black: $n_{i}=1$,
white: $n_{i}=0$.\textbf{ (d)} Effective action $\Gamma(m)$ for
the homogeneous model. From black to light gray: $N=10,20,N_{C},50,100$.
Red curve: Critical value $N_{C}\simeq32$ at which the approximation
of $\Gamma$ becomes non-convex. Empirical results were obtained from
Glauber dynamics simulated for $T=10^{7}$ time steps.\label{fig:TAP}}
\end{figure}

The approximation of $\Gamma$ up to second order of $K_{ij}$ thus
reads

\begin{align}
\Gamma(m) & =\sum_{i=1}^{N}\ln(m_{i})m_{i}+\ln(1-m_{i})(1-m_{i})\label{eq:Gamma_TAP}\\
 & -\frac{\epsilon}{2}\sum_{i\neq j}K_{ij}m_{i}m_{j}\nonumber \\
 & -\frac{\epsilon^{2}}{4}\sum_{i\neq j}K_{ij}^{2}\,m_{i}(1-m_{i})\,m_{j}(1-m_{j})+O(\epsilon^{3}),\nonumber 
\end{align}
where the first term is the entropy of $N$ independent binary variables,
the second yields Curie-Weiss mean-field theory and the last line
is the Onsager reaction term.
\begin{enumerate}
\item[7. ] We now determine the equation of state \prettyref{eq:equation_of_state}.
This will give an expression for the parameters $h_{i}$.
\end{enumerate}
The equation of state therefore is
\begin{align}
h_{i} & =\frac{\partial\Gamma}{\partial m_{i}}\nonumber \\
h_{i} & =\ln(\frac{m_{i}}{1-m_{i}})-\sum_{j}\,\epsilon K_{ij}m_{j}-\frac{\epsilon^{2}}{2}\sum_{j}K_{ij}^{2}\,(1-2m_{i})\,m_{j}(1-m_{j})+O(\epsilon^{3}),\label{eq:equation_of_state_binary}
\end{align}
which can be solved numerically, for example by bisection. The result
of this approximation is shown in \prettyref{fig:TAP}.

The approximation of $Z$ up to second order will lead to a calculation
of several pages to reach the same result. A summary of different
mean-field methods can also be found in \citep{Roudi09,Opper01}.
The original works employed Ising spins $s_{i}\in\{-1,1\}$. We here
instead use binary variables $n_{i}\in\{0,1\}$. Both models are mathematically
identically, because all binary state representations are bijectively
linked.

\subsection*{Inverse problem}

Here we may use the approximation \eqref{Gamma_TAP} of $\Gamma$
to solve the so-called \textbf{inverse problem}, which is finding
the equations for the parameters $h_{i}$ and $J_{ij}$ for given
mean activities $m_{i}$ and covariances $c_{ij}=W_{ij}^{(2)}$. This
problem typically arises in data analysis: We want to construct a
maximum entropy model that obeys the constraints given by the data.
To solve the inverse problem, we can make use of \prettyref{eq:WGamma_inv_Hessians}
and determine the inverse of the covariance matrix
\begin{align*}
\left(W^{(2)}\right)_{ij}^{-1} & =\Gamma_{ij}^{(2)}.
\end{align*}
To determine the $J_{ij}$ it suffices to evaluate this expression
on the off-diagonal $i\neq j$

\begin{align}
i\neq j:\quad\Gamma_{ij}^{(2)} & =\frac{\partial^{2}\Gamma}{\partial m_{i}\partial m_{j}}\nonumber \\
 & =-\epsilon K_{ij}-\frac{\epsilon^{2}}{2}\,K_{ij}^{2}\,(1-2m_{i})\,(1-2m_{j}).\label{eq:TAP_J_ij_inverse}
\end{align}
So given we know the covariance matrix $c_{ij}=W_{ij}^{(2)}$ and
the mean activities $m_{i}$, we may determine $K_{ij}$ from \prettyref{eq:TAP_J_ij_inverse}
as

\begin{align*}
K_{ij}^{\pm} & =-\frac{1}{2\kappa_{ij}}\pm\sqrt{\frac{1}{4\kappa_{ij}^{2}}-\frac{\left[c^{-1}\right]_{ij}}{\kappa_{ij}}}\\
\kappa_{ij} & =\frac{1}{2}(1-2m_{i})(1-2m_{j}),
\end{align*}
where the positive sign yields the correct result $K_{ij}=0$ for
the case of vanishing correlations on the off-diagonal elements of
$c_{ij}$. We then obtain the biases $h_{i}$ from \prettyref{eq:equation_of_state_binary}.

\section{Expansion of cumulants into tree diagrams of vertex functions\label{sec:Tree-level-expansion-Cumulants-Vertex}}

\begin{figure}
\begin{centering}
\includegraphics{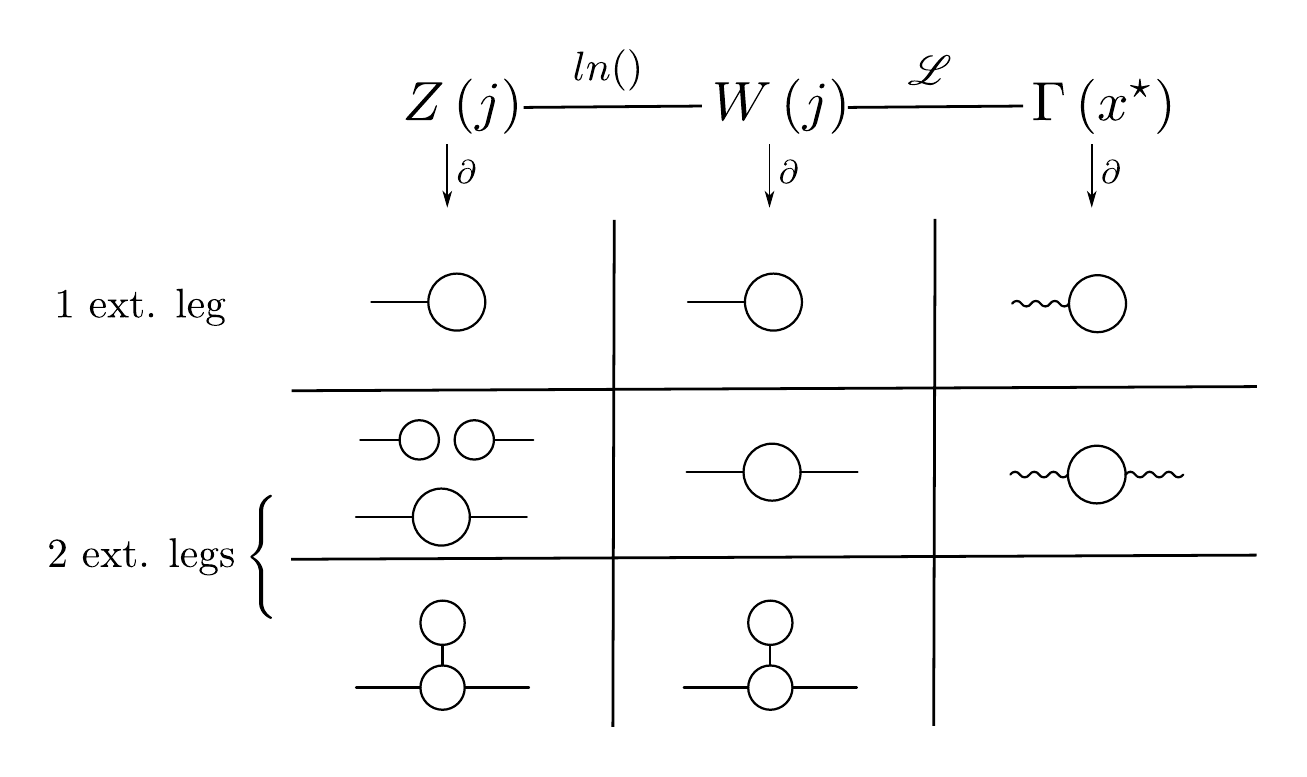}
\par\end{centering}
\caption{\textbf{Graphical summary of the connection between $Z$, $W$ and
$\Gamma$. }Here on the example of a perturbation expansion around
a Gaussian theory with a three-point interaction. The rows correspond
to different orders in the perturbation, the number of three-point
vertices. By the linked cluster theorem, the step from $Z(j)$ to
$W(j)$ removes all diagrams that are disconnected; $W$ only contains
the statistical dependence of the respective variables. Bottom row:
In both cases we get tadpole diagrams appearing as sub-diagrams. These
are perturbative corrections to the first moment, which is identical
to the first cumulant and therefore appear for $Z$ and $W$ alike.
The Legendre transform $\mathcal{L}$ from $W(j)$ to $\Gamma(x^{\ast})$,
which expresses all quantities in terms of the mean value $x^{\ast}=\langle x\rangle(j)$,
removes all diagrams that come about by perturbative corrections to
the mean. This makes sense, because the mean is prescribed to be $x^{\ast}$:
In the Gaussian case, the one-line reducible diagrams are removed,
because the sub-diagram connected with a single line also appears
as a perturbative correction to the mean.\label{fig:Graphical-summary-ZWGamma}}
\end{figure}

In the previous section we have derived an iterative procedure to
construct all contributions to the vertex generating function. In
this section we will show that there is a simple set of graphical
rules that connect the Feynman diagrams that contribute to $\Gamma$
and those of the cumulants, that make up $W$. Graphically, we can
summarize the connection between $Z$, $W$, and $\Gamma$ in \figref{Graphical-summary-ZWGamma}.

In the same line as for the moment and cumulant generating function,
we write $\Gamma$ as a Taylor series, the coefficients of which we
call \textbf{vertex functions} for reasons that will become clear
in the end of this section. These vertex functions are defined as
the $n$-th derivatives of the function $\Gamma$
\begin{align}
\Gamma^{(n_{1},\ldots,n_{N})}(x^{\ast}): & =\partial_{1}^{n_{1}}\cdots\partial_{N}^{n_{N}}\,\Gamma(x^{\ast}).\label{eq:def_vertex_function}
\end{align}
Conversely, we may of course write $\Gamma$ in its Taylor representation
with $\delta x_{i}^{\ast}=x_{i}^{\ast}-x_{0,i}$
\begin{align}
\Gamma(x^{\ast}) & =\sum_{n_{1},\ldots,n_{N}}\frac{\Gamma^{(n_{1},\ldots,n_{N})}(x_{0})}{n_{1}!\ldots n_{N}!}\,\delta x_{1}^{\ast n_{1}}\cdots\delta x_{N}^{\ast n_{N}},\label{eq:Taylor_Gamma}
\end{align}
where $x_{0}$ is an arbitrary point around which to expand. We saw
in the Gaussian case in \secref{Vertex-functions-Gaussian} that the
mean value of the unperturbed Gaussian appeared naturally in the expansion:
the $x^{\ast}$ dependence appeared only on the external legs of the
diagrams in the form $x^{\ast}-x_{0}$.

We will now again use a graphical representation for the Taylor coefficients
that appear in \eqref{def_vertex_function}, where an additional derivative
by $x_{i}^{\ast}$ adds a leg with the corresponding index $i$ to
the vertex $\Gamma^{(n)}$ and similarly for the derivative of $W^{(n)}$.
Without loss of generality, let us assume that we differentiate by
each variable only once and that we can always rename the variables,
so that we differentiate by the first $k$ variables each once. The
general case can be reconstructed from these rules by setting a certain
number of variables equal, as in \secref{Connection-moments-cumulants}.

\begin{center}
\par\end{center}

\begin{fmffile}{k+1_derivatives_of_Gamma}
\fmfset{wiggly_len}{2mm}
\fmfset{thin}{0.75pt} 
	\begin{eqnarray*}
		\frac{\partial}{\partial x^{\star}_{1}}...\frac{\partial}{\partial x^{\star}_{k}} \; \Gamma(x^{\star}) = \; \Gamma^{(k)}_{1,..,k} &=& \qquad \parbox{30mm}{
			\begin{fmfgraph*}(50,50)
				\fmfsurroundn{i}{16} 
				\fmf{wiggly}{i15,v1,i1}
				\fmf{wiggly}{i2,v1,i3}
				\fmf{wiggly}{i5,v1}
				\fmf{phantom}{i4,v1,i6}
				\fmf{phantom}{i7,v1,i8}
				\fmf{phantom}{i9,v1,i10}
				\fmf{phantom}{i11,v1,i12}
				\fmf{phantom}{i13,v1,i14}
				\fmf{phantom}{i16,v1}
				\fmflabel{$1$}{i15}
				\fmflabel{.}{i1}
				\fmflabel{.}{i2}
				\fmflabel{.}{i3}
				\fmflabel{$k$}{i5}
				\fmfv{decor.shape=circle, d.filled=shaded}{v1}
			\end{fmfgraph*}
			} \\ [30pt]
		\frac{\partial}{\partial x^{\star}_{k+1}} \; \parbox{30mm}{
			\begin{fmfgraph*}(50,50)
				\fmfsurroundn{i}{16} 
				\fmf{wiggly}{i15,v1,i1}
				\fmf{wiggly}{i2,v1,i3}
				\fmf{wiggly}{i5,v1}
				\fmf{phantom}{i4,v1,i6}
				\fmf{phantom}{i7,v1,i8}
				\fmf{phantom}{i9,v1,i10}
				\fmf{phantom}{i11,v1,i12}
				\fmf{phantom}{i13,v1,i14}
				\fmf{phantom}{i16,v1}
				\fmflabel{$1$}{i15}
				\fmflabel{.}{i1}
				\fmflabel{.}{i2}
				\fmflabel{.}{i3}
				\fmflabel{$k$}{i5}
				\fmfv{decor.shape=circle, d.filled=shaded}{v1}
			\end{fmfgraph*}
			} &=& \qquad \parbox{30mm}{
			\begin{fmfgraph*}(50,50)
				\fmfsurroundn{i}{16} 
				\fmf{wiggly}{i15,v1,i1}
				\fmf{wiggly}{i2,v1,i3}
				\fmf{wiggly}{i5,v1,i8}
				\fmf{phantom}{i4,v1}
				\fmf{phantom}{i6,v1,i7}
				\fmf{phantom}{i9,v1,i10}
				\fmf{phantom}{i11,v1,i12}
				\fmf{phantom}{i13,v1,i14}
				\fmf{phantom}{i16,v1}
				\fmflabel{$1$}{i15}
				\fmflabel{.}{i1}
				\fmflabel{.}{i2}
				\fmflabel{.}{i3}
				\fmflabel{$k$}{i5}
				\fmflabel{$k\!+\!1$}{i8}
				\fmfv{decor.shape=circle, d.filled=shaded}{v1}
			\end{fmfgraph*}
			}
	\end{eqnarray*}
\end{fmffile}
\begin{center}
\par\end{center}

Analogously we use the graphical representation for the derivatives
of $W$ as
\begin{center}
\par\end{center}

\begin{fmffile}{k+1_derivatives_of_W}
\fmfset{thin}{0.75pt} 
	\begin{eqnarray*}
		\frac{\partial}{\partial j_{k+1}} \; \parbox{30mm}{
			\begin{fmfgraph*}(50,50)
				\fmfsurroundn{i}{16} 
				\fmf{plain}{i15,v1,i1}
				\fmf{plain}{i2,v1,i3}
				\fmf{plain}{i5,v1}
				\fmf{phantom}{i4,v1,i6}
				\fmf{phantom}{i7,v1,i8}
				\fmf{phantom}{i9,v1,i10}
				\fmf{phantom}{i11,v1,i12}
				\fmf{phantom}{i13,v1,i14}
				\fmf{phantom}{i16,v1}
				\fmflabel{$1$}{i15}
				\fmflabel{.}{i1}
				\fmflabel{.}{i2}
				\fmflabel{.}{i3}
				\fmflabel{$k$}{i5}
				\fmfv{decor.shape=circle, d.filled=empty}{v1}
			\end{fmfgraph*}
			} =& \qquad \parbox{30mm}{   
			\begin{fmfgraph*}(50,50)
				\fmfsurroundn{i}{16} 
				\fmf{plain}{i15,v1,i1}
				\fmf{plain}{i2,v1,i3}
				\fmf{plain}{i5,v1,i8}
				\fmf{phantom}{i4,v1}
				\fmf{phantom}{i6,v1,i7}
				\fmf{phantom}{i9,v1,i10}
				\fmf{phantom}{i11,v1,i12}
				\fmf{phantom}{i13,v1,i14}
				\fmf{phantom}{i16,v1}
				\fmflabel{$1$}{i15}
				\fmflabel{.}{i1}
				\fmflabel{.}{i2}
				\fmflabel{.}{i3}
				\fmflabel{$k$}{i5}
				\fmflabel{$k\!+\!1$}{i8}
				\fmfv{decor.shape=circle, d.filled=empty}{v1}
			\end{fmfgraph*}
			}
	\end{eqnarray*}
\end{fmffile}

We already know the relationship of the second derivatives, namely
that the \textbf{Hessians} of $W$ and $\Gamma$ are inverses of one
another (\ref{eq:inverse_W1_Gamma1})

\begin{align}
\Gamma^{(2)}(x^{\ast})\,W^{(2)}(\Gamma^{(1)}(x^{\ast})) & =1\qquad\forall\,x^{\ast}\label{eq:W2Gamma2_inverse}\\
\Gamma^{(2)}(W^{(1)}(j))\,W^{(2)}(j) & =1\qquad\forall\,j\nonumber \\
\sum_{k}\,\Gamma_{ik}^{(2)}W_{kl}^{(2)} & =\delta_{il}.\nonumber 
\end{align}
Graphically, the relation \eqref{W2Gamma2_inverse} can be expressed
as:
\begin{align*}
\Diagram{gpgfcf}
 & =1,
\end{align*}
where the identity operation $1$ must be chosen from the appropriate
space corresponding to $x$. For distributions of an $N$ dimensional
variable this would be the diagonal unit matrix.

In the following we will use subscripts to denote the variables with
respect to which we differentiate, for example
\begin{align*}
\partial_{j_{k}}W & =W_{k}^{(1)}.
\end{align*}

Now let us obtain higher derivatives of \eqref{W2Gamma2_inverse}
with respect to $\frac{\partial}{\partial j_{a}}$: acting on $W^{(2)}$
we add a leg with index $a$, acting on $\Gamma_{ik}^{(2)}$, by the
chain rule, we get $\frac{\partial}{\partial j_{a}}\Gamma_{ik}^{(2)}(W^{(1)}(j))=\sum_{m}\Gamma_{ikm}^{(3)}W_{ma}^{(2)}$,
and, by the product rule, the application to $W_{kl}^{(2)}$ yields
$W_{kla}^{(3)}$, so in total 
\begin{align*}
0 & =\sum_{k,m}\Gamma_{ikm}^{(3)}W_{ma}^{(2)}W_{kl}^{(2)}+\sum_{k}\Gamma_{ik}^{(2)}W_{kla}^{(3)},
\end{align*}
which has the graphical representation:

\begin{center}
\par\end{center}

\begin{fmffile}{treestructure_of_vertices_1}
\fmfset{thin}{0.75pt}
\fmfset{decor_size}{4mm}
	\begin{eqnarray*}
		0 \; = \; \frac{\partial}{\partial j_{a}} \qquad \parbox{30mm}{
			\begin{fmfgraph*}(70,50)
				\fmfleft{i}
				\fmfright{o}
				\fmf{wiggly}{i,v1,v2}
				\fmf{plain}{v2,v3,o}
				\fmflabel{i}{i}
				\fmflabel{$l$}{o}
				\fmfv{label=k, label.angle=-90}{v2}
				\fmfv{decor.shape=circle, d.filled=shaded}{v1}
				\fmfv{decor.shape=circle, d.filled=empty}{v3}
			\end{fmfgraph*}
			} =& \qquad \parbox{30mm}{
			\begin{fmfgraph*}(100,70)
				\fmfleft{i,a}
				\fmfright{o}
				\fmf{wiggly}{i,v1,v2}
				\fmf{plain}{v2,v3,o}
				\fmflabel{i}{i}
				\fmflabel{$l$}{o}
				\fmfv{label=k, label.angle=-90}{v2}
				\fmfv{decor.shape=circle, d.filled=shaded}{v1}
				\fmfv{decor.shape=circle, d.filled=empty}{v3}
				\fmf{wiggly}{v4,v1}
				\fmf{plain}{a,v5,v4}
				\fmflabel{a}{a}
				\fmfv{label=m, label.angle=10}{v4}
				\fmfv{decor.shape=circle, d.filled=empty, d.size=4mm}{v5}
			\end{fmfgraph*}
			} \quad \qquad + \qquad \parbox{30mm}{
			\begin{fmfgraph*}(70,50)
				\fmfleft{i}
				\fmfright{d1,o,d2,a}
				\fmf{wiggly}{i,v1,v2}
				\fmf{plain}{v2,v3,o}
				\fmflabel{i}{i}
				\fmflabel{$l$}{o}
				\fmfv{label=k, label.angle=-90}{v2}
				\fmfv{decor.shape=circle, d.filled=shaded}{v1}
				\fmfv{decor.shape=circle, d.filled=empty}{v3}
				\fmf{plain}{v3,a}
				\fmflabel{a}{a}
			\end{fmfgraph*}
			}
	\end{eqnarray*}
\end{fmffile}
\begin{center}
\par\end{center}

\begin{flushleft}
We may multiply the latter expression by $W_{ib}^{(2)}$ and sum over
all $i$, using \eqref{W2Gamma2_inverse} to see that this operation
effectively removes the $\Gamma^{(2)}$ in the second term to obtain
\begin{align}
0 & =\sum_{i,k,m}\Gamma_{ikm}^{(3)}W_{ma}^{(2)}W_{kl}^{(2)}W_{ib}^{(2)}+W_{bla}^{(3)}.\label{eq:W3_Gamma3}
\end{align}
Graphically:
\par\end{flushleft}

\begin{center}
\par\end{center}

\begin{fmffile}{treestructure_of_vertices_2}
\fmfset{thin}{0.75pt}
\fmfset{decor_size}{4mm}
	\begin{eqnarray*}
	\parbox{30mm}{
		\begin{fmfgraph*}(40,40)
			\fmfsurroundn{i}{3}
			\fmf{plain}{i1,v1,i2}
			\fmf{plain}{i3,v1,}
			\fmflabel{a}{i1}
			\fmflabel{b}{i2}
			\fmflabel{$l$}{i3}
			\fmfv{decor.shape=circle, d.filled=empty}{v1}
		\end{fmfgraph*}
		} =& \; - \; \parbox{30mm}{
		\begin{fmfgraph*}(100,100)
			\fmfsurroundn{i}{3}
			\fmf{wiggly}{v1,v3}
			\fmf{wiggly}{v1,v5}
			\fmf{wiggly}{v1,v7}
			\fmf{plain}{i1,v2,v3}
			\fmf{plain}{i2,v4,v5}
			\fmf{plain}{i3,v6,v7}
			\fmflabel{a}{i1}
			\fmflabel{b}{i2}
			\fmflabel{$l$}{i3}
			\fmfv{label=m, l.a=90}{v3}
			\fmfv{label=i, l.a=-170}{v5}
			\fmfv{label=k, l.a=-10}{v7}
			\fmfv{decor.shape=circle, d.filled=shaded}{v1}
			\fmfv{decor.shape=circle, d.filled=empty}{v2}
			\fmfv{decor.shape=circle, d.filled=empty}{v4}
			\fmfv{decor.shape=circle, d.filled=empty}{v6}
		\end{fmfgraph*}
		}
	\end{eqnarray*}
\end{fmffile}

The latter expression shows that the third order cumulant $W^{(3)}$
can be expressed as a diagram that has so called \textbf{tree structure},
i.e. that does not contain any closed loops. This means that all closed
loops that are contained in the Feynman diagrams of the third cumulant
must be contained in the vertex function $\Gamma^{(3)}$ and in the
lines connecting them, the $W^{(2)}$.

Applying the derivatives by $j$ successively, we see that the left
diagram gets one additional leg, while in the right diagram we can
attach a leg to each of the three terms of $W^{(2)}$ and we can attach
an additional leg to $\Gamma^{(3)}$ that again comes, by the chain
rule, with a factor $W^{(2)}$, so that the tree level structure of
this relation is preserved:

\begin{center}
\par\end{center}

\begin{fmffile}{treestructure_of_vertices_3}
\fmfset{thin}{0.75pt}
\fmfset{decor_size}{4mm}
	\begin{align*}
	0 \; =& \quad \frac{\partial}{\partial j_{c}} \left[ \; \parbox{20mm}{
		\begin{fmfgraph*}(40,40)
			\fmfsurroundn{i}{3}
			\fmf{plain}{i1,v1,i2}
			\fmf{plain}{i3,v1,}
			\fmflabel{a}{i1}
			\fmflabel{b}{i2}
			\fmflabel{$l$}{i3}
			\fmfv{decor.shape=circle, d.filled=empty}{v1}
		\end{fmfgraph*}
		} + \; \parbox{30mm}{
		\begin{fmfgraph*}(70,70)
			\fmfsurroundn{i}{3}
			\fmf{wiggly}{v1,v3}
			\fmf{wiggly}{v1,v5}
			\fmf{wiggly}{v1,v7}
			\fmf{plain}{i1,v2,v3}
			\fmf{plain}{i2,v4,v5}
			\fmf{plain}{i3,v6,v7}
			\fmflabel{a}{i1}
			\fmflabel{b}{i2}
			\fmflabel{$l$}{i3}
			\fmfv{decor.shape=circle, d.filled=shaded}{v1}
			\fmfv{decor.shape=circle, d.filled=empty}{v2}
			\fmfv{decor.shape=circle, d.filled=empty}{v4}
			\fmfv{decor.shape=circle, d.filled=empty}{v6}
		\end{fmfgraph*}
		} \right] \\[30pt]
		=& \; \parbox{20mm}{
		\begin{fmfgraph*}(40,40)
			\fmfleft{b,c}
			\fmfright{l,a}
			\fmf{plain}{a,v1,b}
			\fmf{plain}{c,v1,l}
			\fmflabel{a}{a}
			\fmflabel{b}{b}
			\fmflabel{c}{c}
			\fmflabel{$l$}{l}
			\fmfv{decor.shape=circle, d.filled=empty}{v1}
		\end{fmfgraph*}
		} + \qquad \parbox{30mm}{
		\begin{fmfgraph*}(90,70)
			\fmfleft{dummy1,b,c,dummy2}
			\fmfright{l,a}
			\fmf{wiggly}{v1,v3}
			\fmf{wiggly}{v1,v5}
			\fmf{wiggly}{v1,v7}
			\fmf{plain}{a,v2,v3}
			\fmf{plain}{b,v4,v5}
			\fmf{plain}{c,v4}
			\fmf{plain}{l,v6,v7}
			\fmflabel{a}{a}
			\fmflabel{b}{b}
			\fmflabel{c}{c}
			\fmflabel{$l$}{l}
			\fmfv{decor.shape=circle, d.filled=shaded}{v1}
			\fmfv{decor.shape=circle, d.filled=empty}{v2}
			\fmfv{decor.shape=circle, d.filled=empty}{v4}
			\fmfv{decor.shape=circle, d.filled=empty}{v6}
		\end{fmfgraph*}
		} \; + \; 2 \; \textrm{perm.} \; + \qquad \parbox{30mm}{
		\begin{fmfgraph*}(70,70)
			\fmfleft{b,c}
			\fmfright{l,a}
			\fmf{wiggly}{v1,v3}
			\fmf{wiggly}{v1,v5}
			\fmf{wiggly}{v1,v7}
			\fmf{wiggly}{v1,v9}
			\fmf{plain}{a,v2,v3}
			\fmf{plain}{b,v4,v5}
			\fmf{plain}{c,v8,v9}
			\fmf{plain}{l,v6,v7}
			\fmflabel{a}{a}
			\fmflabel{b}{b}
			\fmflabel{c}{c}
			\fmflabel{$l$}{l}
			\fmfv{decor.shape=circle, d.filled=shaded}{v1}
			\fmfv{decor.shape=circle, d.filled=empty}{v2}
			\fmfv{decor.shape=circle, d.filled=empty}{v4}
			\fmfv{decor.shape=circle, d.filled=empty}{v6}
			\fmfv{decor.shape=circle, d.filled=empty}{v8}
		\end{fmfgraph*}
		} \\[30pt]
	\end{align*}
\end{fmffile}

We may express the intermediate diagram in the last line by using
the diagram for the three point cumulant\begin{fmffile}{treestructure_of_vertices_3_1}
\fmfset{thin}{0.75pt}
\fmfset{decor_size}{4mm}
\begin{align*}
   &
		-\parbox{30mm}{
		\begin{fmfgraph*}(130,70)
			\fmfleft{b,c}
			\fmfright{l,a}
			\fmf{wiggly}{blob1,v3}
			\fmf{wiggly}{blob1,v7}
			\fmf{wiggly, tension=0.9}{blob1,v11}
			\fmf{wiggly}{blob2,v5}
			\fmf{wiggly}{blob2,v9}
			\fmf{wiggly, tension=0.9}{blob2,v12}
			\fmf{plain}{a,v2,v3}
			\fmf{plain}{b,v4,v5}
			\fmf{plain}{c,v8,v9}
			\fmf{plain}{l,v6,v7}
			\fmf{plain, tension=0.9}{v11,v10,v12}
			\fmflabel{a}{a}
			\fmflabel{b}{b}
			\fmflabel{c}{c}
			\fmflabel{$l$}{l}
			\fmfv{decor.shape=circle, d.filled=shaded}{blob1}
			\fmfv{decor.shape=circle, d.filled=shaded}{blob2}
			\fmfv{decor.shape=circle, d.filled=empty}{v2}
			\fmfv{decor.shape=circle, d.filled=empty}{v4}
			\fmfv{decor.shape=circle, d.filled=empty}{v6}
			\fmfv{decor.shape=circle, d.filled=empty}{v8}
			\fmfv{decor.shape=circle, d.filled=empty}{v10}
		\end{fmfgraph*}
		} \\
   \end{align*}
\end{fmffile}
\begin{center}
\par\end{center}

We here did not write the two permutations explicitly, there the derivative
acts on the second cumulants with labels $a$ and $l$. The complete
expression of course contains these two additional terms. By induction
we can show that the diagrams that express cumulants in terms of vertex
functions all have tree structure. We will see a proof of this assertion
in \secref{effective_action}. This feature explains the name \textbf{vertex
functions}: these functions effectively act as interaction vertices.
We obtain the cumulants as combinations of these interaction vertices
with the effective propagator $W^{(2)}$. The special property is
that only tree diagrams contribute. We will see in the next section
that this feature is related to the expression (\ref{eq:pre_Legendre})
in the previous section: Only the right hand side of the expression
contains an integral over the fluctuations $\delta x$. These are
therefore effectively contained in the function $\Gamma$ on the left
hand side. In the following section we will indeed see that fluctuations
are related to the appearance of loops in the Feynman diagrams. The
absence of loops in the Feynman diagrams of $W$ expressed in terms
of the vertex functions can therefore be interpreted as the vertex
functions implicitly containing all these fluctuations. Technically
this decomposition is therefore advantageous: We have seen that in
momentum space, each loop corresponds to one frequency integral to
be computed. Decomposing connected diagrams in terms of vertex functions
hence extracts these integrals; they only need to be computed once.

\subsection{Self-energy or mass operator $\Sigma$\label{sec:self-energy}}

The connection between the two-point correlation function $W^{(2)}$
and the two point vertex function $\Gamma^{(2)}$, given by the reciprocity
relation (\ref{eq:W2Gamma2_inverse}), is special as it does not involve
a minus sign, in contrast to all higher orders, such as for example
(\ref{eq:W3_Gamma3}). The Hessian $W^{(2)}(0)$ is the covariance
matrix, or the full propagator (sometimes called ``dressed'' propagator,
including perturbative corrections), of the system. It therefore quantifies
the strength of the fluctuations in the system so it plays an important
role. One may, for example, investigate for which parameters fluctuations
become large: If the Hessian $\Gamma^{(2)}$ has a vanishing eigenvalue
in a particular direction, fluctuations in the system diverge in the
corresponding direction. Critical phenomena, or second order phase
transitions, are based on this phenomenon. 

In the current section we consider the particular case that the solvable
part of the theory is Gaussian or that fluctuations are approximated
around a stationary point, as will be done in the next section in
the loopwise approximation. In both cases, shown for the perturbation
expansion in \secref{Vertex-functions-Gaussian} and \subref{Example-Vertex-function-phi34},
the Gaussian part of the action also appears (with a minus sign) in
the effective action (\ref{eq:Gamma0_Gauss}), which decomposes as
\begin{eqnarray*}
\Gamma(x^{\ast}) & = & -S_{0}(x^{\ast})+\Gamma_{V}(x).
\end{eqnarray*}
So we may separate the leading order contribution to $\Gamma^{(2)}$
by writing
\begin{align}
\Gamma^{(2)} & =-S_{0}^{(2)}+\Gamma_{V}^{(2)}\label{eq:def_self_energy}\\
 & =:-S_{0}^{(2)}+\Sigma,\nonumber 
\end{align}
where we defined $\Sigma:=\Gamma_{V}^{(2)}$ as the correction term
to the Hessian. In the context of quantum field theory, $\Sigma$
is called the \textbf{self-energy} or \textbf{mass operator}. The
name self-energy stems from its physical interpretation that it provides
a correction to the energy of a particle due to the interaction with
the remaining system. The name ``mass-operator'' refers to the fact
that these corrections affect the second derivative of the effective
action, the constant (momentum-independent) part of which is the particle
mass in a standard $\phi^{4}$ theory.

From (\ref{eq:W2Gamma2_inverse}) then follows that
\begin{align}
1 & =\Gamma^{(2)}W^{(2)}\label{eq:reciprocity_self_energy}\\
 & =\left(-S_{0}^{(2)}+\Sigma\right)W^{(2)}.\nonumber 
\end{align}
We see that hence $(W^{(2)})^{-1}=-S_{0}^{(2)}+\Sigma$, so the full
propagator $W^{(2)}$ results from the inverse of the second derivative
of the bare action plus the self-energy; this explains the interpretation
of $\Sigma$ as an additional mass term: in quantum field theory,
the mass terms typically yield terms that are quadratic in the fields.

In matrix form and multiplied by the propagator $\Delta=\left(-S_{0}^{(2)}\right)^{-1}$
of the free theory from left we get
\begin{align*}
\Delta & =\Delta\left(-S_{0}^{(2)}+\Sigma\right)W^{(2)}\\
 & =\left(1+\Delta\Sigma\right)W^{(2)},
\end{align*}
so that multiplying from left by the inverse of the bracket we obtain
\begin{align}
W^{(2)} & =\left(1+\Delta\Sigma\right)^{-1}\Delta\nonumber \\
 & =\sum_{n=0}^{\infty}\left(-\Delta\Sigma\right)^{n}\Delta\nonumber \\
 & =\Delta-\Delta\Sigma\Delta+\Delta\Sigma\Delta\Sigma\Delta-\ldots,\label{eq:Dyson}
\end{align}
which is a so-called \textbf{Dyson's equation}. These contributions
to $W^{(2)}$ are all tree-level diagrams with two external legs.
Since $W^{(2)}$ contains all connected graphs with two external legs,
consequentially the contributions to $\Sigma$ all must contain two
uncontracted variables and otherwise form connected diagrams which
cannot be disconnected by cutting a single line.

It is interesting to note that if $\Sigma$ has been determined to
a certain order (in perturbation theory or in terms of loops, see
below) that the terms in (\ref{eq:Dyson}) become less and less important,
since they are of increasing order.

The latter property follows from the decomposition in the last line
of (\ref{eq:Dyson}): The expression corresponds to the sum of all
possible graphs that are composed of sub-graphs which are interconnected
by a single line $\Delta$. Hence, these graphs can be disconnected
by cutting a single line $\Delta$. Since $W^{(2)}$ must contain
all connected graphs with two external legs, and the sum already contains
all possible such combinations. No separable components can reside
in $\Sigma$. This follows from the proofs in \secref{Perturbation-expansion-of-Gamma}
and \secref{Loopwise-expansion-vertex-function}, but can also be
seen from the following argument:

$W^{\left(2\right)}$ is composed of all diagrams with two external
legs. Any diagram can be decomposed into a set of components that
are connected among each other by only single lines. This can be seen
recursively, identifying a single line that would disconnect the entire
diagram into two and then proceeding in the same manner with each
of the two sub-diagrams recursively. The remainders, which cannot
be decomposed further, are 1PI by definition and must have two connecting
points. Writing any of the found connections explicitly as $\Delta$,
we see that we arrive at \eqref{Dyson}.

With the help of the Dyson equation \eqref{Dyson}, we can therefore
write the tree-level decomposition of an arbitrary cumulant in \secref{Tree-level-expansion-Cumulants-Vertex}
by replacing the external connecting components, the full propagators
$W^{(2)}$ by \eqref{Dyson}
\begin{align*}
W^{(2)} & =\Diagram{fscfs}
\\
 & \stackrel{(\ref{eq:Dyson})}{=}\Diagram{f}
-\Diagram{f!p{\Sigma}f}
+\Diagram{f!p{\Sigma}f!p{\Sigma}f}
-\ldots.
\end{align*}
In the case of the loopwise expansion, $\feyn{f}=\Delta=(-S^{(2)}(x^{\ast}))^{-1}$
and $\Feyn{p}=\Sigma(x^{\ast})$ are still functions of $x^{\ast}$,
the true mean value. We can therefore express all quantities appearing
in \secref{Tree-level-expansion-Cumulants-Vertex} by explicit algebraic
terms that all depend on $x^{\ast}$, the true mean value. In the
presence of sources $j$, the latter, in turn, follows from the solution
of the equation of state (\ref{eq:equation_of_state}).

\section{Loopwise expansion of the effective action - Tree level\label{sec:Loopwise-expansion-vertex-function}}

We saw in \secref{Perturbation-expansion} that perturbation theory
produces a set of graphs with vertices $V^{(n)}$ of the Taylor expansion
of $V$ and propagators $\Delta=A^{-1}$. If the interaction vertices
$V^{(n)}$ are smaller than $\Delta$ by a small factor $\epsilon$,
this factor is a natural parameter to organize the perturbation series.
In many cases this is not the case and other arrangements of the perturbation
series may have better convergence properties. We will here study
one such reorganization for the case of a strong interaction that
causes non-vanishing mean-values of the stochastic variables. This
is a central tool in quantum field theory and statistical field theory
and will here be used to systematically calculate fluctuations in
classical systems. We here loosely follow \citep[Sec. 6.4]{ZinnJustin96},
\citep[Sec. 2.5]{NegeleOrland98}, and \citep[Sec. 3.2.26]{Kleinert89}.

To illustrate the situation, we again consider the example of the
``$\phi^{3}+\phi^{4}$''-theory with the action 
\begin{align}
S(x)= & l\left(-\frac{1}{2}x^{2}+\frac{\alpha}{3!}x^{3}+\frac{\beta}{4!}x^{4}\right)\label{eq:phi3_4_loop}
\end{align}
with $\beta<0$, $l>0$ and $\alpha$ arbitrary. We now assume that
there is no small parameter $\epsilon$ scaling the non-Gaussian part.
Instead, we may assume that there is a parameter $l$, multiplying
the entire action. Figure \ref{fig:Loopwise-expansion-phi4}a shows
the probability for different values of the source $j$, with a maximum
monotonically shifting as a function of $j$. The mean value $\langle x\rangle$
may be quite far away from $0$ so that we seek a method to expand
around an arbitrary value $x^{\ast}=\langle x\rangle$: The peak of
the distribution can always be approximated by a quadratic polynomial.
One would guess that such an approximation would be more accurate
than the expansion around the Gaussian part $-\frac{1}{2}x^{2}$,
if the peak if far off zero.

In many cases, corrections by fluctuations may be small. We saw a
concrete example in \secref{tanh_network}. We will see in the following
that small fluctuations correspond to $l$ being large. To see this,
we express the cumulant generating function in terms of an integral
over the fluctuations of $x$, as in \eqref{W_as_Z-1}
\begin{align}
\exp\left(W(j)+\ln\Z(0)\right) & =\int dx\,\exp\left(S(x)+j^{\T}x\right),\label{eq:W_as_Z-2}
\end{align}
where we moved the normalization as $\exp(\Z(0))$ to the left hand
side. We expect the dominant contribution to the integral on the right
of \eqref{W_as_Z-2} from the local maxima of $S(x)+j^{\T}x$, i.e.
the points $x_{S}^{\ast}(j)$ at which $S^{(1)}(x_{S}^{\ast})+j=0$
and $S^{(2)}(x_{S}^{\ast})<0$. At these points the action including
the source term is stationary

\begin{align}
\frac{\partial}{\partial x}\left(S(x)+j^{\T}x\right) & \stackrel{!}{=}0\nonumber \\
S^{(1)}(x_{S}^{\ast})+j & =0,\label{eq:stationary_point_S-1-2}
\end{align}
which implicitly defines the function $x_{S}^{\ast}(j)$. Inserted
into the integral, we obtain the lowest order approximation
\begin{align}
W_{0}(j)+\ln\Z(0)= & S(x_{S}^{\ast}(j))+j^{\T}x_{S}^{\ast}(j),\label{eq:lowest_order_W-2}
\end{align}
because the fluctuations of $x$ will be close to $x_{S}^{\ast}$.
The normalization of this distribution will give a correction term,
which, however is $\propto l^{-1}$ as we will see in the next section.
So to lowest order we can neglect it here. In the limit $l\to\infty$
the entire probability mass is concentrated at the point $x=x_{S}^{\ast}$.
The accuracy of this approximation increases with $l$. It corresponds
to our naive mean-field solution (\ref{eq:mf_tanh_network}) in the
problem of the recurrent activity. 

Together with the condition (\ref{eq:stationary_point_S-1-2}), (\ref{eq:lowest_order_W-2})
has the form of a Legendre transform, but with a different sign convention
than in \eqref{def_gamma} and with the inconsequential additive constant
$W(0)$. So to lowest order in the fluctuations, the cumulant generating
function is the Legendre transform of the action. Since we know that
the Legendre transform is involutive for convex functions, meaning
applied twice yields the identity, we conclude with the definition
of $\Gamma$ by \eqref{def_gamma} as the Legendre transform of $W$
that to lowest order in $l$ we have 
\begin{align}
\Gamma_{0}(x^{\ast})-\ln\Z(0) & =-S(x^{\ast})\propto O(l).\label{eq:Gamma0}
\end{align}
(More precisely: $\Gamma_{0}$ is the convex envelope of $-S$, because
the Legendre transform of any function is convex, see \secref{Convexity-W}).
This approximation is called \textbf{tree-level approximation},\textbf{
mean-field} \textbf{approximation}, or \textbf{stationary phase} \textbf{approximation}:
Only the original interaction vertices of $-S$ appear in $\Gamma$.
The name ``tree-level approximation'' can be understood from the
fact that only tree-diagrams contribute to the mean value of $\langle x\rangle(j)$,
as shown in \secref{Equivalence-loopwise-resummation}. We also see
from (\ref{eq:W_as_Z-2}) that we replaced the fluctuating $x$ by
a non-fluctuating mean value $x_{S}$, giving rise to the name ``mean-field''
approximation. In our example in \secref{tanh_network}, the equation
of state (\ref{eq:equation_of_state}) is identical to this approximation
that neglects all fluctuations, cf. (\ref{eq:mf_tanh_network}).

In the following, we want to obtain a systematic inclusion of fluctuations.
We will extend the principle of the previous sections to obtain a
systematic expansion of the fluctuations in Gaussian and higher order
terms around an arbitrary point $x^{\ast}$. In the context of field
theory, this method is known as the background field expansion \citep[section 3.2.26]{Kleinert89},
the formal derivation of which we loosely follow here.

\subsection{Counting the number of loops\label{sec:Counting-loops}}

Before deriving the systematic fluctuation expansion, we will here
make the link between the strength of fluctuations and another topological
feature of Feynman diagrams: their number of loops. For simplicity,
let us first consider a problem with a Gaussian part $-\frac{1}{2}x^{\T}Ax$
and a perturbing potential $V(x)$ of order $x^{3}$ and higher. Let
us further assume that $A$ and $V$ are both of the same order of
magnitude. We introduce a parameter $l$ to measure this scale.

For large $l$, fluctuations of $x$, measured by $\delta x=x-\langle x\rangle$,
are small and we have $\delta x\propto1/\sqrt{l}$. This is because
for small $\delta x$, the quadratic part $-\frac{1}{2}x^{\T}Ax$
dominates over $V(x)$ which is of higher than second order in $x$;
further, because the variance is $\langle\delta x^{2}\rangle=A^{-1}\propto l^{-1}$.
We here seek an expansion around these weak fluctuations of the integral

\begin{align}
W(j) & \propto\ln\int dx\,\exp\left(-\frac{1}{2}x^{\T}\,A\,x+V(x)+j^{\T}x\right)\label{eq:moment_generator_l}\\
 & =\ln\int dx\,\exp\left(-\frac{1}{2}x^{\T}\,l\,a\,x+l\,v(x)+j^{\T}x\right)\nonumber 
\end{align}
around $x=0$, where we defined $a=A/l$ and $v=V/l$ to make the
order of $A$ and $V$ explicit.

Let us first make a simple observation to see that the scale $l$
penalizes contributions from contractions with a high power of $x$.
Since the fluctuations are $\delta x\propto1/\sqrt{l}$, the contribution
of a diagram whose summed power of $x$ in all vertices is $x^{n}$
will be $\propto l^{-n/2}$ (for $n$ even), because each contraction
of a pair of $x$ yields a factor $l^{-1}$. This can be seen from
the substitution $\sqrt{l}x\equiv y$ (neglecting the change of the
determinant $l^{-\frac{N}{2}}$, which just yields an inconsequential
additive correction to $W(j)$)
\begin{align*}
W(j) & \propto\ln\int dy\,\exp\left(-\frac{1}{2}y^{\T}\,a\,y+l\,v(\frac{y}{\sqrt{l}})+j^{\T}\frac{y}{\sqrt{l}}\right)
\end{align*}
and then considering the form of the contribution of one graph (assuming
$j=0$) of the form $\sum_{n_{1}+\ldots+n_{k}=n}\,l\,\frac{v^{(n_{1})}}{n_{1}!}\cdots l\,\frac{v^{(n_{k})}}{n_{k}!}\,\langle\frac{y^{n}}{l^{\frac{n}{2}}}\rangle\propto l^{k-\frac{n}{2}}$.
Since each contraction $\langle yy\rangle\propto a^{-1}=O(1)$ corresponds
to one propagator, i.e. one line in the graph, we can also say that
the contribution is damped by $l^{k-n_{\Delta}}$, where $n_{\Delta}$
is the number of lines in the graph and $k$ is the number of vertices.

\begin{figure}
\begin{centering}
\begin{minipage}[t]{0.3\columnwidth}%
\selectlanguage{english}%
a)

\selectlanguage{american}%
\vspace*{.3cm}
\begin{fmffile}{name_a_tobechanged_68} 
\begin{fmfgraph*}(40, 50) 
\fmfleft{i1,i2}
\fmfright{o1,o2,o3}
\fmf{plain}{i2,v1,i1}
\fmf{plain}{v1,o2}
\fmflabel{$V^{(3)}$}{v1}
\fmfdotn{v}{1}
\end{fmfgraph*}
\hspace*{\fill}
\begin{fmfgraph*}(40,50)
\fmfright{r}
\fmfleft{l}
\fmftop{t}
\fmfbottom{b}
\fmf{plain}{l,v1,t}
\fmf{plain, label=$\Delta$}{b,v1}
\fmf{plain}{v1,r} 
\fmflabel{$j$}{b}
\fmfdot{v1}
\end{fmfgraph*}
\end{fmffile}

\begin{align*}
n_{V} & =2\\
n_{\Delta}^{\mathrm{int.}} & =0\\
n_{L} & =0\\
n_{j} & =1
\end{align*}
\end{minipage}\hspace*{\fill}%
\begin{minipage}[t]{0.3\columnwidth}%
\selectlanguage{english}%
b)

\selectlanguage{american}%
\vspace*{.3cm}
\begin{fmffile}{name_b_tobechanged_17} 
\begin{fmfgraph*}(100, 50)
\fmfleftn{l}{3}
\fmfrightn{r}{4}
\fmffreeze
\fmfshift{(.3w,0.)}{l2}
\fmfshift{(-.2w,.2h)}{r2}
\fmfshift{(-.25w,0.)}{r4}
\fmf{plain}{l1,l2,l3}
\fmf{plain, label=$\Delta$}{l2,r2}
\fmf{plain, label=$\Delta$}{r2,r1}
\fmf{plain}{r3,r2,r4}
\fmflabel{$V^{(3)}$}{l2}
\fmfv{l=$V^{(4)}$, l.a=-30, l.d=.1h}{r2}
\fmflabel{$j$}{r1}
\fmfdot{l2,r2}
\end{fmfgraph*}
\end{fmffile} 

\begin{align*}
n_{V} & =2\\
n_{\Delta}^{\mathrm{int.}} & =1\\
n_{L} & =0=n_{\Delta}^{\mathrm{int.}}-n_{V}+1\\
n_{j} & =1
\end{align*}
\end{minipage}\hspace*{\fill}%
\begin{minipage}[t]{0.3\columnwidth}%
\selectlanguage{english}%
c)

\selectlanguage{american}%
\vspace*{.3cm}
\begin{fmffile}{name_c_tobechanged_10} 
\begin{fmfgraph*}(100, 50)
\fmfleftn{l}{3} \fmfrightn{r}{4} \fmffreeze \fmfshift{(.3w,0.)}{l2} \fmfshift{(-.2w,.2h)}{r2} \fmfshift{(-.25w,0.)}{r4} \fmf{plain, label=$\Delta$}{l2,r2} \fmf{plain, label=$\Delta$}{r2,r1} \fmf{phantom}{l1,l2} \fmf{plain, label=$\Delta$}{l2,l2} \fmf{plain,left, label=$\Delta$}{r2,r2} 
\fmfv{l=$V^{(4)}$, l.a=-30, l.d=.1h}{r2} \fmfv{l=$V^{(3)}$, l.a=-90, l.d=.1h}{l2}
\fmflabel{$j$}{r1}
\fmfdot{l2,r2}
\end{fmfgraph*}
\end{fmffile}

\begin{align*}
n_{V} & =2\\
n_{\Delta}^{\mathrm{int.}} & =3\\
n_{L} & =2=n_{\Delta}^{\mathrm{int.}}-n_{V}+1\\
n_{j} & =1
\end{align*}
\end{minipage}
\par\end{centering}

\caption{\textbf{Stepwise construction of a connected diagram.} Steps of constructing
a connected graph of $n_{V}$ vertices with $n_{j}$ external legs.
\textbf{a} Assignment of external legs to vertices, requiring $n_{\Delta}^{\mathrm{ext.}}=n_{j}$
external lines. \textbf{b} Connection of all vertices into one component,
requiring $n_{\Delta}^{\mathrm{int.}}=n_{V}-1$ lines. \textbf{c}
Contraction of all remaining free legs of the vertices. The number
of loops in the diagrams in b and c is $n_{L}=n_{\Delta}^{\mathrm{int}}-n_{V}+1$.\label{fig:Loopwise-expansion-Steps}}
\end{figure}
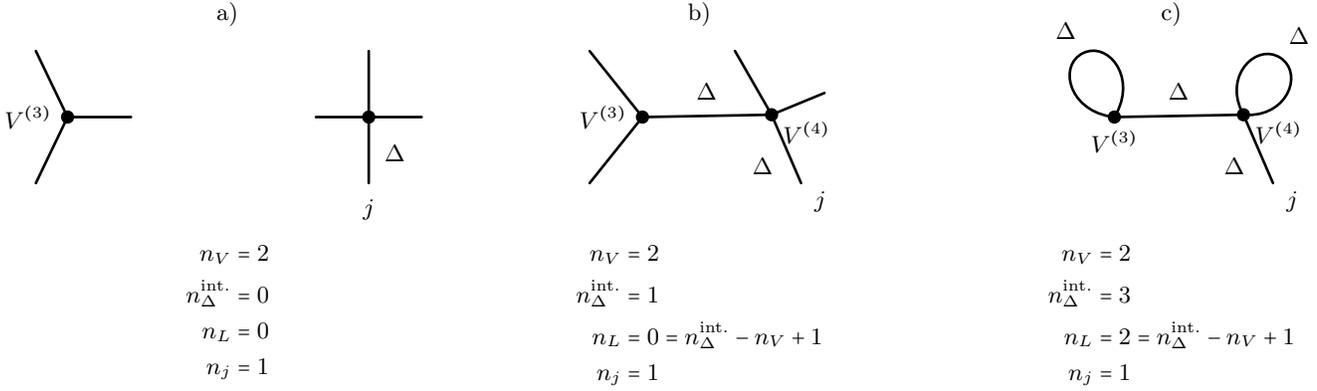

We can see this relation also directly on the level of the graphs,
illustrated in \figref{Loopwise-expansion-Steps}. The parameter $l$
scales the propagators and the vertices in the Feynman graphs differently.
The propagator is $\Delta=(la)^{-1}=\frac{1}{l}a^{-1}\propto l^{-1}$,
a vertex $lv^{(n)}\propto l$.

To make the link to the topology of the graph, we construct a connected
diagram with $n_{V}$ vertices and $n_{j}$ external legs. We first
connect these legs to vertices, each yielding a contribution $\propto\frac{1}{l}$
due to the connecting propagator. We now need to connect the $n_{V}$
vertices among each other so that we have a single connected component.
Otherwise the contribution to $W$ would vanish. This requires at
least $n_{\Delta}^{\mathrm{int},\mathrm{min}}=n_{V}-1$ internal lines,
each coming with one factor $l^{-1}$. By construction, the formed
graph so far has no loops, so $n_{L}=0$. Now we need to contract
all remaining legs of the vertices. Each such contraction requires
an additional propagator, coming with a factor $l^{-1}$ and necessarily
produces one additional loop in the graph, because it connects to
two vertices that are already connected. We therefore see that the
number of loops equals
\begin{align*}
n_{L} & =n_{\Delta}^{\mathrm{int}}-n_{V}+1.
\end{align*}
The prefactor of a graph is hence
\begin{align*}
l^{n_{V}-n_{\Delta}^{\mathrm{int}}}\,l^{-n_{j}} & =l^{1-n_{L}-n_{j}}.
\end{align*}
This shows two things: First, noticing that the number of external
legs equals the order of the cumulant, cumulants are damped by the
factor $l^{-n_{j}}$ in relation to their order; this term just stems
from the $n_{j}$ external legs, each of which being connected with
a single propagator. Second, for a given order of the cumulant to
be calculated, we can order the contributions by their importance;
their contributions diminish with the number $n_{L}$ of the loops
in the corresponding graphs. The latter factor stems from the amputated
part of the diagram \textendash{} the part without external legs.

\begin{figure}
\begin{centering}
\includegraphics{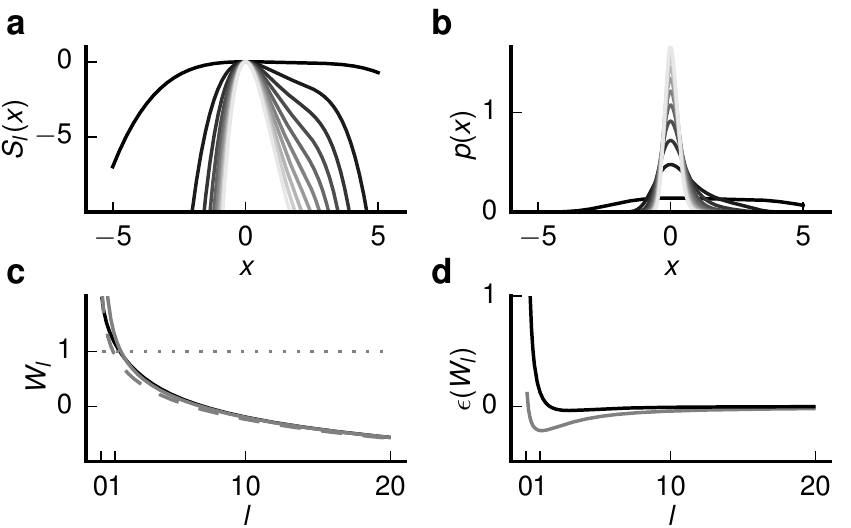}
\par\end{centering}
\caption{\textbf{Loopwise expansion for the action ``$\phi^{3}+\phi^{4}$''
theory. a} Action $S_{l}(x)=l\left(\frac{1}{2}x^{2}+\frac{\alpha}{3!}x^{3}+\frac{\beta}{4!}x^{4}\right)$,
from black ($l=0.1$ to light gray $l=20$). \textbf{b} Probability
$p(x)=e^{S_{l}(x)-W_{l}}$. \textbf{c} Normalization given by $W_{l}=\ln\,\int dx\,\exp(S_{l}(x))$
from numerical integration (black) and zero loop approximation ($W_{l}=1$,
gray dotted), one-loop approximation (gray dashed), and two-loop approximation
(gray solid) (see exercises). \textbf{d} Error of one-loop (gray),
and two-loop approximation (black). Other parameters: $\alpha=\frac{3}{2}$,
$\beta=-1$.\label{fig:Loopwise-expansion-phi34}}
\end{figure}

The loopwise approximation can be illustrated by a one-dimensional
integral shown in \figref{Loopwise-expansion-phi34} that is calculated
in the exercises: The loop corrections converge towards the true value
of the integral for $l\gg1$. The two-loop approximation has a smaller
error in the range $l\gg1$ than the one-loop approximation.

\subsection{Loopwise expansion of the effective action - Higher numbers of loops\label{sec:Loopwise-gamma}}

In this section, we will use the loopwise expansion of \secref{Counting-loops}
to systematically calculate the corrections to $\Gamma$. To lowest
order we already know from \eqref{Gamma0} that $\Gamma_{0}(x^{\ast})=-S(x^{\ast})+\ln\Z(0)$.
With the general definition \eqref{def_gamma} of $\Gamma$, \eqref{pre_Legendre}
takes the form
\begin{align}
\exp\left(-\Gamma(x^{\ast})+\ln\Z(0)\right) & =\int\,d\delta x\,\exp\left(S(x^{\ast}+\delta x)+j^{\T}\delta x\right).\label{eq:Gamma_int_general}
\end{align}
To lowest order in the fluctuations, we set $\delta x=0$, leading
to the same result as (\ref{eq:Gamma0}). We now set out to derive
an iterative equation to compute $\Gamma$, where the iteration parameter
is the number of loops in the diagrams. We use the equation of state
\eqref{equation_of_state} to replace $j(x^{\ast})$ in the latter
equation to obtain
\begin{align}
\exp\left(-\Gamma(x^{\ast})+\ln\Z(0)\right) & =\int d\delta x\,\exp\left(S(x^{\ast}+\delta x)+\Gamma^{(1)\T}(x^{\ast})\,\delta x\right).\label{eq:Gamma_int_dx}
\end{align}
Now let us expand the fluctuations of $x$ around $x^{\ast}$. We
perform a Taylor expansion of the action around $x^{\ast}$ 
\begin{align}
S(x^{\ast}+\delta x) & =S(x^{\ast})+S^{(1)}(x^{\ast})\,\delta x+\frac{1}{2}\delta x^{\T}S^{(2)}(x^{\ast})\delta x+R(x^{\ast},\delta x).\label{eq:expansion_S}
\end{align}
Here all terms in the expansion higher than order two are contained
in the remainder term $R(x^{\ast},\delta x$). Inserting \eqref{expansion_S}
into \eqref{Gamma_int_general}, we get 

\begin{align}
\exp\left(-\Gamma(x^{\ast})-S(x^{\ast})+\ln\Z(0)\right)= & \int d\delta x\,\exp\Big(\big(S^{(1)}(x^{\ast})+\Gamma^{(1)\T}(x^{\ast})\big)\,\delta x+\frac{1}{2}\delta x^{\T}S^{(2)}(x^{\ast})\delta x+R(x^{\ast},\delta x)\Big),\label{eq:pre_Gamma_fluct}
\end{align}
where we sorted by powers of $\delta x$ on the right side and moved
the term $S(x^{\ast})$, which is independent of $\delta x$, to the
left. Since by \eqref{Gamma0} to lowest order $\Gamma_{0}-\ln\Z(0)=-S$
we now define the corrections due to fluctuations on the left hand
side as 
\begin{align}
\Gammafl(x^{\ast}) & :=\Gamma(x^{\ast})+S(x^{\ast})-\ln\Z(0).\label{eq:def_Gammafl}
\end{align}
The first term on the right hand side of \eqref{pre_Gamma_fluct}
can therefore be written as $\frac{\partial}{\partial x^{\ast}}\left(S(x^{\ast})+\Gamma(x^{\ast})\right)\equiv\Gammafl^{(1)}(x^{\ast})$
(since $\ln\Z(0)$ is independent of $x^{\ast}$), so that we obtain
the final result
\begin{align}
\exp\left(-\Gammafl(x^{\ast})\right)= & \int d\delta x\,\exp\left(\frac{1}{2}\delta x^{\T}S^{(2)}(x^{\ast})\delta x+R(x^{\ast},\delta x)+\Gammafl^{(1)\T}(x^{\ast})\,\delta x\right).\label{eq:Gammafl_int}
\end{align}

The latter expression allows us to again use the reorganization of
the loopwise expansion (\secref{Counting-loops}) for the calculation
of $\Gamma$ by \eqref{Gammafl_int}, sorting the different terms
by their importance in contributing to the fluctuations.

Comparing \eqref{moment_generator_l} and \eqref{Gammafl_int}, we
identify the terms
\begin{align*}
S^{(2)}(x^{\ast}) & \equiv-A\\
R(x^{\ast},\delta x) & \equiv V(\delta x)\\
\frac{\partial\Gammafl}{\partial x^{\ast}}(x^{\ast}) & \equiv j.
\end{align*}
We can therefore think of \eqref{Gammafl_int} as an effective quadratic
theory with free part $A$ given by $-S^{(2)}$, a small perturbing
potential $V$ given by $R$ and a source $j$ given by $\partial\Gammafl/\partial x^{\ast}$.
From this identification by the linked cluster theorem (\secref{General-proof-linked-cluster}),
all connected diagrams that are made up of the propagators (lines)
$\Delta=A^{-1}=\left(-S^{(2)}(x^{\ast})\right)^{-1}$, vertices $V(x)=R(x^{\ast},x)$
and ``external lines'' $j=\frac{\partial\Gammafl^{\T}}{\partial x^{\ast}}(x^{\ast})$
contribute to $\Gammafl$; the latter are of course unlike the external
lines appearing in $W$, since $\frac{\partial\Gammafl^{\T}}{\partial x^{\ast}}$
corresponds to s sub-diagram, as we will see below.

We now seek an approximation for the case that the integral is dominated
by a narrow regime around the maximum of $S$ close to $x^{\ast}$
in the same spirit as in \secref{Counting-loops}. This is precisely
the case, if the exponent has a sharp maximum; formally we may think
of an intrinsic scale $l\gg1$ present in the action. We will introduce
this scale as a parameter $l$ and rewrite \eqref{Gammafl_int} as
\begin{align}
-l\gammafl(x^{\ast}) & =\ln\int d\delta x\,\exp\left(l\big(\frac{1}{2}\delta x^{\T}s^{(2)}(x^{\ast})\delta x+r(x^{\ast},\delta x)+\gammafl^{(1)\T}(x^{\ast})\,\delta x\big)\right),\label{eq:gammafl_l}
\end{align}
where we defined $s^{(2)}:=\frac{S^{(2)}}{l}$, $r:=\frac{R}{l}$
and $\gammafl:=\frac{\Gammafl}{l}$. As before, we will use $l$ as
an expansion parameter. We remember from \secref{Counting-loops}
that a diagram with $n_{L}=n_{\Delta}-n_{V}+1$ loops has a prefactor
$l^{n_{V}-n_{\Delta}}=l^{1-n_{L}}$. So the overall contribution to
the integral diminishes with the number of loops.

Let us first imagine the term $l\gammafl^{(1)}\,\delta x$ was absent
on the right hand side. The integral would then generate all connected
Feynman diagrams with the propagator $\frac{1}{l}\left[-s^{(2)}\right]^{-1}$
and the vertices in $lr$. Due to the logarithm, by the linked cluster
theorem, only connected diagrams contribute in which all legs of all
vertices are contracted by propagators. The $l$-dependent factor
of each such diagram would hence be $l^{n_{V}-n_{\Delta}}=l^{1-n_{L}}$,
which counts the number of loops $n_{L}$ of the diagram. Due to the
prefactor $l$ on the left hand side of \eqref{gammafl_l}, the contribution
of a graph with $n_{L}$ loops to $\gammafl$ comes with a factor
$l^{-n_{L}}$.

To find all contributing graphs, our reasoning will now proceed by
induction in the number of loops and we successively compute
\begin{align*}
 & \gammafl^{0},\gammafl^{1},\gammafl^{2},\ldots
\end{align*}
with the given number of loops in the superscript as $\gammafl^{n_{L}}$.

To zero-loop order, we already know that $\Gamma-\ln\Z(0)=-S$, so
$\gammafl^{0}=0$. Proceeding to all one-loop contributions to $\gammafl^{1}$,
we can therefore drop the term $\gammafl^{0(1)}$. Since the vertices
in $R$, by construction, have three or more legs, they only yield
connected diagrams with two or more loops. The only contribution we
get is hence the integral over the Gaussian part of the action, i.e.
\begin{align}
\gammafl^{1}(x^{\ast})= & -\frac{1}{l}\ln\int d\delta x\,\exp\left(\frac{1}{2}\delta x^{\T}ls^{(2)}(x^{\ast})\delta x\right)\label{eq:gammafl_Gauss}\\
= & -\frac{1}{l}\ln\left(\sqrt{\left(2\pi\right)^{N}\det(-S^{(2)}(x^{\ast})^{-1})}\right),\nonumber \\
\Gammafl^{1}(x^{\ast})=l\gammafl^{1}(x^{\ast})= & \frac{1}{2}\ln\left(\left(2\pi\right)^{-N}\det(-S^{(2)}(x^{\ast}))\right),\nonumber 
\end{align}
where $N$ is the dimension of $x$ and the last step follows, because
$\det(A)^{-1}=\prod_{i}\lambda_{i}^{-1}=\det(A^{-1})$. We now see
that the one-loop correction grows as $O(\ln(l))$, which is smaller
than $O(l)$ of the zeroth order \eqref{Gamma0}, a posteriori justifying
our lowest order approximation. 

In the context of quantum mechanics, the approximation to one-loop
order is also called \textbf{semi-classical approximation}, because
it contains the dominant quantum fluctuation corrections if the system
is close to the classical limit; in this case $\hbar$ plays the role
of the expansion parameter $l^{-1}$ \citep[see also][]{ZinnJustin96}.

As in \secref{Tree-level-expansion-Cumulants-Vertex}, we denote the
function $\Gamma(x^{\ast})$ by a hatched circle and each derivative
adds one external leg to the symbol, so that the term $\gammafl^{(1)}=\partial_{x^{\ast}}\gammafl$
is denoted by:

\begin{center}
\begin{align*}
\partial_{x^{\ast}}\gammafl & =\Diagram{gp}
\end{align*}
\par\end{center}

We will now make the iteration step. Assume we have calculated the
contributions to $\gammafl^{n_{L}}$ up to loop-order $n_{L}$. The
integral in \eqref{gammafl_l} produces contributions at loop order
$n_{L}+1$ of two different types:
\begin{enumerate}
\item \label{enu:All-vacuum-diagrams}All vacuum diagrams (no external legs)
made up of $n_{V}$ vertices in $lr$ and $n_{\Delta}$ propagators
$\left(-ls^{(2)}\right)^{-1}$ with $n_{L}+1=n_{\Delta}-n_{V}+1$.
\item \label{enu:All-reducible-diagrams}All diagrams made of a subgraph
of $n_{L_{1}}$ loops composed of $n_{V}$ vertices from $lr$ and
$n_{\Delta}$ propagators $\left(-ls^{(2)}\right)^{-1}$ with $n_{L_{1}}=n_{\Delta}-n_{V}+1$
and the graphs of loop order $n_{L_{2}}\le n_{L}$ already contained
in $\gammafl^{n_{L}}$, so that $n_{L_{1}}+n_{L_{2}}=n_{L}+1$: The
term $l\gammafl^{(1)\T}\delta x=l\,\sum_{a=1}^{N}\frac{\partial\gammafl}{\partial x_{a}^{\ast}}\delta x_{a}$
allows the contraction of the $\delta x_{a}$ by the propagator to
some other $\delta x_{l}$ belonging to a vertex. For the example
of $n_{L_{1}}=1$ and $n_{L_{2}}=n_{L}$ one such contribution would
have the graphical representation:
\end{enumerate}
\begin{center}
\par\end{center}

\begin{figure}
\selectlanguage{english}%
\begin{minipage}[t]{0.45\columnwidth}%
\begin{flushright}
subgraph made of $\Delta$, $r$
\end{flushright}%
\end{minipage}\hfill%
\begin{minipage}[t]{0.45\columnwidth}%
subgraph of $n_{L_{2}}<n_{L}$ loops\linebreak{}
in $\gamma_{\mathrm{fl}}^{(n_{L})}$%
\end{minipage}

\selectlanguage{american}%
\noindent\begin{minipage}[t]{1\columnwidth}%
\begin{center}
\begin{fmffile}{tmp_8}
\begin{fmfgraph*}(100, 50)
\fmftopn{t}{3}
\fmfbottom{b1}
\fmffreeze
\fmfshift{(.2w, -.4h)}{t1}
\fmfshift{(0.w, -.5h)}{t2}
\fmfshift{(0.w, -.4h)}{t3}
\fmftop{t4}
\fmfdot{t1}
\fmffreeze
\fmf{plain}{t1,t2}
\fmf{plain, label=$(-ls^{(2)})^{-1}$}{t2,t3}
\fmf{plain, label=$(-ls^{(2)})^{-1}$}{t1,t1}
\fmf{dashes}{t4,t2,b1}
\fmfblob{.2w}{t3}
\fmfv{label=$n_{L_2}$, l.a=0, l.d=.4w}{t3}
\fmfv{label=$l\frac{r^{(3)}}{3!}$, l.a=45, l.d=.1h}{t1}
\fmfv{label=$(-1)l$, l.a=150, l.d=.15w}{t3}
\end{fmfgraph*}
\end{fmffile}
\end{center}%
\end{minipage}

\begin{minipage}[t]{0.5\columnwidth}%
\selectlanguage{english}%
\begin{center}\qquad
$n_{L_1}=1$
\end{center}\selectlanguage{american}%
\end{minipage}\hfill%
\begin{minipage}[t]{0.45\columnwidth}%
\selectlanguage{english}%
$n_{L_{2}}$ loops\selectlanguage{american}%
\end{minipage}
\end{figure}

Since the left portion of the diagram must have one or more loops
and the factor $l^{-1}$ from the connecting propagator $(-l\,s^{(2)})^{-1}$
cancels with the factor $l$ from $l\,\gammafl^{(1)}$, we see that
we only need diagrams with $n_{L}$ or less loops on the right. So
the iteration is indeed closed: We only need in the $n_{L}+1$ step
diagrams that we already calculated.
\begin{flushleft}
We see that a derivative $\partial_{x_{k}^{\ast}}$ attaches one leg
with index $k$. The terms contained in $\gammafl$ are diagrams,
where the legs of all vertices are contracted by propagators. The
derivative by $x^{\ast}$ may act on two different components of such
a diagram: a vertex $S^{(n)}(x^{\ast})=l\,s^{(n)}(x^{\ast}),$ $n>2$,
or a propagator $\Delta(x^{\ast})=\left(-l\,s^{(2)}(x^{\ast})\right)^{-1}$;
this is because both depend on $x^{\ast}$. Note that the $x^{\ast}$
dependence of these terms is the point around which the expansion
is performed. The typical portion of such a diagram around a vertex
$s^{(n)}$ therefore has the form
\begin{align}
 & \cdots S_{1\cdots n}^{(n)}(x^{\ast})\,\Pi_{i=1}^{n}\Delta_{i\,k_{i}}(x^{\ast})\cdots,\label{eq:part_of_diagram}
\end{align}
where each of the legs $1,\ldots,n$ of the vertex $S^{(n)}$ are
connected to a single propagator $\Delta$. The other end of each
of these propagators is denoted by the indices $k_{1},\ldots,k_{n}$.
Without loss of generality, we assume ascending indices of the first
vertex. Applying the derivative $\partial_{x_{a}^{\ast}}$ to this
diagram, the indicated portion (\ref{eq:part_of_diagram}) will lead
to the contributions
\par\end{flushleft}

\begin{flushleft}
\begin{eqnarray}
 &  & \cdots\partial_{x_{a}^{\ast}}\left\{ S_{1\cdots n}^{(n)}(x^{\ast})\,\Pi_{i=1}^{n}\Delta_{i\,k_{i}}(x^{\ast})\right\} \cdots\label{eq:portion_graph}\\
 & = & \cdots S_{1\cdots n\,a}^{(n+1)}(x^{\ast})\,\Pi_{i=1}^{n}\Delta_{i\,k_{i}}(x^{\ast})\cdots\nonumber \\
 &  & +\cdots S_{1\cdots n}^{(n)}(x^{\ast})\,\sum_{j=1}^{n}\left\{ \Pi_{i\neq j}\Delta_{i\,k_{i}}(x^{\ast})\right\} \partial_{x_{a}^{\ast}}\Delta_{j\,k_{j}}(x^{\ast})\cdots\nonumber 
\end{eqnarray}
The first term in the second last line adds one leg $a$ to the vertex,
therefore converting this vertex from an $n$-point to an $n+1$ point
vertex. To see the graphical representation of the last line, we rewrite
the derivative $\partial_{x_{a}^{\ast}}\Delta_{m_{j}k_{j}}(x^{\ast})$
by expressing the propagator $\Delta=-\left(S^{(2)}\right)^{-1}$,
differentiating $\Delta\,S^{(2)}=-\mathbf{1}$ and treat $\Delta$
as a matrix in its pair of indices. Differentiating the identity yields
\par\end{flushleft}

\begin{eqnarray*}
\partial x_{a}^{\ast}\left\{ \Delta\,S^{(2)}\right\}  & = & 0\\
\left(\partial_{x_{a}^{\ast}}\Delta\right)S^{(2)}+\Delta\underbrace{\partial_{x_{a}^{\ast}}S^{(2)}}_{S^{(3)}} & = & 0\\
\partial_{x_{a}^{\ast}}\Delta_{kl} & = & \left(\Delta\,S_{\circ a\circ}^{(3)}\,\Delta\right)_{kl},
\end{eqnarray*}
showing that the derivative of the propagator is transformed into
a three point vertex that has one leg labeled $a$ and is, by the
pair of propagators, connected to the remaining part of the diagram.
\begin{flushleft}
Thus the differentiation of the portion of the graph in \eqref{portion_graph},
for the example of $n=3$, takes on the diagrammatic representation
\par\end{flushleft}

\begin{eqnarray*}
 &  & \partial_{x_{a}^{\ast}}\ldots\Diagram{\vertexlabel^{1}\\
 & fdf\vertexlabel^{2}\\
 & \vertexlabel_{3}fu
}
\ldots=\ldots\Diagram{\vertexlabel^{1} &  & \vertexlabel^{2}\\
fd & fu\\
\vertexlabel_{3}fu & gd\vertexlabel_{a}
}
\ldots+\ldots\,\Diagram{ & \vertexlabel^{1}\\
\vertexlabel^{a}g & fv\\
 & fdf\vertexlabel_{2}\\
 & \vertexlabel_{3}fu
}
\ldots+2\text{ perm.}
\end{eqnarray*}

The meaning of the last expression is the conversion of the connecting
line between $m_{j}$ to $k_{j}$ into a connecting line between $m_{j}$
and $k_{j}$ to one leg of the three-point vertex $S^{(3)}$ that
in addition has another leg $a$. 

We first note that the latter contribution comes with a minus sign,
due to the minus sign on the left hand side of \eqref{gammafl_l},
while the vacuum diagrams produced by step \enuref{All-vacuum-diagrams}
comes with a plus sign. Second, we realize that the contributions
of the terms \enuref{All-reducible-diagrams} have the property of
being \textbf{one-line reducible} (also called \textbf{one-particle
reducible}), which means that the diagram can be disconnected by cutting
a single line, namely the line which connects the two subgraphs. Third,
we see that the power in $l$ of the latter contribution is $l^{-(n_{L_{1}}+n_{L_{2}}-1+1)}=l^{-(n_{L}+1)}$,
because the factor $l^{-1}$ in the connecting propagator and the
factor $l$ in $l\partial_{x}\gammafl^{\T}\delta x$ cancel each other.
So the power is the same as a $n_{L}+1$ loop contribution constructed
from step \enuref{All-vacuum-diagrams}. In conclusion we see, that
the graphs constructed by step \enuref{All-reducible-diagrams} cancel
all one-particle reducible contributions constructed by \enuref{All-vacuum-diagrams},
so that only \textbf{one-line irreducible} (or \textbf{one-particle
irreducible},\textbf{ 1PI}) graphs remain in $\gammafl$.

Coming back to our initial goal, namely the expansion of the vertex
generating function in terms of fluctuations centered around the mean
value, we see again how the term $l\partial_{x}\gammafl\delta x$
successively includes the contribution of fluctuations to the effective
value of the source $j_{\mathrm{fl}}(x^{\ast})=l\partial_{x}\gammafl$
that is required to reach the desired mean value $x^{\ast}=\langle x\rangle$
in the given approximation of $\gammafl$. In conclusion we have found
\begin{align*}
\Gamma(x^{\ast}) & -\ln\Z(0)=-S(x^{\ast})+\Gammafl(x^{\ast}),\\
\Gammafl(x^{\ast}) & =-\frac{1}{2}\ln\left(\left(2\pi\right)^{-N}\det(-S^{(2)}(x^{\ast}))\right)-\sum_{\mathrm{1PI}}\in\text{vacuum graphs}(\Delta,R),\\
\Delta(x^{\ast}) & =-\left[S^{(2)}(x^{\ast})\right]^{-1}.
\end{align*}
where the contribution of a graph to $\Gammafl$ declines with the
number of its loops.

If we are interested in the $n$-point vertex function, we may directly
calculate it from the diagrams with $n$ legs $\Feyn{g}$. We see
from the consideration above in \eqref{portion_graph} that the differentiation
by $x^{\ast}$ may attach a leg either at a vertex contained in $R$
or it may insert a three-point vertex into a propagator.

The combinatorial factor of a diagram contributing to $\Gamma^{(n)}$,
with $n$ external legs $\Feyn{g}$, is the same as for the diagrams
contributing to $W^{(n)}$ with $n$ external legs $j$: Since an
external line $j$ in a contribution to $W$ also connects to one
leg of an interaction vertex, just via a propagator, the rules for
the combinatorial factor must be identical. In the following example
we will see how to practically perform the loopwise approximation
for a concrete action.

\subsection{Example: $\phi^{3}+\phi^{4}$-theory\label{sub:Example_phi34_loopwise}}

Suppose we have the action
\begin{align*}
S_{l}(x) & =l\left(-\frac{1}{2}x^{2}+\frac{\alpha}{3!}x^{3}+\frac{\beta}{4!}x^{4}\right).
\end{align*}
with a parameter $l>0$ and possibly $l\gg1$, so that fluctuations
are small.

We start with the zero loop contribution, which by \eqref{def_Gammafl},
is
\begin{align}
\Gamma^{0}(x^{\ast})-\ln\Z(0) & =-S_{l}(x^{\ast})=l\left(\frac{1}{2}x^{\ast2}-\frac{\alpha}{3!}x^{\ast3}-\frac{\beta}{4!}x^{\ast4}\right).\label{eq:Gamma_0_phi34}
\end{align}
To obtain the corrections due to fluctuations collected in $\Gammafl$
according to \eqref{Gammafl_int}, we need to determine the effective
propagator $\left(-S_{l}^{(2)}\right)^{-1}$ as 
\begin{align}
S_{l}^{(2)}(x^{\ast}) & =l\left(-1+\alpha x^{\ast}+\frac{\beta}{2}x^{\ast2}\right)\label{eq:S2_phi4-1-1}\\
\Delta(x^{\ast})=-\left(S_{l}^{(2)}(x^{\ast})\right)^{-1} & =\frac{1}{l\left(1-\alpha x^{\ast}-\frac{\beta}{2}x^{\ast2}\right)}.\nonumber 
\end{align}
The one-loop correction is therefore given by the Gaussian integral
(\ref{eq:gammafl_Gauss}) appearing in \eqref{Gammafl_int}, leading
to
\begin{align}
\Gammafl^{1}(x^{\ast}) & =\ln\sqrt{\frac{-S_{l}^{(2)}(x^{\ast})}{2\pi}}=\frac{1}{2}\ln\left(\frac{l\left(1-\alpha x^{\ast}-\frac{\beta}{2}x^{\ast2}\right)}{2\pi}\right)\label{eq:Gammafl1_phi34-1}
\end{align}
The interaction vertices are
\begin{align}
\frac{1}{3!}S_{l}^{(3)}(x^{\ast}) & =\frac{1}{3!}\left(\alpha l+\beta lx^{\ast}\right)\label{eq:three_point-1}\\
\frac{1}{4!}S_{l}^{(4)} & =\frac{1}{4!}\beta l\nonumber \\
S_{l}^{(>4)} & =0.\nonumber 
\end{align}

Suppose we are only interested in the correction of the self-consistency
equation for the average $\langle x\rangle$, given by the solution
to the equation of state (\ref{eq:equation_of_state}), $\partial\Gamma/\partial x^{\ast}=0$
in the absence of fields. We have two possibilities: Either we calculate
the vertex function $\Gammafl^{1}$ to first order, given by (\ref{eq:Gammafl1_phi34-1})
and then take the derivative. This approach yields
\begin{align}
\frac{\partial\Gammafl^{1}}{\partial x} & =\frac{1}{2}\,\frac{l\,(-\alpha-\beta x^{\ast})}{l\left(1-\alpha x^{\ast}-\frac{\beta}{2}x^{\ast2}\right)}.\label{eq:Gammafl_1_dx_correction-1}
\end{align}
Alternatively, we may calculate the same contribution directly. We
therefore only need to consider those 1PI diagrams that have one external
leg (due to the derivative by $x$) and a single Gaussian integral
(one loop). The only correction is therefore a tadpole diagram including
the three-point vertex (\ref{eq:three_point-1}) and the propagator
(\ref{eq:S2_phi4-1-1}) 
\begin{align*}
\frac{\partial\Gammafl^{1}}{\partial x} & =-3\cdot\Diagram{gf0flflu}
\\
\\
 & =-3\left(-S_{l}^{(2)}\right)^{-1}\frac{1}{3!}S_{l}^{(3)}\\
 & =\frac{1}{2}\,\frac{\alpha l+\beta lx^{\ast}}{l\left(-1+\alpha x^{\ast}+\frac{\beta}{2}x^{\ast2}\right)},
\end{align*}
where the combinatorial factor is $3$ (three legs to choose from
the three point vertex to connect the external leg to and $1/3!$
stemming from the Taylor expansion of the action), which yields the
same result as (\ref{eq:Gammafl_1_dx_correction-1}). Both corrections
are of oder $\mathcal{O}(1)$ in $l$. So in total we get at $1$
loop order with $-S_{l}^{(1)}(x^{\ast})=l\left(x^{\ast}+\frac{\alpha}{2!}x^{\ast2}+\frac{\beta}{3!}x^{\ast3}\right)$
the mean value $x^{\ast}$ as the solution of
\begin{align*}
j & =\Gamma^{(1)}\stackrel{\text{1 loop order}}{\simeq}-S^{(1)}(x^{\ast})+\Gammafl^{(1)}(x^{\ast})\\
 & =l\left(x^{\ast}-\frac{\alpha}{2!}x^{\ast2}-\frac{\beta}{3!}x^{\ast3}\right)+\frac{1}{2}\,\frac{\alpha+\beta x^{\ast}}{-1+\alpha x^{\ast}+\frac{\beta}{2}x^{\ast}{}^{2}}.
\end{align*}
The tree level term is here $\mathcal{O}(l)$, the one loop term $\mathcal{O}(1)$.
The two-loop corrections $\mathcal{O}(l^{-1})$ are calculated in
the exercises. The resulting approximations of $\Gamma$ are shown
in \figref{Loopwise-expansion-phi4}.

\subsection{Appendix: Equivalence of loopwise expansion and infinite resummation\label{sec:Equivalence-loopwise-resummation}}

To relate the loopwise approximation to the perturbation expansion,
let us assume a setting where we expand around a Gaussian
\begin{align}
S(x) & =l\big(-\frac{1}{2}x^{\T}Ax+\epsilon V(x)\big).\label{eq:action_quadratic}
\end{align}
To relate the two approximations, we now have both expansion parameters,
$l$ and $\epsilon$. Here $\epsilon$ just serves us to count the
vertices, but we will perform an expansion in $l$. For the tree-level
approximation (\ref{eq:Gamma0}), the equation of state (\ref{eq:equation_of_state})
takes the form
\begin{align*}
j\stackrel{(\ref{eq:equation_of_state})}{=}\frac{\partial\Gamma_{0}(x^{\ast})}{\partial x^{\ast}} & \stackrel{(\ref{eq:Gamma0})}{=}-\frac{\partial S(x^{\ast})}{\partial x^{\ast}}\\
 & =l\big(A\,x^{\ast}-\epsilon V^{(1)}(x^{\ast})\big).
\end{align*}
We may rewrite the last equation as $x^{\ast}=A^{-1}j/l+A^{-1}\epsilon V^{(1)}(x^{\ast})$
and solve it iteratively 
\begin{align}
x_{0}^{\ast} & =A^{-1}j/l\label{eq:iterative_tree_level}\\
x_{1}^{\ast} & =A^{-1}j/l+A^{-1}\epsilon\,V^{(1)}(\underbrace{A^{-1}j/l}_{\equiv x_{0}})\nonumber \\
x_{2}^{\ast} & =A^{-1}j/l+A^{-1}\epsilon\,V^{(1)}(\underbrace{A^{-1}j/l+A^{-1}\epsilon\,V^{(1)}(A^{-1}j/l)}_{\equiv x_{1}})\nonumber \\
 & \vdots\nonumber 
\end{align}
Diagrammatically we have a tree structure, which for the example (\ref{eq:phi3_4_loop})
and setting $\beta=0$ would have $\epsilon V(x)=\frac{\alpha}{3!}x^{3}$
so $\epsilon V^{(1)}(x)=3\cdot\frac{\alpha}{3!}x^{2}$, where the
factor $3$ can also be regarded as the combinatorial factor of attaching
the left leg in the graphical notation of the above iteration
\begin{align*}
x^{\ast} & =\Diagram{f\vertexlabel^{j/l}}
\quad+\quad\Diagram{ & fu\vertexlabel^{j/l}\\
f\\
 & fd\vertexlabel^{j/l}
}
\quad+2\cdot\quad\Diagram{ & fu\vertexlabel^{j/l}\\
f\\
 & fd\\
 &  & fd & fu\vertexlabel^{j/l}\\
 &  &  & fd\vertexlabel^{j/l}
}
\quad+\quad\Diagram{ &  &  & fu\vertexlabel^{j/l}\\
 &  & fu & fd\vertexlabel^{j/l}\\
 & fu\\
f\\
 & fd\\
 &  & fd & fu\vertexlabel^{j/l}\\
 &  &  & fd\vertexlabel^{j/l}
}
\quad+\ldots,
\end{align*}
justifying the name ``tree-level approximation''. We see that we
effectively re-sum an infinite number of diagrams from ordinary perturbation
theory. We may also regard $x^{\ast}=W^{(1)}$ and hence conclude
that $W$ in this approximation corresponds to the shown graphs, where
an additional $j$ is attached to the left external line.

This property of resummation will persist at higher orders. It may
be that such a resummation has better convergence properties than
the original series.

We may perform an analogous computation for any higher order in the
loopwise approximation. We will exemplify this here for the \textbf{semi-classical
approximation} or \textbf{one-loop corrections}. To this end it is
easiest to go back to the integral expression \eqref{gammafl_Gauss}
in the form
\begin{align}
\Gammafl^{1}(x^{\ast}) & =-\ln\int d\delta x\,\exp\left(\frac{1}{2}\delta x^{\T}\,S^{(2)}(x^{\ast})\,\delta x\right)\label{eq:one_loop_integral}
\end{align}
and assume an action of the form \eqref{action_quadratic}. So we
have $S^{(2)}(x)=-lA+\epsilon lV^{(2)}(x)$ and hence \eqref{one_loop_integral}
takes the form
\begin{align*}
\Gammafl^{1}(x^{\ast}) & =-\ln\int d\delta x\,\exp\left(-\frac{1}{2}\delta x^{\T}\,lA\,\delta x+\frac{\epsilon}{2}\,\delta x^{\T}\,lV^{(2)}(x^{*})\,\delta x\right).
\end{align*}
We may consider $\Delta=l^{-1}A^{-1}=\Feyn{f}$ as the propagator
and the second quadratic term as the interaction $\frac{\epsilon}{2}\,\delta x^{\T}\,lV^{(2)}(x)\,\delta x=\Feyn{fufd}$.
Due to the logarithm we only get connected vacuum diagrams. A connected
diagram with $k$ vertices and all associated $\delta x$ being contracted
necessarily has a ring structure. Picking one of the vertices and
one of its legs at random, we have $k-1$ identical other vertices
to choose from and a factor $2$ to select one of its legs to connect
to. In the next step we have $2\,(k-2)$ choices to that we finally
arrive at
\begin{align*}
 & 2^{k-1}(k-1)!\,\underbrace{\Diagram{fufd & f & fufd\\
fv & f &  & fv\\
fdfu &  & fdfu
}
}_{k=4\text{ vertices}}.
\end{align*}
Since each vertex comes with $\frac{\epsilon}{2}$ and we have an
overall factor $\frac{1}{k!}$, we get 
\begin{align}
\Gammafl^{1}(x^{\ast}) & =\underbrace{-\frac{1}{2}\ln\,(\left(2\pi\right)^{N}\det(lA)^{-1})}_{\text{const.}(x^{\ast})}-\frac{1}{2}\,\sum_{k=1}^{\infty}\frac{\epsilon^{k}}{k}\tr\,\underbrace{(A^{-1}V^{(2)}(x^{\ast}))^{k}}_{k\text{ terms }V^{(2)}},\label{eq:resum_1loop}
\end{align}
where the latter term is meant to read (on the example $k=2$) $\tr\,A^{-1}V^{(2)}A^{-1}V^{(2)}=\sum_{i_{1}i_{2}i_{3}i_{4}}A_{i_{1}i_{2}}^{-1}V_{i_{2}i_{3}}^{(2)}A_{i_{3}i_{4}}^{-1}V_{i_{4}i_{1}}^{(2)}$
etc, because the propagator $A_{ik}^{-1}$ contracts corresponding
$\delta x_{i}$ and $\delta x_{k}$ associated with the terms $\delta x_{i}V_{ik}^{(2)}\delta x_{k}$.

We make three observations:
\begin{itemize}
\item In each term, the vertices form a single loop.
\item We get a resummation of infinitely many terms from perturbation theory,
had we expanded the action around some non-vanishing $x^{\ast}$.
\item The latter term in (\ref{eq:resum_1loop}) has the form of the power
series of $\ln(1-x)=\sum_{n=1}^{\infty}\frac{x}{n}$, so we can formally
see the result as $\ln(-S^{(2)})=\ln(lA-l\epsilon V^{(2)})=\ln lA+\ln(1-A^{-1}V^{(2)})=\ln lA+\sum_{k=1}^{\infty}\frac{(A^{-1}V^{(2)})^{k}}{k}$.
Further one can use that $\det(-S^{(2)})=\Pi_{i}\lambda_{i}$ with
$\lambda_{i}$ the eigenvalues of $-S^{(2)}$ and hence $\ln\det(M)=\sum_{i}\ln\lambda_{i}=\tr\ln(M)$,
because the trace is invariant under the choice of the basis.
\item The $x^{\ast}$-dependence in this case is only in the $V^{(2)}(x^{\ast})$.
$A$ instead is independent of $x^{\ast}$. A correction to the equation
of state would hence attach one additional leg to each term in these
factors, converting $V^{(2)}\to V^{(3)}$
\begin{align*}
 & \,\Diagram{fufd & f & fufd\\
fv & f &  & fv\\
fdfu &  & fdfu\\
fs0gv
}
\\
\end{align*}
\end{itemize}

\subsection{Appendix: Interpretation of $\Gamma$ as effective action\label{sec:effective_action}}

We can provide a formal reasoning, why $\Gamma$ is called ``effective
action''. We here follow \citep[section 16.2]{Weinberg05_II}. To
this end let us define the cumulant generating function $W_{\Gamma}$
\begin{align}
\exp(l\,W_{\Gamma,l}(j)) & :=\int\,dx\,\exp\Big(l\,\big(-\Gamma(x)+j^{\T}x\big)\Big),\label{eq:def_W_Gamma}
\end{align}
where we use the effective action $\Gamma$ in place of the action
$S$. We also introduced an arbitrary parameter $l$ to rescale the
exponent. The quantity $W_{\Gamma,l}$ does not have any physical
meaning. We here introduce it merely to convince ourselves that $\Gamma$
is composed of all one-line irreducible diagrams. As in \secref{Loopwise-gamma},
it will serve us to organize the generated diagrams in terms of the
numbers of loops involved.

For large $l\gg1$, we know from \secref{Loopwise-gamma} that the
dominant contribution to the integral on the right side of \eqref{def_W_Gamma}
originates from the points at which
\begin{align*}
\frac{\partial}{\partial x}\left(-\Gamma(x)+j^{\T}x\right) & \stackrel{!}{=}0\\
\frac{\partial\Gamma}{\partial x} & =j,
\end{align*}
which is the equation of state (\ref{eq:equation_of_state}) obtained
earlier for the Legendre transform. In this limit, we obtain the approximation
of \eqref{def_W_Gamma} as
\begin{align*}
W_{\Gamma,l\to\infty}(j) & =\sup_{x}\,j^{\T}x-\Gamma(x),
\end{align*}
which shows that $W_{\Gamma,l}$ approaches the Legendre transform
of $\Gamma$. Since the Legendre transform is involutive (see \secref{Vertex-generating-function}),
we conclude that $W_{\Gamma,l\to\infty}\to W(j)$ becomes the cumulant
generating function of our original theory. This view explains the
name \textbf{effective action}, because we obtain the true solution
containing all fluctuation corrections as the $x$ that minimizes
$\Gamma$ in the same way as we obtain the equations of motion of
a classical problem by finding the stationary points of the Lagrangian.

The property of one-line irreducibility now follows from \secref{Loopwise-gamma}
that to lowest order in $l$ only tree level diagrams contribute:
The zero loop approximation of an ordinary action replaces $\Gamma_{0}(x^{\ast})-\ln\Z(0)=-S(x^{\ast})$,
which contains all vertices of the original theory. The equation of
state, as shown in \secref{Equivalence-loopwise-resummation}, can
be written as all possible tree-level diagrams without any loops.

Applied to the integral \eqref{def_W_Gamma}, which, at lowest order
is the full theory including all connected diagrams with arbitrary
numbers of loops, we see that all these contributions are generated
by all possible tree level diagrams composed of the components of
$\Gamma$. Expanding $\Gamma(x^{\ast})$ around $x_{0}$ we get from
the equation of state (\ref{eq:equation_of_state})

\begin{align*}
j & =\underbrace{\Gamma^{(2)}(x_{0})}_{=(W^{(2)})^{-1}}(x^{\ast}-x_{0})+\sum_{k=3}^{\infty}\frac{1}{k-1!}\Gamma^{(k)}(x_{0})(x^{\ast}-x_{0})^{k-1}.
\end{align*}
We can therefore solve the equation of state in the same iterative
manner as in (\ref{eq:iterative_tree_level}) with $\delta x:=x^{\ast}-x_{0}$
\begin{align*}
\delta x_{i}^{0} & =W_{ik}^{(2)}\,j_{k}\\
\delta x_{i}^{1} & =W_{ik}^{(2)}\,j_{k}+\frac{1}{2!}\Gamma_{ikl}^{(3)}W_{kn}^{(2)}W_{lm}^{(2)}\,j_{n}j_{m}\\
 & \vdots
\end{align*}

The connections in these diagrams, in analogy to \eqref{Gammafl_int},
are made by the effective propagator $(\Gamma^{(2)})^{-1}=W^{(2)}$
(following from \eqref{W2Gamma2_inverse}), which are lines corresponding
to the full second cumulants of the theory. The vertices are the higher
derivatives of $\Gamma$, i.e. the vertex functions introduced in
\eqref{def_vertex_function}. This view is therefore completely in
line with our graphical decomposition developed in \secref{Tree-level-expansion-Cumulants-Vertex}
and again shows the tree-level decomposition of $W$ into vertex functions:
Here we have the explicit expansion of $\delta x=W^{(1)}$. This in
turn means that the components of $\Gamma$ can only be those diagrams
that are one-line irreducible, i.e. that cannot be disconnected by
cutting one such line, because otherwise the same diagram would be
produced twice.

\subsection{Loopwise expansion of self-consistency equation\label{sec:Loopwise-expansion-self-consistency}}

We here come back to the example from \secref{tanh_network} and want
to obtain a loopwise approximation for the one-dimensional self-consistency
equation

\begin{align}
x & =\underbrace{J_{0}\phi(x)}_{=:\psi(x)}+\mu+\xi.\label{eq:self_cons_tanh_loop}
\end{align}
We need to construct the moment-generating function. We define a function
$f(x)=x-\psi(x)-\mu$ so that we may express $x$ as a function of
the noise realization $x=f^{-1}(\xi)$. We can define the moment-generating
function
\begin{align}
Z(j) & =\langle\exp(j\,\underbrace{f^{-1}(\xi)}_{x})\rangle_{\xi}\nonumber \\
 & =\big\langle\int\,dx\,\delta(x-f^{-1}(\xi))\,\exp(j\,x)\big\rangle_{\xi},\label{eq:mom_gen_self_consistent}
\end{align}
where in the last step we introduced the variable $x$ explicitly
to get a usual source term.

Since the constraint is given by an implicit expression of the form
$f(x)=\xi$, with $f(x)=x-\psi(x)$ we need to work out what $\delta(f(x))$
is, which follows from substitution as 
\begin{align}
\int\,g(x)\,\delta(\underbrace{f(x)}_{=:y})\,dx & =\int\,g(f^{-1}(y))\,\delta(y)\,\frac{1}{\frac{dy}{dx}}\,dy=\int\,g(f^{-1}(y))\,\delta(y)\,\frac{1}{f^{\prime}(f^{-1}(y))}\,dy=\frac{g(f^{-1}(0))}{f^{\prime}(f^{-1}(0))}\nonumber \\
\delta(f(x)-\xi)\,f^{\prime}(x) & \to\delta(x-f^{-1}(\xi)),\label{eq:delta_f_rule-1}
\end{align}
We can therefore rewrite (\ref{eq:mom_gen_self_consistent}) as 
\begin{align}
Z(j) & \stackrel{(\ref{eq:delta_f_rule-1})}{=}\big\langle\int\,dx\,f^{\prime}(x)\,\delta(f(x)-\xi)\,\exp(j\,x)\big\rangle_{\xi},\label{eq:Z_with_func_det}
\end{align}
which satisfies $Z(j)=0$ as it should. We now resolve the Dirac $\delta$
constraint by the introduction of an \textbf{auxiliary field} $\tilde{x}$
and represent the Dirac $\delta$ in Fourier domain as 
\begin{align*}
\delta(x) & =\frac{1}{2\pi i}\,\int_{-i\infty}^{i\infty}\,e^{\tx\,x}\,d\tx.
\end{align*}
We get
\begin{align}
Z(j) & =\int_{-\infty}^{\infty}\,dx\,\int_{-i\infty}^{i\infty}\frac{d\tx}{2\pi i}\,(1-\psi^{\prime}(x))\,\exp\big(\tx\,(x-\psi(x))-\mu\,\tx+j\,x\big)\,\underbrace{\langle\exp(-\tx\xi)\rangle_{\xi}}_{\equiv Z_{\xi}(-\tx)=\exp(\frac{D}{2}\tx^{2})},\label{eq:Z_self_con_pre-1}
\end{align}
where we identified the moment-generating function of the noise in
the underbrace and inserted $f^{\prime}=1-\psi^{\prime}$. We notice
that $\mu$ couples to $\tx$ in a similar way as a source term. We
can therefore as well introduce a source $\tj$ and remove $\mu$
from the moment generating function

\begin{align}
Z(j,\tj) & :=\int_{-\infty}^{\infty}\,dx\,\int_{-i\infty}^{i\infty}\frac{d\tx}{2\pi i}\,(1-\psi^{\prime}(x))\,\exp(S(x,\tx)+jx+\tj\tx)\label{eq:Z_l_unsymm-1}\\
S(x,\tx) & :=\tx\,(x-\psi(x))+\frac{D}{2}\tx^{2}.\nonumber 
\end{align}
In determining the solution in the presence of $\mu$, we need to
ultimately set $\tj=-\mu$.

We see from this form a special property of the field $\tx$: For
$j=0$ and $\tilde{j}\in\mathbb{R}$ arbitrary we have due to normalization
of the distribution $Z(0,\tilde{j})=1=\mathrm{const.}(\tilde{j})$.
Consequently
\begin{align*}
\langle\tx^{n}\rangle\big|_{j=0} & \equiv\frac{\partial^{n}Z(0,\tilde{j})}{\partial\tilde{j}^{n}}=0\quad\forall\tilde{j}.
\end{align*}
We hence conclude that there cannot be any diagrams in $W(j)$ with
only external legs $\tj$.

To obtain the loopwise expansion of $\Gamma(x^{\ast},\tx^{\ast})$,
we can start at the lowest order. To lowest order we have \eqref{Gamma0}
and therefore get the pair of equations
\begin{align}
j & =-\frac{\partial S(x^{\ast},\tx^{\ast})}{\partial x}=-\tx^{\ast}\psi^{\prime}(x^{\ast})\label{eq:eq_state_tanh_j}\\
\tj & =-\frac{\partial S(x^{\ast},\tx^{\ast})}{\partial\tx}=x^{\ast}-\psi(x^{\ast})+D\,\tx^{\ast}.\label{eq:eq_state_tanh_til_j}
\end{align}
The first equation, for $j=0$ allows the solution $\tx^{\ast}=\langle\tx\rangle=0$,
which we know to be the true one from the argument above. Inserted
into the second equation, we get with $\tj=-\mu$
\begin{align*}
x^{\ast} & =\underbrace{\psi(x^{\ast})}_{J_{0}\phi(x^{\ast})}+\mu,
\end{align*}
which is in line with our naive solution (\ref{eq:self_consistent_mean_tanh}).

To continue to higher orders, we need to determine the effective propagator
from the negative inverse Hessian of $S$, which is
\begin{align}
S^{(2)}(x^{\ast},\tx^{\ast}) & =\left(\begin{array}{cc}
-\tx\,\psi^{(2)}(x^{\ast}) & 1-\psi^{(1)}(x^{\ast})\\
1-\psi^{(1)}(x^{\ast}) & D
\end{array}\right).\label{eq:Hessian_S_tanh}
\end{align}
From the general property that $\langle\tx\rangle=0$, we know that
the correct solution of the equation of state must expose the same
property. So we may directly invert (\ref{eq:Hessian_S_tanh}) at
the point $\tx^{\ast}=\langle\tx\rangle=0$, which is
\begin{align*}
\Delta(x^{\ast},0)=(-S^{(2)}(x^{\ast},0))^{-1} & =\left(\begin{array}{cc}
\frac{D}{(1-\psi^{\prime}(x^{\ast}))^{2}} & -\frac{1}{1-\psi^{\prime}(x^{\ast})}\\
-\frac{1}{1-\psi^{\prime}(x^{\ast})} & 0
\end{array}\right)=:\left(\begin{array}{cc}
\feyn{fVfA} & \feyn{fV}\\
\feyn{fA} & 0
\end{array}\right).
\end{align*}
We see that to lowest order hence $\langle\tx^{2}\rangle=0$, indicated
by the vanishing lower right entry. In the graphical notation we chose
the direction of the arrow to indicate the contraction with a variable
$x$ (incoming arrow) or a variable $\tx$ (outgoing arrow). 

We conclude from the argument that $\langle\tx^{2}\rangle=0$ that
the correction to the self-energy has to vanish as well $\Sigma_{xx}=0$.
This can be seen by writing the second cumulant with (\ref{eq:def_self_energy})
as
\begin{align}
W^{(2)} & =\left(\Gamma^{(2)}\right)^{-1}\label{eq:W_2_self_consistent}\\
 & =(-S^{(2)}+\Sigma)^{-1}\nonumber \\
 & =\left[-\left(\begin{array}{cc}
0 & 1-\psi^{(1)}(x^{\ast})\\
1-\psi^{(1)}(x^{\ast}) & D
\end{array}\right)+\left(\begin{array}{cc}
\Sigma_{xx} & \Sigma_{x\tx}\\
\Sigma_{\tx x} & \Sigma_{\tx\tx}
\end{array}\right)\right]^{-1}.\nonumber 
\end{align}
In order for $W_{\tj\tj}^{(2)}$ to vanish, we need a vanishing $\Sigma_{xx}$,
otherwise we would get an entry $W_{\tj\tj}^{(2)}\propto\Sigma_{xx}$.

The interaction vertices, correspondingly, are the higher derivatives
of $S$. Due to the quadratic appearance of $\tx$, we see that no
vertices exist that have three or more derivatives by $\tx$. Because
$\tx^{\ast}=0$, we see that all $\tx$ must be gone by differentiation
for the vertex to contribute. The only vertex at third order therefore
is
\begin{align*}
S_{\tx xx}^{(3)} & =-\frac{1}{2!}\psi^{(2)}(x^{\ast})=\Diagram{\vertexlabel^{\tx}fV & fuV & \vertexlabel^{x}\\
 & fdV & \vertexlabel^{x}
}
\quad,
\end{align*}
where the factor $1/2!$ stems from the Taylor expansion due to the
second derivative of $S_{\tx xx}^{(3)}$ by $x$. We observe that
at arbitrary order $n$ we get
\begin{align*}
S_{\tx x^{n-1}}^{(n)} & =-\frac{1}{n-1!}\psi^{(n-1)}(x^{\ast}).
\end{align*}
 We are therefore ready to apply the loopwise expansion of the equation
of state. 

We know from the general argument above that we do not need to calculate
any loop corrections to (\ref{eq:eq_state_tanh_j}), because we know
that $\tx^{\ast}\equiv0$. We obtain the one-loop correction to the
equation of state (\ref{eq:eq_state_tanh_til_j}) from the diagram
with one external $\tx$-leg
\begin{align*}
\frac{\partial\Gammafl}{\partial\tx}=-\Diagram{gVf0flVfluVf0}
 & =(-1)\underbrace{(-\frac{1}{2!}\psi^{(2)}(x^{\ast}))}_{S_{\tx xx}^{(3)}}\,\underbrace{\frac{D}{(1-\psi^{\prime}(x^{\ast}))^{2}}}_{\left(-S^{(2)}\right)_{xx}^{-1}}\\
 & =\frac{D}{2}\,\frac{J_{0}\phi^{(2)}(x^{\ast})}{(1-J_{0}\phi^{(1)}(x^{\ast}))^{2}}.
\end{align*}
Note that there is no combinatorial factor $3$ here, because there
is only one variable $\tx$ to choose for the external leg. So together
with the lowest order (\ref{eq:eq_state_tanh_til_j}) we arrive at
the self-consistency equation for $x^{\ast}$
\begin{align}
\tj=-\mu= & -\frac{\partial S(x^{\ast},\tx^{\ast})}{\partial\tx}-\Diagram{gVf0flVfluVf0}
\nonumber \\
\nonumber \\
x^{\ast}= & J_{0}\phi(x^{\ast})+\mu+\frac{D}{2}\,\frac{J_{0}\phi^{(2)}(x^{\ast})}{(1-J_{0}\phi^{(1)}(x^{\ast}))^{2}}.\label{eq:eq_state_1loop}
\end{align}
Comparing to (\ref{eq:one_loop_tanh}), we see that we have recovered
the same correction. But we now know that the additional term is the
next to leading order systematic correction in terms of the fluctuations.
Also, we may obtain arbitrary higher order corrections. Moreover,
we are able to obtain corrections to other moments, such at the variance
by calculating the self-energy (see exercises). From our remark further
up we already know that $\Sigma_{xx}\equiv0$, so that diagrams
\begin{align*}
0\equiv & \Diagram{gAf0flVfluAf0gV}
\quad,\\
\end{align*}
which have two external $x$-legs need to vanish and therefore do
not need to be calculated.

\section{Loopwise expansion in the MSRDJ formalism}

We now want to apply the loopwise expansion developed in \secref{Loopwise-expansion-vertex-function}
to a stochastic differential equation, formulated as an MSRDJ path
integral. We remember from \secref{Loopwise-expansion-vertex-function},
that the loopwise expansion is a systematic perturbative technique,
whose expansion parameter is the number of loops in the diagrams,
which we showed to measure the fluctuations in the system.

\subsection{Intuitive approach\label{sub:Intuitive-approach_MSR_Loop}}

Before embarking on this endeavor in a formal way, we would like to
present the naive approach of obtaining an expansion for small fluctuations
for a stochastic differential equation in the limit of weak noise,
i.e. for weak variance of the driving noise $W_{W}^{(2)}(0)\ll1$
in \eqref{MSR_Z_tilde_J}, where $W_{W}(j)=\ln\,Z_{W}(j)$ is the
cumulant generating functional of the stochastic increments. We want
to convince ourselves that to lowest order, the formal approach agrees
to our intuition.

To illustrate the two approaches, we consider the stochastic differential
equation
\begin{align}
dx(t) & =f(x(t))\,dt+dW(t)\label{eq:SDE_loop}
\end{align}
which has action in the general (possibly non-Gaussian noise) case
\begin{align*}
S[x,\tilde{x}] & =\tilde{x}^{\T}(\partial_{t}x-f(x))+W_{W}(-\tilde{x}).
\end{align*}
The naive way of approximating \eqref{SDE_loop} for small noise is
the replacement of the noise drive by its mean value $\langle dW\rangle(t)=\frac{\delta W_{W}(0)}{\delta j_{W}(t)}dt\equiv W_{W,t}^{(1)}(0)dt=:\bar{W}(t)\,dt$
(using that the derivative $W_{W,t}^{(1)}(0)$ is the mean stochastic
increment at time $t$). This yields the ODE
\begin{align}
\partial_{t}x & =f(x(t))+\bar{W}(t).\label{eq:zero_loop_MSR_naive}
\end{align}
We will now check if the lowest order loopwise expansion yields the
same result. To this end we use \eqref{Gamma0}, i.e. $\Gamma_{0}[x,\tilde{x}]=-S[x,\tilde{x}]$
and obtain the pair of equations from the equation of state \eqref{equation_of_state}
\begin{align*}
-\frac{\delta S[x,\tilde{x}]}{\delta x(t)}=\frac{\delta\Gamma_{0}[x,\tilde{x}]}{\delta x(t)} & =j(t)\\
-\frac{\delta S[x,\tilde{x}]}{\delta\tilde{x}(t)}=\frac{\delta\Gamma_{0}[x,\tilde{x}]}{\delta x(t)} & =\tilde{j}(t).
\end{align*}
The explicit forms of these equations with \eqref{SDE_loop} is
\begin{align}
\left(\partial_{t}+f^{\prime}(x^{\ast}(t))\right)\,\tilde{x}^{\ast}(t) & =j(t),\label{eq:eq_state_zero_loop}\\
\partial_{t}x^{\ast}(t)-f(x^{\ast}(t))-W_{W,t}^{(1)}(-\tilde{x}^{\ast}) & =-\tilde{j}(t),\nonumber 
\end{align}
where in the first line we used integration by parts to shift the
temporal derivative to $\tilde{x}^{\ast}$, assuming negligible boundary
terms at $t\to\pm\infty$. In the absence of external fields $j=\tilde{j}=0$,
the second equation is hence identical to the naive approach \eqref{zero_loop_MSR_naive},
if $\tilde{x}^{\ast}\equiv0$. The first equation indeed admits this
solution. Interestingly, the first equation has only unstable solutions
if and only if the linearized dynamics for $x$ (which is $(\partial_{t}-f^{\prime}(x^{\ast}))\delta x$
see below) is stable and vice versa. So the only finite solution of
the first equation is the vanishing solution $\tilde{x}\equiv0$.

We have anticipated the latter result from the perturbative arguments
in \secref{Vanishing-response-loops}: to all orders the moments of
$\tilde{x}$ vanish for $j=0$. We hence know from the reciprocity
relationship \eqref{inverse_W1_Gamma1} between $W^{(1)}[j,\tilde{j}]=(x^{\ast},\tilde{x}^{\ast})$
and $\Gamma^{(1)}[x^{\ast},\tilde{x}^{\ast}]=(j,\tilde{j})$ that
the minimum of $\Gamma[x,\tilde{x}]$ must be attained at $\tilde{x}^{\ast}=\langle\tilde{x}\rangle=0$
for any value of $\tilde{j}(t)$. Solving the equations of state (\ref{eq:eq_state_zero_loop})
with a non-zero $j(t)$, this property ceases to be valid, in line
with \eqref{eq_state_zero_loop}. A vanishing source $j$, however,
does not pose any constraint to the applicability to physical problems,
since $j$ does not have any physical meaning. The freedom to choose
a non-zero $\tilde{j}$ in \eqref{eq_state_zero_loop}, on the contrary,
is useful, because it appears as the inhomogeneity of the system (see
\secref{Response-function-in}) and hence allows us to determine the
true mean value of the fields in the presence of an external drive
to the system.

Continuing the naive approach to include fluctuations, we could linearize
the \eqref{SDE_loop} around the solution $x^{\ast}$, defining $\delta x(t)=x(t)-x^{\ast}(t)$
with the resulting SDE for the fluctuation $\delta x$
\begin{align}
d\delta x(t) & =f^{\prime}(x^{\ast}(t))\,\delta x(t)\,dt+dW(t)-\bar{W}(t)dt.\label{eq:SDE_delta_x_loop}
\end{align}
The equation is linear in $\delta x$ and the driving noise $dW(t)-\bar{W}(t)dt$,
by construction, has zero mean. Taking the expectation value of the
last equation therefore shows that, for stable dynamics, $\langle\delta x(t)\rangle$
decays to $0$, so $\delta x$ has zero mean (is a centered process).
Its second moment is therefore identical to the second cumulant, for
which \eqref{SDE_delta_x_loop} yields the differential equation
\begin{align}
(\partial_{t}-f^{\prime}(x^{\ast}(t)))(\partial_{s}-f^{\prime}(x^{\ast}(s)))\langle\delta x(t)\delta x(s)\rangle & =\delta(t-s)\,W_{W,t}^{(2)},\label{eq:cov_loop_intuitive}
\end{align}
where we used that the centered increments $dW(t)-\bar{W}(t)$ are
uncorrelated between $t\neq s$ and hence have the covariance $\delta(t-s)W_{W,t}^{(2)}\,dt\,ds$.

We now want the see if we get the same result by the formal approach.
We may therefore determine the Hessian $\Gamma_{0,t,s}^{(2)}[x^{\ast},\tilde{x}]$,
the inverse of which, by \eqref{W2Gamma2_inverse}, is the covariance
$W^{(2)}$
\begin{align*}
\Gamma_{0,t,s}^{(2)}[x^{\ast},\tilde{x}^{\ast}] & \equiv\frac{\delta^{2}\Gamma_{0}}{\delta\{x,\tilde{x}\}(t)\delta\{x,\tilde{x}\}(s)}\\
 & =\left(\begin{array}{ccc}
0 & \qquad & \delta(t-s)\,(\partial_{t}+f^{\prime}(x^{\ast}))\\
\delta(t-s)\,(-\partial_{t}+f^{\prime}(x^{\ast})) & \qquad & -\delta(t-s)\,W_{W,t}^{(2)}(0)
\end{array}\right),
\end{align*}
where the top left entry $\tilde{x}^{\ast}(t)\,f^{\prime}(x^{\ast})\delta(t-s)$
vanishes, because we evaluate the Hessian at the stationary point
with $\tilde{x}^{\ast}\equiv0$ and we used that the noise is white,
leading to $\delta(t-s)$ in the lower right entry. We may therefore
obtain the covariance matrix as the inverse, i.e. $W^{(2)}=\left[\Gamma^{(2)}\right]^{-1}$
in the sense
\begin{align*}
\diag(\delta(t-u)) & =\int\,\Gamma_{0,t,s}^{(2)}\,W_{s,u}^{(2)}\,ds,\\
W_{t,s}^{(2)}=\frac{\delta^{2}W}{\delta\{j,\tilde{j}\}(t)\delta\{j,\tilde{j}\}(s)} & =\left(\begin{array}{cc}
\llangle x(t)x(s)\rrangle & \llangle x(t)\tilde{x}(s)\rrangle\\
\llangle\tilde{x}(t)x(s)\rrangle & \llangle\tilde{x}(t)\tilde{x}(s)\rrangle
\end{array}\right)
\end{align*}
leading to the set of four differential equations
\begin{align*}
\delta(t-u) & =(\partial_{t}+f^{\prime}(x^{\ast}(t)))\,\llangle\tilde{x}(t)x(u)\rrangle\\
0 & =-(\partial_{t}+f^{\prime}(x^{\ast}(t)))\,\llangle\tilde{x}(t)\tilde{x}(u)\rrangle\\
0 & =(-\partial_{t}+f^{\prime}(x^{\ast}(t)))\,\llangle x(t)x(u)\rrangle-W_{W,t}^{(2)}(0)\,\llangle\tilde{x}(t)x(u)\rrangle\\
\delta(t-u) & =(-\partial_{t}+f^{\prime}(x^{\ast}(t)))\,\llangle x(t)\tilde{x}(u)\rrangle-W_{W,t}^{(2)}(0)\,\llangle\tilde{x}(t)\tilde{x}(s)\rrangle.
\end{align*}
For stable dynamics of $x$, the operator in the second equation
is necessarily unstable, because the temporal derivative has opposite
sign. The only admissible finite solution is therefore the trivial
solution $\llangle\tilde{x}(t)\tilde{x}(u)\rrangle\equiv0$. The last
equation therefore rewrites as
\begin{align*}
\delta(t-u) & =(-\partial_{t}+f^{\prime}(x^{\ast}(t)))\,\llangle x(t)\tilde{x}(u)\rrangle.
\end{align*}
Applying the operator $(-\partial_{u}+f^{\prime}(x^{\ast}(u)))$ to
the third equation and using the last identity we get
\begin{align*}
(\partial_{t}-f^{\prime}(x^{\ast}(t)))\,(\partial_{u}-f^{\prime}(x^{\ast}(u)))\,\llangle x(t)x(u)\rrangle & =\delta(t-u)\,W_{W,t}^{(2)}(0),
\end{align*}
which is the same result as obtained by the intuitive approach in
\eqref{cov_loop_intuitive}.

So to lowest order in the loopwise expansion, we see that the naive
approach is identical to the systematic approach. Up to this point
we have of course not gained anything by using the formal treatment.
Going to higher orders in the loopwise expansion, however, we will
obtain a systematic scheme to obtain corrections to the naive approach.
The fluctuations of $\delta x$ obviously could change the mean of
the process. This is what will, by construction, be taken into account
self-consistently.

\subsection{Loopwise corrections to the effective equation of motion}

In the following, we want to use the loopwise expansion to approximate
the average value of the stochastic variable $x$. Let us assume that
it fulfills the stochastic differential equation
\begin{equation}
dx+x\,dt=J\phi\left(x\right)\,dt+dW\left(t\right),\label{eq:Langevin_equation}
\end{equation}
where 
\[
\phi\left(x\right)=J\left(x-\alpha\frac{x^{3}}{3!}\right)
\]
and $dW$ is white noise with
\[
\left\langle dW\left(t\right)\right\rangle =0,\ \left\langle dW\left(t\right)dW\left(t^{\prime}\right)\right\rangle =D\delta_{tt^{\prime}}\,dt.
\]
The fix points of this ODE in the noiseless case (i.e. $D=0$) are
\[
x_{0}:=0,\ x_{\pm}:=\sqrt{3!\frac{J-1}{\alpha J}},\text{ for }\quad\frac{J-1}{\alpha J}>0.
\]
The trivial fix point $x_{0}$ is stable as long as $J<1$ and the
fix points $x_{\pm}$ are stable for $J>1$. For $\alpha<0$ and $0<J<1$
or $\alpha>0$ and $J<1$, the nontrivial fix points exist, but are
unstable. In other words: If $\alpha>0$, the system becomes bistable
if the level of excitation is high enough and if $\alpha<0$, it explodes
for too high excitation.

Due to the fluctuations, the average $\lim_{t\rightarrow\infty}\left\langle x\right\rangle \left(t\right)$
will deviate from $x_{0}$. We will determine this deviation in the
following.

\begin{figure}
\begin{centering}
\includegraphics{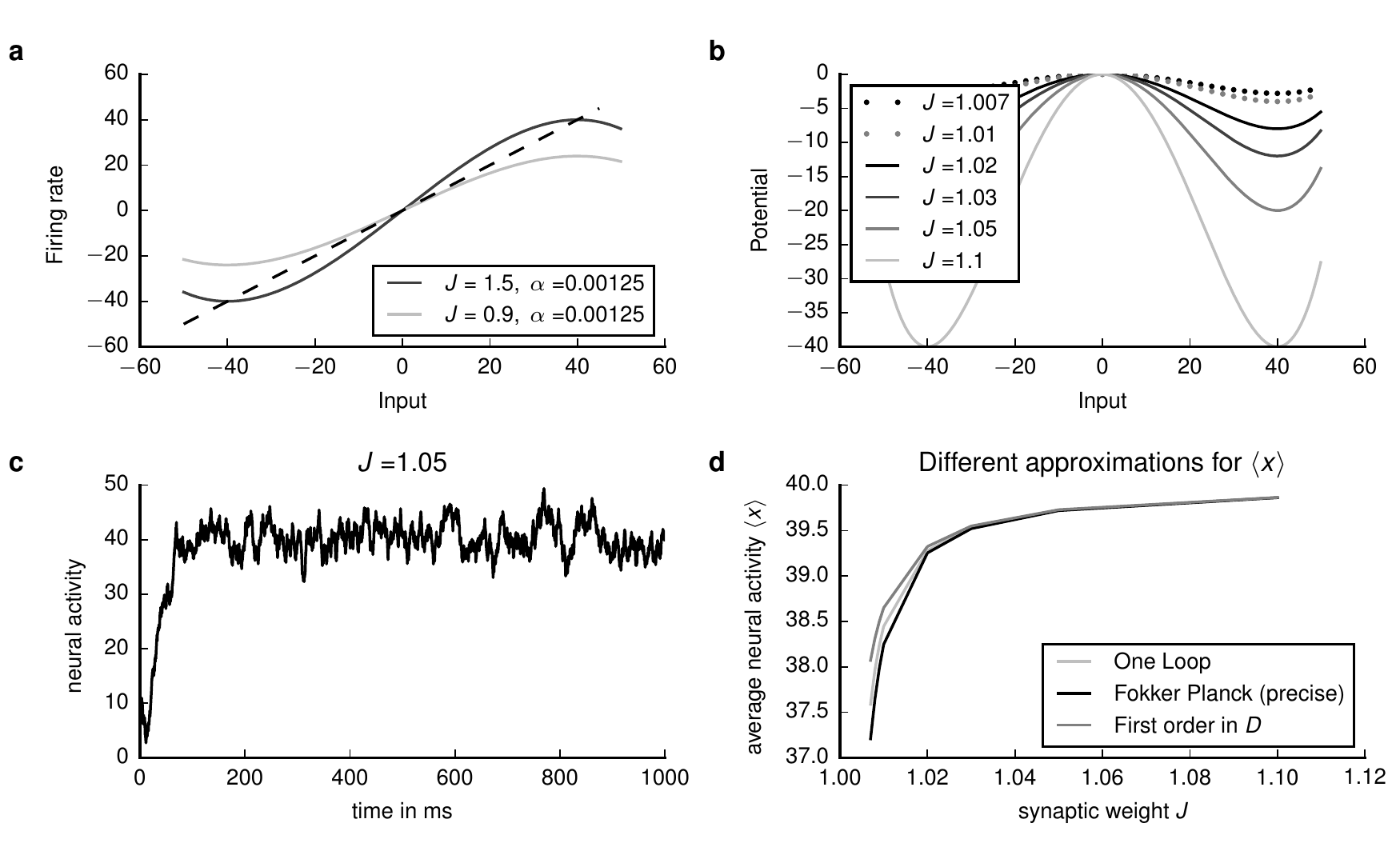}
\par\end{centering}
\caption{\textbf{a} Transfer function for the symmetry-broken phase and the
phase with $\langle x\rangle=0$. \textbf{b} Potential including leak
term in the symmetry-broken phase with different synaptic weights,
but with $x_{0}=\pm40$ always. \textbf{c} One realization of the
stochastic process. \textbf{d} Deviation of the fix point value for
the cases of b depending on the strength of the non-linearity $\alpha$
calculated (numerically) exact by solving the Fokker-Planck equation
of \eqref{Langevin_equation} in the stationary case, in the one-loop
approximation and by expanding the one-loop solution in $D$ to first
order (which amounts to expanding the Fokker-Planck-solution in $D$).\label{fig:OUP_loop_figure}}
\end{figure}
For this, we need the action of the stochastic ODE
\[
S[x,\widetilde{x}]=\tx^{\T}\left[\left(\partial_{t}+1-J\right)x+\frac{\alpha J}{3!}x^{3}\right]+\frac{D}{2}\tx^{\T}\tx.
\]
We now want to calculate the vertex-generating function successively
in different orders of numbers of loops in the Feynman diagrams. To
lowest order (\ref{eq:Gamma0}) we have
\[
\Gamma_{0}\left[x^{\ast},\tx^{\ast}\right]=-S\left[x^{\ast},\tx^{\ast}\right]
\]
We know from the general proof in \secref{Vanishing-response-field}
and from the remarks in \subref{Intuitive-approach_MSR_Loop} that
the true mean value of the response field needs to vanish $\tx^{\ast}=0$.
The true mean value $x^{\ast}$ is, so far, unknown. We have to determine
it by solving the equation of state
\[
\left(\begin{array}{c}
\frac{\partial}{\partial x^{\star}}\\
\frac{\partial}{\partial\widetilde{x}^{\star}}
\end{array}\right)\Gamma\left[x^{\star},\tx^{\star}\right]=\left(\begin{array}{c}
j\\
\tj
\end{array}\right).
\]
One of the equations will just lead to $\tx=0$. To convince ourselves
that this is indeed so, we here calculate this second equation as
well:
\begin{gather*}
\frac{\partial\Gamma}{\partial x^{\ast}}\left[x^{\ast},\tx^{\ast}\right]=-\left(-\partial_{t}+1-J\right)\tx^{\ast}\left(t\right)-\frac{\alpha J}{2}\left(x^{\ast}\left(t\right)\right)^{2}\tx^{*}\left(t\right)+\mathcal{O}\left(\text{loop corrections}\right)=j\\
\frac{\partial\Gamma}{\partial\widetilde{x}^{\ast}}\left[x^{\ast},\tx^{\ast}\right]=-\left(\partial_{t}+1-J\right)x^{\ast}\left(t\right)-\frac{\alpha J}{3!}\left(x^{\ast}\left(t\right)\right)^{3}-D\tx\left(t\right)+\mathcal{O}\left(\text{loop corrections}\right)=\tj.
\end{gather*}
Next, we will be concerned with the stationary solution and drop the
time derivatives. This makes it much easier so determine the one-loop-contributions.
They consist of a three-point-vertex at which are attached one external
amputated line, $x^{*}$ or $\widetilde{x}^{*}$, and two ``normal''
lines, associated with $\delta x$ or $\delta\widetilde{x}$, which
are contracted. The only nonzero three-point-vertices in our theory
are
\begin{align}
\frac{1}{3!}\,\frac{\delta^{3}S}{\delta x\left(s\right)\delta x\left(t\right)\delta x\left(u\right)} & =\frac{1}{3!}\delta\left(t-s\right)\delta\left(t-u\right)J\alpha\widetilde{x}^{\ast}\left(t\right)\label{eq:vertex_MSR_loop_time}\\
\frac{1}{2!}\,\frac{\delta^{3}S}{\delta x\left(s\right)\delta x\left(t\right)\delta\widetilde{x}\left(u\right)} & =\frac{1}{2!}\delta\left(t-s\right)\delta\left(t-u\right)J\alpha x^{\ast}\left(t\right).\nonumber 
\end{align}
According to the rules derived in \secref{Feynman-rules-Fourier},
in Fourier domain these read
\begin{align*}
\frac{1}{3!}\,\frac{\delta^{3}S}{\delta X\left(\omega\right)\delta X\left(\omega^{\prime}\right)\delta X\left(\omega^{\prime\prime}\right)} & =\frac{1}{2\pi}\frac{1}{3!}\delta\left(\omega+\omega^{\prime}+\omega^{\prime\prime}\right)\,J\alpha\widetilde{x}^{\ast}\\
\frac{1}{2!}\,\frac{\delta^{3}S}{\delta X\left(\omega\right)\delta X\left(\omega^{\prime}\right)\delta\tX\left(\omega^{\prime\prime}\right)} & =\frac{1}{2\pi}\frac{1}{2!}\delta\left(\omega+\omega^{\prime}+\omega^{\prime\prime}\right)\,J\alpha x^{\ast}
\end{align*}
 For the term $\frac{\partial\Gamma}{\partial\widetilde{x}^{\ast}}$,
we have
\begin{align}
\nonumber \\
\feyn{!{gV}{\widetilde{x}^{\ast}}f0!{flV}{\delta x}!{fluV}{\delta x}f0f0}.\label{eq:one_loop_deriv_x_tilde_star}\\
\nonumber 
\end{align}
For the term $\frac{\partial\Gamma}{\partial x^{\star}}$, this leads
to the one-loop diagrams

\begin{align}
3\cdot\feyn{!{gA}{x^{\ast}}f0!{flV}{\delta x}!{fluV}{\delta x}f0f0+f0!{gA}{x^{\ast}}f0!{flV}{\delta x}!{fluA}{\delta\widetilde{x}}f0f0},\label{eq:One_loop_deriv_x_star}\\
\nonumber 
\end{align}
where the last diagram vanishes in the Ito-convention because it includes
a response function starting and ending at the same vertex.

To determine the values of these diagrams, we need to know the propagator,
which is the inverse of the second derivative of the action $S^{\left(2\right)}$.
Limiting ourselves first to the stationary case, this can be achieved
by going into Fourier space. However, let us first note, how $S^{\left(2\right)}$
looks like in the general case:
\begin{equation}
S_{t,s}^{(2)}\left[x^{\ast},\widetilde{x}^{\ast}\right]=\left(\begin{array}{cc}
J\alpha\widetilde{x}^{\ast}\left(t\right)x^{\ast}\left(t\right) & \partial_{s}+1-J+\frac{J\alpha}{2}\left(x^{\ast}\left(t\right)\right)^{2}\\
\partial_{t}+1-J+\frac{J\alpha}{2}\left(x^{\ast}(t)\right)^{2} & D
\end{array}\right)\,\delta\left(t-s\right).\label{eq:S2_MSR}
\end{equation}
With the abbreviations $-m:=1-J+\frac{J\alpha}{2}\left(x_{0}^{\ast}\right){}^{2}$
and $\tilde{D}:=J\alpha x_{0}^{\ast}\widetilde{x}_{0}^{\ast}$, in
Fourier domain, this becomes for $\widetilde{x}^{\ast}\left(t\right)=\widetilde{x}_{0}^{\ast}$,
$x^{\ast}\left(t\right)=x_{0}^{\ast}$

\[
S_{\omega^{\prime}\omega}^{(2)}(x_{0}^{\ast},\widetilde{x}_{0}^{\ast})=\left(\begin{array}{cc}
\tilde{D} & -i\omega-m\\
i\omega-m & D
\end{array}\right)\,\delta\left(\omega-\omega^{\prime}\right).
\]
The inverse of this matrix, the propagator then becomes
\begin{align*}
\Delta(x,\widetilde{x}_{0}^{\ast})\left(\omega^{\prime},\omega\right) & =\left(-S_{\omega^{\prime}\omega}^{(2)}\left[x_{0}^{\ast},\widetilde{x}_{0}^{\ast}\right]\right)^{-1}\\
 & =-\frac{1}{\tilde{D}D-\left(\omega^{2}+m^{2}\right)}\,\left(\begin{array}{cc}
D & i\omega+m\\
-i\omega+m & \tilde{D}
\end{array}\right)\,\delta\left(\omega-\omega^{\prime}\right).
\end{align*}
Let us assume that $\widetilde{x}_{0}^{\star}=0$ - we will see later
that this is a consistent assumption. Then $\tilde{D}=0,$ so the
propagator is given by
\[
\Delta(x_{0}^{\ast},\widetilde{x}_{0}^{\ast})\left(\omega^{\prime},\omega\right)=\left(\begin{array}{cc}
\frac{D}{\omega^{2}+m^{2}} & \frac{1}{-i\omega+m}\\
\frac{1}{i\omega+m} & 0
\end{array}\right)\,\delta\left(\omega-\omega^{\prime}\right).
\]
Comparing to \eqref{Z_J_OUP}, we see that the propagator is, of course,
of the same form as in the Gaussian case, since the loopwise approximation
is an approximation around a local maximum. The back transform to
time domain with \eqref{free_covariance-1-1} and \eqref{free_response}
therefore reads
\begin{equation}
\Delta\left[x_{0}^{\ast},\widetilde{x}_{0}^{\ast}\right]\left(t^{\prime},t\right)=\left(\begin{array}{cc}
-\frac{D}{2m}\,e^{m\left|t-t^{\prime}\right|} & \Theta\left(t-t^{\prime}\right)\,\exp\left(m\left(t^{\prime}-t\right)\right)\\
\Theta\left(t^{\prime}-t\right)\exp\left(m\left(t-t^{\prime}\right)\right) & 0
\end{array}\right).\label{eq:prop_loop_time}
\end{equation}
In other words: If $\widetilde{x}_{0}^{\star}=0$, the response functions
are (anti-)causal. That means that the contributions of the two last
diagrams in \eqref{One_loop_deriv_x_star} vanish. 

With these results, we may evaluate the first diagram of \eqref{One_loop_deriv_x_star}.
Due to the two Dirac $\delta$ in the interaction vertex in time domain,
this is easiest done in time domain. The diagram \eqref{one_loop_deriv_x_tilde_star}
results in
\begin{align*}
\\
 & \feyn{!{gA}{\tx^{\ast}}f0!{flV}{\delta x}!{fluV}{\delta x}f0f0}\\
\\
= & \frac{1}{2!}\,S_{\tx(t)x(s)x(u)}\Delta_{x(s)x(u)}\\
= & \frac{1}{2!}\,\iint\,ds\,du\:\delta(t-s)\delta(t-u)\,J\alpha\,x_{0}^{\ast}\,\frac{-D}{2m}\,e^{m\left|t-u\right|}\\
= & \frac{-J\alpha D}{4m}\,x_{0}^{\ast}.
\end{align*}
The second diagram of \eqref{One_loop_deriv_x_star} vanishes, because
the response functions at equal time points vanish. The first diagrams,
by the linear dependence of the interaction vertex \eqref{vertex_MSR_loop_time}
on $\tx^{\ast}$ has the value

\begin{align*}
3\cdot\feyn{!{gA}{x^{\star}}f0!{flV}{\delta x}!{fluV}{\delta x}f0f0} & =\frac{3}{3!}\,\frac{-D}{2m}\,J\alpha\tx_{0}^{\ast}=\frac{-J\alpha D}{4m}\,\tx_{0}^{\ast},
\end{align*}
which vanishes as well for $\tx^{\ast}=0$, showing that this value
is a consistent solution for the loopwise correction. 

Inserted into the equation of state this yields the self-consistency
equation for the mean value $x^{\ast}$ 
\begin{equation}
\left(1-J\right)x_{0}^{\ast}+\frac{\alpha J}{3!}\left(x_{0}^{\ast}\right)^{3}+\frac{1}{4}J\alpha x_{0}^{\ast}D\frac{1}{1-J+\frac{J\alpha}{2}\left(x_{0}^{\ast}\right){}^{2}}=0.\label{eq:x_star_one_loop_x_tilde_zero}
\end{equation}
We can check our result by solving the Fokker-Planck equation for
the system, which gives a (numerically) exact solution for the fixpoints
of \eqref{Langevin_equation}. This is shown in \figref{OUP_loop_figure}.

\subsection{Corrections to the self-energy and self-consistency}

We may determine corrections to the self-energy by forming all 1PI
diagrams with two external amputated legs. We can use these terms
to obtain corrections to the second moments of the process by obtaining
the inversion:
\begin{align*}
\left(-S^{(2)}[x^{\ast},\tx^{\ast}]+\Sigma\right)\,\Delta & =1.
\end{align*}
With (\ref{eq:S2_MSR}) and $\tx\equiv0$ we obtain the set of coupled
differential equations
\begin{align}
\int\,dt^{\prime}\left(\begin{array}{cc}
0 & (\partial_{t}+m)\,\delta(t-t^{\prime})+\Sigma_{x\tx}(t,t^{\prime})\\
(-\partial_{t}+m)\,\delta(t-t^{\prime})+\Sigma_{\tx x}(t,t^{\prime}) & -D\,\delta(t-t^{\prime})+\Sigma_{\tx\tx}(t,t^{\prime})
\end{array}\right)\,\left(\begin{array}{cc}
\Delta_{xx}(t^{\prime},s) & \Delta_{x\tx}(t^{\prime},s)\\
\Delta_{\tx x}(t^{\prime},s) & 0
\end{array}\right) & =\diag(\delta(t-s)).\label{eq:self_energy_delta}
\end{align}
We may write (\ref{eq:self_energy_delta}) explicitly to get two linearly-independent
equations
\begin{align*}
(\partial_{t}+m)\,\Delta_{\tx x}(t,s)+\int_{s}^{t}\,dt^{\prime}\,\Sigma_{x\tx}(t,t^{\prime})\,\Delta_{\tx x}(t^{\prime},s) & =\delta(t-s),\\
(-\partial_{t}+m)\,\Delta_{xx}(t,s)+\int_{-\infty}^{t}dt^{\prime}\,\Sigma_{\tx x}(t,t^{\prime})\Delta_{xx}(t^{\prime},s)-D\,\Delta_{\tx x}(t,s)+\int_{s}^{\infty}\Sigma_{\tx\tx}(t,t^{\prime})\Delta_{\tx x}(t^{\prime},s) & =0.
\end{align*}
Compared to (\ref{eq:def_A_OUP}), we may interpret (\ref{eq:self_energy_delta})
as describing a linear stochastic differential-convolution equation
\begin{align}
(\partial_{t}-m)\,y & =\int dt^{\prime}\,\Sigma_{\tx x}(t,t^{\prime})\,y(t^{\prime})+\eta(t),\label{eq:interpretation_lin_diffeq}
\end{align}
where the noise $\eta$ is Gaussian and has variance
\begin{align*}
\langle\eta(t)\eta(s)\rangle & =D\,\delta(t-s)-\Sigma_{\tx\tx}(t,s).
\end{align*}
The self-energy terms therefore have the interpretation to define
a linear process that has the same second order statistics as the
full non-linear problem. This is consistent with the self-energy correcting
the Gaussian part and therefore the propagator of the system.

\subsection{Self-energy correction to the full propagator}

Instead of calculating the perturbative corrections to the second
cumulants directly, we may instead compute the self-energy first and
obtain the corrections to the covariance function and the response
function from Dyson's equation, as explained in \secref{self-energy}.
This is possible, because we expand around a Gaussian solvable theory.

The bare propagator of the system is given by \eqref{prop_loop_time}.
We have Dyson's \eqref{Dyson} in the form
\begin{eqnarray*}
W^{(2)} & = & \Delta-\Delta\Sigma\Delta+\Delta\Sigma\Delta\Sigma\Delta-\ldots.
\end{eqnarray*}
So we need to compute all 1PI diagrams that contribute to the self-energy
$\Sigma$. If we restrict ourselves to one-loop corrections. At this
loop order, we get three diagrams with two amputated external legs
\begin{align}
\nonumber \\
-\Sigma_{\tx\tx}(t,s) & =2\cdot\Diagram{\vertexlabel_{\tx^{\ast}(t)}gVf0!{fl}{\Delta_{xx}}!{flu}{\Delta_{xx}}f0gA\vertexlabel_{\tx^{\ast}(s)}}
\label{eq:Sigma_tx_x-1}\\
\nonumber \\
\nonumber \\
 & =2\cdot\frac{1}{2!}\big(\frac{J\alpha x^{\ast}}{2!}\big)^{2}\,\big(\Delta_{xx}(t,s)\big)^{2}\nonumber \\
\nonumber \\
-\Sigma_{\tx x}(t,s) & =2\cdot2\cdot2\cdot\Diagram{\vertexlabel_{\tx^{\ast}(t)}gVf0!{fl}{\Delta_{xx}}!{fluV}{\Delta_{\tx x}}f0gV\vertexlabel_{x^{\ast}(s)}}
\quad+\quad3\cdot\quad\text{\ensuremath{\Diagram{\vertexlabel^{x^{\ast}(s)}\quad gdA & f0!{fl}{\Delta_{xx}}fluf0\\
\vertexlabel^{\tx^{\ast}(t)}\quad guV
}
}}\label{eq:Sigma_tx_x}\\
\nonumber \\
 & =2\cdot2\cdot2\cdot\frac{1}{2!}\big(\frac{J\alpha x^{\ast}}{2!}\big)^{2}\,\Delta_{xx}(t,s)\Delta_{\tx x}(t,s)+\delta(t-s)\,3\,\frac{J\alpha}{3!}\,\Delta_{xx}(t,t)\nonumber \\
\nonumber 
\end{align}

(The combinatorial factors are as follows. For $\Sigma_{\tx\tx}$:
$2$ $2$ possibilities to connect the inner lines of the diagram
in the loop, directly or crossed. For $\Sigma_{\tx x}$: $2$ vertices
to choose from to connect the external $\tx$; $2$ legs to choose
from at the other vertex to connect the external $x$; $2$ ways to
connect the internal propagator to either of the two $x$-legs of
the vertex. The latter factor $1/2!$ stems from the repeated appearance
of the interaction vertex). The factor $3$ in the last diagram comes
from the $3$ possibilities to select one of the three $x$-legs of
the four-point interaction vertex. We cannot construct any non-zero
correction to $\Sigma_{xx}$ due to the causality of the response
function $\Delta_{x\tx}$.

We may now use Dyson's equation to compute the corrections to the
covariance and the response function. We get
\begin{align}
W^{(2)} & =\Delta-\Delta\Sigma\Delta+\ldots\label{eq:Dyson_matrix_MSRDJ}\\
 & =\left(\begin{array}{cc}
\Diagram{fVfA}
 & \Diagram{fV}
\\
\Diagram{fA}
 & 0
\end{array}\right)\nonumber \\
 & +\left(\begin{array}{cc}
\Diagram{fVfA}
 & \Diagram{fV}
\\
\Diagram{fA}
\quad & 0
\end{array}\right)\left(\begin{array}{cccc}
0 &  &  & \Diagram{gAf0!{fl}{}!{fluA}{}f0gA}
+\Diagram{gdA & f0!{fl}{}fluf0\\
guV
}
\\
\\
\\
\Diagram{gVf0!{fl}{}!{fluV}{}f0gV}
+\Diagram{gdA & f0!{fl}{}fluf0\\
guV
}
 &  &  & \Diagram{gVf0!{fl}{}!{flu}{}f0gA}
\end{array}\right)\left(\begin{array}{cc}
\Diagram{fVfA}
 & \Diagram{fV}
\\
\Diagram{fA}
 & 0
\end{array}\right)\nonumber \\
 & -\ldots,\nonumber 
\end{align}
where we suppressed the combinatorial factors for clarity (they need
to be taken into account, of course). Performing the matrix multiplication,
we may, for example, obtain the perturbation correction to $W_{xx}^{(2)}$,
the upper left element. We hence get the corrected covariance
\begin{eqnarray*}
W_{xx}^{(2)} & = & \Diagram{fVfA}
\\
 &  & +\Diagram{fVfAf0!{fl}{}!{fluA}{}f0fA}
+\Diagram{fVf0!{fl}{}!{fluV}{}f0fVfA}
+\Diagram{fdV\\
 & fdA & f0flfluf0\\
 & fuV
}
+\Diagram{ &  & fuA\\
f0flfluf0 & fuV\\
 & fdA
}
\\
\\
 &  & +\Diagram{fVf0!{fl}{}!{flu}{}f0fA}
\\
\\
 &  & -\ldots.
\end{eqnarray*}
We see that the contribution of the diagram \eqref{Sigma_tx_x} that
is non-symmetric under exchange of $t\leftrightarrow s$ that contributes
to $\Sigma_{\tx x}$ appears in a symmerized manner in the covariance,
as it has to be. This result is, of course, in line with \eqref{F_contrib_I}
above. The practical advantage of this procedure is obvious: We only
need to compute the self-energy corrections once. To get the response
functions $W_{\tx x}^{(2)}$, we would just have to evaluate the off-diagonal
elements of the matrix \eqref{Dyson_matrix_MSRDJ}.

\subsection{Self-consistent one-loop}

We may replace the bare propagators that we use to construct the self-energy
diagrams by the solutions of (\ref{eq:self_energy_delta}). We then
obtain a self-consistency equation for the propagator. Calculating
the correction to the self-energy to first order in the interaction
strength, we get the diagram
\begin{align*}
\Diagram{\vertexlabel^{\tx(t)}fs0 & gdV & f0flVfluVf0\\
\vertexlabel^{x(s)}fs0 & guA
}
 & =3\cdot\frac{\alpha J}{3!}\,\Delta_{xx}(t,t)\,\delta(t-s),
\end{align*}
where we plugged in the full propagator $\Delta_{xx}(t,t)$, which,
at equal time points, is just the variance of the process. In the
interpretation given by the effective equation (\ref{eq:interpretation_lin_diffeq}),
this approximation hence corresponds to
\begin{align*}
(\partial_{t}-m)\,y & =\frac{\alpha}{2}\,\Delta_{xx}(t,t)\,y(t)+\eta(t)\\
 & =3\cdot\frac{\alpha}{3!}\,\langle y(t)y(t)\rangle\,y(t)+\eta(t).
\end{align*}
The last line has the interpretation that two of the three factors
$y$ are contracted, giving $3$ possible pairings. This approximation
is known as the \textbf{Hartree-Fock approximation} or \textbf{self-consistent
one-loop approximation}.

\subsection{Appendix: Solution by Fokker-Planck equation}

The Fokker-Planck equation corresponding to the stochastic differential
equation (\ref{eq:Langevin_equation}) reads \citep{Risken96}

\begin{align*}
\tau\partial_{t}\,\rho\left(x,t\right) & =-\partial_{x}\left(f(x)-\frac{D}{2}\partial_{x}\right)\,\rho\left(x,t\right),\\
f(x) & =-x+J\left(x-\frac{\alpha}{3!}x^{3}\right)
\end{align*}
As we are interested in the stationary case, we set the left hand
side to $0$. This leads to
\begin{equation}
\left(f(x)-\frac{D}{2}\partial_{x}\right)\,\rho_{0}\left(x\right)=\varphi=\mathrm{const.}\label{eq:Fokker_Planck_current_zero}
\end{equation}
Since the left hand side is the flux operator and since there are
neither sinks nor sources, the flux $\varphi$ must vanish in the
entire domain, so the constants $\varphi\equiv0$. The general solution
of (\ref{eq:Fokker_Planck_current_zero}) can be constructed by variation
of constants
\begin{align*}
\partial_{x}\,\rho_{0}(x) & =\frac{2}{D}\,f(x)\,\rho_{0}(x)\\
\rho_{0}(x) & =\exp\,\Big(\frac{2}{D}\int^{x}f(x^{\prime})\,dx^{\prime}\Big)\\
 & =C\,\exp\,\Big(\frac{2}{D}\,\Big(\frac{(J-1)}{2}\,x^{2}-\frac{J\alpha}{4!}x^{4}\Big)\Big),
\end{align*}

where the choice of the lower boundary amounts to a multiplicative
constant $C$ that is fixed by the normalization condition
\begin{align*}
1= & \int\,\rho_{0}(x)\,dx.
\end{align*}
Therefore, the full solution is given by
\[
\rho_{0}\left(x\right)=\frac{\exp\,\Big(\frac{2}{D}\int_{0}^{x}f(x^{\prime})\,dx^{\prime}\Big)}{\int_{-\infty}^{\infty}\exp\left(\frac{2}{D}\left(\int_{0}^{x}f(x^{\prime})dx^{\prime}\right)\right)dx}.
\]

\section{Nomenclature}

We here adapt the nomenclature from the book by Kleinert on path integrals
\citep{Kleinert89}. We denote as $x$ our ordinary random variable
or dynamical variable, depending on the system. Further we use
\begin{itemize}
\item $p(x)$ probability distribution
\item $\langle x^{n}\rangle$ $n$-th moment 
\item $\llangle x\rrangle$ $n$-th cumulant
\item $S(x)\propto\ln\,p(x)$ action
\item $-\frac{1}{2}x^{\T}Ax$ quadratic action
\item $S^{(n)}$ $n$-th derivative of action
\item $\Delta=A^{-1}$ or $\Delta=\left(-S^{(2)}\right)^{-1}$inverse quadratic
part of action, propagrator
\item $Z(j)=\langle\exp(j^{\T}x)\rangle$ moment generating function{[}al{]}
or partition function
\item $W(j)=\ln\,Z(j)$ cumulant generating function{[}al{]} or generating
function of connected diagrams
\item $\Gamma[y]=\sup_{j}\,j^{\T}y-W[j]$ generating function{[}al{]} of
vertex function or one-particle irreducible diagrams
\item $\Gamma_{0}=-S$: zero loop approximation of $\Gamma$
\item $\Gammafl$: fluctuation corrections to $\Gamma$
\item $\Sigma=\Gammafl^{(2)}$ self-energy 
\end{itemize}

\begin{acknowledgments}
We would like to thank Tobias K\"uhn, Jannis Sch\"ucker, and Sven
G\"odeke for contributions to the preparation of this collection
of material and to Christian Keup and Sandra Nestler for typesetting
many of the Feynman diagrams within these notes. This work was partly
supported by the Helmholtz association: Helmholtz Young investigator's
group VH-NG-1028; HBP - The Human Brain Project SGA2 (2018-04-01 -
2020-03-30); Juelich Aachen Research Alliance (JARA); the ERS RWTH
Seed fund ``Dynamic phase transitions in cortical networks''.
\end{acknowledgments}

\end{document}